\pdfoutput=1
\documentclass[12pt]{book}
\usepackage[lflt]{floatflt}
\usepackage{latexsym}
\usepackage{amsfonts}
\usepackage{amsmath}
\usepackage{amssymb}

\usepackage{graphicx}
\usepackage{epsfig}
\usepackage{mathrsfs}
\usepackage{fancyhdr}
\usepackage{makeidx} 
\usepackage[vcentermath,enableskew]{youngtab}
 
\fancypagestyle{plain}{
\fancyhf{}

\fancyfoot[LE,RO]{\thepage}
}
\makeatletter
    \def\cleardoublepage{\clearpage\if@twoside \ifodd\c@page\else%
        \hbox{}%
        \thispagestyle{empty}
        \newpage%
        \if@twocolumn\hbox{}\newpage\fi\fi\fi}
\makeatother

\newcommand{\I}{\imath}

\newcommand{\ID}{\mathbf{1}}

\newcommand{\CP}{\mathbb{C}\mathrm{P}}
\newcommand{\CAS}{\mathrm{C}_2}

\newcommand{\calL}{{\mathcal{L}}}

\newcommand{\be}{\begin{equation}}
\newcommand{\ee}{\end{equation}}
\newcommand{\bea}{\begin{eqnarray}}
\newcommand{\eea}{\end{eqnarray}}
\newcommand{\med}{\frac{1}{2}}

\newcommand{\Tr}{{\rm Tr}}
\newcommand{\trm}{{ t}}

\newcommand{\la}{\langle}
\newcommand{\ra}{\rangle}
\newcommand{\Yp}{\hat{Y}}
\newcommand{\lpl}{\hat{\mathcal{L}}}
\newcommand{\cte}{\frac{4 \pi}{L+1}}
\newcommand{\clebsch}{\mathtt{C}}
\newcommand{\real}{\mathbb{R}}

\newcommand{\vlm}{\overrightarrow{\mathbf{c_1}}}
\newcommand{\field}{\mathbf{\Phi}}
\newcommand{\partition}{\mathcal{Z}}
\newcommand{\probability}{\mathcal{P}}
\newcommand{\ccv}{\mathbf{c}}

\newcommand{\cteA}{{\mathbf{\mathtt{A}}}}
\newcommand{\cteB}{\mathbf{\mathtt{B}}}
\newcommand{\cteC}{\mathbf{\mathtt{C}}}
\newcommand{\cteD}{\mathbf{\mathtt{D}}}
\newcommand{\cteAA}{{\mathbf{\mathsf{A}}}}
\newcommand{\cteBB}{\mathbf{\mathsf{B}}}
\newcommand{\cteCC}{\mathbf{\mathsf{C}}}
\newcommand{\cteDD}{\mathbf{\mathsf{D}}}
\newcommand{\eater}{\hat{D}}
\newcommand{\proj}{\mathcal{P}}
\newcommand{\identy}{\mathbf{1}}
\newcommand{\gen}{{\mathcal{J}}}
\newcommand{\PP}{\mathrm{P}}
\newcommand{\metric}{\mathcal{G}}
\newcommand{\Stwo}{{s}^{s^2}}
\newcommand{\Diag}{\mathbf{\Lambda}}
\newcommand{\ijtp}{\em{Int. J. Mod. Phys.}}
\newcommand{\cqg}{\it{Class. and Quant. Grav.}}
\newcommand{\cmp}{\it{Commun. Math Phys.}} 
\newcommand{\ijmpa}{\it{Int. J. Mod. Phys.} {\bf{A}}}
\newcommand{\hepth}{{\tt{hep-th/}}} 
\newcommand{\heplat}{{\tt{hep-lat/}}}
\newcommand{\npps}{{\it{Nucl. Phys.}}}
\newcommand{\jgp}{\it{J.Geom.Phys.}}
\newcommand{\appol}{\it{Acta Phys. Polon. }}

\newcommand{\JHEP}{{\it{JHEP }}}
\newcommand{\condmat}{\tt{cond-mat}} 
\newcommand{\comppc}{{\it{Comput. Phys. Comm.}}} 
\newcommand{\PLett}{{\it{ Phys. Lett.}}}

\newcommand{\logo}{
\begin{floatingfigure}[l]{5mm}{
     \includegraphics[width=2cm]{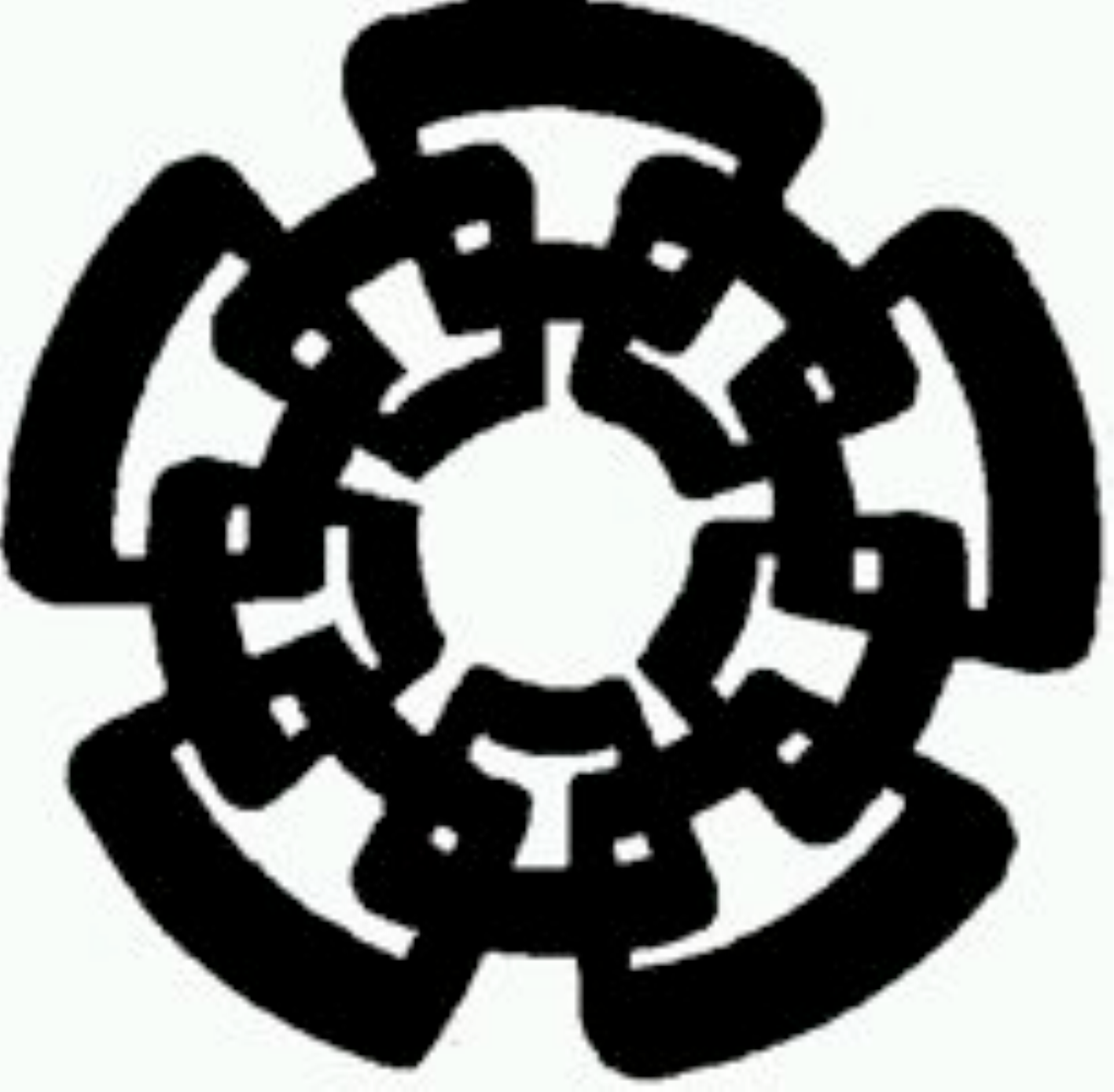}}
\end{floatingfigure}   
}
 
\title{Fuzzy Scalar Field Theories: Numerical and Analytical Investigations}

\author{Julieta Medina Garc\'{\i}a  \\ \  \\  
        Advisors: \\
       Dr. Denjoe O'Connor \\
       Dr. Wolfgang Bietenholz}

\begin{document}

\thispagestyle{empty}
\vspace{-1.0in} 
\rule{-1.0in}{0.3in}
\begin{minipage}{170mm}
\begin{center}
\parbox{20mm}{\logo}\\[3ex]
{\Large \sf \bf{\hspace{-0.5cm} CENTRO DE INVESTIGACION Y DE \\
ESTUDIOS AVANZADOS DEL
\\  INSTITUTO POLITECNICO NACIONAL}}
\end{center}
\end{minipage}
\\[5ex]
\begin{center}
{\LARGE \sf DEPARTAMENTO DE F{I}SICA}\\[10ex]

{\huge {\sf{\bf{Teor\'{\i}as de Campo Escalares Difusas:}}}}\\[1ex]
{\Large {\sf{{\bf Investigaciones Anal\'{\i}ticas y Num\'ericas }}}}\\[1ex] 
{\large {\sf{ Tesis que presenta}}}\\[7ex]
{\Large Julieta Medina Garc\'{\i}a }\\[7ex]
 {\sf{Para obtener el grado de:}}\\[4ex]
{\Large {\sf{DOCTORA EN CIENCIAS}}}\\[2ex]
 {\large \sf{En la especialidad de}}\\[2ex]
{\Large { \sf{ F\'{\i}sica}}}\\

\vfill
{\Large Directores: Dr. Denjoe O'Connor \\
                  Dr.Wolfgang Bietenholz
}\\[4ex]
\end{center}
{\large {\sf{M\'exico, D.F. \hfill April, 2006}}}

\pagebreak
\thispagestyle{empty}
.
\thispagestyle{empty}
\vspace{-1.0in} 
\rule{-1.0in}{0.3in}
\begin{minipage}{170mm}
\begin{center}

\parbox{20mm}{\logo}\\[3ex]
{\Large \sf \bf{\hspace{-0.5cm} CENTRO DE INVESTIGACION Y DE \\
ESTUDIOS AVANZADOS DEL
\\  INSTITUTO POLITECNICO NACIONAL}}
\end{center}
\end{minipage}
\\[5ex]
\begin{center}

{\LARGE \sf PHYSICS DEPARTMENT}\\[10ex]

{\huge {\sf{\bf{Fuzzy Scalar Field Theories:}}}}\\[1ex]
{\huge {\sf{{\bf Numerical and Analytical Investigations }}}}\\[1ex] 
{\large {\sf{ Thesis submmited by}}}\\[7ex]
{\Large Julieta Medina Garc\'{\i}a }\\[7ex]
 {\sf{In order to obtain the}}\\[4ex]
{\Large {\sf{Doctor in Science}}}\\[2ex]
 {\large \sf{degree, speciality in}}\\[2ex]
{\Large { \sf{ Physics}}}\\

\vfill
{\Large Advisors: Dr.\ Sc.\  Denjoe O'Connor \\
                  Dr.\ Sc.\ Wolfgang Bietenholz
}\\[4ex]
\end{center}
{\large {\sf{M\'exico City \hfill April, 2006}}}

\pagebreak
\thispagestyle{empty}
.
\pagenumbering{roman} \setcounter{page}{1}
\tableofcontents
\thispagestyle{plain}{

\chapter*{Acknowledgement}
It is with sincere appreciation that I wish here to acknowledge my debt 
to all those who contributed so much to making this study possible.

I want to thank to CONACYT for supporting me with a Ph.D. scholarship.

I am very indebted to the CINVESTAV which made available a grant-in-aid.

To the {\em Dublin Institute for Advanced Studies} where I did a predoctoral 
stay, I am indebted for many courtesies which 
facilitated my research including financial support.

To  the {\em Institut f\"{u}r Physik, Humboldt Universit\"{a}t zu Berlin}
for allowing me do a stay there.

To my advisors Dr. Denjoe O'Connor  and Dr. Wolfgang Bietenholz  for all of their guidance, encouragement, and support,
I am grateful  for their careful review of my thesis, 
and his valuable comments on my work.

I wish to thank all my former colleagues, especially Idrish Huet, Fernando Garc\'{\i}a, 
Pavel Castro, Rodrigo Delgadillo and Juan Aguilar. 

To people in the Physics Department, especially to Dr. Gabriel L\'opez Castro for his help and 
support, to Dr. Jos\'e Mendez.
To my friends from  the Physics Department, Xavier Amador, Carlos Soto, Alfredo L\'opez,  Sendic Estrada, Sara Cruz and Luz del Carmen Cort\'es. 

To people belonging to the {\em fuzzy gang }: Xavier Martin, Brian Dolan, Professor Balachandran, 
Marco Panero and Badis Ydri.
A special acknowledgement goes to Frank Hofheinz and Jan Volkholz for introducing me to parallel 
computing and invaluable comments on lattice topics. 
}
\pagebreak

\thispagestyle{plain}{
I would like to thank all of those people who provided technical support,
especially to Ronan Cunniffe, system administrator from {\em DIAS} 
during my stay there.

Additionally, I wish to thank all of secretariat staff of the Physics Department, especially to 
Patricia Villar, Ma. Eugenia L\'opez and Flor Ib\'a\~nez.

All my friends deserve my special thanks for bringing joy into my life.

My deepest thanks go to my parents, for their loving care and encouragement and for giving me 
the foundation from which to build my life. I am also grateful to my sister Gabriela.

Finally, I owe a very special thanks to my husband Amilcar for his unfailing love and support 
through it all. His understanding attitude towards my work and confidence in me has been 
essential for the completion of this thesis.
}           
\thispagestyle{empty}{
\chapter*{Resumen}
Este trabajo est\'a dedicado al estudio de Teorias de Campo (QFT) en {\em espacios difusos}.

Los espacios difusos son aproximaciones al \'algebra de funciones de un espacio continuo
 por medio de un \'algebra matricial finita. En el l\'{\i}mite de matrices infinitamente grandes la 
aproximaci\'on es exacta.

Una caracteristica atractiva de esta aproximaci\'on es que muestra de una manera 
transparente como son preservadas las propiedades geom\'etricas del espacio continuo.

En el estudio del r\'egimen no perturbativo de QFT los espacios difusos proveen una posible alternativa a la red como
m\'etodo de regularizaci\'on. Esta tesis est\'a dividida en dos partes:

\begin{enumerate}
 \item Realizamos la simulaci\'on Monte Carlo de la teor\'{\i}a $\lambda \phi^4$ 
en un espacio Euclideano de tres dimensiones. La regularizaci\'on se compone por 
una esfera difusa de dos dimensiones, $S^2_F$, para las direcciones espaciales 
m\'as una red convencional para la direcci\'on temporal. 
Se identifica el diagrama de phase de este modelo.
Adem\'as de las fases desordenada y ordenada uniforme usuales encontramos 
una tercera fase de ordenamiento no uniforme. Ello indica la existencia del fen\'omeno conocido como mezclamiento UV-IR en el r\'egimen de acoplamiento fuerte.

 \item Como segundo punto presentamos un an\'alisis geom\'etrico de una teor\'{\i}a 
escalar general en una esfera difusa de cuatro dimensiones, $S^4_F$.
Una aproximaci\'on  para $S^4$ es de un interes especial dado que $S^4$ es la subs-
tituci\'on natural de $\real^4$ 
en estudios de QFT Euclideana. 
\end{enumerate}
}
\pagebreak

\thispagestyle{empty}{
Sin embargo la versi\'on difusa de $S^4$ no puede obtenerse mediante la cuantizaci\'on del espacio cl\'asico.
   
El problema es {\em rodeado} definiendo una teor\'{\i}a escalar en un espacio {\em m\'as grande}, 
que es $\CP^3$ el cu\'al es de dimension seis. Incluye grados de libertad de  $S^4$ m\'as otros 
que no lo son. Esos grados de libertad extras se eliminan din\'amicamente mediante un 
m\'etodo probabil\'{\i}stico. El an\'alisis de las estructuras geom\'etricas nos permite 
interpretar a este procedimiento como una reducci\'on de Kaluza-Klein de $\CP^3$ a $S^4$.

}                            
\chapter*{Abstract}

This work is devoted to the study of Quantum Field Theories (QFT) on {\em fuzzy spaces}.

Fuzzy spaces are approximations to the algebra of functions of a continuous
space by a finite matrix algebra. In the limit of infinitely large matrices the formulation is exact.

An attractive feature of this approach is that it transparently shows 
 how the geometrical properties of the continuous space are preserved.

In the study of the non-perturbative regime of QFT,  fuzzy spaces provide a 
possible alternative to the lattice
as a regularisation method.
The thesis is divided into two parts:
\begin{enumerate}
 \item We perform  Monte Carlo simulations of a $\lambda \phi^4$ theory 
       on a $3$-dimensional Euclidean space. The regularisation consist of replacing space by a 
       fuzzy $2$-dimensional sphere,  namely $S^2_F$, and Euclidean time by
       a conventional lattice. 
       We identify  the phase diagram of this model. In addition to the usual 
       disordered and uniform ordered 
       phases we find a third phase of non-uniform ordering.
       This indicates the existence of the phenomenon called UV-IR mixing in the 
       strong coupling regime.
 \item Second we present a geometrical analysis of the scalar field theory on a 
       $4$-dimensional fuzzy sphere, $S^4_F$.
       An approximation to $S^4$ is of special interest since $S^4$  is 
       the natural replacement of  $\real^4$ in studies of Euclidean QFT. 
       Nevertheless a fuzzy version of $S^4$ cannot be achieved by quantisation of the classical
       space.  
\end{enumerate}   

\pagebreak

\thispagestyle{empty}{
       The problem is circumvented by defining a scalar theory on a {\em larger} space, 
       $\CP^3$ which is $6$-dimensional. It includes degrees of freedom related to $S^4$ 
       plus others beyond $S^4$. Those extra degrees of freedom are dynamically 
       suppressed through a probabilistic method.
       The analysis of the geometrical structures allows us to interpret this procedure
       as a Kaluza-Klein reduction of  $\CP^3$ to $S^4$.
}


\pagenumbering{arabic} \setcounter{page}{1}
\chapter{Introduction}
\label{introduction}

Physics works best when there is a good interaction between experiment
and theory. Unfortunately, for many of the interesting questions
that arise, either it is impossible to perform the appropriate experiments
or they are too costly. This is where the power of computer simulations
plays an important role. In some sense, the computer simulation plays the role of
the experiment. In modern particle physics, the experimental tools
are large accelerators and the theories are typically quantum field
theories. One of these is that of the strong interactions known as
Quantum Chromodynamics. It is very difficult to extract some predictions
from this theory as they fall in a non-perturbative regime and many
physicists have resorted to computer simulations to extract the physical
predictions.

Similarly, many of the more speculative ideas emerging in physics
involve strongly interacting field theories, some of these have
novel features such as non-commutativity of the space-time coordinates.
This type of structure is also suggested by string theory.
The work of this thesis is dedicated to developing
non-perturbative techniques adequate to these non-commutative
theories and hopefully to string theory.

Fuzzy spaces are included in the wider framework of non-commutative geometry. 

The idea of involving non-commutativity into Physics dated from the middle of the last century, 
nevertheless the substantial development has taken place in the last few years. 
There are several reasons why studying non-commutative (NC) spaces has become so popular in the physics community.
Although our interest in the study of fuzzy spaces is related to the study of Quantum Field Theories
 as we will discuss later, we would like to mention some other motivations for the study of NC 
geometry in Physics.

Many interesting phenomena in Physics have been discovered by extensions, 
therefore generalising commutative spaces into non-commutative spaces seems a natural extension.  In this spirit, generalising  commutative spaces to non-commutative spaces seems motivated.
Non-commutativity can be incorporated into many  branches of Physics like
Gravitational Theories, Condensed Matter Physics and Quantum Field Theories. 
The first attempts to involve NC theories in Quantum Gravity date from the last decade \cite{classical-Gravity-on-fuzzy}. 
Fuzzy spaces can be found in String Theories (with D-Branes) under certain conditions 
--- see Ref.\cite{castelino}-\cite{Ho-Ramgoolam}. 
In Condensed Matter Physics it was found that the  Quantum Hall Effect can be  formulated in terms of 
non-commutative coordinates where a magnetic background field $B$ is related to the non-commutative parameter 
\cite{Quantum-Hall-effect}-\cite{Campbell-Kaminsky}.

One of the open problems in  Field Theory is the existence of non well defined finite quantities: the
{\em divergences}. The regularisation procedure modifies the Field Theory  to remove those divergences.
The three well establish methods to the date are the dimensional regularisation, 
Pauli-Villars regularisation and the lattice regularisation \cite{Zinn-Justin}. The first two methods are
for exclusive application at small coupling regimes.
Regarding our motivation, we plan to test the feasibility of 
fuzzy spaces as a regularisation scheme in Quantum Field Theories. It should work, as the lattice procedure, 
at any regime.

Suppose we want to study  QFT through the path integral formalism 
on a given space. If we want to access the non-perturbative regime it is necessary to {\em discretise} the 
space in order to get a finite number of degrees of freedom. The standard method is to approximate the space  by discrete points  --- a lattice ---  representing the space and then calculate 
the observables over that set of points. This simple idea has generated some of the most successful theories  
in Physics, Lattice Field Theories, a review of them can be found in Ref. \cite{LFT}.

On the lattice the continuous translational and rotational symmetry is explicitly broken. 
This certainly is a disadvantage in models where these symmetries are important.
This takes us to search an alternative method that preserves these symmetries.

Before proposing an alternative one has to ask  which is the information necessary to describe an arbitrary 
space. The answer to this question was found by Connes and others  in the context of 
non-commutative geometry \cite{Connes_book}. It is known that it is possible to reconstruct a manifold\footnote{Spaces are included in this more general notion of {\em manifold}. For our purposes we work only with spaces.} 
$\mathcal{M}$ 
if we have the algebra $\mathcal{A}$ of functions over $\mathcal{M}$, a Hilbert space, $\mathcal{H}$, and a 
differential operator able to specify the geometry  (in  Ref.~\cite{Connes_book} that operator is the Dirac 
operator, $\mathcal{D}$, although for scalar theories as those studied in the present thesis it was conjectured in \cite{Frohlich} that the 
Laplace-Beltrami operator $\Delta$ is enough to specify the geometry).
Then, instead of discretising directly $\mathcal{M}$ by means of points we can {\em discretise} the triplet 
$\left(\mathcal{A},\mathcal{H},\mathcal{D} \right)$ and here  the fuzzy spaces enter: they are 
essentially discretisations at algebraic level.
If we want to obtain a finite number of degrees of freedom --- namely, the coefficients 
in the expansion of a function in the algebra basis --- 
the algebra has to be finite dimensional, i.e.\ a 
matrix algebra of dimension $N$, $Mat_N$, and as a consequence we have a non-commutative algebra. 
The elements in the algebra act on a finite dimensional version of the Hilbert space, $\mathcal{H}_N$, and 
an appropriate version of $\mathcal{D}$ is needed.
In the limit $N \longrightarrow \infty$ (called the commutative limit) we have to recover $\mathcal{M}$.

Summarising, the fuzzy discretisation\footnote{We denote it as {\em fuzzification}.} consist in the replacement:
\[
 \left(\mathcal{A},\mathcal{H},\mathcal{D} \right) \longrightarrow \left(Mat_N,\mathcal{H}_N,\mathcal{D}_N \right).
\]

The term ``fuzzy'' originates from the following observation: since in the fuzzy space the coordinates will be 
matrices and they do not commute, this will mean in the Quantum Mechanics spirit that the notion of points 
does not exist, i.e.\  the space turns fuzzy.

The seminal work on fuzzy spaces is due to Madore; in his work \cite{Madore} a fuzzy approximation of a 
two-dimensional sphere is constructed. Since then there exists a large compendium of fuzzy literature, 
e.g. Refs. \cite{Ramgoolam_0105006}-\cite{starprod_CPN}. 
Most of  fuzzy spaces have been constructed based on the following observation: 
If we quantise a classical phase space we obtain a finite dimensional Hilbert space. 
This implies that the candidates to be quantised are manifolds of finite volume which have a symplectic structure.
Co-adjoint orbits of Lie groups fall into this class. A didactic example of them are the complex projective spaces,
$\CP^n$, they are $2n$-dimensional spaces that can be defined as $SU(n)$ orbits. 
A discussion of its fuzzy version is given in Ref.~\cite{starprod_CPN}.
The family of $\CP^n_F$ is especially interesting since $\CP^2_F \cong S^2_F$; $S^4_F$ and $S^2_F$ can be obtained
form $\CP^3_F$ (see Ref. \cite{3sphere}).

Once we count with a fuzzy version of a space, the next step in our program is to define 
a Field Theory on it, e.g. Refs.~\cite{fuzzy-physics}-\cite{pcastro}. Then we need to construct fuzzy versions of Laplacians, Dirac operators, etc.
The solid mathematical background of the fuzzy approach makes it  easy to identify such fuzzy versions.
Of course one has to check whether the proposed theory reproduces the continuum theory.

Field Theories on fuzzy spaces share a generic property of Field Theories on general non-commutative  spaces 
called the UV-IR mixing \cite{Gubser:2000cd}. The UV-IR mixing was originally discovered in perturbative calculations Refs.~\cite{UV-IRNC-complex-scalar-field}-\cite{delgadillo_thesis}.
In  non-commutative spaces we have two kinds of diagrams, those that reduce to commutative diagrams 
and those   diagrams     without a  commutative counterpart, these  are divergent at low momenta.

Simulations on fuzzy spaces  are a relatively  recent topic,  see Refs.~\cite{xavier}-\cite{Das-Digal-Govindarajan}.

In the present thesis we concentrate on the study of Scalar Field Theories on fuzzy spaces.
We cover two important remarks in the fuzzy program:
\begin{enumerate}
 \item Test the feasibility as a discretisation method through a numerical simulation.
 \item Show that the fuzzy approach allows for a transparent geometrical analysis of Field Theories. 
\end{enumerate}

As a pilot study we describe a numerical study of the $\lambda \phi^4$ model
on the 3 dimensional Euclidean space $S^2 \otimes T$
by means of the Metropolis algorithm.

Our regularisation consists of
\begin{itemize}
\item {\em the fuzzy sphere} $S_{F}^{2}$ for the spatial coordinates 
\item a conventional lattice with periodic boundary conditions for the time direction.
\end{itemize}
   
The organisation of this first part of this thesis is the following:
In chapter \ref{section2} we present a detailed description of both discretisation schemes,
 emphasising the advantages of each method.
In chapter \ref{chapter-phases-characterisation} we present the characterisation of the phases in the model; we 
dedicate part of this chapter to the description of some technical aspects related to 
thermalisation problems in Monte Carlo simulations.
In chapter \ref{chapter4} we identify the phase diagram of the model analysing the scaling 
behaviour of the critical lines.
In chapter \ref{discussion} we present the discussion of our results. 
The key point of this analysis is the behaviour of the triple point under different limits. 
It reveals that in the thermodynamic limit $N \longrightarrow \infty$ is it not possible to recover 
the Ising universality class due to the dominance of a phase that breaks the rotational symmetry spontaneously. 
We find that the UV-IR mixing predicted in the perturbative regime of the model appears in the strong 
coupling regime as well.   
There are ways to remove those divergences. In the context of the scalar field theory $\lambda \phi^4$
 on the fuzzy sphere this is done by a suitable choice of the action \cite{matrix_phi4_models}.  The 
UV-IR mixing exists in the non-perturbative regime e.g. Refs.~\cite{BHN}-\cite{Procs} where it was detected  as 
a {\em matrix } or {\em striped} phase which has no counterpart in the commutative theory.

As a second point in this thesis we present an analytical part. We study a Scalar Field Theory on $S^4_F$.
$S^4$ is a special $4$-dimensional curved space, taking its radius to infinity, we arrive at $\real^4$. $S^4$
 is not a phase space, hence its ``construction'' involves some complications 
that are explained in chapter \ref{construction-S4}. 
To solve this problem we allow $S^4$ to fluctuate into a {\em larger} space, this is $\CP^3$. 
We present a review of the construction of $\CP^3$ as $SU(4)\cong Spin(6)$ orbit.
Nevertheless for {\em moving} on $S^4$, it is enough to preserve rotations in $5$ dimensions.
We find that $\CP^3$ can also be constructed as a $Spin(5)$ orbit, but 
demanding this less restrictive symmetry we construct a {\em squashed} $\CP^3$. 
For our purposes we demonstrate that this construction has more advantages since it allows us to identify  
$\CP^3$ as a fibre bundle over $S^4$ with  $S^2$ as the fibre.
We start chapter \ref{geometry_for_CP3} with a short review of
\cite{scalar-FT-S4}. There, a Scalar Field Theory on an {\em squashed} $\CP^3_F$ is defined. 
Then a penalisation method for all the non $S^4$ modes is introduced. 
This probabilistic method introduces an apparently ``artificial'' parameter $h$, 
such that  $h$ positive and large makes the non $S^4$ configurations improbable. 
Now we give an interpretation to this parameter through a geometrical analysis of the proposed model.
Using coherent state techniques it is possible to extract the geometry. 
At the end we are able to ``visualise'' the penalisation method as a Kaluza Klein reduction of $\CP^3$ to $S^4$. 
$h$ is interpreted in terms of the radius of the fibre $S^2$.	             

\part{Simulations of the $\lambda \phi^4$ Model
on the Space $S^2_F \times S_1$ }
\label{simulation-part}

\chapter{Generalities of the  method}
\label{path-int-chapter}
\section{Path integrals and functional integrals}
The functional integral provides a powerful tool to study Quantum Field Theories. 
It can be thought of as a generalisation of the path integral formalism in Quantum Mechanics 
introduced by R.~P.\ Feynman in the late 40's (see e.g. \cite{Zinn-Justin} and \cite{Feynman-Hibbs}).
The crucial idea behind the path integral is the superposition law. 
If we want to calculate the transition amplitude for going from an initial state at $\tau'$ 
to a final one at time $\tau''$, one has to consider a superposition of all possible paths.

To state this in a mathematical form, let us suppose the initial state at $\tau'$  to be 
denoted by $\vert \psi(x') \ra$ and at time
$\tau''$ we have $\vert \psi(x'') \ra$. Then the transition amplitude is given by
\be
  \la \psi(x'')\vert \mathcal{U}(\tau'',\tau')\vert \psi(x') \ra
 \label{trans-amplitude}
\ee
where 
\be
\mathcal{U}(\tau'',\tau')= e^{-\imath \mathcal{H} (\tau''-\tau')/ \hbar}, \label{time-evol-operators}
\ee
${\mathcal H}$ is the Hamiltonian of the system, which we assume to be time independent.

We start {\em slicing} the time interval $\left[ \tau',\tau'' \right]$ into $N$ 
subintervals of duration $\epsilon=\tau_{i+i}-\tau_i$ as in
{\bf figure \ref{paths-graph}}.
\begin{center} 
   \includegraphics[width=4.0in]{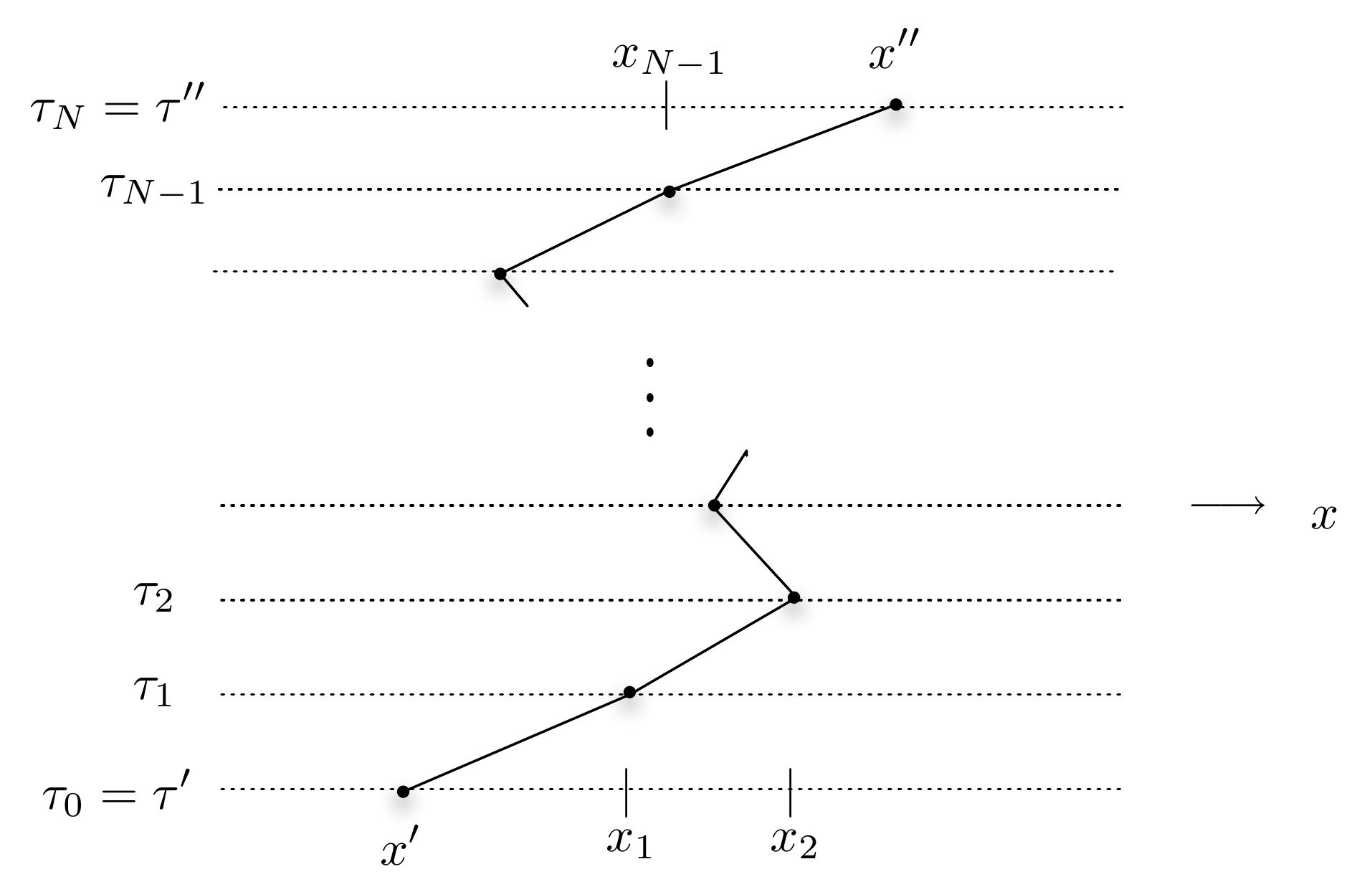}
\end{center}
\begin{figure}[h]
 \vspace{-0.35in}
 \caption{Schematic representation of the path.}
\label{paths-graph}
\end{figure}

The time evolution operator can be broken into intervals:
\bea
 e^{-\imath \mathcal H (\tau''-\tau')/ \hbar}&=& 
   e^{-\frac{\imath}{\hbar} \mathcal H (\tau_N -\tau_{N-1}+\tau_{N-1}-\cdots -\tau_2+\tau_2-\tau_{1}+\tau_{1}-\tau_0)}, \nonumber \\
 &=&\left( e^{-\imath \mathcal{H} \epsilon/ \hbar} \right)^N. 
\label{time-evolution-sliced}
\eea
where $\tau'=\tau_0$ and $\tau''=\tau_N$. \\
$\mathcal{H}$ has the general form $\mathcal{H}= \mathcal{H}_0+\mathcal{V}$, where ${\cal H}_0 =\frac{ p^2 }{2m}$.
For $\epsilon \to 0$
\be
  e^{-\imath \mathcal H (\tau''-\tau')/ \hbar} \approx
  \left( e^{-\imath \mathcal{H}_0 \epsilon/ \hbar} e^{-\imath \mathcal{V} \epsilon/ \hbar} \right)^N, \label{exp_H_by_Trotter}
\ee
where we have used Trotter's formula. Eq.~(\ref{exp_H_by_Trotter}) holds if $\mathcal{H}_0$ and $\mathcal{V}$ are semibounded.
The next trick is to insert between each term  $e^{-\imath \mathcal{H}_0 \epsilon/ \hbar} e^{-\imath \mathcal{V} \epsilon/ \hbar}$  the set of complete states 
\be
\int dx_i \vert \psi(x_i)\ra \la \psi(x_i) \vert = \mathbf{1},
\ee
then eq.~(\ref{trans-amplitude}) can be written as the product of $N$ terms 
{\footnotesize
\hspace{-0.5cm}
\bea
    \la \psi(x'')\vert \mathcal{U}(\tau'',\tau')\vert \psi(x') \ra& \approx & 
 \int dx_{N-1} \int dx_{N-2} \cdots  \int dx_{1}  
  \la \psi(x_N)\vert e^{-\imath \mathcal{H}_0 \epsilon/ \hbar} e^{-\imath \mathcal{V} \epsilon/ \hbar}\vert \psi(x_{N-1}) \ra 
  \nonumber \\
         & &  \times       \la \psi(x_{N-1})\vert e^{-\imath \mathcal{H}_0 \epsilon/ \hbar}
    e^{-\imath \mathcal{V} \epsilon/ \hbar}\vert \psi(x_{N-2}) \ra \nonumber \\
  & & \vdots \nonumber \\
  & & \times \la \psi(x_{1})\vert e^{-\imath \mathcal{H}_0 \epsilon/ \hbar} e^{-\imath \mathcal{V} \epsilon/ \hbar}\vert \psi(x_{0}) \ra. \nonumber
\eea
}

Suppose $\mathcal{V}$ depends on the position $X$ and $\mathcal{H}_0$ depends on the momentum $P$. It is possible to demonstrate that 
(see e.g. Refs.~\cite{Zinn-Justin},\cite{Jaffe}-\cite{Peskin}):

\be
 \la \psi(x_{i+1})\vert  e^{\imath \mathcal{H}_0 \epsilon / \hbar}e^{\imath \mathcal{V} \epsilon / \hbar} \vert  \psi(x_{i})\ra \approx 
\int \frac{dp_{i+1}}{2\pi \hbar}e^{\frac{\imath}{\hbar}\left( p_{i+1}(x_{i+1}-x_i)- \epsilon \mathcal{H}(p_{i+1}, \med(x_{i+1}+x_{i})) \right) }. \label{sliced-x_1+1}
\ee
In eq.~(\ref{sliced-x_1+1}) we note that the argument in the exponential can be written as
$$\frac{\imath}{\hbar} \epsilon\left( p_{i+1}\frac{x_{i+1}-x_i}{\epsilon}-  \mathcal{H}(p_{i+1}, \med(x_{i+1}+x_{i})) \right),
$$
where we recognise a discrete version of Lagrangian in the interval $ \left[\tau_{i} ,\tau_{i+1}\right]$ times the duration of the interval $\epsilon$, i.e., the action in such an interval.
Taking the product of the $N$-terms of the type in eq.~(\ref{sliced-x_1+1}) and taking $N\to \infty$ 
we arrive at  
\be
    \la \psi(x')\vert \mathcal{U}(\tau'',\tau')\vert \psi(x') \ra=\int_{x(\tau')=x'}^{x(\tau'')=x''}
  \left[ D x(\tau) \right]e^{\imath S\left[x \right] /\hbar} \label{path_integral}
\ee
where $S\left[ x \right]$ is the action
\be
 S\left[ x \right]= \int_{\tau'}^{\tau''} L(x, \dot{x}) d \tau,
\ee
and $L$ is the Lagrangian of the system.
$x(\tau)$ is a path that interpolates between $x'$ and $x''$, $ \left[ D x(\tau) \right]$ is the functional measure,
therefore ``$\int_{x'}^{x''}
  \left[ D x(\tau) \right]$'' denotes the integral over all paths between $x(\tau')=x'$ and
$x(\tau'')=x''$.

An important remark is that from eq.(\ref{path_integral}) we can beautifully recover the least action principle 
noting that in the limit, $\hbar\to 0$, the path of minimum action dominates the integral since 
the phase $e^{\imath S/\hbar}$ of any path away from this fluctuates rapidly and different contributions cancel.

The generalisation to quantum fields is a straightforward
generalisation of eq.~(\ref{path_integral}). 
But before introducing its expression we would like to remark that using path-integral methods 
it is common to give the action an imaginary time 
in order to simplify the calculations --- the weight in the path integral is an exponential with real argument, which is easier to handle numerically --- and then return to a real action at the end. This can be done if the Osterwalder-Schrader axioms hold (see Refs.~\cite{Osterwalder}-\cite{Roepstorff}).

Besides simplification purposes, in imaginary time the paths away from the classical path are 
exponentially suppressed. This makes the path integral  to converge much better than the phase rotation.
This is crucial for numerical studies since it allows us to have reliable results with a relatively
modest statistics.

There is a deeper  consequence of considering an imaginary time: it 
allows us to establish a connection to Statistical Mechanics.

\subsection{From Euclidean time to real time}
In complex analysis, a branch of mathematics, analytic continuation is a technique to be used in the domain 
of definition of a given analytic function.
We can apply such techniques here to go from real time $\tau$ to the imaginary time $\trm$ called the {\em Euclidean time}.
(For a formal treatment see Ref.~\cite{Jaffe}).

Imaginary time and spatial coordinates play equivalent r\^oles. 
For example, in real time 
 the D'Alembertian operator is given by:
\be
  \Box =\frac{\partial^2}{\partial \tau^2}-\frac{\partial^2}{\partial x_1^2}
   -\frac{\partial^2}{\partial x_2^2}-\frac{\partial^2}{\partial x_3^2}. \nonumber
\ee
Under the Euclidean prescription we have
$$
  \Delta =-\frac{\partial^2}{\partial x_0^2}-\frac{\partial^2}{\partial x_1^2}
   -\frac{\partial^2}{\partial x_2^2}-\frac{\partial^2}{\partial x_3^2},
$$
where we set $x_0=\trm$.

In Quantum Mechanics we consider the possible particle positions at each time,
given by functions $x(t)$ (or $\vec x(t)$ in $d=3$),
and the path integral integrates over all these functions, i.e.\ over all
possible particle paths (with the given end-point).
This reproduces the canonical Quantum Mechanics, but space and time
are not treated in the same way.
 In field theory, one does treat them in the same manner and
introduces a functions of any space-time point
$x = (\vec x ,t)$, which are denoted as fields. The simplest
case is a neutral scalar field, where this field values are real,
$\phi(x) \in \real$. The assignement of a field value in each
space-time point $x$ is called a configuration, and it takes
over the role of paths in Quantum Mechanics. Consequently, the functional
integral now runs over all field configurations, $\int \left[ D \phi \right]$.
The Lagrangian $L$ is now the integral of a Lagrangian density at each point $\vec{x}$, 
$\mathcal{L}(\phi (x), \partial_{\mu} \phi (x))$, and the action is obtained by  an integral over the space-time volume, $S = \int d^4 x  \mathcal{L}$.

In imaginary time, the analog of eq.~(\ref{path_integral}) for quantum fields is given by\footnote{We set $\hbar=1$.}
\be
    \la \phi''\vert \mathcal{T}(\trm'',\trm')\vert \phi' \ra=\int
  \left[D \phi(x,\trm ) \right] e^{-S \left[ \phi \right]}. \label{functional_integral_ima_time}
\ee

In the Euclidean formulation the time evolution becomes a transfer
matrix, $\mathcal{T}(\trm'',\trm')$, and in the case that $\mathcal{H}$ is time independent  eq.~(\ref{time-evol-operators}) becomes
$e^{-{\cal H}t}$. Furthermore the quantum partition function
$\Tr \left(e^{-\beta {\cal H}}\right)$ becomes the functional integral over paths
that are periodic in Euclidean time of period $\beta$, where
$\beta={1\over k_B T}$ with $k_B$ Boltzmann's constant.  There is also a
second interpretation of the resulting functional integral as a
functional integral in statistical field theory. Here one considers
the Euclidean action as the energy functional of an analog statistical
mechanical system with $k_B T=1$.  As is conventional in lattice field
theory it is the latter analogue that will be used in this thesis.
Then eq.~(\ref{functional_integral_ima_time}) describes a statistical system in  equilibrium.

\subsection{Expectation values}
The expectation values of an observable $F$, denoted $\la F \ra$,
 can be calculated as follows:
\be 
   \left\langle F\right\rangle=\frac{1}{\partition}\int \left[D\phi\right] F(\phi)e^{-S[\phi]}.\label{def-expectation-value}
\ee
where 
\be
\partition = \int \left[ D \phi\right] e^{-S\left[\phi\right]} \label{def-partition-function}
\ee
is the partition function.

The  integration in eq.~(\ref{def-expectation-value}) involves all the possibles 
configurations in the functional space.

The problem is how to measure (or estimate) the value in (\ref{def-expectation-value}).
Here is where the {\em importance sampling methods} enters. The most popular approach is the Monte Carlo Method 
(see appendix \ref{small-description-MC} for a brief description).
The main idea of this method is that we can estimate (\ref{def-expectation-value}) considering a representative sample of (independent) configurations.

The way to produce the {\em representative} samples is through random moves to explore the search space. 

In this thesis we use a variant of the Monte Carlo method called the Metropolis algorithm \cite{Metropolis}
to estimate the expectation values of the observables defined in chapter \ref{section2}.

\chapter{A review of the 2 dimensional $\lambda \phi^4$ model on a fuzzy sphere}
\label{2-dim-model-chapter}
We devote this section to a review of some aspects of the $2$-dimensional $\lambda \phi^4$ model on a fuzzy sphere. 
We will discuss generic properties of fuzzy spaces by means of the most studied example: the fuzzy sphere.
We will  show that the fuzzy sphere can retain the exact rotational 
symmetry. 
 
\section{The 2 dimensional $\lambda \phi^4$ model}
A quite general scalar field theory on a $2$ dimensional sphere is given by the action
\be
  \Stwo(\phi):=\int_{S^2}\left[ \med \phi(x) \frac{\calL^2}{R^2} \phi(x)+ V\left[ \phi(x) \right]
                              \right] R^2 d \Omega \label{2-dim-continous_action-gen}
\ee
where $ d \Omega=\sin\theta d\theta d \varphi$,  
$\phi(x)$ is a neutral scalar field on the sphere. It depends 
on the coordinates $x_i(\theta, \varphi)$ which satisfy:
\be
  \sum_{i=1}^3 x_i^2=R^2, \label{sphere_equation}
\ee
where $R$ is the radius of the sphere. 
$\calL^2= \sum_{i}^3 \calL_i^2$, and $\calL_i$ are the angular momentum
operators.
$V\left[ \phi(x) \right]$ is the potential of the model.

Eq.~(\ref{sphere_equation}) describes $S^2$ embedded in $\real^3$.

\section{The fuzzy sphere}
\label{fuzzy_sphere_section}
To obtain a {\em fuzzy} version of a continuous space
we have to replace
the algebra of the continuous space by a sequence of matrix
algebras of dimension $N$, ${\tt{Mat}}_{N} $. 

In the case of the {\em fuzzy sphere},
the permitted values of  $N$ are $L+1$ where $L$ is
the largest angular momentum (the cutoff), which can take the values $L=0,1,2,\cdots$ .
The coordinates $x_i$ are elements in the algebra of functions of $S^2$, $C^{\infty}(S^2)$. They are replaced by
the {\em coordinate operators}, $X_i$, which are defined as $X_i= 2R
\frac{L_i}{\sqrt{N^2-1}}$, where  $L_i, i=1,2,3$, are the  $SU(2)$
generators in the $N=(L+1)$-dimensional irreducible
representation.
  
The {\em coordinate operators} satisfy the constraint
\be
   \sum_{i=1}^3 X_i^2= R^2  \cdot 1 \!\! 1 \ , \label{fuzzy_sphere_equation}
\ee
which can be interpreted as a matrix equation for a sphere, the
analog to eq.~(\ref{sphere_equation}). Note that the operators $X_i$ do not commute,
\be
   \left[X_i,X_j  \right]=\imath \epsilon_{ijk} \frac{2
     R}{\sqrt{N^2-1}}X_k. \label{fuzzy_commutator}
\ee

Following  the above prescription,
the scalar field is represented by
a hermitian matrix $\Phi$ of dimension $N$. Just as in the
standard case where  $\phi$ can
be expressed as a polynomial in the coordinates $x_i$, its {\em fuzzy version} $\Phi$ can be
written as a polynomial in the {\em fuzzy coordinates}.
The differential operators $\calL_i \cdot$ are
replaced by $\left[L_i, \cdot \right]$ and the integral over $S^2$ is
replaced by the trace. Summarising the above:

\bea
   x_i\in C^{\infty}(S^2)  & \longrightarrow & X_i \in Mat_{N},  
                                                        \label{replace_fuzzy_1}\\
   \phi(x)\in C^{\infty}(S^2)  & \longrightarrow & \Phi \in Mat_{N},  
                                                        \label{replace_fuzzy_2}\\
   \calL_i  \phi(x)            & \longrightarrow & \left[ L_i, \Phi \right],  
                                                        \label{replace_fuzzy_3} \\
  R^2 \int_{S^2} \phi(x) d \Omega & \longrightarrow & \frac{4 \pi R^2}{N} 
                                                        \Tr
                                                        \left(\Phi \right), 
                                                        \label{replace_fuzzy_4} \\
  \calL^2 \cdot& \longrightarrow &\lpl^2 \cdot :=  \sum_{i=1}^3
                    \left[L_i, \left[L_i, \cdot  \right] \right]. \label{replace_fuzzy_5}
\eea

Note that $\phi \in \real$ implies that $\Phi$ is hermitian and the
                    choice of normalisation in eq.~(\ref{replace_fuzzy_4})
                    ensures that the integral of the unit function
                    equals the trace of the unit matrix, i.e.

\be
 \frac{4 \pi R^2}{N}  \Tr   1 \!\! 1 \ = 4 \pi R^2  =  R^2 \int_{S^2}d \Omega.
  \label{int_volume_fuzzy}
\ee

Rotations  on the {\em fuzzy sphere} are performed by the adjoint action of
an element $U$ of $SU(2)$ in the dimension $N$ unitary irreducible
representation. $U$  has the general form
$U=e^{\imath \omega_i L_i}$. The coordinate operators are then rotated
as
\be
  U X_i U^{\dagger}= {\mathtt{R}}_{ij} X_j, \quad  \mathtt{R} \in SO(3)
\ee
and  the field transforms as
\be
  \Phi \longrightarrow \Phi'=U \Phi U^{\dagger} \label{rotation_fuzzy}.
\ee

\subsection{Limits of the fuzzy sphere}\label{limits-fuzzy-sphere}
Following \cite{steinacker}, for the spatial part of our model (the fuzzy sphere) we have:
\begin{itemize}
 \item The {\em commutative sphere limit} $S^2$:
   \be  N \longrightarrow \infty, \quad 
       R \mbox{ fixed}.  \label{commutative-sphere-limit}
   \ee 
  \item The {\em Moyal plane limit}  $\real_{\Theta}^2$
    \be  N \longrightarrow \infty, \,
       R^2=\frac{N \Theta}{2}, \quad \Theta \mbox{ constant}.\label{moyal-plane-limit}
    \ee
  \item The {\em commutative flat limit} $\real^2$
    \be  N \longrightarrow \infty, \,
       R \propto N^{\med(1-\epsilon )},   \quad 1>\epsilon >0.  \label{commutative-flat-limit}
    \ee
\end{itemize}
The limit given by eq.~(\ref{commutative-sphere-limit}) arises naturally from the fact that $N \longrightarrow \infty$  recovers $ C^{\infty}(S^2)$.

A short way to deduce eqs.~(\ref{moyal-plane-limit})-(\ref{commutative-flat-limit})  is the following:

Considering the  north pole on the fuzzy sphere where $X_3^2 \sim  R^2 1 \!\! 1 \ $,  we can re-scale the coordinate $X_3^2$ to  $\frac{X_3^2}{R^2} \sim   1 \!\! 1 \ $, and re-write the commutation relation (\ref{fuzzy_commutator}) at the  north pole  as 
\be
   \left[X_1,X_2  \right]=\imath  \frac{2
     R}{\sqrt{N^2-1}}X_3 \approx  \imath \frac{2
     R^2}{N}\frac{X_3}{R}. \label{fuzzy_commutator-at-north-pole}
\ee
We propose $R$ as a function in $N$. For the non-commutative plane we have:
\be
    \left[X_1,X_2  \right]= \imath \Theta, \label{Moyal-commutator}
\ee
then, comparing  eq.~(\ref{fuzzy_commutator-at-north-pole}) to (\ref{Moyal-commutator}) we obtain  $\frac{2
     R^2}{N}=\Theta$.

We define the exponent $\epsilon$ in the relation  
\be
  R^2 \propto N^{1-\epsilon}, \quad 1>\epsilon\ge0.
\ee
 If $\epsilon =0$ we have $\Theta=const.$ For $\epsilon >0$ the commutator (\ref{fuzzy_commutator-at-north-pole}) vanishes if $N\longrightarrow \infty$.  
 Note that in this limit we also require the commutator given by  eq.~(\ref{fuzzy_commutator}) to vanish 
and this requirement is immediately satisfied  for $\epsilon >0$.

\subsection{The scalar action on the fuzzy sphere}
The next step is to define our field theory on the fuzzy sphere.
Implementing the replacements given by eqs.~(\ref{replace_fuzzy_1})-(\ref{replace_fuzzy_5}) in 
eq.~(\ref{2-dim-continous_action-gen}) we arrive at the following expression,
\be
 \Stwo \left[ \Phi \right] = \frac{4 \pi R^2 }{N} 
                            \Tr \Big(
                            \frac{1}{2} \Phi \frac{\lpl^2}{R^2} 
                            \Phi +
                            V\left[ \Phi \right]
                           \Big) \ . \label{action_fuzzy_sphere-2d}
\ee

Eq.~(\ref{action_fuzzy_sphere-2d}) is valid for any potential  $V\left[ \Phi \right]$.
For testing purposes it is convenient to select a simple model.
In Ref.~\cite{Garcia-Martin-OConnor} the $\lambda \phi^4$ model on a fuzzy sphere was studied,
where the action is written as
\begin{equation}\label{eq:accion}
  S[\Phi] = \Tr \left[ a  \Phi {\calL^2} \Phi + b \Phi^{2} + c
  \Phi^{4} \right]. 
\end{equation}
$\Phi$ is a Hermitian matrix of size $N$.
 After a suitable rescaling, the parameters $b$ and
$c$ become the mass squared and the self-coupling,
respectively.

In Ref.~\cite{Garcia-Martin-OConnor}  $\phi$ was rescaled to fix $a=1$.
The model in eq.~(\ref{eq:accion}) was also studied in Ref.~\cite{xavier} but in terms of a different convention of  parameters
\be 
a=\frac{4\pi}{N}, \quad b=a r R^2, \quad  c=a\lambda R^2. \label{convention-xavier}
\ee

\section{Numerical results on the two dimensional model}
\begin{figure}[!h]
\begin{center}
\includegraphics[width=0.70 \textwidth,angle=-90]{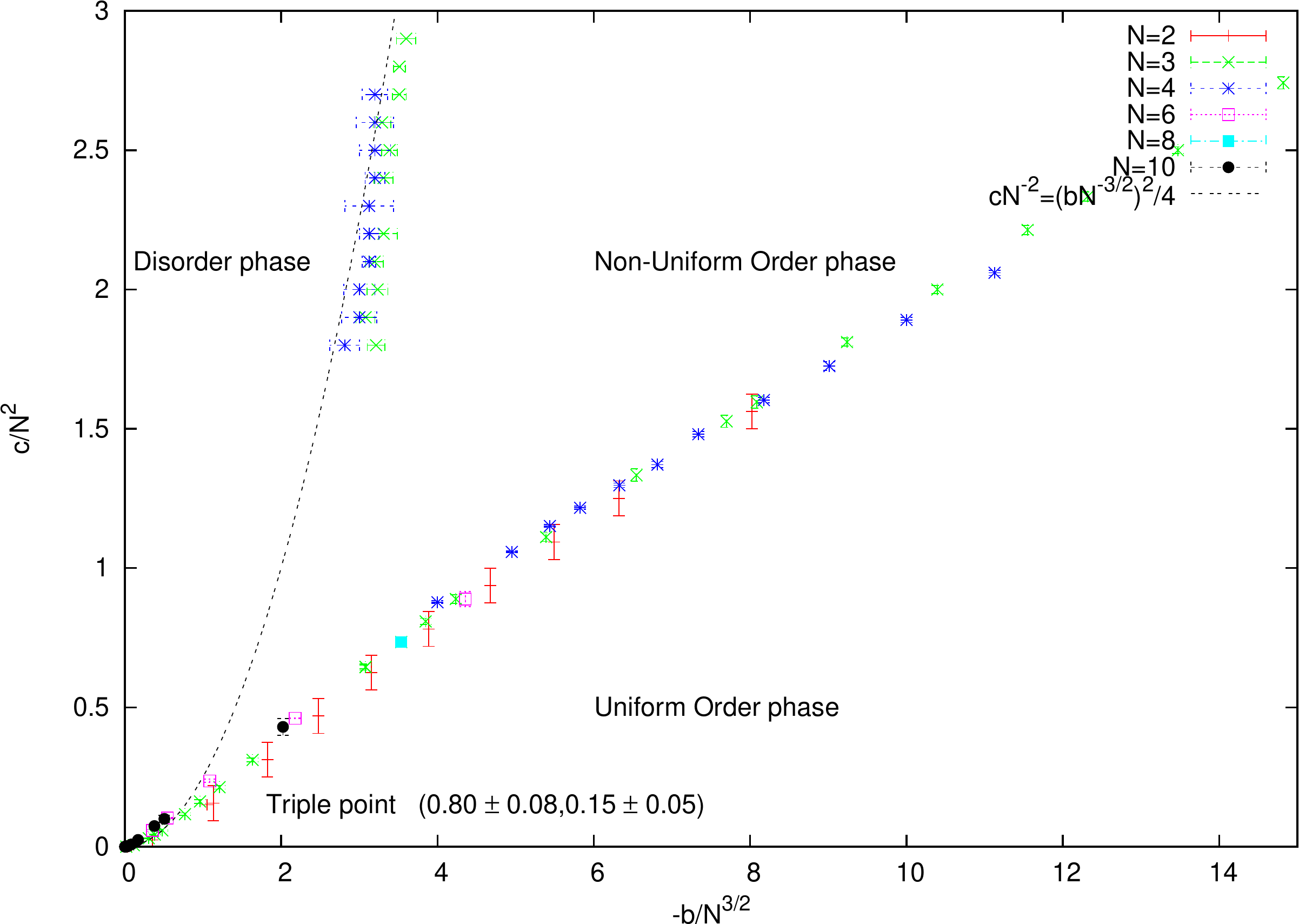}   
\caption{Phase diagram obtained from Monte Carlo simulations of the
  model (\protect\ref{eq:accion}) in Ref.~\cite{Garcia-Martin-OConnor}.}
\label{fig:diagram}
\end{center}
\end{figure}

The model in eq.~(\ref{eq:accion}) has been studied  numerically by several authors ---see  Refs.~\cite{xavier},\cite{Garcia-Martin-OConnor}-\cite{Das-Digal-Govindarajan}. 
We follow those results in  Refs.~\cite{Garcia-Martin-OConnor}, which
are summarised in {\bf figure \ref{fig:diagram}}.\footnote{We thank the authors of  Refs.~\cite{Garcia-Martin-OConnor} for their permission to reproduce the graph here.}

{\bf Figure \ref{fig:diagram}} shows the existence of three phases:
\begin{itemize}
 \item Disordered
 \item Non-uniform ordered
 \item Uniform ordered
\end{itemize}
It is remarkable that the three coexistence lines collapse under the same re-scaling in $N$ for the axes
$\frac{-b}{N^{3/2}}$ vs. $\frac{c}{N^2}$.
Two coexistence lines are:
\begin{itemize}
 \item Disordered - Ordered uniform: 
  \be
    b=-0.35\sqrt{N} c. \label{eq-phase-trans-2-dim-1}
  \ee
 \item Disordered - Non-uniform ordered:
  \be
    \frac{c}{N^2}=\frac{(bN^{-3/2})^2}{4}. \label{eq-phase-trans-2-dim-2}
  \ee
\end{itemize}
The triple point is given by the intersection of the three critical lines 
\begin{equation}
(b_{_T},c_{_T})=(-0.15 N^{3/2},0.8 N^{2}). \label{triple-point-fuzzy-sphere}
\end{equation}

The transition in eq.~(\ref{eq-phase-trans-2-dim-2}) is only valid for large values of $c$.
Therefore eq.~(\ref{triple-point-fuzzy-sphere}) is not the intersection of
eqs.~(\ref{eq-phase-trans-2-dim-1})-(\ref{eq-phase-trans-2-dim-2}).

Although we discretised the spatial part of our three dimensional model by means of a {\em fuzzy sphere},
our simulations will show  that the properties differ from  those of the two dimensional model in eq.~(\ref{eq:accion}).
In certain limits the $\lambda \phi^4$ theory on the fuzzy sphere can emerge as a limit of the $3$-dimensional model
in eq.~(\ref{action_1}).

The model in eq.~(\ref{eq:accion}) depends on the parameters $N,\ R,\ m^2$ and $\lambda$ 
---or equivalently it depends on $N,\ a,\ b$ and $c$, see eq~.(\ref{convention-xavier})---
but it effectively depends only on two out of  three parameters $a,\ b$ and $c$.
For the three dimensional model in addition to the parameters in the $2$-dimensional model we have
as parameters the number of lattice sites, $N_t$ and the lattice spacing $\Delta t$.
We will fix in section \ref{details_parameters} $N_t=N$  and the model will effectively 
depend on four parameters.

The first question is if the three dimensional model has the phase of  non-uniform ordering. 
We will see in chapter \ref{chapter-phases-characterisation} that the answer to this question is yes.
\chapter{Description of the model}
\label{section2}

In this Chapter we present the discretisation of the $3$-dimensional model composed
by a $2$-dimensional sphere plus a Euclidean time direction.

We first recall the results of chapter \ref{2-dim-model-chapter} for the 
discretisation of the continuous model. We will apply them to the $3$-dimensional model.
After performing  the discretisation in the time direction we present the model to be studied by Monte Carlo techniques. In section \ref{Definitions-of-the-observables} we present the observables and a brief description of their meaning.

\section{Regularisation of the action}

It is convenient to consider the Euclidean version of the model. 

As it was remark in chapter \ref{path-int-chapter}, the main advantage of working in this formalism is that it allows to establish a connection to Statistical Physics and the functional integral converges with a relatively modest statistics.

The model to regularise is
{\small
\be
  S(\phi):=\int_{S^1} d t \int_{S^2}\left[ \med \phi(x,t) \left(\frac{\calL^2}{R^2}-\partial_t^2 
                                                   \right) \phi(x,t)+
                              \frac{m^2}{2}\phi^2(x,t)+\frac{\lambda}{4}\phi^4(x,t)         
                        \right] R^2 d \Omega \label{continous_action}
\ee
}
$\phi(x,t)$ is a neutral scalar field on the sphere. It depends on time (euclidean) and
on the coordinates $x_i(\theta, \varphi)$ satisfying eq.~(\ref{sphere_equation}),
where $R$ is the radius of the sphere. 

We consider the integral over Euclidean time on a compact version, $S_1$, which 
has circumference $T$.

First we will explain how to discretise the {\em spatial}  directions
and then we will perform the discretisation in time direction.

\subsection{Discretising by a fuzzy sphere}

Let us consider the spatial part of the action given in eq.\ (\ref{continous_action}),
\be
  s(\phi,t):= \int_{S^2}\left[ \med \phi(x,t)
                      \left( \frac{\calL^2}{R^2}-\partial_t^2 \right)
                                  \phi(x,t)
             +                \frac{m^2}{2}\phi^2(x,t)
           +\frac{\lambda}{4}\phi^4(x,t)         
                        \right] R^2 d \Omega. \label{continous_action_spatial}
\ee

Implementing the replacements given by eqs.~(\ref{replace_fuzzy_1})-(\ref{replace_fuzzy_5}) in 
eq.~(\ref{continous_action_spatial}) we arrive at
\be
 s\left[ \Phi ,t\right] = \frac{4 \pi R^2 }{N} 
                            \Tr \Big(
                            \frac{1}{2} \Phi(t)
                            \left( \frac{\lpl^2}{R^2}  -\partial_{t}^2
                            \right) \Phi(t) 
                            + \frac{m^2}{2} \Phi^2(t)
                    +\frac{\lambda}{4} \Phi^4(t)
                           \Big) \ . \label{action_fuzzy_sphere}
\ee

Then, the action (\ref{continous_action}) ``discretised'' in the spatial
directions is:

\be
S \left[ \Phi\right] = \frac{4  \pi R^2 }{N} \int_{S_1} d t \, 
                            \Tr \Big[
                            \frac{1}{2} \Phi\left(t\right)
                             \left( \frac{\lpl^2}{R^2}-\partial_t^2 \right)
                    \Phi\left(t\right)  
                    + \frac{m^2}{2} \Phi^2(t)
                    +\frac{\lambda}{4} \Phi^4(t)
                           \Big] \ . \label{action_0}
\ee
The model given by eq.~(\ref{action_0}) has the exact rotation  symmetry of model
(\ref{continous_action}) since any rotation on the sphere is allowed
and the action (\ref{action_0}) is invariant under uniform rotations
given by eq.~(\ref{rotation_fuzzy}).

\subsection{Discretisation  of the time direction }
\label{Discretization-of-the-time-direction}
To discretise the time direction we take a set of $N_{\trm}$
equidistant points, then $T=N_{\trm} \Delta t $.

The changes to implement in eq.~(\ref{continous_action}) are:
\bea
    \int_{S_1} dt     & \longrightarrow &  \sum_{\trm=1}^{N_{\trm}} \Delta t , \label{replace_lattice_1} \\
    \partial_t \phi(x,t) & \longrightarrow & \frac{\phi(x,t+\Delta t) -\phi(x,t)}{\Delta t}.
                                                  \label{replace_lattice_3}
\eea  
We arrive at
\small{
\bea
S \left[ \Phi\right] = \frac{4  \pi R^2 }{N}  \Delta t \sum_{\trm=1}^{N_{\trm}} 
                            \Tr   \Big[&& \hspace{-5mm}
                            \frac{1}{2R^2} \Phi\left(t\right)
                             \lpl^2 \Phi\left(t\right) +  \med 
                    \left( \frac{\Phi(t+\Delta t)  -\Phi(t)}{\Delta t}  \right)^2 \nonumber \\ 
& & \hspace{-5mm}
                    + \frac{m^2}{2} \Phi^2(t)
                    +\frac{\lambda}{4} \Phi^4(t)
                           \Big]. \label{action_1}
\eea
}
One configuration $\Phi$ corresponds to a set of matrices
$\{ \Phi (t) \}$,  for $\trm=1, \dots ,N_{\trm}$.

Alternatively we can write down eq.~(\ref{action_1}) in terms of the constants $\cteA$, $\cteD$, $\cteB$ and $\cteC$ defined in eqs.~(\ref{cte-A})-(\ref{cte-C}):
\bea
  \cteA &=& \frac{2  \pi \Delta t }{N}, \label{cte-A} \\
  \cteD &=& \frac{2  \pi R^2 }{N \Delta t}, \label{cte-D} \\
  \cteB &=& \frac{2  \pi R^2 m^2  \Delta t}{N}, \label{cte-B} \\
  \cteC &=& \frac{ \pi R^2 \lambda \Delta t }{N}.  \label{cte-C}
\eea
Then the action reads:
\be
S \left[ \Phi\right] =  \sum_{\trm=1}^{N_{\trm}} 
                            \Tr \Big[ 
                            \cteA \Phi\left(t\right)
                             \lpl^2 \Phi\left(t\right) +  \cteD
                    \left( \Phi(t+\Delta t) -\Phi(t)  \right)^2
                    + \cteB \Phi^2(t)
                    +\cteC \Phi^4(t) \Big] \label{action_two}.
\ee

\section{Decomposition of the field}
\label{Decomposition-of-the-field}
As we mentioned in the previous section, we are representing a
configuration of the field in our model by a set of matrices 
$\{ \Phi (t) \}$, for $t=1, \dots ,N_{\trm}$. Every element in this
set can be expanded in the {\em polarisation tensor basis}
\be
   \Phi(t)= \sum_{l=0}^{N-1} \sum_{m=-l}^l c_{lm}(t) \Yp_{lm},   \label{expansion_field}
\ee
where $c_{lm}(t)$ are $N^2$ coefficients.
The polarisation tensors $\Yp_{lm}$ are $N \times N$ matrices
that are the analog of the spherical harmonics,
$Y_{lm}(\theta,\varphi)$. Details about the polarisation tensors are presented
in  appendix \ref{app-polarisation}.

At the end, the quantities of  interest can be expressed as expectation
values or averages over the configurations.
The expectation value
of the observable $F(\Phi)$  was defined in eq.~(\ref{def-expectation-value}):
\be
  \la F \ra =\int \left[ D \Phi \right] F (\Phi)\frac{e^{-S\left[\Phi\right]}}{\partition} 
\label{def-expectation-value-2}
\ee
where $\partition= \int \left[ D \Phi\right] e^{-S\left[\Phi\right]}$ is the partition function.
Quantities of interest will be

\bea
 & &  \la \Phi(\trm) \ra , \label{av1} \\
 & &  \la \Phi(\trm) \Phi(\trm') \ra  \label{correlator} 
\eea
where eq.~(\ref{av1}) is a {\em condensate} and eq.~(\ref{correlator}) is a
 correlation function. These  can be mapped to standard correlation
functions by replacing the $ \Yp_{lm} $ by $Y_{lm}(\theta,\varphi)$. 
We can reduce the expressions
(\ref{av1})-(\ref{correlator})  to combinations of the expectation values of the coefficients $c_{lm}(t)$
introduced in eq.~(\ref{expansion_field})
\bea
 & &  \la c_{lm}(\trm) \ra , \label{correlator1}  \\
 & &  \la c_{lm}^*(\trm) c_{l'm'}(\trm')\ra . \label{correlator2} 
\eea
Now it is convenient to compute the quantities (\ref{correlator1})-(\ref{correlator2})
 after a Fourier transform in (Euclidean) time.

Following Ref.~\cite{delgadillo_thesis}, the complete Fourier decomposition of the
field is given by

\be
   \Phi(t):= \sum_{l,m}\sum_{k=0}^{N_{\trm}-1} c_{lm}(k)
                      e^{\imath \frac{2 \pi k t}{N_{\trm}}}  
                      \Yp_{lm},   \label{expansion_fourier}
\ee
where
\bea
    c_{lm}(k)&:=& \frac{1}{N_{\trm}}\sum_{\trm}  
                   e^{-\imath \frac{2 \pi k t}{N_{\trm}}} 
                   \underbrace{\frac{4 \pi}{N}\Tr \left(
                               \Yp_{lm}^{\dagger} \Phi(t) \right) }. \label{coeff-fourier} \\
                & & \hspace{1.4in} c_{lm}(t)  \nonumber
\eea
It will sometimes prove convenient to define
\be
 \Phi(k)= \frac{1}{2 \pi N_{\trm}}\sum_{\trm} 
  e^{-\imath \frac{2 \pi k t}{N_{\trm}}}\Phi(t).  \label{expansion_fourier2}
\ee

In this space the correlator (\ref{correlator2}) is diagonal,
\be
  \la c_{lm}^*(k) c_{l'm'}(k') \ra = G_{lm}(k)\delta_{kk'}
                                     \delta_{l l'}\delta_{m m'}. \label{space_correlator}
\ee
$ G_{lm}(k)$ is the Green function in  momentum space. Here the term
``momentum space'' is used to include both angular momentum $(l,m)$ and
frequency $k$.

\section{The different limits}
\label{limits-of-the-model}
For all our simulations we are interested in taking
the thermodynamic limit $N \longrightarrow \infty$.

For the time direction, we are interested on taking  $N_{\trm} \longrightarrow \infty $. 

We now set to  $\Delta t=1$. Now, the tricky part is how to relate the parameter $R$, the radius of the spheres, and $N$, the dimension of the matrices.

For the limits of the spatial part of our model we follow section \ref{limits-fuzzy-sphere}.

\section{Definitions of the observables}
\label{Definitions-of-the-observables}
We define {\em the field averaged over the time lattice as:}
\be
   \field:=\frac{1}{N_{\trm}}\sum_{\trm}\Phi(t). \label{averaged_field}
\ee

The average over the time lattice of the coefficients $c_{lm}$ are:
\be
   \ccv_{lm}:=\frac{1}{N_{\trm}}\sum_{\trm} c_{lm}(\trm). \label{averaged_coeff}
\ee

This picks out the zero frequency component of $\Phi$, i.e.\ eqs.~(\ref{averaged_field})-(\ref{averaged_coeff})
 are particular cases of the equations
(\ref{expansion_fourier})-(\ref{expansion_fourier2})  when $k=0$.

Some particular cases in eq.~(\ref{averaged_coeff}) are
\bea
   \ccv_{00}&:=&\frac{\sqrt{4 \pi}}{N} \Tr  \field, \label{averaged_coeff-00}\\
   \ccv_{1m}&:=&\frac{4 \pi}{N} \Tr\left( \Yp_{1,m}^{\dagger} \field \right), 
   \label{averaged_coeff-1m}
\eea
where $\Yp_{1,m}$ are given in eqs.~(\ref{Yp11})-(\ref{Yp1-1}) of appendix \ref{app-polarisation}.

\subsection{Order parameters}
\label{definitions-order-parameters}
We want to measure the contributions of different modes to the configuration
$\field$. For this purpose we need a control parameter. This turns out  to be
the sum 

$ \vert \ccv_{lm}
\vert^2$, this quantity was called  {\em the full power of the field} in Ref.~\cite{xavier} and it represents the {\em norm} of the field $\field$; it can be calculate as :
\be
   \varphi_{all}^2:=\sum_{l,m} \vert \ccv_{lm} \vert^2=\frac{4 \pi}{N}
                \Tr\left( \field^2\right).  \label{full_power_field}
\ee

Although $\la \varphi_{all}^2   \ra $ cannot play the r\^ole of an order
parameter, we will show that it is useful to localise the region where 
 the phases split into disordered and ordered. We expect $\la {\varphi}_{all}^2
\ra \sim 0 $ in the disordered phase and $\la {\varphi}_{all}^2
\ra \gg 0 $ in the ordered phase. 

To distinguish the contributions from the different modes to eq.~(\ref{full_power_field}) we define the  quantity:
\be
  \varphi_l:=  \sqrt{ \sum_{m=-l}^{l} \vert c_{l,m} \vert^2}.
   \label{phi_l}
\ee

We can re-write eq.~(\ref{full_power_field}) in terms of the quantities in eq.~(\ref{phi_l})
\be
   \varphi_{all}^2:=\sum_{l} \varphi_l^2.  \label{full_power_field-2}
\ee
In the disordered phase we expect $\la \varphi_l \ra \approx 0$ for all $l$.

Studying the contributions of the different
modes to $\la \varphi_{all}^2 \ra$ can provide more information about the phases.
If $\la \varphi_l \ra \gg 0$ for $l>0$ it indicates that the rotational symmetry is broken.
In our simulations we measure quantities related to the lowest modes: the zero mode for $l=0$
and the first mode for $l=1$ as representative of those modes where the rotational symmetry is broken.

Choosing the particular case $l=0$ in eq.~(\ref {phi_l}) we have

\be
   \varphi_0:=  \vert \ccv_{00}  \vert.
     \label{parameter0}
\ee
For $m^2<0$, if the contribution  of the fuzzy kinetic term to the action is not negligible 
we can expect the kinetic term to select the zero mode as the leading one,
$\la \varphi_0^2 \ra \cong \la \varphi_{all}^2 \ra$.  As a consequence $\la \varphi_0 \ra \gg 0$ in the uniform 
ordered phase;
$\la \varphi_l \ra$ is expected to be close to zero in the  disordered phase.

Its corresponding  susceptibility is defined as:

\be
  \chi_0:= \la \varphi_0^2 \ra - \la \varphi_0 \ra ^2. \label{sus_zero_mode}
\ee

As the contribution of the kinetic term to the action reduces compared to the potential contribution we can
expect the system can undergo the condensation of higher modes.
Let us consider the p-wave contribution to $\Phi$,  i.e.\ the contribution
of the $l=1$ mode.
Using  $\ccv_{1m}$,  $m=1,0,-1$ we introduce a 3-dimensional vector, 
\[
\vlm:= \left( \begin{array}{c}
                            c_{1,1} \\
                            c_{1,0}  \\
                            c_{1,-1}  
                        \end{array}
                  \right) .
\]

With this vector we can define the order parameter $\varphi_1$, as
a particular case $l=1$ in eq.~(\ref {phi_l}) we have
\be
  \varphi_1:=  \sqrt{ \sum_{m=-1}^{1} \vert c_{1,m} \vert^2}:=\vert \vlm
  \vert  \label{phi_1}
\ee

and its {\em susceptibility}, $\chi_1$:
\be
    \chi_1:= \la {\varphi_1}^2 \ra -  \la {\varphi_1} \ra^2. \label{sus_first_mode}
\ee

Following Ref.~\cite{xavier} the ordered non-uniform  phase is then characterised by  $\la \varphi_1^2 \ra
\gg0$. Note, however, that due to  fluctuations  we will always have $\la
\varphi_1^2 \ra >0$, so we
have to specify {\em how large} it has to be. We will give more details of how to characterise this phase in
the next section.

We can include contributions of the remaining modes generalising
(\ref{phi_1}) and (\ref{sus_first_mode}). In practice the study of
the first two modes should be enough to understand the behaviour of
the system.

\subsection{Energy and specific heat}

The internal energy is defined as:
\be
  E(m^2,\lambda):=\la S \ra,  \label{energy}
\ee
and the specific heat takes the form
\be
  C(m^2,\lambda):= \la  S^2\ra - \la  S \ra^2. \label{specific_heat}
\ee
These terms correspond to the usual definitions
$E(m^2,\lambda) =-\frac{1}{\partition}\frac{\partial \partition}{\partial \beta}$
and $ C(m^2,\lambda)=\frac{\partial E}{\partial \beta} $
where $\partition$ is the partition function.\footnote{$\beta$ is proportional to the inverse of the temperature $T$, i.e.\ $\beta=\frac{1}{k_B T}$, where $k_B$ is the Boltzmann constant.}

We separate the action (\ref{action_two}) into its four contributions:
\bea
S_1 \left[ \Phi\right] &=& \cteA \sum_t
                            \Tr \Big(
                            \Phi\left(t\right) \lpl^2
                    \Phi\left(t\right)  \Big), \label{fuzzy_contribution}
\\
S_2 \left[ \Phi\right] &=& \cteD \sum_t
                            \Tr \Big(  \Phi(t+1) -\Phi(t)  \Big)^2,\label{temporal_contribution}
\\
    S_3 \left[ \Phi\right] &=& \cteB \sum_t
                            \Tr \Big(  \Phi^2(t)\Big),  \label{mass_contribution}  \\
    S_4 \left[ \Phi\right] &=& \cteC \sum_t
                            \Tr \Big(  \Phi^4(t)\Big), \label{quartic_contribution}
\eea
where $\cteA,\cteB,\cteC,\cteD$ were defined in  eqs.~(\ref{cte-A})-(\ref{cte-C}).

The corresponding expectation values to eqs.~(\ref{fuzzy_contribution})-(\ref{quartic_contribution}) are
\bea
  E_1(m^2,\lambda)&:=&\la S_1 \ra,  \label{energy1}   \\
  E_2(m^2,\lambda)&:=&\la S_2 \ra,  \label{energy2}   \\
  E_3(m^2,\lambda)&:=&\la S_3 \ra,  \label{energy3}   \\
  E_4(m^2,\lambda)&:=&\la S_4 \ra.  \label{energy4}   
\eea

\subsection{Dimensionless parameters.}  
\label{details_parameters}
\begin{itemize}
 \item Eq.~(\ref{action_1}) is written in terms of the following parameters: a general temporal lattice spacing $\Delta t$ , the radius of the sphere $R$, the dimension of the matrices $N$, the number of points in the lattice $N_{\trm}$,the mass squared $m^2$ and the self-coupling $\lambda$.  
 \item In order to simplify the simulations, we use the freedom to re-scale the field $\Phi$ to fix the value of one of the constants given in eqs.~(\ref{cte-A})-(\ref{cte-C}). For our simulations we fix $A=2\pi$ --- see chapter \ref{discussion} for more details. We defined the dimensionless parameters:
\bea
       \bar{R} & = & \frac{R}{\Delta t}, \label{R-prime} \\
     \bar{m}^2 & = & (\Delta t)^2 m^2,   \label{m-prime}  \\
 \bar{\lambda} & = & \Delta t \lambda \label{lambda-prime}.
\eea
 \item In all our simulations for the $3$-dimensional model $N_{\trm}$ was taken equal to $N$.
\end{itemize}

\chapter{Description of the different phases in the model}
\label{chapter-phases-characterisation}
In this chapter we characterise the different phases present in this model.
According to the values of $\bar{\lambda}$ relative to a critical value $\bar{\lambda}_T$   we will see that we can divide the space of parameters into two regions. In both cases we can subdivide according to values of $\bar{m^2}$:

\begin{enumerate}
     \item $\bar{\lambda}_T > \bar{\lambda} > 0$.
          \begin{enumerate}             \item For $\bar{m^2}  < \bar{m^2_c} $ we have a uniform ordering (Ising type).            \item For $\bar{m^2}  > \bar{m^2_c} $ we have the disordered phase.         \end{enumerate}  
    \item $\bar{\lambda} > \bar{\lambda_T} $. 
          \begin{enumerate}
             \item For $\bar{m^2}  < \bar{m^2_c} $ we have a non-uniform ordering.
             \item For $\bar{m^2}  > \bar{m^2_c} $ we have the disordered phase.
           \end{enumerate}
\end{enumerate}

\section{Behaviour of the system for $\bar{\lambda}_T > \bar{\lambda} > 0$}
In the previous chapter we defined the observable $ \varphi_{all}^2$ called the {\em full power of the field}. It provides a {\em control parameter} since it represents a  norm of the matrix $ \field $. We will see that $ \la \varphi_{all}^2 \ra \approx 0$ defines a disordered phase while  $ \la \varphi_{all}^2 \ra \gg 0$ defines a kind of ordered regime. But  to describe the type of ordering it is necessary to study the contribution from separate modes to $\la \varphi_{all}^2 \ra$.
 In  {\bf figure \ref{ordermix-N12-l2-R4} } we present a typical case  for  $\bar{\lambda}_T > \bar{\lambda}$ where we show the partial contributions from the zero and first mode to $ \la \varphi_{all}^2  \ra$.

 \begin{center} 
   \includegraphics[width=3.6in]{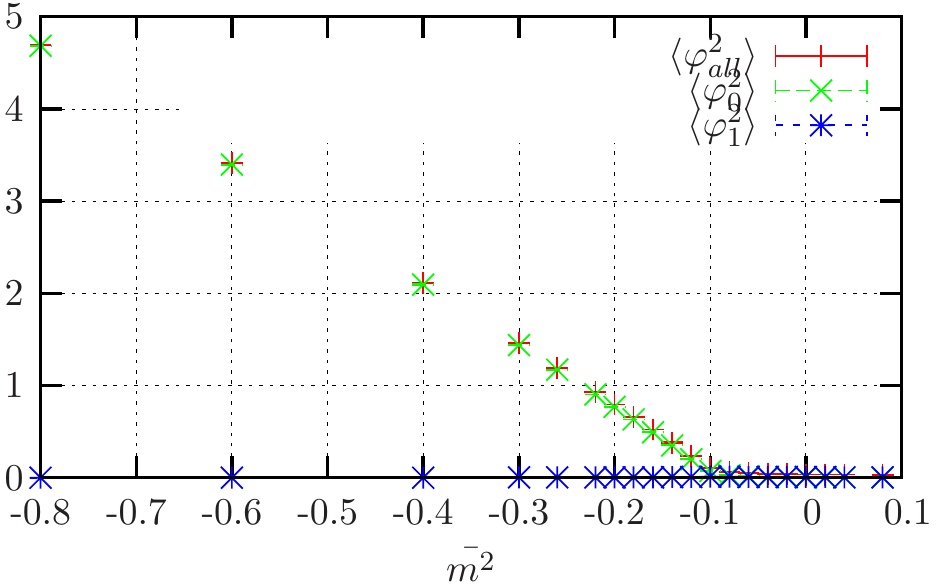}
\end{center}
\begin{figure}[h]
 \vspace{-0.35in}
 \caption{  $\la \varphi_{all}^2 \ra$, $\la \varphi_0^2 \ra$ and $\la \varphi_1^2 \ra$
   vs. $\bar{m^2}$ at $\bar{\lambda}=0.17$, 
    $\bar{R}=4$, $N=12$.}
\label{ordermix-N12-l2-R4}
\end{figure}

From {\bf figure \ref{ordermix-N12-l2-R4}} we can observe  that for $\bar{m^2}>-0.1$: $\varphi_{all}^2\approx0$, $\bar{m^2}<-0.1$: $\varphi_{all}^2\sim \varphi_0^2>0$, so the dominant mode turns out to be the zero mode.

\begin{center} 
   \includegraphics[width=3.6in]{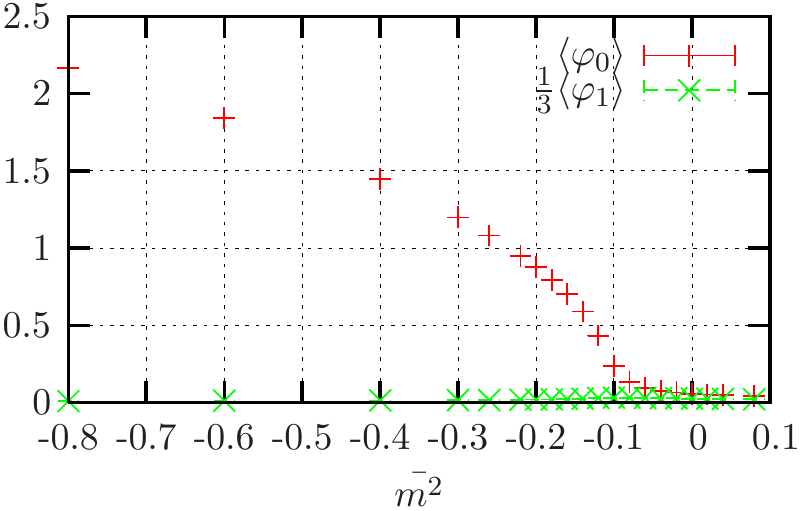}
\end{center}
\begin{figure}[h]
 \vspace{-0.35in}
 \caption{$\la \varphi_0 \ra$ and $\la \varphi_1 \ra$ vs. $\bar{m^2}$ at $\bar{\lambda}=0.17$, 
    $\bar{R}=4$, $N=12$.}
\label{order-N12-l2-R4}
\end{figure}
{\bf Figure \ref{order-N12-l2-R4} } shows the order parameters $\la \varphi_0 \ra$ and  $\la \varphi_1 \ra$. 
In order to determine  precisely where the phase transition occurs the standard way  is to search the maximum in the susceptibility, in this case  $\chi_0$ since the zero mode is  relevant for this phase transition.

\begin{figure}[h]
\begin{center}
  \includegraphics[width=4.6in]{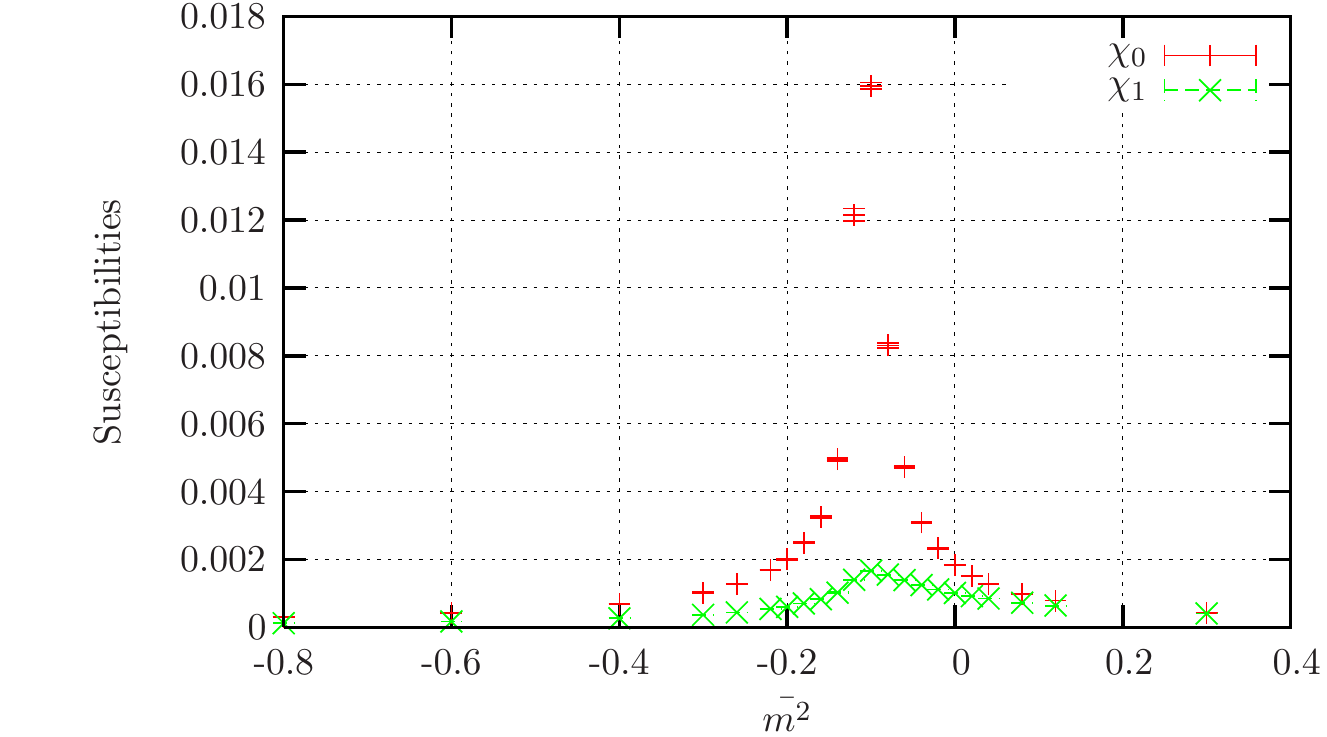}
 \end{center}
 \vspace{-0.2in}
 \caption{The susceptibilities $\chi_0$ and $\chi_1$, in eqs.~(\ref{sus_zero_mode}) and (\ref{sus_first_mode}), at $\bar{\lambda}=0.17$, $\bar{R}=4$, $N=12$}
\label{sus-mix-N12-l2-R4}
\end{figure}
$\chi_0$ peaks at $\bar{m^2}=-0.1$ and in {\bf figure \ref{sus-mix-N12-l2-R4}  } we can observe that the susceptibility associated to the first mode reveals a small response too.

{\bf Figure \ref{contrib-N12-l2-R4}} shows the internal energy for the same parameters as in {\bf figures \ref{ordermix-N12-l2-R4}-\ref{sus-mix-N12-l2-R4}}.
 \begin{center}
  \includegraphics[width=4.6in]{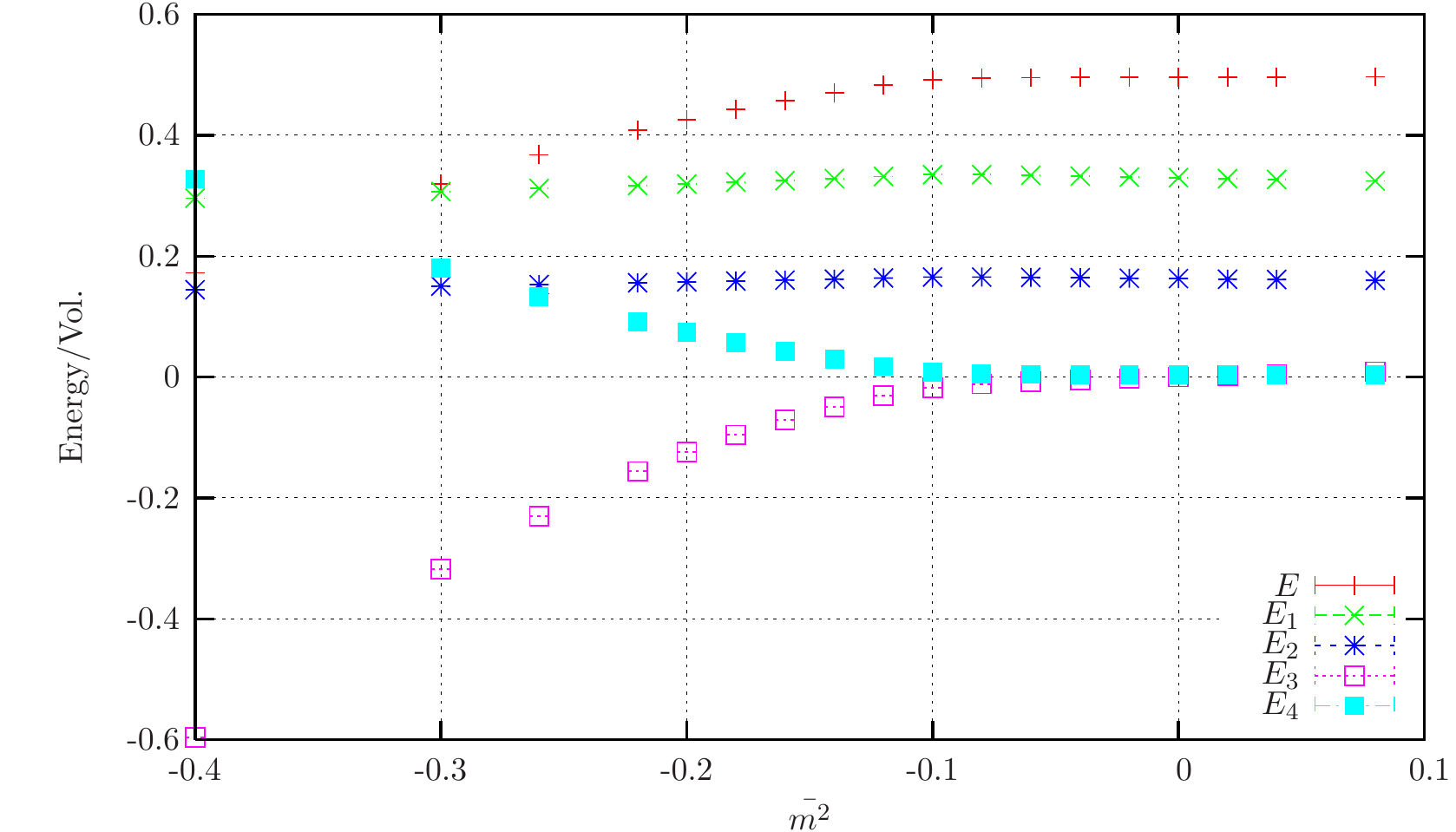}
 \end{center}
\begin{figure}[h]
 \vspace{-0.2in}
 \caption{ Internal energy density $E$, see eq.~(\ref{energy}), and its partial contributions, given in eqs.~(\ref{energy1})-(\ref{energy4}), at $\bar{\lambda}=0.17$, $\bar{R}=4$, $N=12$.}
\label{contrib-N12-l2-R4}
\end{figure}
We observe in {\bf figure  \ref{contrib-N12-l2-R4}} that the leading contribution for $\bar{m^2}>-0.1$ to the internal energy $E$, is the one that comes from the kinetic fuzzy term in  eq.~(\ref{energy1}). 
Note that the phase transition occurs at $\bar{m^2}$ where the potential contributions $E_3$ and $E_4$ deviate from zero.

{\bf Figure \ref{sus-heat-N16-l2-R4} }shows an archetypical behaviour of the specific heat for $\bar{\lambda}_T > \bar{\lambda}$.

 \begin{center}
  \includegraphics[width=4.6in]{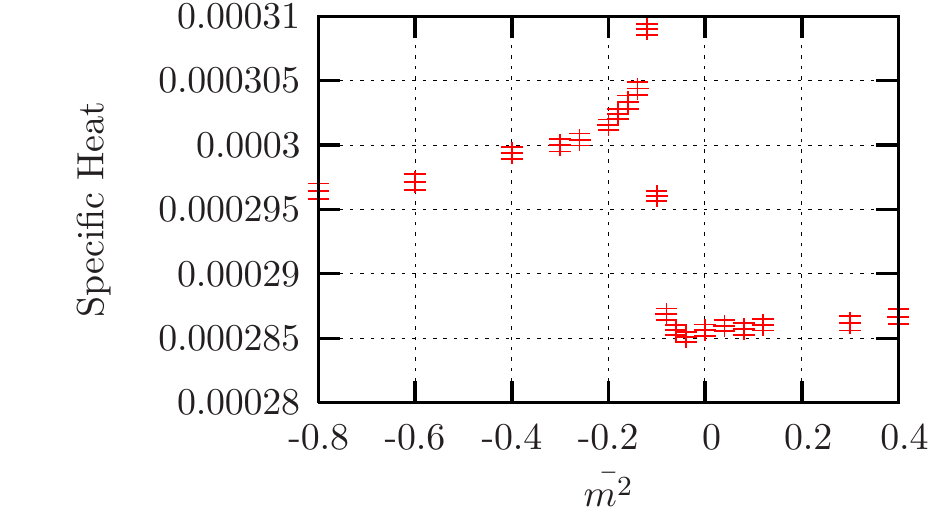}
 \end{center}
\begin{figure}[h]
 \vspace{-0.1in}
 \caption{Specific heat per volume at $\bar{\lambda}=0.17$, $\bar{R}=4$, 
           $N=12$. It follows from eq.~(\ref{int_volume_fuzzy}) that for 
           this values of the parameters the volume is the constant 
           $4 \pi R^2 \times  N_{\trm}= 192 \pi$.}
\label{sus-heat-N16-l2-R4}
\end{figure}
The specific heat provides an alternative criterion to the susceptibilities  to determine  where the 
phase transition occurs. It provides information about the order of the phase transition.
We prefer at this point to follow the specific heat criterion because it  is a more universal quantity 
that does not distinguish the dominant mode. Since we expect that as we increase 
$\bar{\lambda}$ the dominant modes are higher than the zero and first mode, we cannot ensure that
in that region the susceptibilities related to such modes $\chi_0$ and $\chi_1$ give a reliable prediction 
to the critical point. 
If we follow the criteria of the susceptibilities we have to take into account which mode is the dominant one. 
Both criteria are theoretically supposed to detect the same phase transition at the same parameters. 
For $\bar{\lambda}<\bar{\lambda}_T $ where the dominant mode in the ordered phase is the zero mode,
this is confirmed comparing {\bf figure \ref{sus-mix-N12-l2-R4}} to {\bf figure \ref{sus-heat-N16-l2-R4}} 
since the susceptibility of the dominant mode, $\chi_0$, and the specific heat peaks around the same value in $\bar{m^2}$.
We can also observe in  {\bf figure \ref{sus-heat-N16-l2-R4}} that there is a smaller response in the 
susceptibility of the first mode, $\chi_1$.
For $\bar{\lambda}<\bar{\lambda}_T$ we will see that this situation is different since the susceptibilities of the 
non-dominant modes do not peak at the phase transition.

We call the value of $\bar{m^2}$ where the specific heat peaks  $\bar{m^2_c}$, and for 
{\bf figure \ref{sus-heat-N16-l2-R4}} $\bar{m^2_c}=-0.12\pm 0.02$.\footnote{For practical purposes we had to estimate the error by referring to the spacing of the $\bar{m^2}$ values that we simulated.}
In the case that for $\bar{m^2} \ge \bar{m^2_c}$ the kinetic term is not leading, 
it appears as a {\em small} shift between both peaks. This happens when  $\bar{R}$ is 
big enough to have $\frac{N}{\bar{R}}$ {\em small}, $\bar{\lambda}_T > \bar{\lambda}$. 
Another observation is that in {\bf figure \ref{sus-mix-N12-l2-R4}} the error bars are 
smaller than in the case of the specific heat from  {\bf figure \ref{sus-heat-N16-l2-R4}}. 
The reason is that in general more statistics is necessary for the specific heat than for 
the susceptibilities.
This phase transition is of second order as it is shown in {\bf figure \ref{sus-heat-N16-l2-R4}}

\subsection{Thermalisation with respect to the observables}
In this section we want to present the typical behaviour of the observables for $\bar{m^2}>\bar{m^2_c}$ and  $\bar{m^2}<\bar{m^2_c}$.

\paragraph{$\bar{m^2} > \bar{m^2_c}$}
\ \\ \
First we discuss some aspects of the  thermalisation procedure.
We define the thermalisation time as the number of Monte Carlo steps necessary for an 
observable to stabilise around one value independently of the starting conditions.
If $\bar{m^2}> \bar{m^2_c}$ we are in the disordered phase that is characterised by 
the property that the coefficients $\ccv_{lm}$ in eq.~(\ref{averaged_coeff}) 
are in average near to zero. We assume that the thermalisation of the coefficients $\ccv_{lm}$ is similar, and
we check if the coefficient  $\ccv_{00}$ thermalises.
{\bf Figure \ref{ther-hot-m0-l2-N12-R4}} shows the thermalisation of the action and the coefficient\footnote{During the run the values stored were $\ccv_{00}$ and from them we can trivially calculate $\varphi_{0}$. We prefer to present the histories and histogram of $\ccv_{00}$ rather than $\varphi_{0}$ in order to check if the samples are symmetric under $\ccv_{00}\longrightarrow -\ccv_{00} $.} $\ccv_{00}$ for the point $\bar{m^2}=0$ in the {\bf figures \ref{ordermix-N12-l2-R4}- \ref{sus-heat-N16-l2-R4}}.

\centerline{
 \includegraphics[width=2.9in]{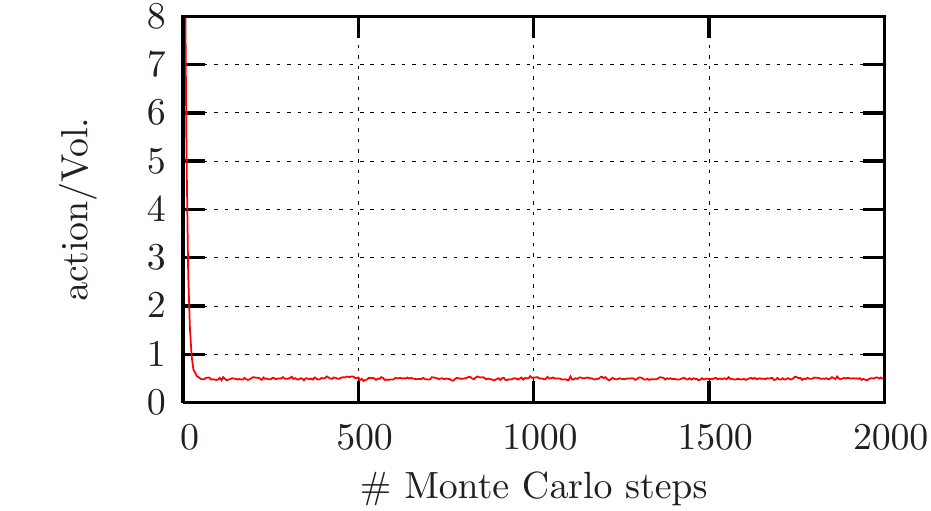}
 \includegraphics[width=2.9in]{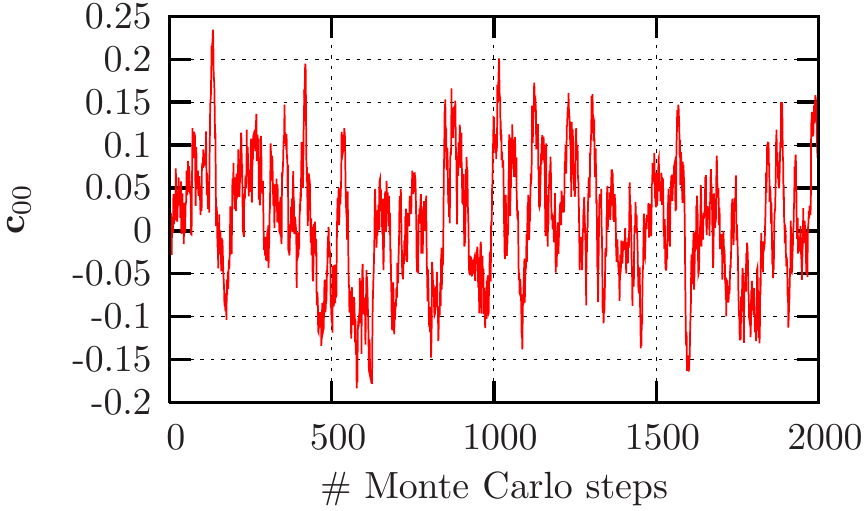}}
\begin{figure}[h]
 \vspace{-0.2in}
 \caption{ Example of thermalisation of the action and the coefficient $\ccv_{00}$
            at $\bar{m^2}=0$,
 $\bar{\lambda}=0.17$, $\bar{R}=4$, $N=12$.  }
\label{ther-hot-m0-l2-N12-R4}
\end{figure}

We chose a {\em hot start }\footnote{The starting configuration is a vector of hermitian matrices 
filled in with random numbers. For more details see appendix \ref{small-description-MC}.} 
for {\bf figure  \ref{ther-hot-m0-l2-N12-R4} }. 
We simulate the same parameters with a {\em cold start }\footnote{The starting configuration is a vector of matrices proportional to the unit.} and we obtained results in agreement within the statistical errors.
The thermalisation time for $\ccv_{00}$ in {\bf figure \ref{ther-hot-m0-l2-N12-R4}} is estimated to $1500$ Monte Carlo steps.

After the thermalisation we begin the measurement procedure. {\bf Figure \ref{ther-hot-c00-m0-l2-N12-R4}} shows the histograms\footnote{The area is normalised to $1$. The number of bins is $500$.} for the observables and parameters used in the simulations of {\bf figure \ref{ther-hot-m0-l2-N12-R4}}:

\centerline{
 \includegraphics[width=2.5in]{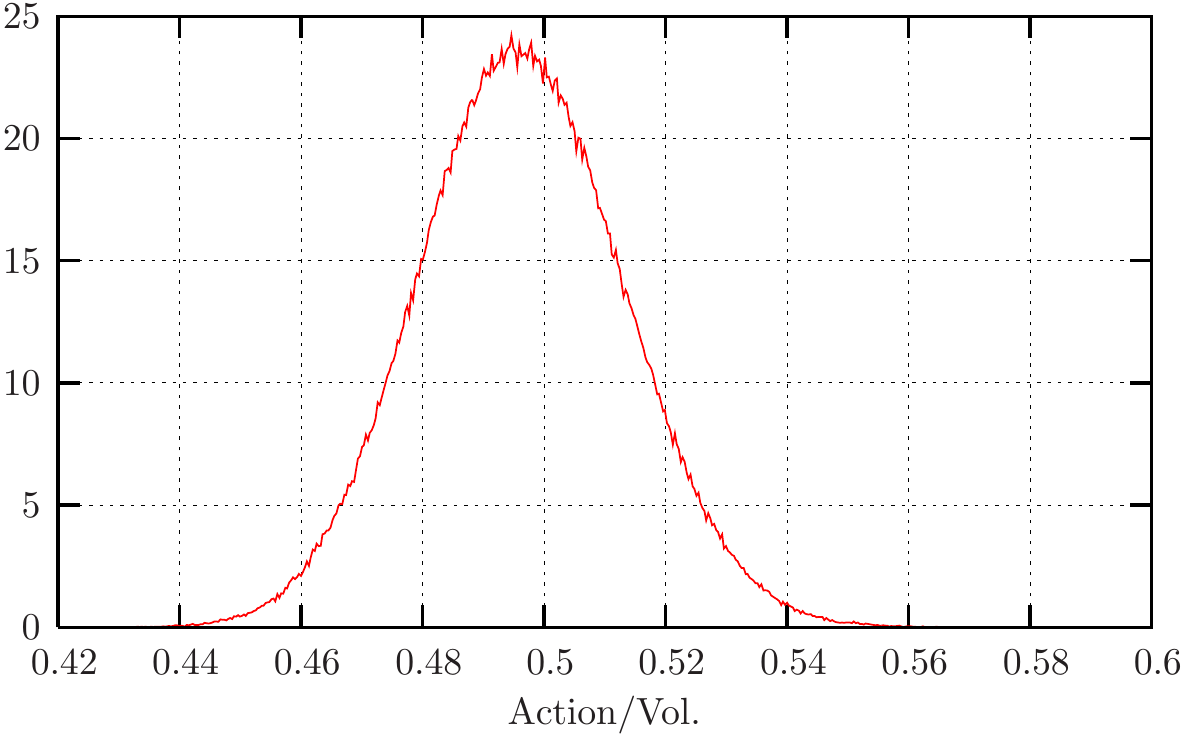}
 \hspace{0.3cm}
 \includegraphics[width=2.5in]{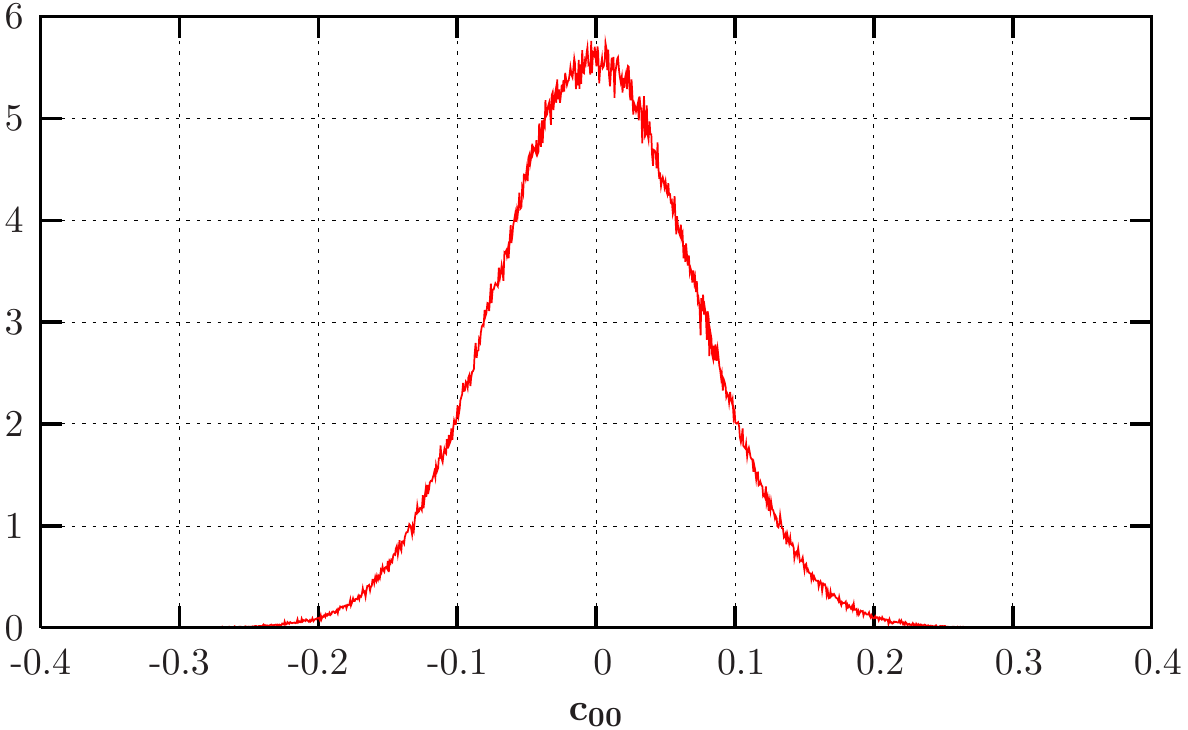}}
\begin{figure}[h]
 \vspace{-0.33in}
 \caption{ Histogram of the of the action and the coefficient $\ccv_{00}$
 at $\bar{m^2}=0$,
 $\bar{\lambda}=0.17$, $\bar{R}=4$, $N=12$.}
\label{ther-hot-c00-m0-l2-N12-R4}
\end{figure}

\vspace{-0.3in}
\paragraph{$\bar{m^2}<\bar{m^2_c}$}
\ \\ \

If $\bar{m^2}>\bar{m^2_c}$ we are in the ordered regime, for $\bar{\lambda}<\bar{\lambda}_T$ we have  uniform 
ordering characterised by the property that in the  expansion (\ref{averaged_field}) all coefficients $\ccv_{lm}$ for $l>0$ average zero. Again it is sufficient to check if the coefficient  $\ccv_{00}$ thermalise.
{\bf Figure \ref{ther-hot-m0p2-l2-N12R4}} shows the thermalisation of the action and the parameter $\ccv_{00}$ for the point $\bar{m^2}=-0.2$ in the {\bf figures \ref{ordermix-N12-l2-R4}- \ref{sus-heat-N16-l2-R4}}.

\centerline{
 \includegraphics[width=2.8in]{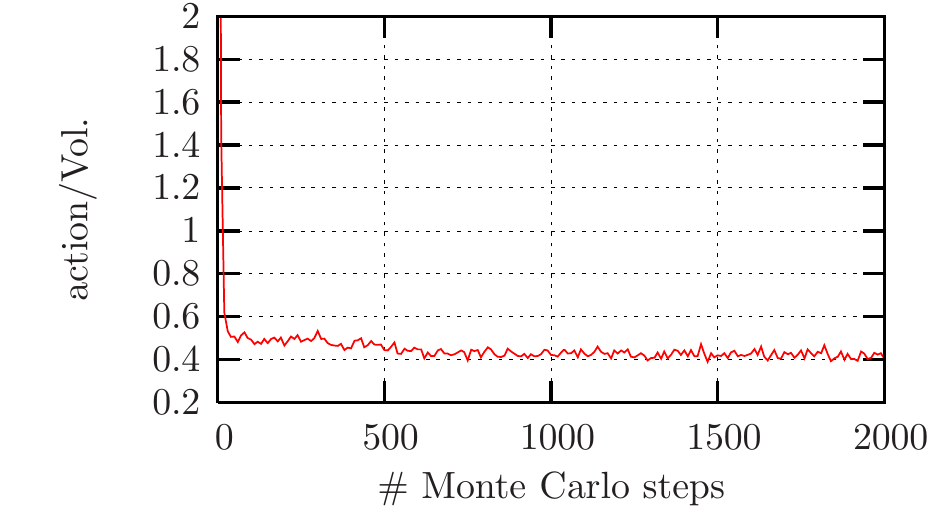}
 \includegraphics[width=2.8in]{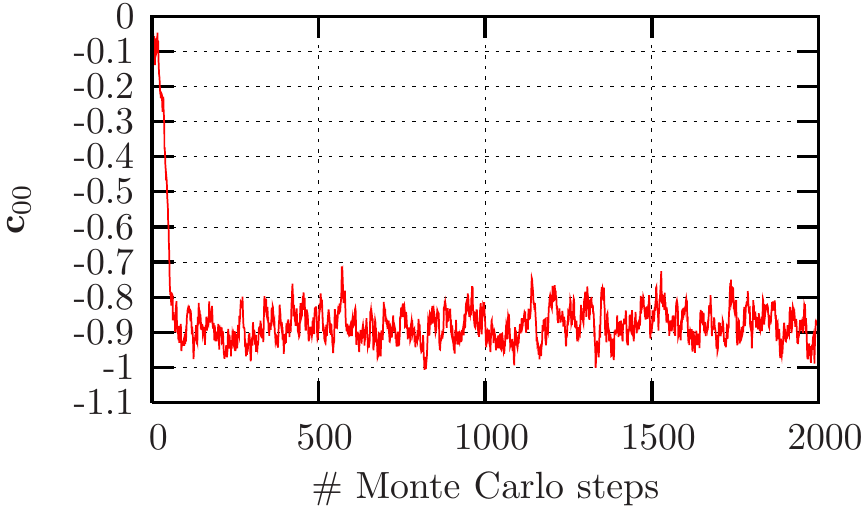}}
\begin{figure}[h]
 \vspace{-0.33in}
 \caption{ Example of the thermalisation of the action and the coefficient $\ccv_{00}$
  at $\bar{m^2}=-0.2$,
 $\bar{\lambda}=0.17$, $\bar{R}=4$, $N=12$.}
\label{ther-hot-m0p2-l2-N12R4}
\end{figure}
We chose a {\em hot start } for {\bf figure \ref{ther-hot-m0p2-l2-N12R4}}. After $1500$ Monte Carlo steps the value of $\ccv_{00}$ oscillates around $0.9$ and the energy per unit of volume fluctuates around $0.43$.
We also simulated at the same parameters with  a {\em cold start }. We obtained the same results within the statistical error but the thermalisation time decreases by more than  $50 \% $ as we 
can observe in {\bf figure \ref{ther-cold-m0p2-l2-N12R4}}.

\centerline{
 \includegraphics[width=2.8in]{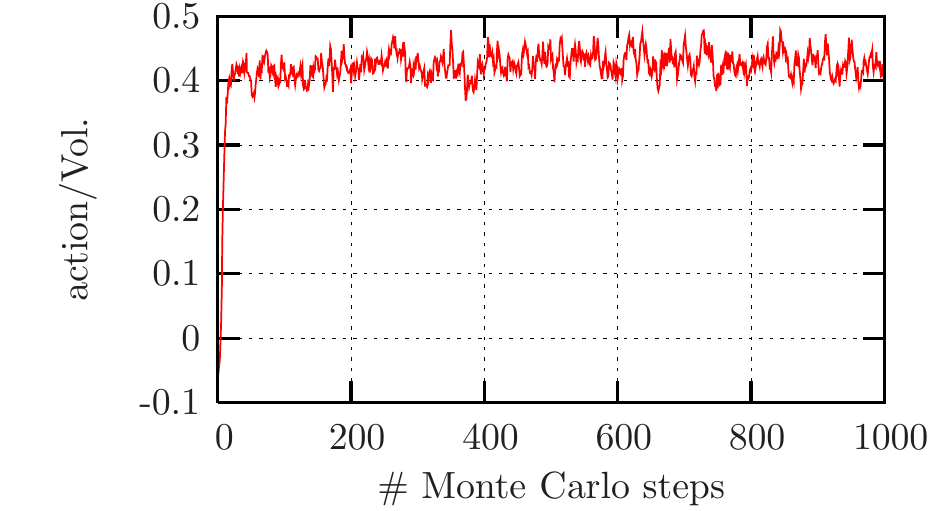}
 \includegraphics[width=2.8in]{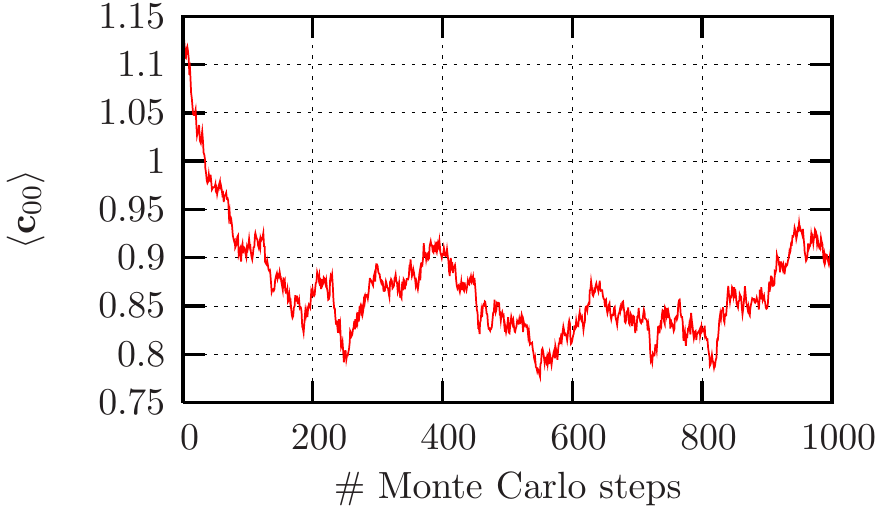}}
\begin{figure}[h]
 \vspace{-0.30in}
 \caption{ Example of thermalisation with {\em cold} starting conditions of the action and the coefficient $\ccv_{00}$
  at $\bar{m^2}=-0.2$,
 $\bar{\lambda}=0.17$, $\bar{R}=4$, $N=12$.}
\label{ther-cold-m0p2-l2-N12R4}
\end{figure}

Finally, after the thermalisation procedure we measure the expectation value of the observables. We present the histograms for the same parameters as in {\bf figure \ref{ther-hot-m0p2-l2-N12R4}}:

\centerline{
 \includegraphics[width=2.7in]{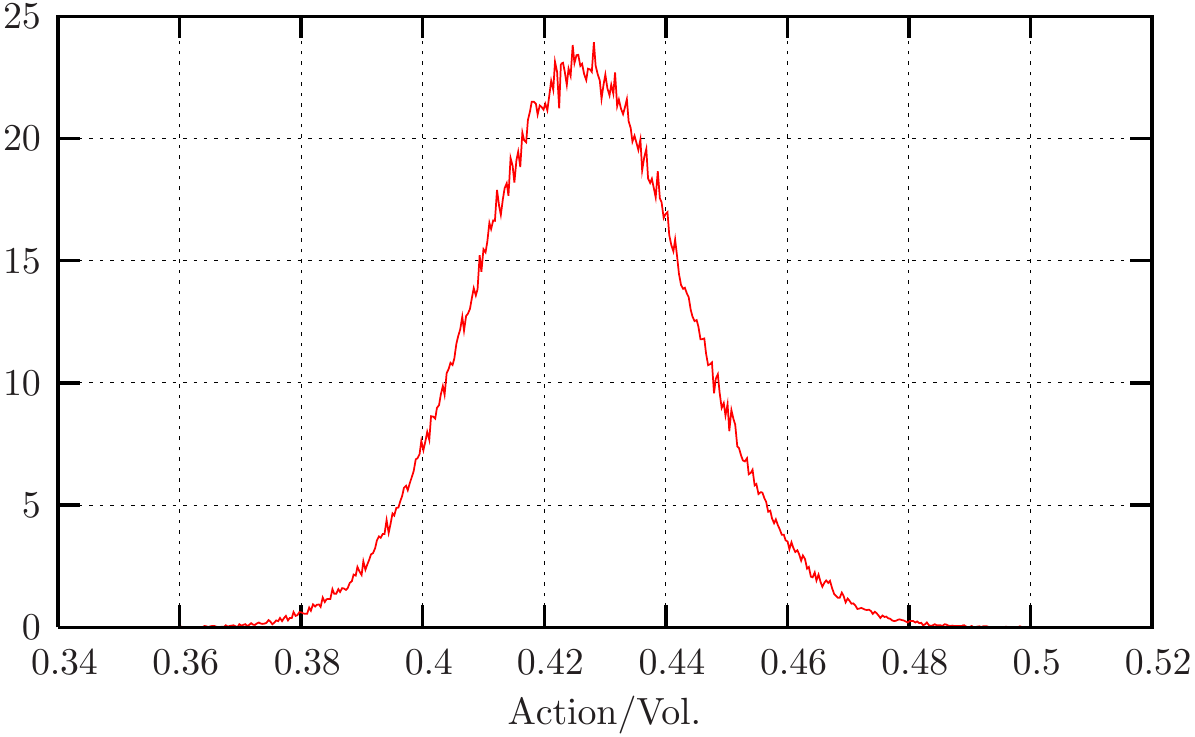}
 \hspace{.3cm}
 \includegraphics[width=2.7in]{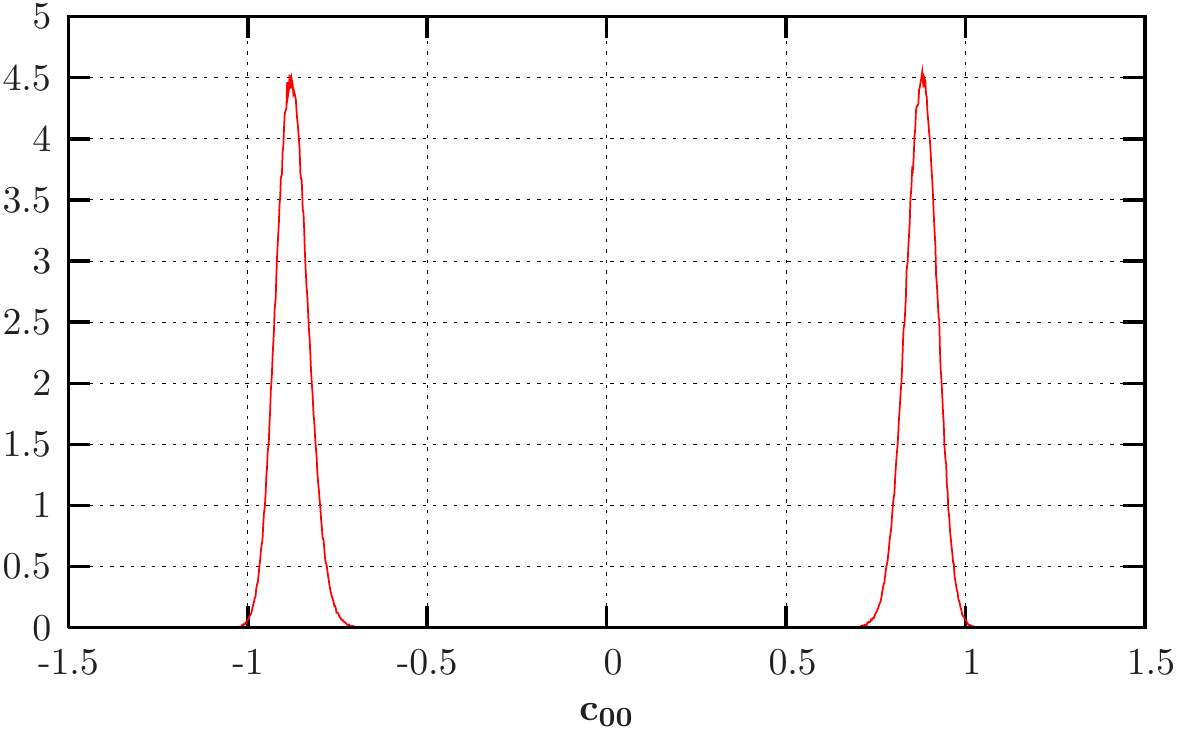}}
\begin{figure}[h]
 \vspace{-0.3in}
 \caption{ Histograms of the action and the coefficient $\ccv_{00}$ at $\bar{m^2}=-0.2$,
 $\bar{\lambda}=0.17$, $\bar{R}=4$, $N=12$.  }
\label{ther-hot-m0-l2-N12R4}
\end{figure}
The peaks in the probability distribution of $\ccv_{00}$ are approximately located at 
$\sqrt{\frac{4\pi \vert \bar{m}^2\vert}{N \bar{\lambda}}}$.
In {\bf figure \ref{ther-hot-m0-l2-N12R4}} the peaks are approximately located at $\ccv_{00}=\pm0.95$ 
and $\sqrt{\frac{4\pi \times0.2}{12 \times0.17}}=1.12$.
As we move forward for a more negative $\bar{m^2}$ this prediction is more accurate. This is shown in
{\bf figure \ref{ther-hot-m2-l2-N12R4}}. 
\begin{center}
 \includegraphics[width=3.6in]{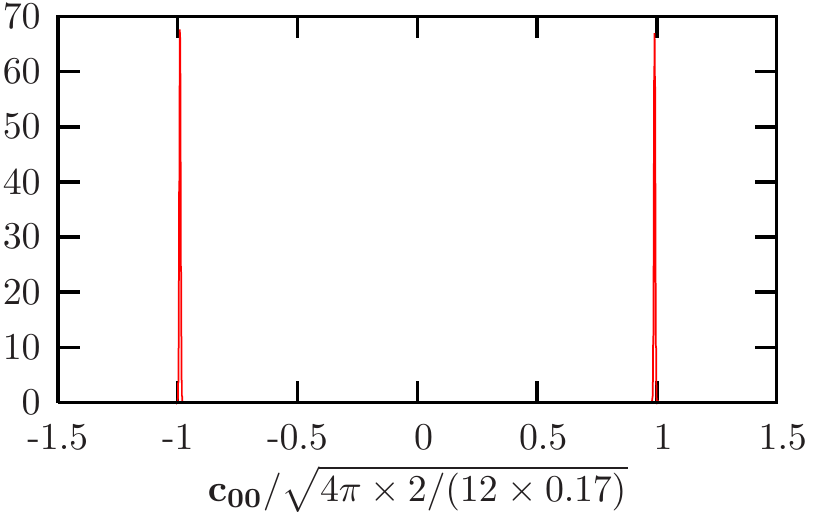}
\end{center}
\begin{figure}[h]
 \vspace{-0.4in}
 \caption{ Histogram of coefficient $\ccv_{00}$ at $\bar{m^2}=-2$,
 $\bar{\lambda}=0.17$, $\bar{R}=4$, $N=12$. 
 We re-scale the $x$-axis by the factor $\sqrt{\frac{4\pi \vert\bar{m}^2 \vert}{N \bar{\lambda}}}$ 
 which in this case takes the value $3.54$.}
\label{ther-hot-m2-l2-N12R4}
\end{figure}

\section{Behaviour of the system for $\bar{\lambda} > \bar{\lambda}_T $}
For this region of the phase diagram we observe two phases: for $\bar{m^2}> \bar{m^2_c}$  we have the disordered phase  characterise by $\langle \varphi_{all}^2  \rangle \approx 0$, for $\bar{m^2}<\bar{m^2_c}$  we have the ordered phase characterised by $\langle \varphi_{all}^2  \rangle>0$. 
For $\bar{m^2} \ll \bar{m^2_c}$ there are thermalisation problems,
we will discuss these difficulties in section \ref{thermalization-problems}.

In {\bf Figure \ref{ordermix-N12-l15-R8} } we present the partial contributions from the zero and first mode in $ \la \varphi_{all}^2  \ra$.
\begin{center} 
   \includegraphics[width=3.6in]{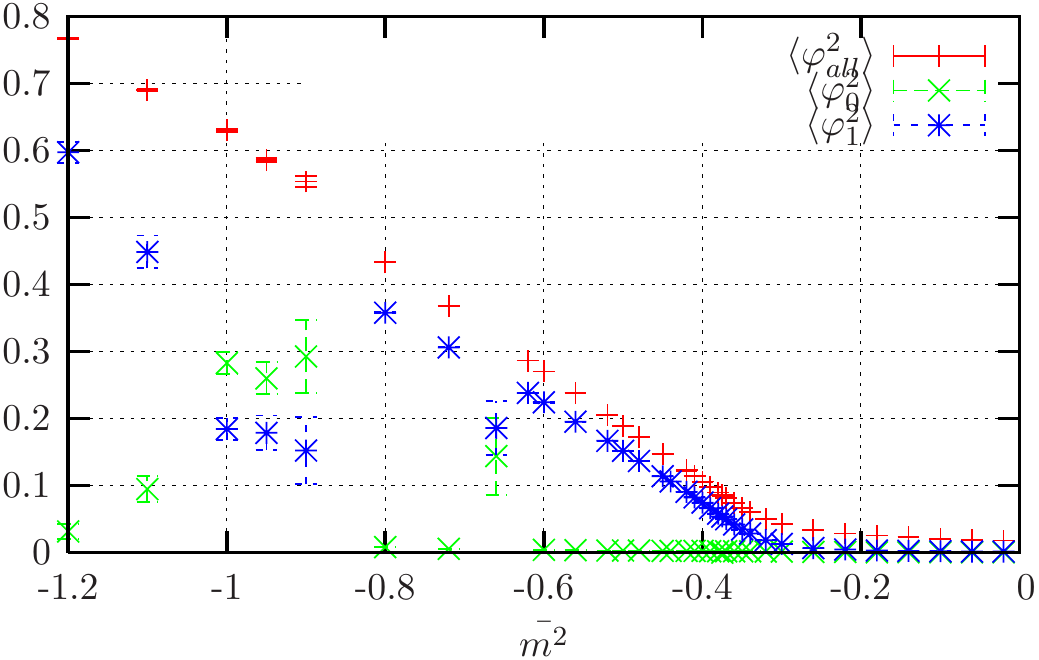}
\end{center}
\begin{figure}[h]
 \vspace{-0.4in}
 \caption{  $\la {\varphi_{all}^2} \ra$, $\la \varphi_0^2 \ra$ and $\la \varphi_1^2 \ra$
   vs. $\bar{m^2}$ at $\bar{\lambda}=1.25$, 
    $\bar{R}=8$, $N=12$.}
\label{ordermix-N12-l15-R8}
\end{figure}
We observe in {\bf figure \ref{ordermix-N12-l15-R8}}  $\la \varphi_{all}^2 \ra \approx 0$ for $\bar{m^2} >-0.3$ while  for $\bar{m^2} < -0.3$ we have $\la \varphi_{all}^2 \ra > 0$ and for $-0.6> \bar{m^2} > -0.1$ we observe $\la{\varphi_{all}}^2 \ra \sim \la \varphi_1^2 \ra$. 

\begin{center} 
   \includegraphics[width=3.6in]{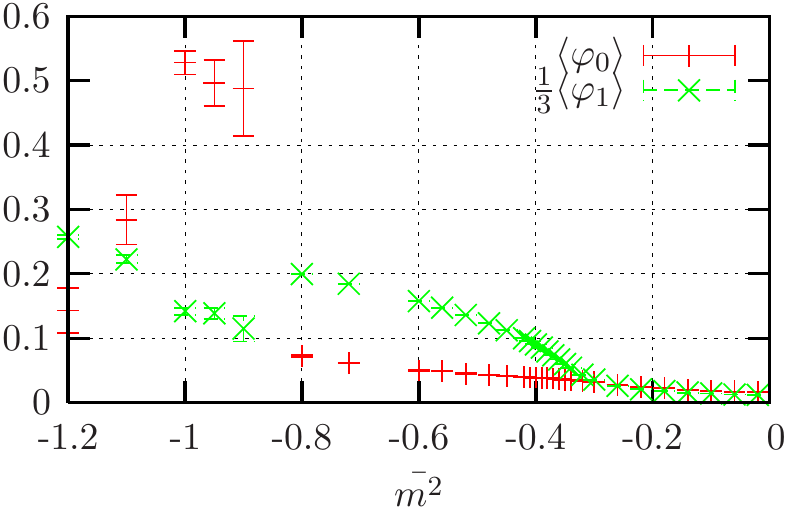}
\end{center}
\begin{figure}[h]
 \vspace{-0.35in}
 \caption{$\la \varphi_0 \ra$ and $\la \varphi_1 \ra$ vs. $\bar{m^2}$ at $\bar{\lambda}=1.25$, 
    $\bar{R}=8$, $N=12$.}
\label{order-N12-l15-R8}
\end{figure}

For those values of $\bar{m^2}$ where $\bar{m^2}<-0.6$ we can observe a kind of irregularity in the quantities $\la \varphi_0^2 \ra$ and $\la \varphi_1^2 \ra$ in {\bf figure \ref{ordermix-N12-l15-R8}}  and in $\varphi_0$ and $ \varphi_1$ in {\bf figure \ref{order-N12-l15-R8} } since they do not grow monotonously. We will come back to this point at section \ref{thermalization-problems}. For  the moment we focus on the region where the observables behave smoothly, i.e.\ $\bar{m^2}>-0.6$.

\begin{figure}[h]
\begin{center}
  \includegraphics[width=4.6in]{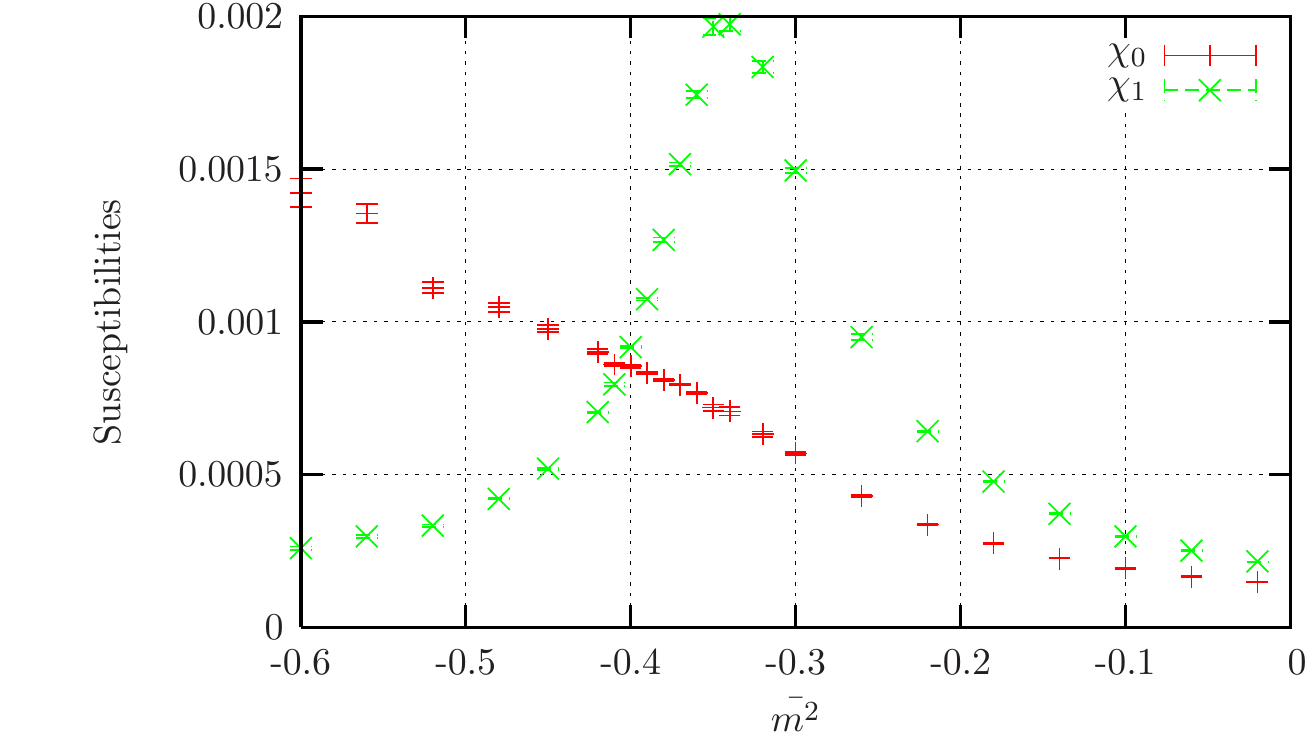}
 \end{center}
 \vspace{-0.3in}
 \caption{ Susceptibilities  $\chi_0$  and $\chi_1$  at $\bar{\lambda}=1.25$, $\bar{R}=8$, $N=12$.}
\label{sus-mix-N12-l15-R8}
\end{figure}

In {\bf figure  \ref{sus-mix-N12-l15-R8}} we can observe that $\chi_1$ indicates a phase transition for $\bar{m^2} \simeq -0.35$ while $\chi_0$ cannot detect it since  $\chi_0$ keeps growing as $m^2$ decreases. We conjecture that $\chi_0$ should peak for some value of $\bar{m^2}<-0.6$.

 \begin{center}
  \includegraphics[width=4.6in]{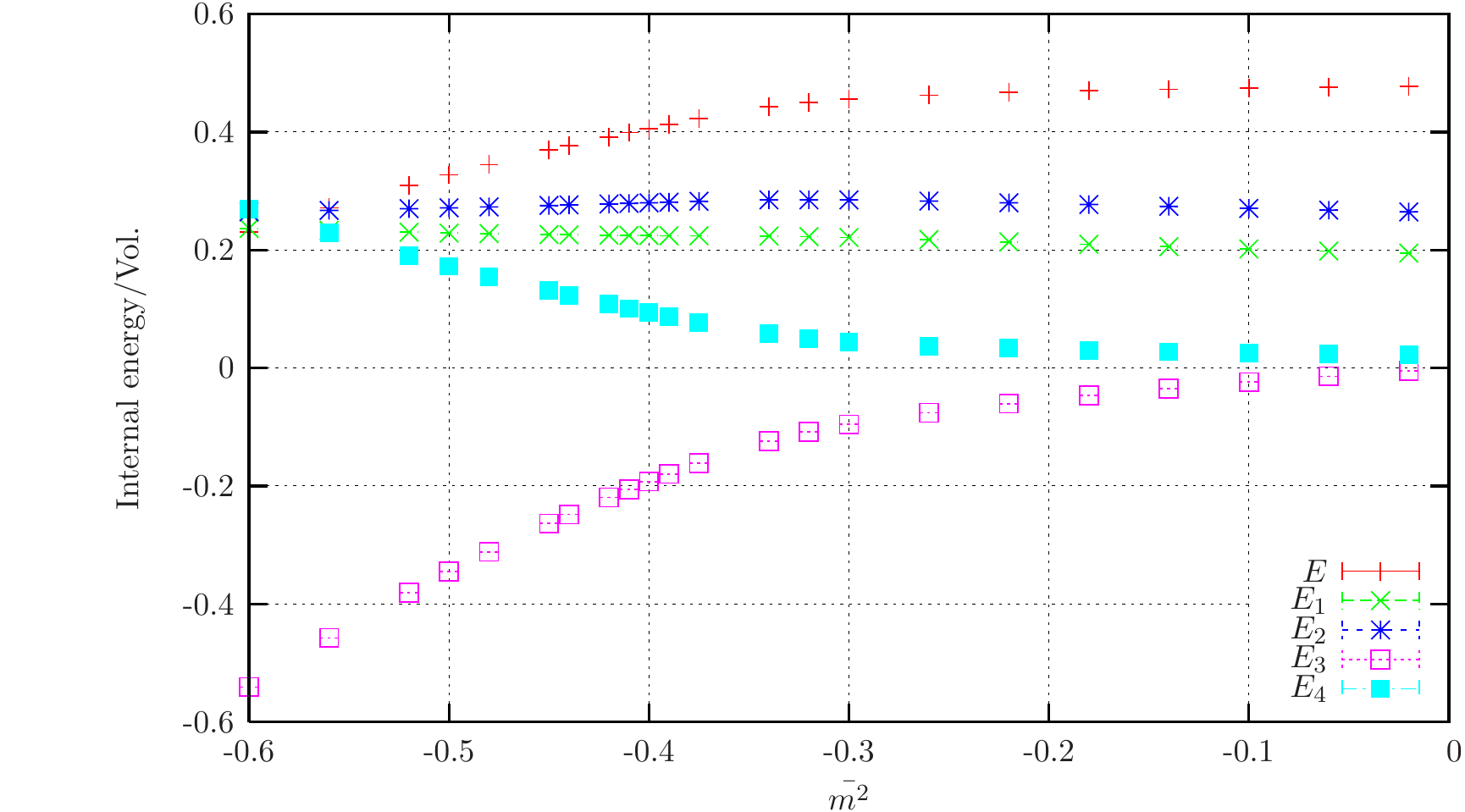}
 \end{center}
\begin{figure}[h]
 \vspace{-0.3in}
 \caption{ Internal energy $E$ of eq.~(\ref{energy}) and its partial contributions eqs.~(\ref{energy1})-(\ref{energy4}) at $\bar{\lambda}=1.25$, $\bar{R}=8$, $N=12$.}
\label{contrib-N12-l15-R8}
\end{figure}

The specific heat in {\bf figure \ref{sus-heat-N12-l15-R8}} indicates a phase transition at $\bar{m^2}=-0.37\pm 0.02$.
 \begin{center}
  \includegraphics[width=4.6in]{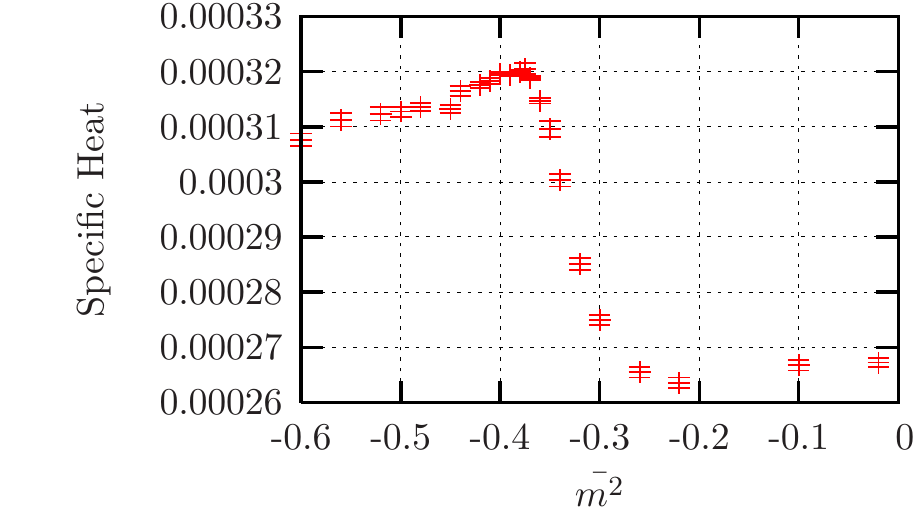}
 \end{center}
\begin{figure}[h]
 \vspace{-0.3in}
 \caption{ Specific heat at $\bar{\lambda}=1.25$, $\bar{R}=8$, $N=12$.}
\label{sus-heat-N12-l15-R8}
\end{figure}
As we can observe comparing  {\bf{ figure \ref{sus-mix-N12-l15-R8}}} to {\bf{ figure \ref{sus-heat-N12-l15-R8}}}, there is a small difference in the critical value of $\bar{m}^2$ predicted by the susceptibility of the zero mode $\chi_0$ and the one given by the specific heat. We conjecture that this is due to a finite volume effect, nevertheless both criteria are qualitatively the same. For more details see appendix \ref{appendix-aside-results}.

\vspace{0.5in}
\subsection{Thermalisation problems}
\label{thermalization-problems}
In this section we want to sketch the thermalisation problems.

The program was designed to perform an arbitrary number of independent\footnote{If we choose a {\em hot start}, they have different starting configurations.}
simulations in every run, $n_{sim}$. In the case of  {\bf figure \ref{ther-hot-m0p66-l15-N12-R8}} we performed ten independent simulations,
the first three of them with a {\em cold start} while for the 
last seven simulations we chose a {\em hot start}.

\begin{center}
 \includegraphics[width=3.3in]{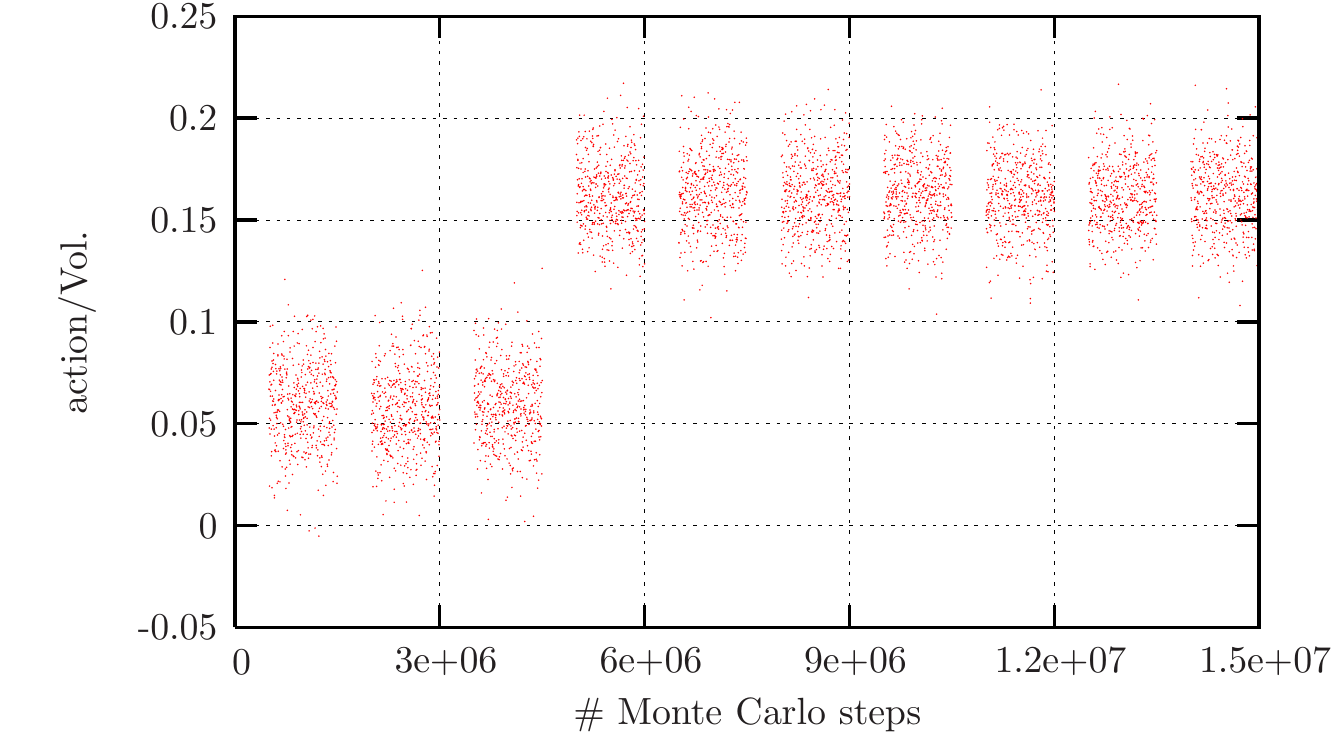}
 \includegraphics[width=3.3in]{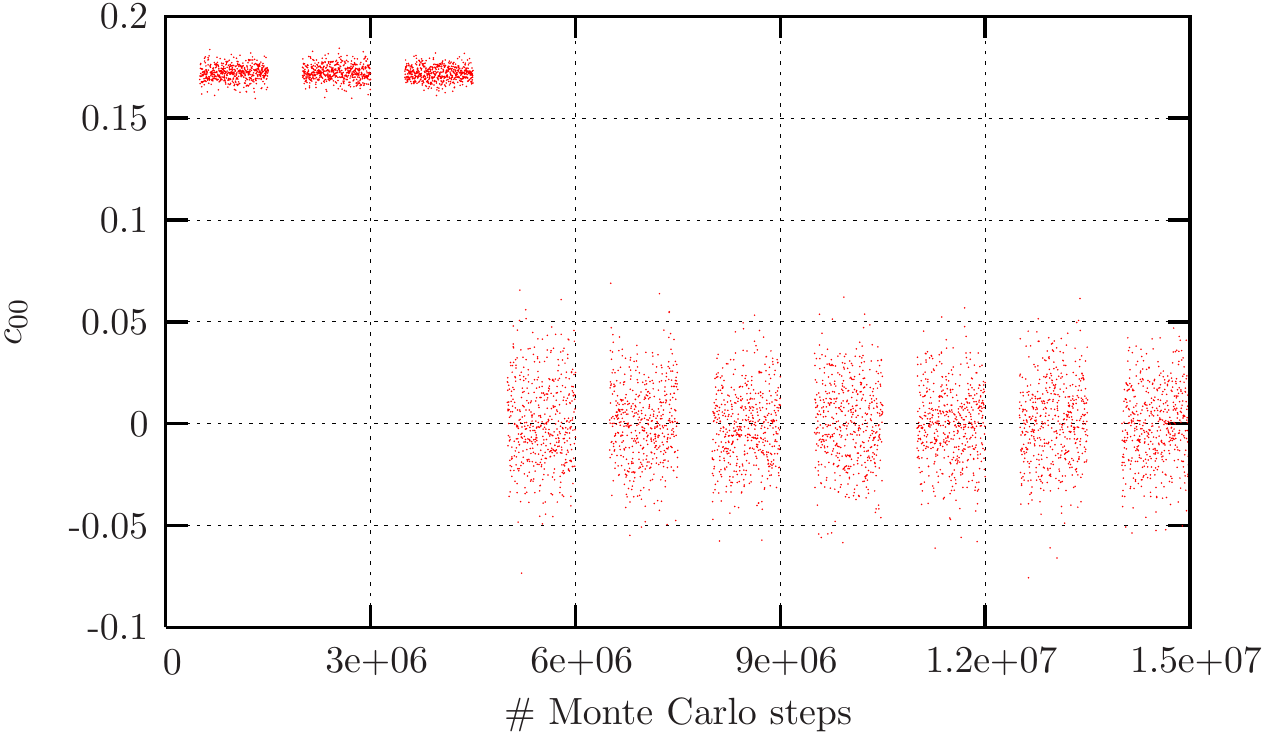}
\end{center}
\begin{figure}[h]
 \vspace{-0.10in}
 \caption{ History of the action and the coefficient $c_{00}$ for  different starting conditions 
  at $\bar{m^2}=-0.66$,
 $\bar{\lambda}=1.25$, $\bar{R}=8$, $N=12$.  }
\label{ther-hot-m0p66-l15-N12-R8}
\end{figure}
We can observe in {\bf figure \ref{ther-hot-m0p66-l15-N12-R8}} that the expectation value of the energy and 
the coefficient $c_{00}$  depend on the starting conditions. 
The interpretation of this phenomenon is that the effective potential
of the system has several local minima with barriers large enough to suppress tunnelling between them. 
For different starting conditions the system gets trapped in one of those minima. 
As a consequence we have  {\em thermalisation  problems} (or  practical ergocidity problems in the algorithm).  
They should disappear at an infinitely large Monte Carlo time, $T_{MC}$. 
In  {\bf figure \ref{ther-hot-m0p66-l15-N12-R8}} we have a new simulation  every $1,500,000$ Monte Carlo 
steps from which the first $500,000$ steps where taken as thermalisation time.
 
The large error bars in the observables at $\bar{m^2}=-0.66$ in {\bf figures \ref{ordermix-N12-l15-R8}} and 
{\bf \ref{order-N12-l15-R8}} can be explained because the different simulations give different results. 
For {\em cold starts} the action of the system oscillates around $0.05$ -- this is the value of the energy in 
 the absolute minima --
and the trace $\ccv_{00}$ of the sampled 
configurations fluctuate around $0.17$; for {\em hot starts} the energy of the system oscillates around $0.18$ 
and the trace $\ccv_{00}$ of the sampled configurations fluctuate around zero.


{\bf Figure \ref{mix-hot-m0p66-l15-N12-R8}} shows the histograms corresponding to {\bf figure \ref{ther-hot-m0p66-l15-N12-R8}}. 

\centerline{
 \includegraphics[width=2.8in]{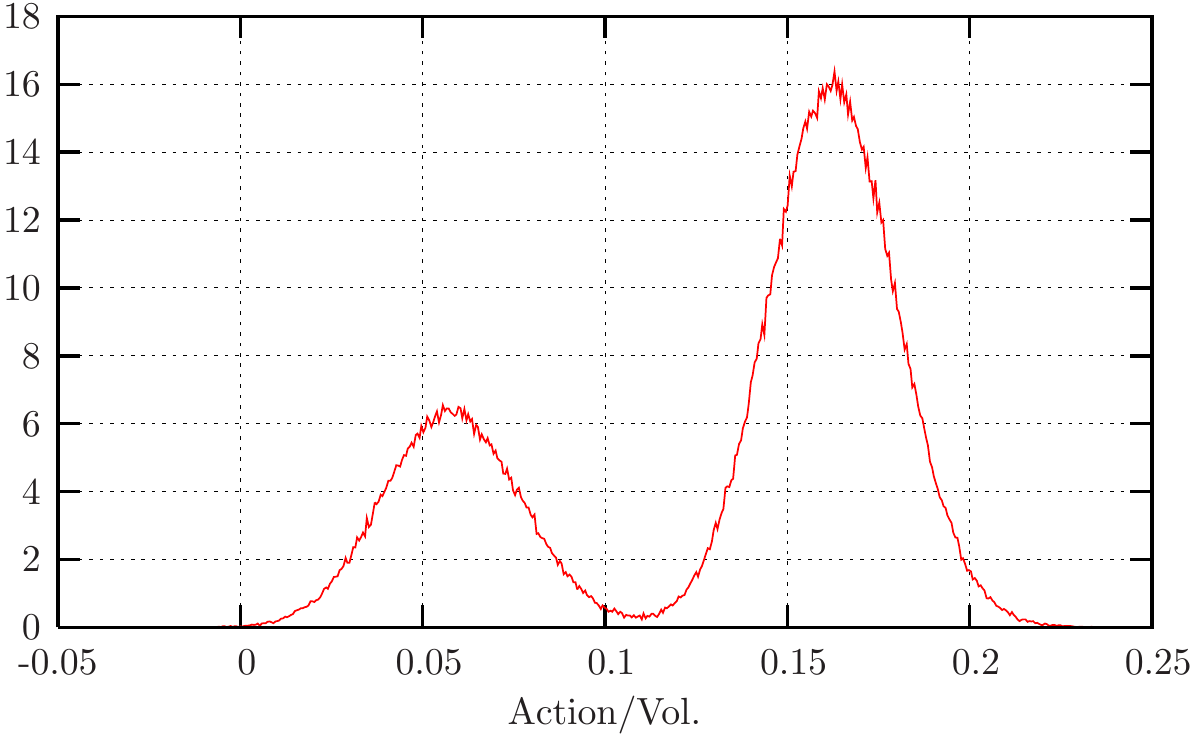}
 \includegraphics[width=2.8in]{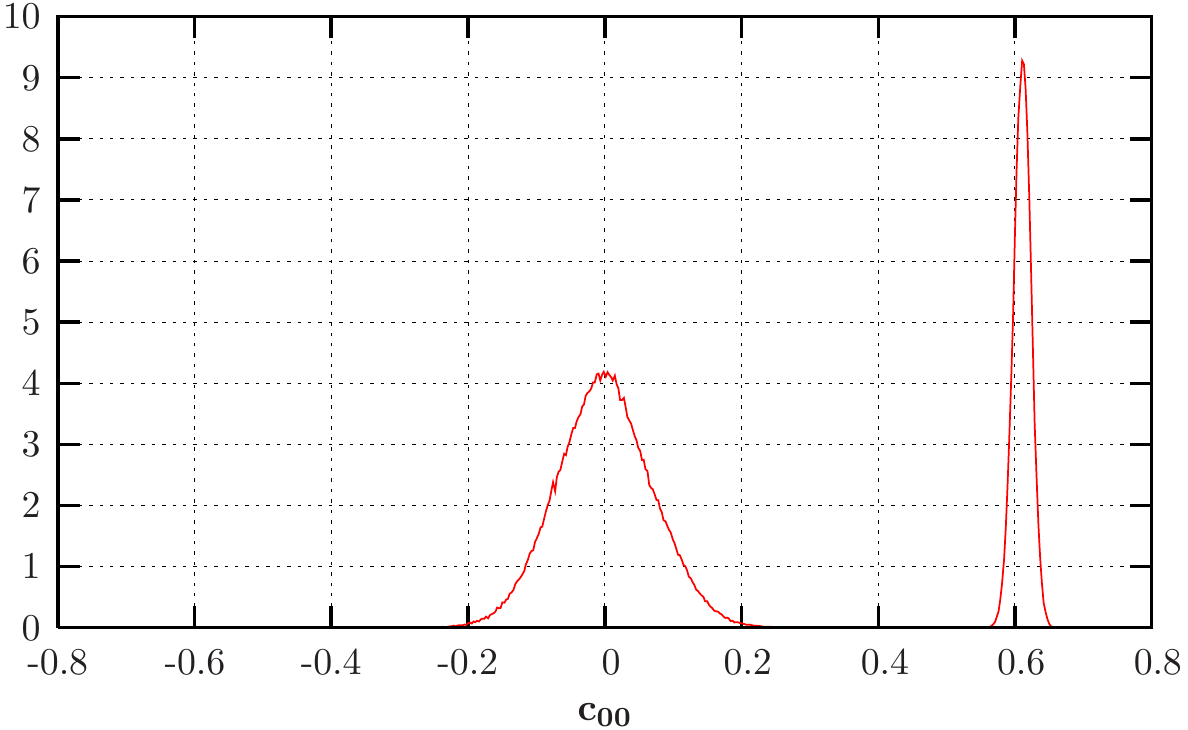}}
\begin{figure}[h]
 \vspace{-0.0in}
 \caption{ Histogram of the action and the coefficient $\ccv_{00}$
  at $\bar{m^2}=-0.66$,
 $\bar{\lambda}=1.25$, $\bar{R}=8$, $N=12$.}
\label{mix-hot-m0p66-l15-N12-R8}
\end{figure}

\section{Estimating the maximal number of minima}
It is possible to observe, from the Monte Carlo time evolution of the observable $\ccv_{00}$ in 
eq.~(\ref{averaged_coeff-00}) that there are no thermalisation problems and that its probability 
distribution has several peaks.

\begin{center}
  \includegraphics[width=3.8in]{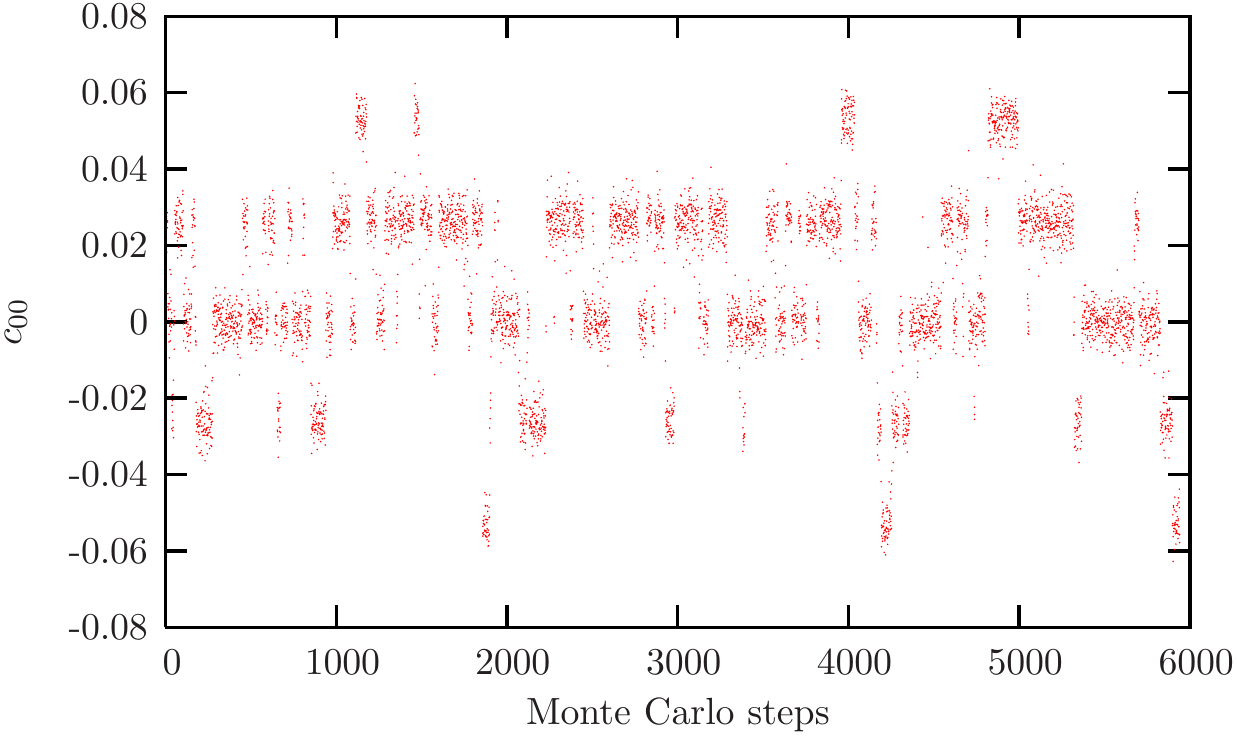}
 \end{center}
\begin{figure}[h]
 \vspace{-0.2in}
 \caption{ History of the coefficient $\ccv_{00}$ for $\bar{m}^2=0.3$ at $\bar{\lambda}=0.75$ $\bar{R}=16$, $N=12$.}
\label{history-order-N12-m-0p30-l9-R16}
\end{figure}

The history  corresponding of the action for the same parameters in
{\bf figure \ref{history-order-N12-m-0p30-l9-R16}} is the following:
\begin{center}
  \includegraphics[width=3.8in]{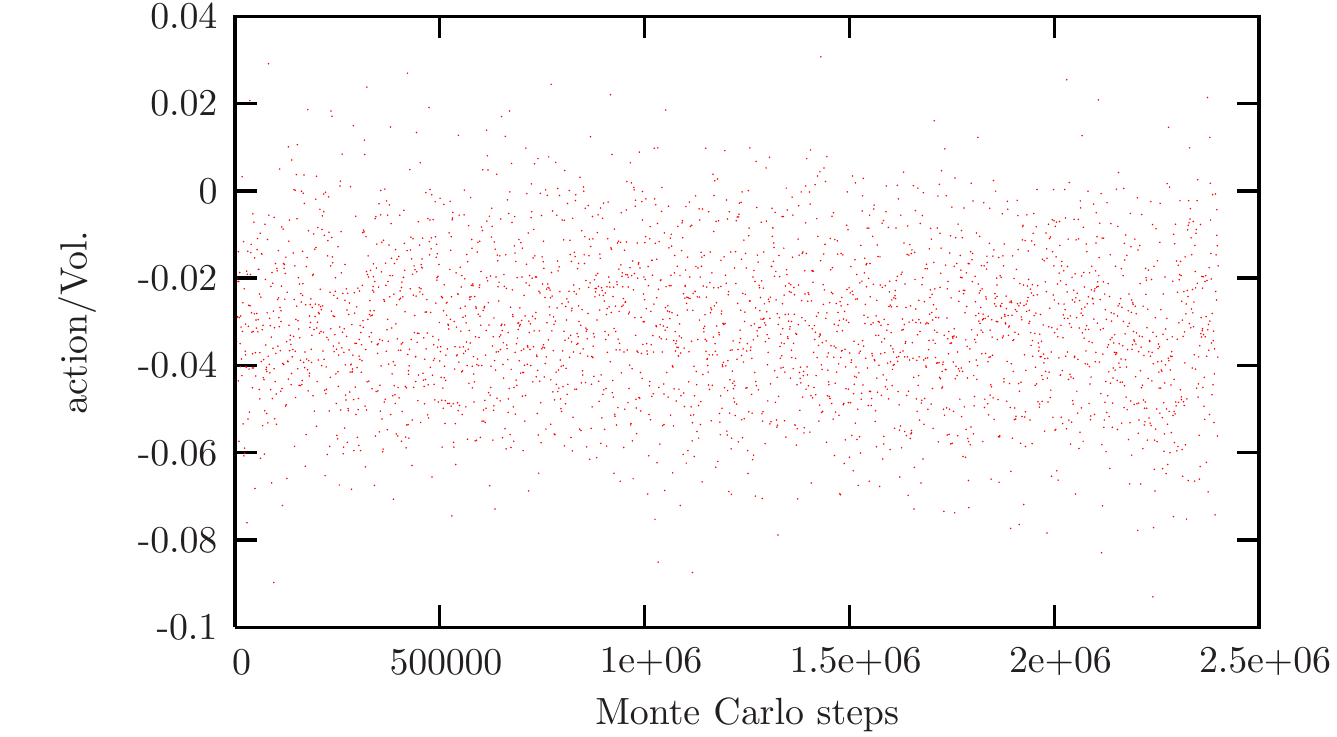}
 \end{center}
\begin{figure}[h]
 \vspace{-0.3in}
 \caption{ History of the action for $\bar{m}^2=-0.30$ at $\bar{\lambda}=0.75$ $\bar{R}=16$, $N=12$}
\label{history-action-N12-m-0p30-l9-R16}
\end{figure}
In this case the fluctuations were large enough to jump from one minimum to
another. Nevertheless the configurations sampled belong to different subspaces of the spaces of configurations characterised by the different values of $\ccv_{00}$  it is shown in {\bf figure \ref{history-order-N12-m-0p30-l9-R16}}.

The histogram corresponding to {\bf figure \ref{history-order-N12-m-0p30-l9-R16}} is presented in {\bf figure \ref{histogram-0p30}}.
\begin{center}
  \includegraphics[width=3.8in]{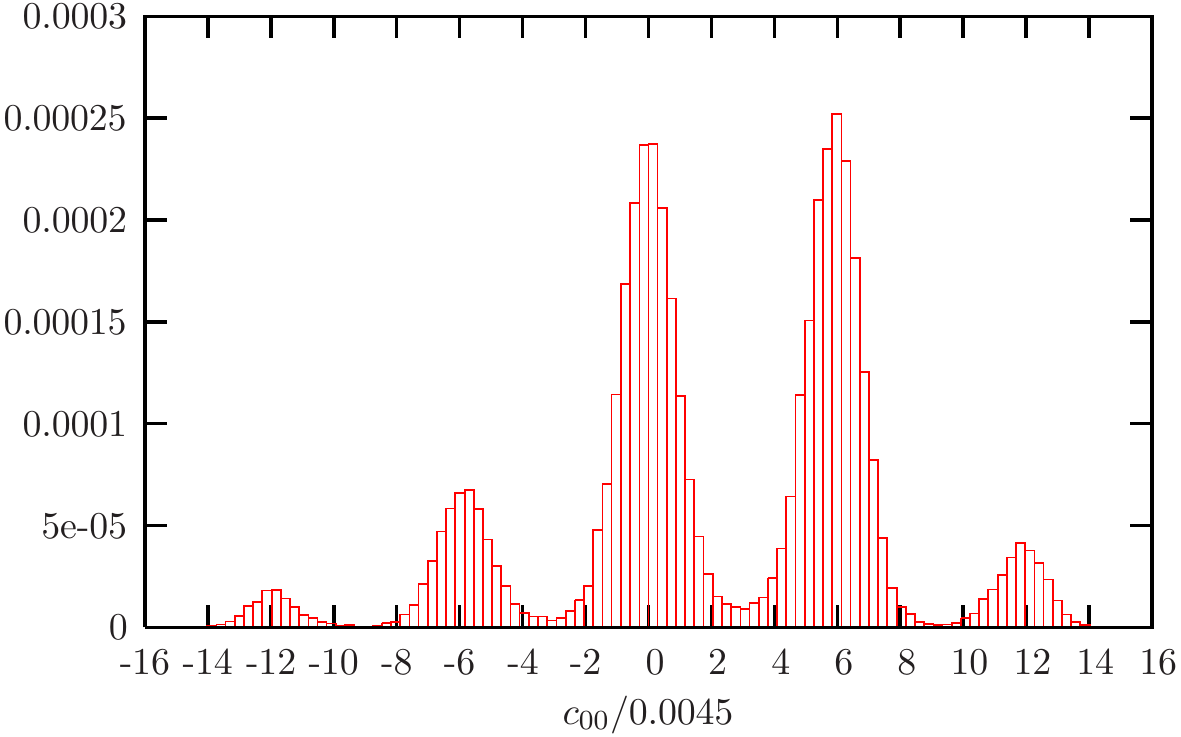}
 \end{center}
\begin{figure}[h]
 \vspace{-0.3in}
 \caption{ Histogram of the coefficient $\ccv_{00}$ for $\bar{m}^2=-0.30$ at $\bar{\lambda}=0.75$ $\bar{R}=16$, $N=12$.
          We divided the $x$-axis by the factor 
          ${ \frac{\sqrt{4 \pi \vert \bar{m}^2 \vert/ (N\bar{\lambda}) }}{N N_{\trm}}  } \simeq0.0045$ to emphasise that the peaks are around integer values.}
\label{histogram-0p30}
\end{figure}

The estimated value for the triple point for $N=12$, $\bar{R}=16$ is 
$\left(\bar{\lambda_T},\bar{m}^2_T \right)= \left(0.25,-0.062  \right)$,
the parameters simulated for {\bf figures \ref{history-order-N12-m-0p30-l9-R16}}-{\bf \ref{history-action-N12-m-0p30-l9-R16}} are in the region $\lambda> \lambda_T$ (i.e.\ in the ordered non-uniform phase). For this region the leading contribution comes from configurations characterised by a angular momentum $l>0$ as we can check from the expansion of the norm of the field eq.~(\ref{full_power_field}):
$\ \varphi_{all}^2=\varphi_0^2 +\varphi_1^2 + \cdots$ 
For the parameters simulated for {\bf figure \ref{history-order-N12-m-0p30-l9-R16}} the contributions to  $\la \varphi_{all}^2 \ra$ split as follows $\la \varphi_{all}^2\ra \simeq 0.30 \simeq 0.014+0.23+ \cdots$.

The several maxima in {\bf figure \ref{histogram-0p30}} reflect the existence of several minima in the effective action.
If $\lambda \gg \lambda_T$ it is expected that the potential provides the leading contribution to the action, turning into a  pure potential model. Then, if  $\lambda$ is large enough, for this region of parameters the minima in the effective action are given by the minima in the potential.

\section{The equilibrium configurations}
The maximal number of local minima for our model can be obtained from  the maximal number of local minima 
for the two-dimensional model.
In Ref.~\cite{xavier} it was conjectured  that the minima in the 2-dimensional {\em pure potential model} are given by the disjoint orbits:
\be
  O_n=\{ \frac{-m^2}{\lambda}U^{\dagger}\left(\mathbf{1}_n \oplus \mathbf{1}_{2s+1-n} U \vert
           U \in U(2s+1)/[U(n) \times U(2s+1-n)] \right) \}, \label{disjoint-orbits-fuzzy-sphere}
\ee
where $n \le s+\med$ and $\mathbf{1}_n$ are $n\times n$  matrices ($2s+1=N$), i.e.\ we have $N$ disjoint orbits.

In our case that we have  $N_{\trm}$ lattice points. Note that in the expression 
(\ref{disjoint-orbits-fuzzy-sphere}) the constant $\frac{-m^2}{\lambda}$ 
is in terms of the parameters of the $2$-dimensional ---see eqs.~(\ref{eq:accion})-(\ref{convention-xavier}).
For large values of $\bar{\lambda}$ we can establish their  ``equivalence'' in the three dimensional model 
via eq.~(\ref{prediction-trans-large-lambda-2d}) ---see section \ref{collapse-large-lambda}.
We conjecture that the minima in the $3$-dimensional model are at 
\be
 \{ \Phi (t) \}_ {\trm=1}^{N_{\trm}}, \quad
 \Phi (t)=\sqrt{\frac{\vert \bar{m}^2 \vert}{N\bar{\lambda}}}U^{\dagger} \Diag U \label{minima-3dim}
\ee
where $\Diag$ is a diagonal matrix with eigenvalues $1$ or $-1$.

To prove the expression (\ref{minima-3dim}) we analyse  $\ccv_{00}$.

The coefficient $\ccv_{00}$ --see eq.~(\ref{averaged_coeff-00})-- can be expressed as
\be
    \ccv_{00}= \frac{\sqrt{4 \pi}}{N N_{\trm}} 
    \Tr \left[ \sum_{\trm=1}^{N_{\trm}}\Phi\left(t\right) \right] \label{c-00-v2}
\ee
From expression (\ref{minima-3dim}) we have $N N_{\trm}+1$ possible values
\be
 \sqrt{\frac{N\bar{\lambda}}{\bar{m}^2}} \sum_{\trm=1}^{N_t} \Tr(\Phi(t))= 
  -N_{\trm} N, -N N_{\trm}+2, \dots, N N_{\trm}-2, N N_{\trm}, \label{integer-values-trace-phi}
\ee
then
\be
  { \frac{\sqrt{4 \pi \vert \bar{m}^2 \vert/ (N\bar{\lambda}) }}{N N_{\trm}}  }\times \ccv_{00}= 
    \Tr \left[ \sum_{\trm=1}^{N_{\trm}}\Phi\left(t\right) \right] \label{c-00-v3}
\ee
takes integer values as in {\bf figure \ref{histogram-0p30}} where $N \times N_{\trm}-2k=-12,-6,0,6,12$.
 
The maximal number of minima is $N \times N_{\trm} +1$. 
For $N=12=N_{\trm}$ as in  {\bf figure \ref{histogram-0p30}} the maximal number of peaks is $145$
but we just observe $5$ of them.

We can discuss intuitively why we cannot observe the maximal number of peaks.
There is a single way to obtain the value
$ \sqrt{\frac{N\bar{\lambda}}{\bar{m}^2}} \sum_{\trm=1}^{N_t} \Tr(\Phi(t))
N \times N_{\trm}$ in eq.~(\ref{integer-values-trace-phi})
--all the matrices $\Phi (\trm)$, $\trm=1,\cdots,N_{\trm}$ should be proportional to the identity.
To get in eq.~(\ref{integer-values-trace-phi}) the value $N  N_{\trm}-2$ we should have 
$\Phi (\trm_0)\propto U^{\dagger} Diag(1,1,\cdots,1,-1) U$,  
$\Phi (\trm)\propto U^{\dagger}\identy U$ for $\trm \neq \trm_0$.
Since  $\sum_{\trm=1}^{N_t} \Tr(\Phi(t))$
is invariant under interchange of the lattice points and interchange of the eigenvalues on each
matrix $\Tr(\Phi(t))$ there are $N  N_{\trm}$ ways to obtain  the value $N  N_{\trm}-2$ in 
eq.~(\ref{integer-values-trace-phi}).
This give us an rough notion of the volume of the potential minima. Those when 
eq.~(\ref{integer-values-trace-phi}) are around $0$  have a larger volume than those where
 $\Tr(\Phi(t))= N  N_{\trm}$,
they are the most probable to fall  in a simulation with  {\em hot} starting conditions. 

Since the number of local minima is larger
the measurement problems are more severe than in the $2$-dimensional case, see e.g.\ 
Ref~\cite{xavier}-\cite{Garcia-Martin-OConnor}.
The tunnelling between different minima depends on the size of its potential barrier  
and the size of the fluctuations.

\begin{figure}[h!]
\begin{center}
  \includegraphics[width=4.2in]{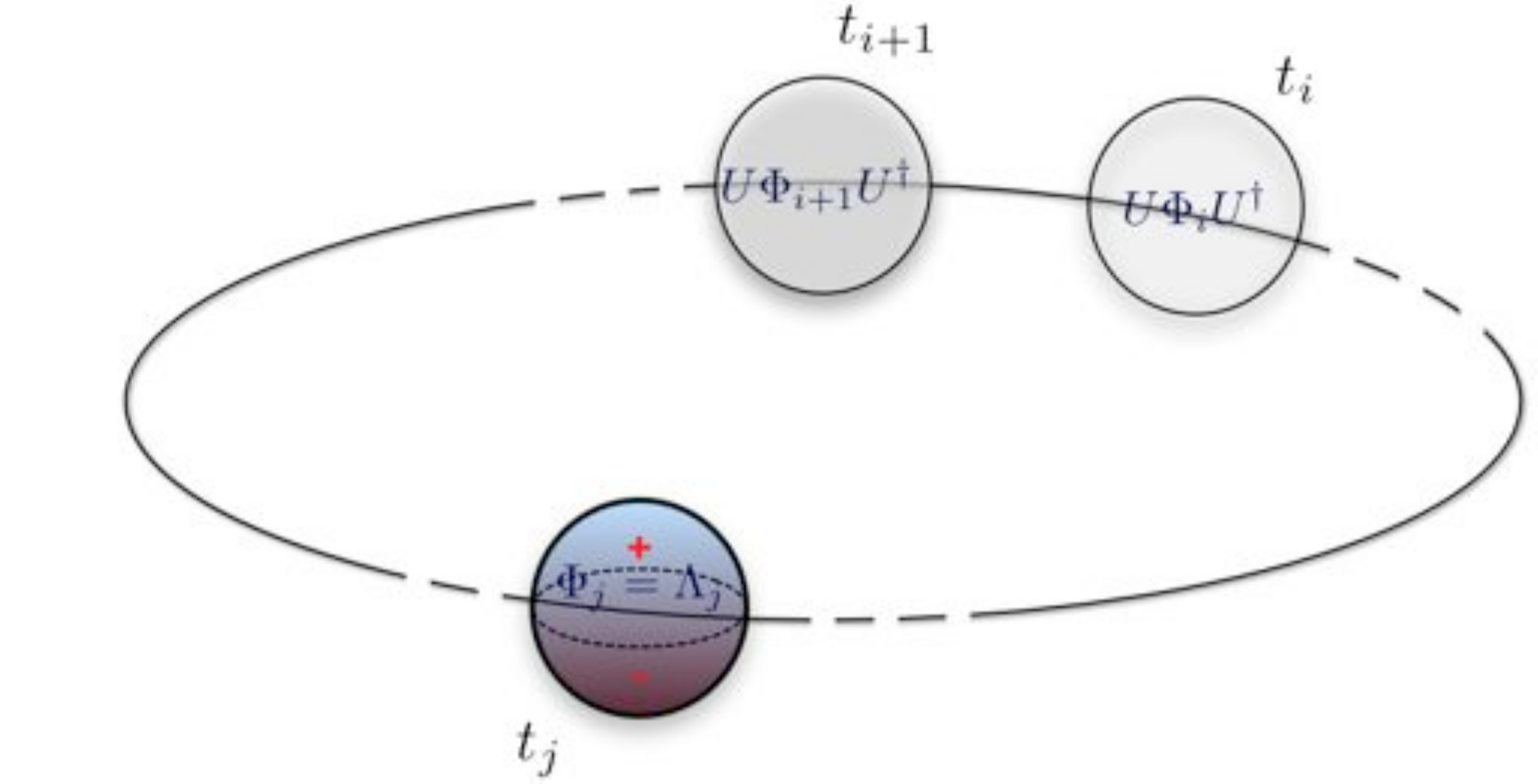}
 \end{center}
 \caption{Schematic view of the equilibrium configuration.}
\label{one-in-the-chain-diagonal}
\end{figure}

In general 
we can diagonalise only one of the $N$ matrices  $\{ \Phi(\trm_i) \}_{i=1}^N$.
This is represented in {\bf figure \ref{one-in-the-chain-diagonal}} where the matrix $\Phi(\trm_j)$ is diagonalised
though a rotation $U \in SU(2)$ to $\Diag_j$.
For $\{ \Phi(\trm_i) \}_{i=1}^{N_{\trm}}$, $i\neq j$ the matrices are not diagonal.

$\Phi(\trm_j)\propto \Diag_j$ it can be map to  the continuum to $f(\theta,\varphi)$
\be
 \Phi(\trm_j)\propto \Diag_j = \sum_{l=0}^{N-1} f_l \Yp_{l0}\longrightarrow f(\theta,\varphi)=\sum_{l=0}^{N-1} f_l Y_{l0}(\theta,\varphi).
\ee

 This is schematically shown in {\bf figure \ref{one-in-the-chain-diagonal}} with the signs\footnote{$Y_{l0}(\theta,\varphi)>0$ for $\frac{\pi}{2}>\theta>0$ and $Y_{l0}(\theta,\varphi)<0$ for $\pi>\theta > \frac{\pi}{2}$.}  "$\pm$" in red on  $\Phi(\trm_j)$.

\pagebreak

\vspace{0.2in}

\begin{figure}[h]
 \vspace{0.0in}
\begin{center}
  \includegraphics[width=4.2in]{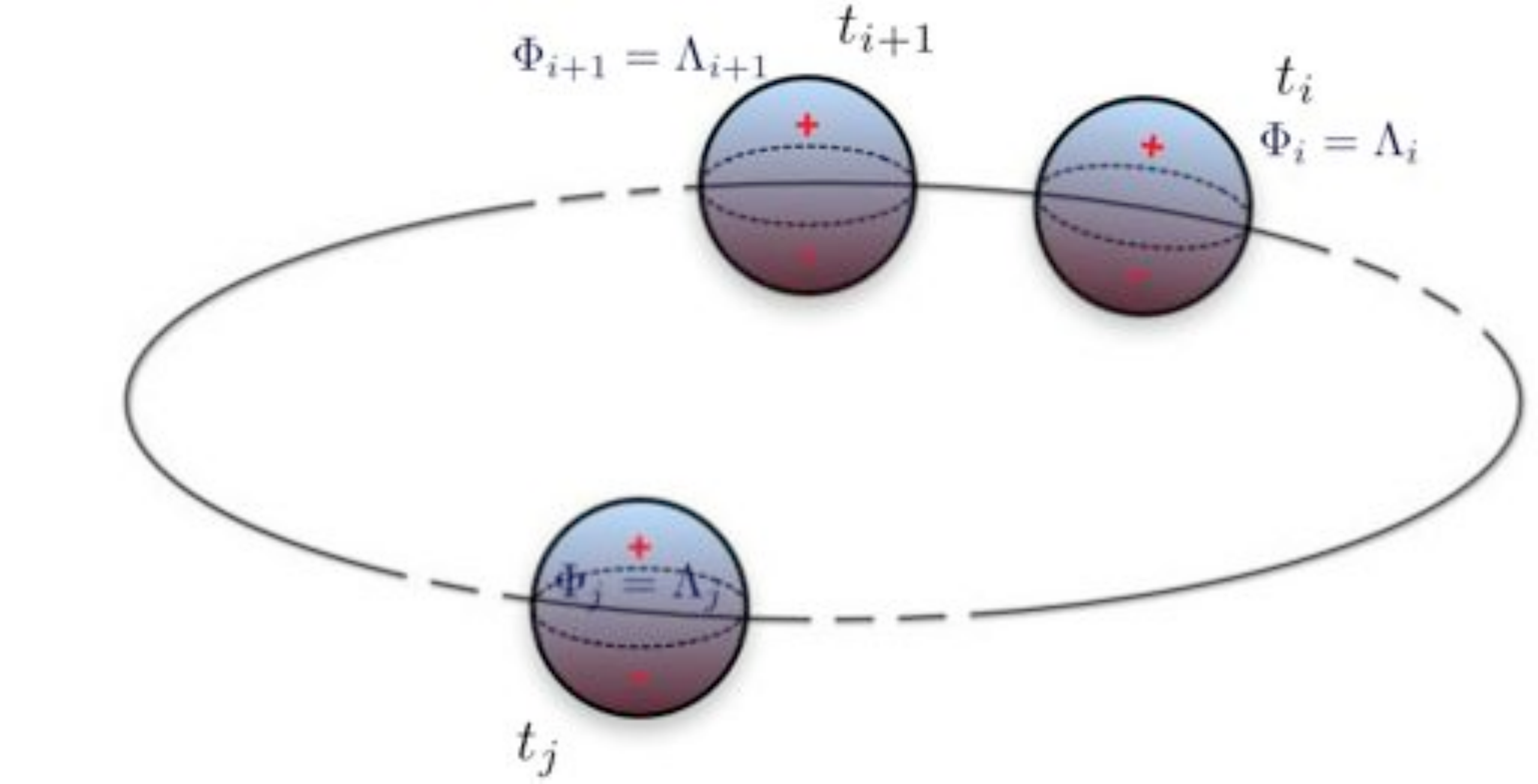}
 \end{center}
 \caption{Schematic view of the equilibrium configuration when the fuzzy kinetic term is negligible.}
\label{all-in-the-chain-diagonal}
\end{figure}

Note that the model in eq.~(\ref{action_1}) can be brought to one of diagonal matrices if the 
contribution of fuzzy kinetics term in 
eq.~(\ref{fuzzy_contribution}) is negligible (see Ref.~\cite{Zinn-Zuber}-\cite{Matytsin}).
We  have the  picture it {\bf figure \ref{all-in-the-chain-diagonal}}.

\chapter{The scaling behaviour}
\label{chapter4}
In this chapter we focus on finding out the dependence of the transition curves and 
triple point on the parameters of the system.

\section{Phase transition disordered to ordered-uniform}
\label{ph-tr-dis-ord-uni}
First we explore the transition curve for $N=12, \bar{R}=8$ and we consider
several values for $\bar{\lambda}$.

 \begin{center}
  \includegraphics[width=4.2in]{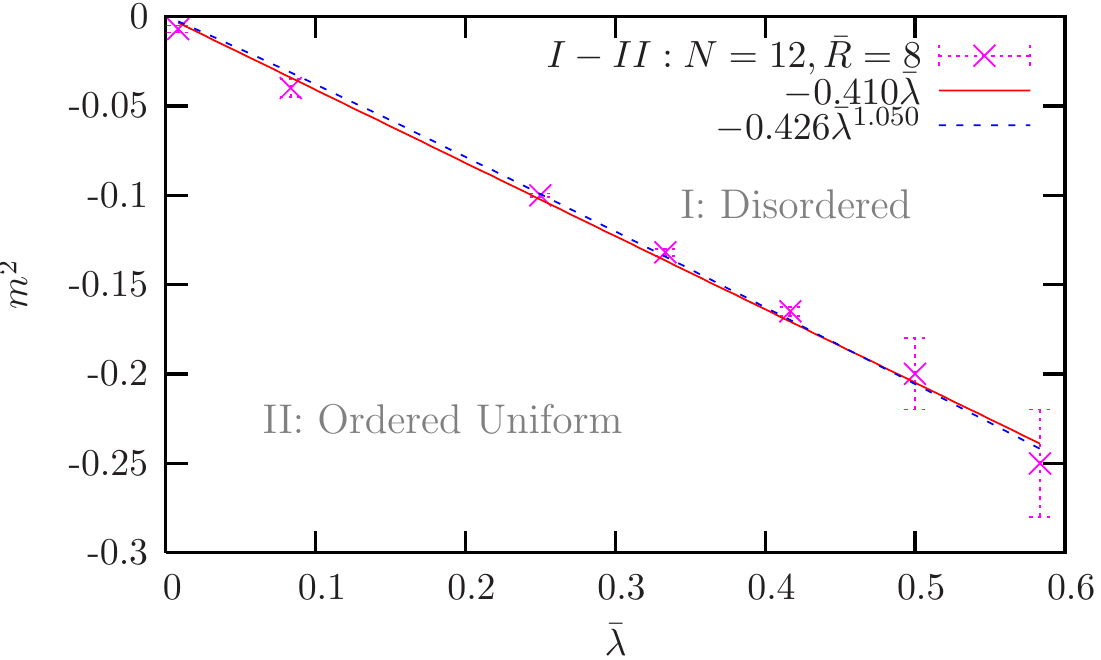}
 \end{center}
\begin{figure}[h]
 \vspace{-0.4in}
 \caption{Transition curve from the disordered to ordered-uniform phase for $N=12$.}
\label{tc1-2-N12-R8}
\end{figure}
We  observe in {\bf figure \ref{tc1-2-N12-R8}} that the transition curve is a line 
that crosses the origin, given by the equation
\be
   \bar{m^2} =  -(0.410 \pm 0.007) \bar{\lambda}. \label{eq-tc1-2-N12-R8}
\ee 
In addition we tried a fit of the form ${m}^2=a \bar{\lambda}^{b}$ and we
found the expression
\be
   \bar{m^2} = -(0.426\pm0.021) \bar{\lambda}^{(1.050 \pm 0.060)}.  \label{other-fit-tc1-2-N12-R8}
\ee 
Eqs.~(\ref{eq-tc1-2-N12-R8}) and (\ref{other-fit-tc1-2-N12-R8}) are in agreement. 
We choose a linear fit for our transition curves.

For general values of $N$ and $\bar{R}$ we propose a transition line between the
disordered phase and the ordered uniform phase of the form

\be
 \bar{m^2}= f_1(N,\bar{R}) \bar{\lambda}. \label{general-form-tc-I-II}
\ee

To identify the slope $f_1(N,\bar{R})$ our strategy is the following:
\begin{enumerate}
  \item First we extract the dependence on $N$ varying $N$ and keeping
  $\bar{R}$ fixed. What we would expect is of the form
  \be
    \bar{m^2}= f_1(\bar{R})N^{\delta_1} \bar{\lambda}. \label{general-form-tc-I-II-part1}
  \ee   
  \item Then, if $\delta_1$ does not depend on $\bar{R}$ we extract the
  dependence on $\bar{R}$ proposing 
  \be
    f_1(\bar{R})=const. \bar{R}^{\delta_2}. \label{general-form-tc-I-II-part2}
  \ee  
  \item Note that eq.(\ref{general-form-tc-I-II}) would  still be
  valid if both sides of the expression are multiplied by a common
  factor. We will use this factor to stabilise the triple point to a
  fixed value. This procedure will be explained in section
  \ref{estimation_triplepointsection}.
\end{enumerate}

The slope $f_1(N,\bar{R})$ could have a more complicated dependence on $N$ or $\bar{R}$, nevertheless
we will prove the choice $f_1(N,\bar{R})=const. N^{\delta_1} \bar{R}^{\delta_2} $ is a good ansatz.

\paragraph{Step 1}
\ \newline
Now, to find the collapse on $N$, we keep $\bar{R}=8$ and we consider $N=8,12,16,23,33$.

 \begin{center}
  \includegraphics[width=3.8in]{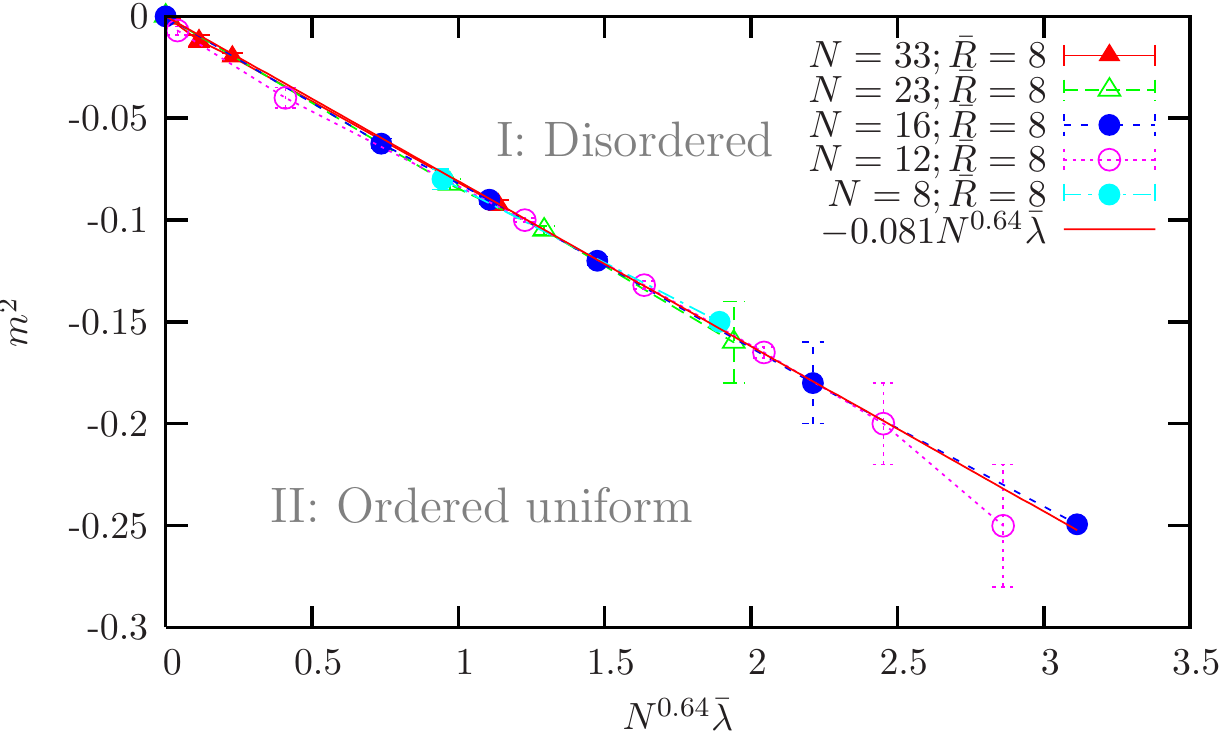}
 \end{center}
\begin{figure}[h]
 \vspace{-0.3in}
 \caption{Transition curve from the ordered-uniform phase to the ordered non-uniform phase for $\bar{R}=8$.}
\label{coll-N-R8-I-II}
\end{figure}
In {\bf figure \ref{coll-N-R8-I-II}} we re-scale the $x$-axis by a factor of $N^{0.64}$. 
We will see in section \ref{estimation_triplepointsection}
this factor stabilises the value of the triple point in $N$.

The equation of the fit of {\bf figure \ref{coll-N-R8-I-II} } is:

\be
    \bar{m^2} =   -0.081 N^{0.64} \bar{\lambda}.  \label{eqcoll2}
\ee

The exponent on $N$, $\delta_1=0.64$ is optimal for this
fit; to compare with another exponent we show the same data on the
next {\bf figure \ref{g2-coll-1-2-R8}  } for  $\delta_1=0.5$:

\begin{center}
  \includegraphics[width=3.8in]{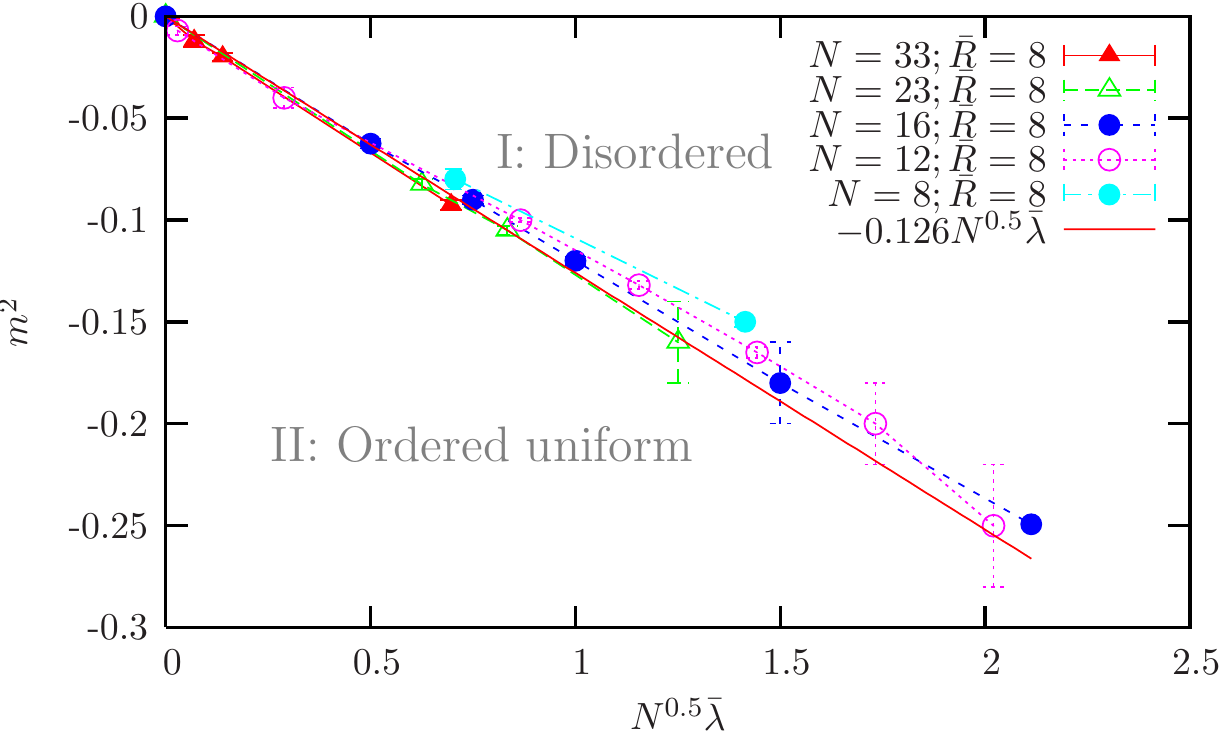}
 \end{center}
\begin{figure}[h]
 \vspace{-0.3in}
 \caption{Transition curve from the ordered-uniform phase to the ordered non-uniform phase for 
          $\bar{R}=8$ for ${\delta_1=0.5}$.}
\label{g2-coll-1-2-R8}
\end{figure}

As we observed comparing {\bf figure  \ref{coll-N-R8-I-II}} to {\bf figure \ref{g2-coll-1-2-R8}}, 
the fit for the exponent ${\delta_1=0.64}$ is better and we estimate the error on $\delta_1$ to
  the value $0.2$ for $\bar{R}=8$, this is $\delta_1(\bar{R}=8)=0.64\pm0.2$.

The next case $\bar{R}=4$ in {\bf figure \ref{coll2-32}} seems to confirm the value for the exponent, $\delta_1(\bar{R}=4)=0.64\pm0.2$.
 \begin{center}
  \includegraphics[width=3.8in]{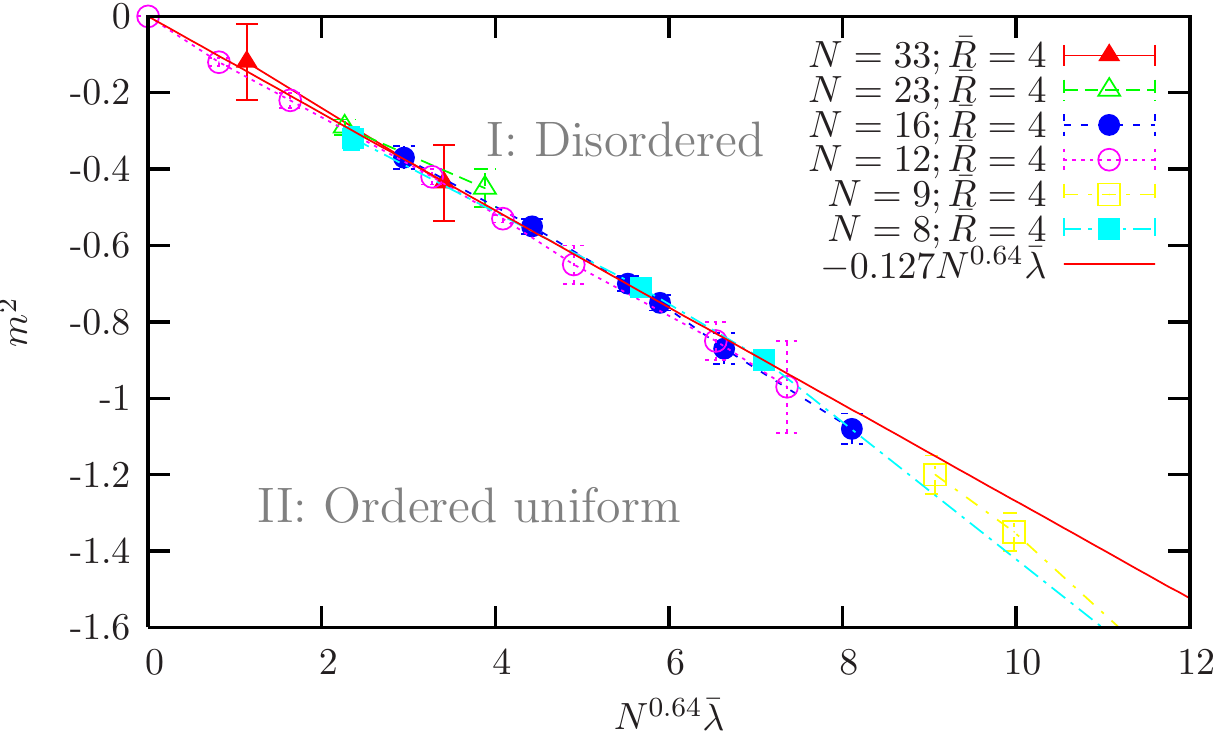}
 \end{center}
\begin{figure}[h]
 \vspace{-0.3in}
 \caption{Transition curve between the disordered phase and the ordered-uniform phase for $\bar{R}=4$.}
\label{coll2-32}
\end{figure}

We illustrate the case  $\bar{R}=16$ in {\bf figures \ref{coll-R16}}-{\bf \ref{coll2-R16}}.
 \begin{center}
  \includegraphics[width=3.8in]{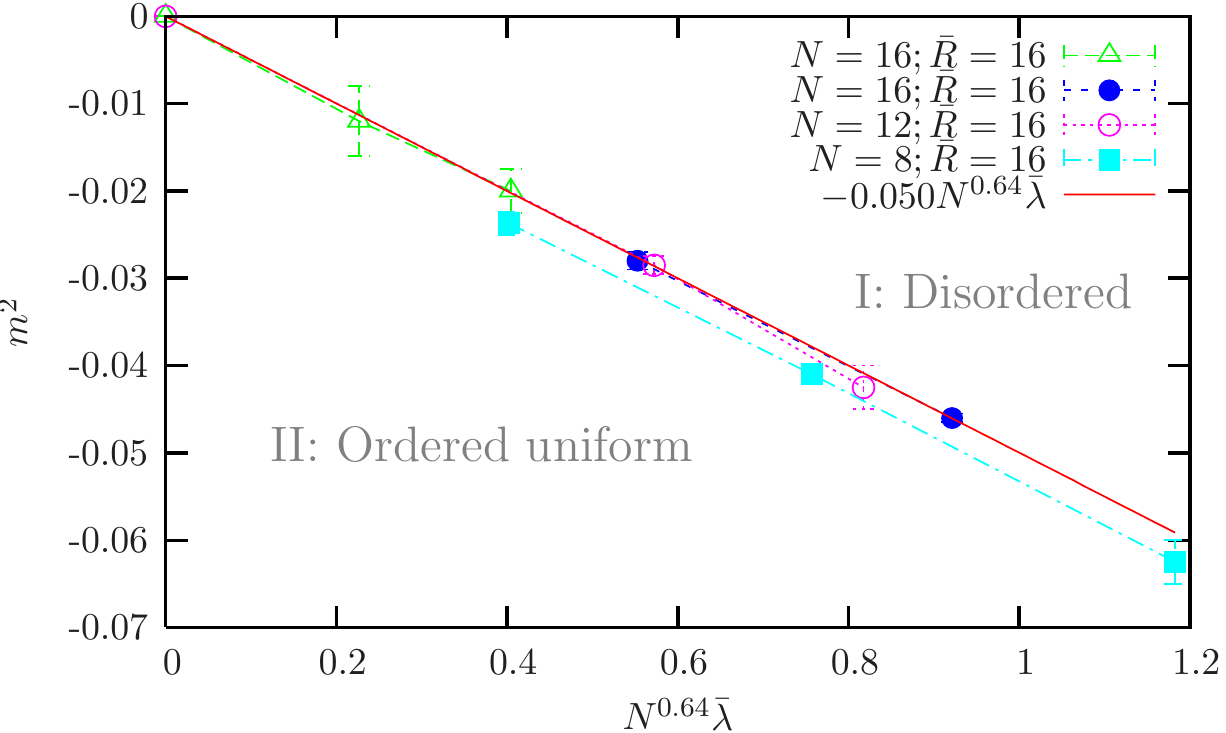}
 \end{center}
\begin{figure}[h]
 \vspace{-0.3in}
 \caption{Transition curve between the disordered phase and the ordered-uniform phase for 
          $\bar{R}=16$ for $\delta_1=-0.64$.}
\label{coll-R16}
\end{figure}

 \begin{center}
  \includegraphics[width=3.8in]{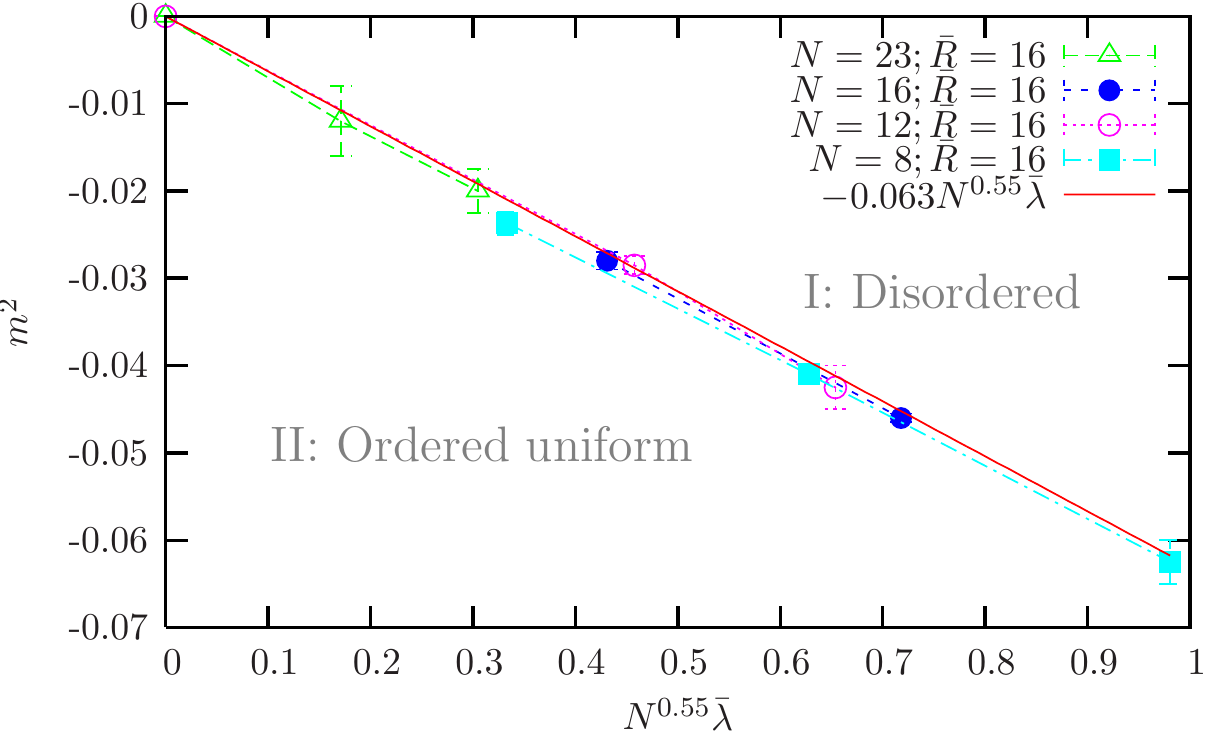}
 \end{center}
\begin{figure}[h]
 \vspace{-0.1in}
 \caption{Transition curve between the disordered phase and the ordered-uniform phase for 
          $\bar{R}=16$ and the optimal exponent $\delta_1=0.55$.}
\label{coll2-R16}
\end{figure}

We have evidence to believe the exponent $\delta_1$ depends on $\bar{R}$. But for the moment we content ourselves with choosing $\delta_1$ as $0.64$ and we consider the error as $0.30$, this is  $\delta_1=0.64\pm 0.3$.

Now that we have the collapse on $N$ we can study the coefficient
$f_1(\bar{R})$ in eq.~(\ref{general-form-tc-I-II}).

\paragraph{Step 2}
\ \newline
\begin{center}
  \includegraphics[width=4.2in]{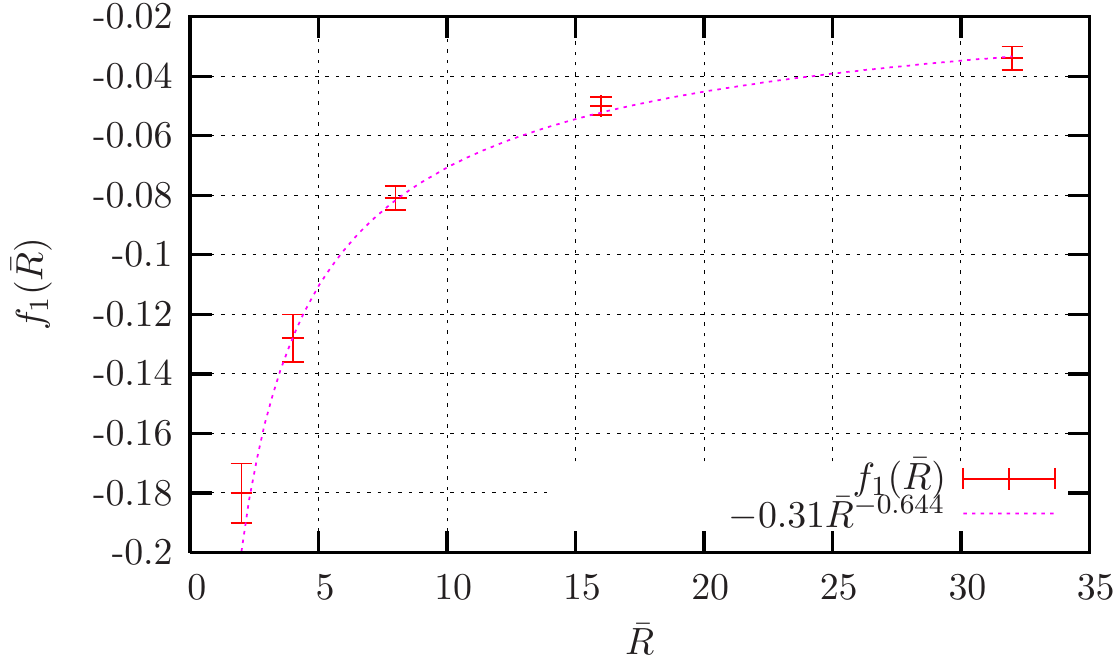}
 \end{center}
\begin{figure}[h]
 \vspace{-0.3in}
 \caption{Coefficients $f_1(\bar{R})$ for $\bar{R}=2,4,8,16,32$ and the fit in eq.~(\ref{fit-f_1-R}).}
\label{g0}
\end{figure}

\pagebreak

In the previous {\bf figure \ref{g0}} we estimate the errors on the fit of $f_1(\bar{R})$:
\be
  f_1(\bar{R})=(-0.31\pm0.1 )\bar{R}^{-(0.64\pm0.1 )}. \label{fit-f_1-R}
\ee   

Then the equation for the transition curve from the disordered phase
to the ordered-uniform phase is:
\be
   \bar{m}^2 = ( -0.31\pm0.1) \frac{N^{0.64\pm0.3}}{\bar{R}^{0.64\pm0.1}} \bar{\lambda}. 
    \label{eq-collapse-uno-dos-v0}
\ee
In terms of the constants $\cteA,\cteB,\cteC$ and $\cteD$  in  eqs.~(\ref{cte-A})-(\ref{cte-C}) we re-write\footnote{ $\bar{\lambda}=\frac{2 \cteC}{\cteD}$, $\bar{m^2}= \frac{\cteB}{\cteD}$, $N\bar{R}^{-1}=\frac{2 \pi}{\sqrt{\cteA \cteD}}$, and $N \bar{R}^2=2\pi \frac{\cteD}{\cteA^2}$} eq.~(\ref{eq-collapse-uno-dos-v0}) as
\be
   \cteB \simeq  -2.01 \frac{ \cteC}{({\cteA \cteD})^{0.32}}. \label{eq-collapse-uno-dos-v2}
\ee

 Checking the collapse on $\bar{R}$ for $N=12$ we obtain {\bf figure \ref{coll-on-R-N12}}:
 \begin{center}
  \includegraphics[width=4.2in]{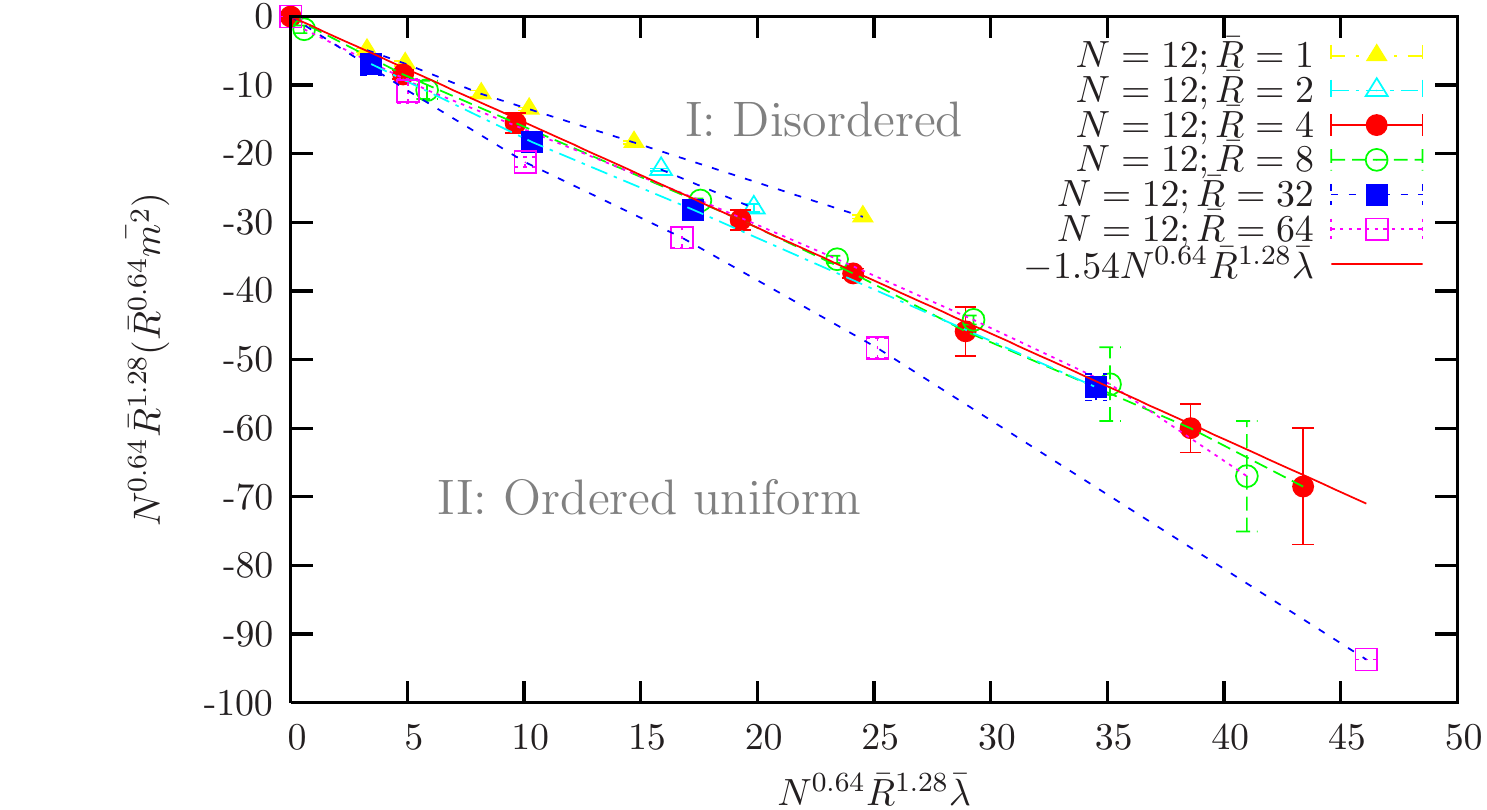}
 \end{center}
\begin{figure}[h]
 \caption{Transition curve from the disordered to ordered-uniform phase for $N=12$.}
\label{coll-on-R-N12}
\end{figure}
We observed for $\bar{R}=64$ -- where $\frac{N}{\bar{R}}=0.1875 \ll 1$ -- that the collapse
given by eq.~(\ref{eq-collapse-uno-dos}) is not good.

Now we can check the collapse on $\bar{R}$ for $N=23$, for  $\frac{N}{\bar{R}} > 1$, 
see {\bf figure \ref{coll-on-R23-N23}}.
\begin{center}
  \includegraphics[width=3.8in]{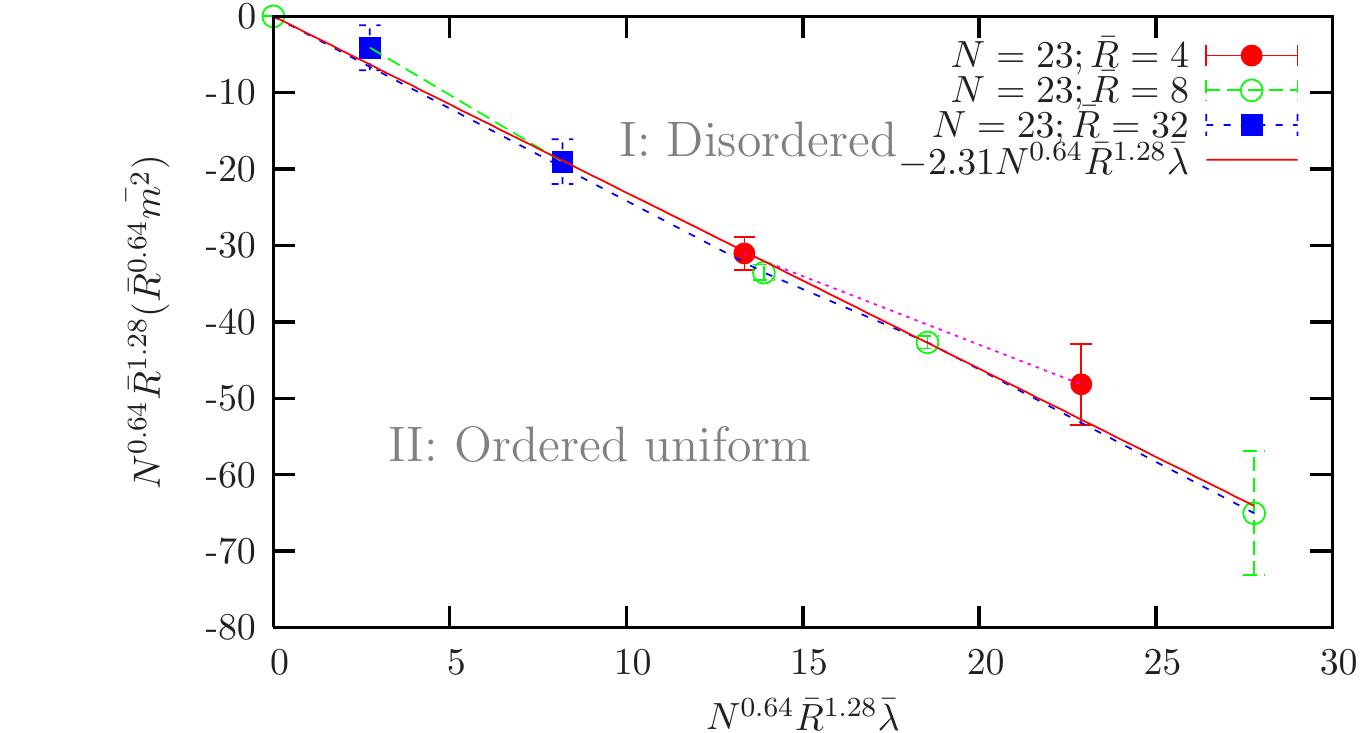}
 \end{center}
\begin{figure}[h]
 \vspace{-0.4in}
 \caption{Transition curve from the disordered to ordered-uniform phase for $N=23$.}
\label{coll-on-R23-N23}
\end{figure}

Finally the collapse of data in eq.~(\ref{eq-collapse-uno-dos}) considering all
our data is shown in {\bf figure \ref{coll-uno-dos-all}}.

 \begin{center}
  \includegraphics[width=5.6in]{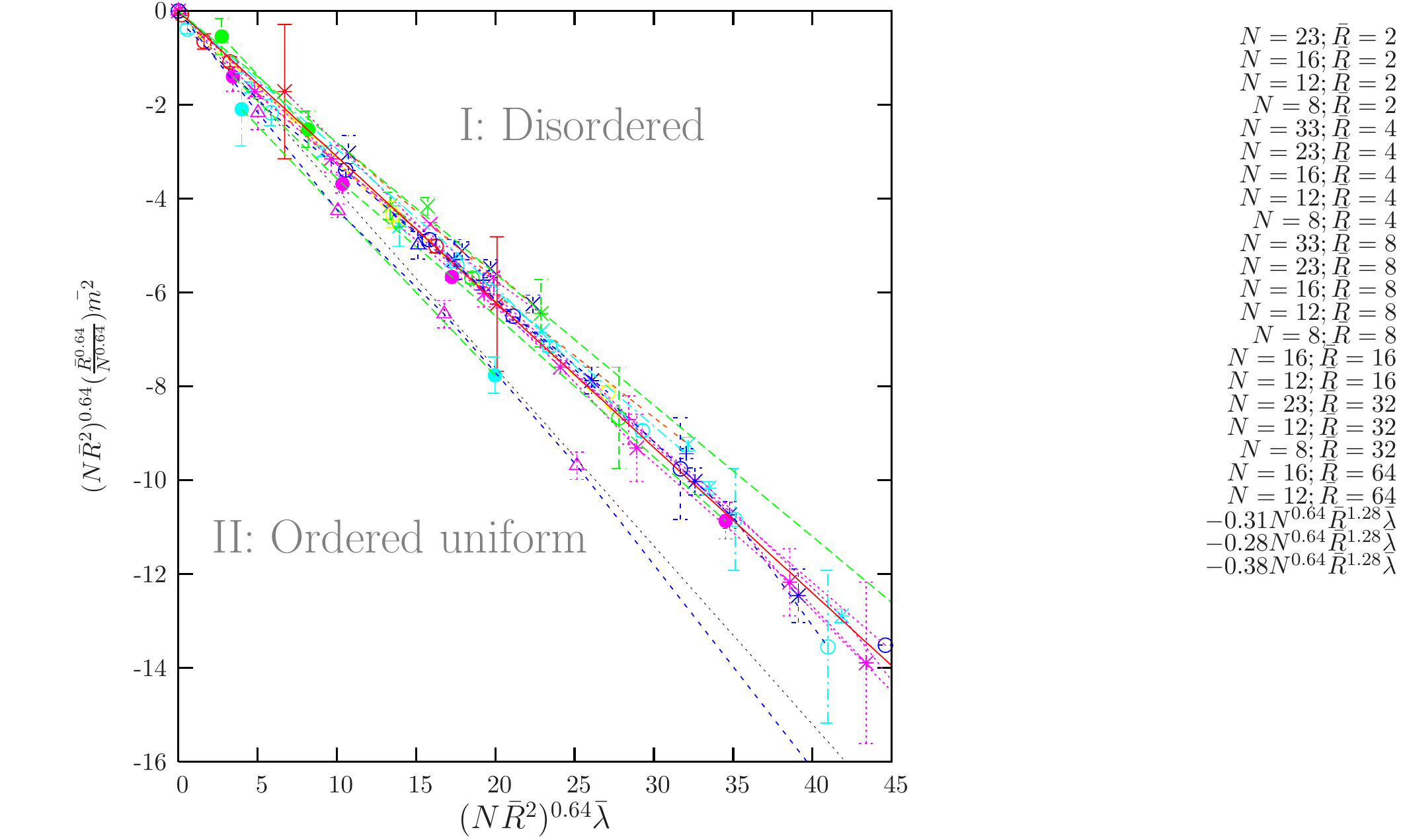}
 \end{center}
\begin{figure}[h]
 \vspace{-0.4in}
 \caption{Collapse of data for the disorder to the order-uniform phase transition.}
\label{coll-uno-dos-all}
\end{figure}
We observe that for  $\frac{N}{\bar{R}}<0.375$ and
$\frac{N}{\bar{R}}>8$
the collapse is not good\footnote{The case $N=23$, $\bar{R}=4$ could fall in this
  class, but is not possible to conclude due to the poor
  resolution for these points.}.

We conclude that the range where eq.~(\ref{eq-collapse-uno-dos}) gives a good
approximation is $\frac{N}{\bar{R}} \in \left(0.375,8 \right)$. 
If we want to include transition points for $\frac{N}{\bar{R}} \notin \left(0.375,8 \right)$
we have to reconsider $f_1(N,\bar{R})$ defined in eq.~(\ref{general-form-tc-I-II})
as a more complicated function on $\frac{N}{\bar{R}}$.
 For the moment we exclude the data that are not in the interval $\left(0.375,8 \right)$ and we present the {\bf figure \ref{coll-uno-dos}}.

 \begin{center}
  \includegraphics[width=5.5in]{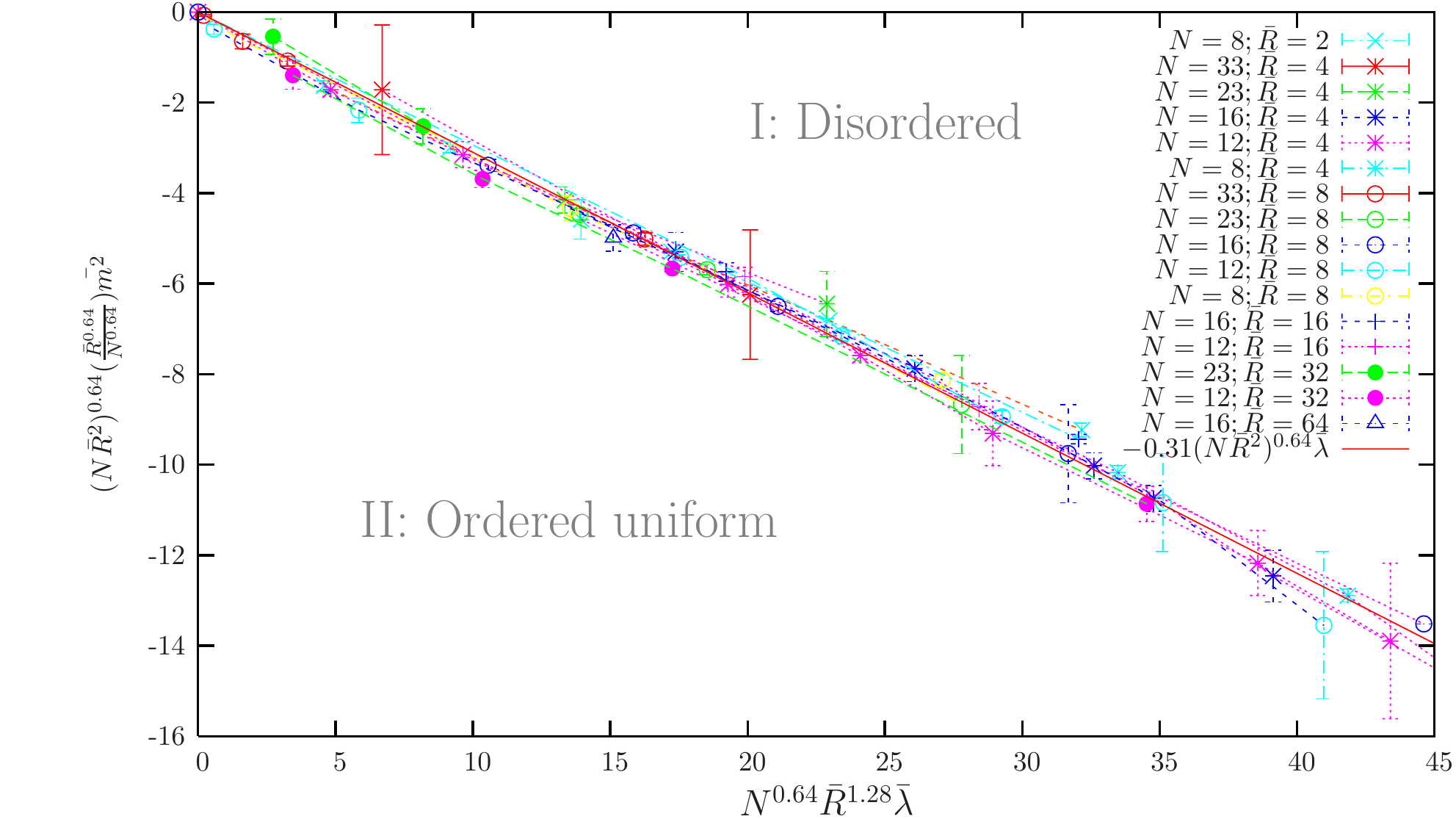}
 \end{center}
\begin{figure}[h]
 \vspace{-0.0in}
 \caption{Transition curve from the disordered phase to the ordered-uniform phase.}
\label{coll-uno-dos}
\end{figure}

To conclude the present section, the expression of the collapse in {\bf figure \ref{coll-uno-dos}} is

\be 
\framebox[1.1\width]{$ \displaystyle
   \bar{m}_c^2 = ( -0.31\pm0.1) \frac{N^{0.64\pm0.3}}{\bar{R}^{0.64\pm0.1}} \bar{\lambda}. $}
    \label{eq-collapse-uno-dos}
\ee
\vspace{0.5in}

\pagebreak

\section{Phase transition  disordered to ordered non-uniform}
\label{section-transition-disordered-to-ordered-non-uniform}
First we explore this transition  curve for $N=12, \bar{R}=8$. 
\begin{center}
  \includegraphics[width=4.6in]{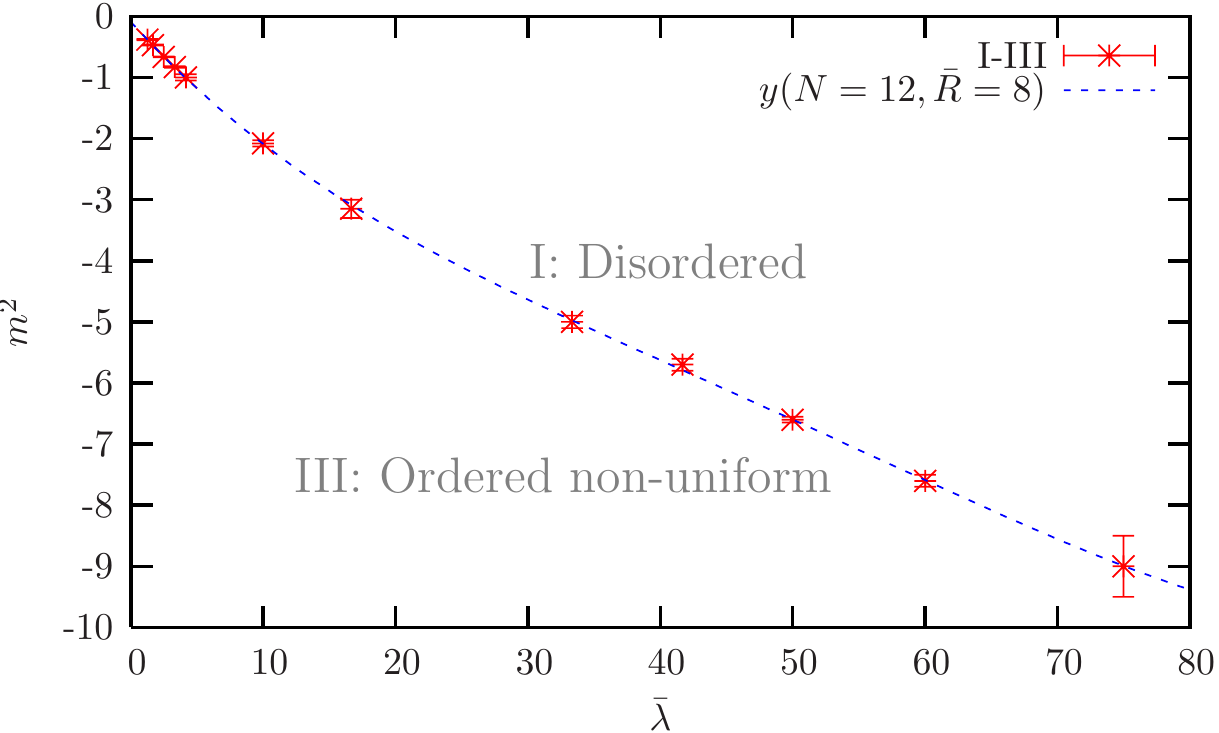}
 \end{center}
\begin{figure}[h]
 \vspace{-0.3in}
 \caption{Transition curve from the disordered phase to the ordered non-uniform phase for $N=12,\bar{R}=8$.}
\label{tc-curve-N12-R8}
\end{figure}
We observe that the transition curve shows curvature. The most natural fit we can propose 
is a polynomial  where the coefficients are functions that depend on $\bar{R}$ and $N$.
In {\bf figure \ref{tc-curve-N12-R8} } we used a polynomial fit of
degree $4$:

\bea
 y(N=12,\bar{R}=8) &=& -0.09025 -0.2389 \bar{\lambda} 
                                     +  0.0045
				     \bar{\lambda}^2  \nonumber \\
				     & &    -6.19 \times 10^{-5} \bar{\lambda}^3 
                                     +  3.12 \times 10^{-7}\bar{\lambda}^4.
\eea
In the {\bf figure \ref{tc-curve-N12-R8}} we covered a large range of
 $\bar{\lambda}$. 
Nevertheless, to predict the triple point we can concentrate on  {``small''} values 
of $\bar{\lambda}$ but above  $\bar{\lambda}_{triple}$. For this range of values the 
transition curve can be approximated by a polynomial of a smaller degree. 
In  {\bf figure \ref{tc-line-N12-R8} } we present a linear fit for an interval of 
{\bf figure \ref{tc-curve-N12-R8} }:
 \begin{center}
  \includegraphics[width=3.8in]{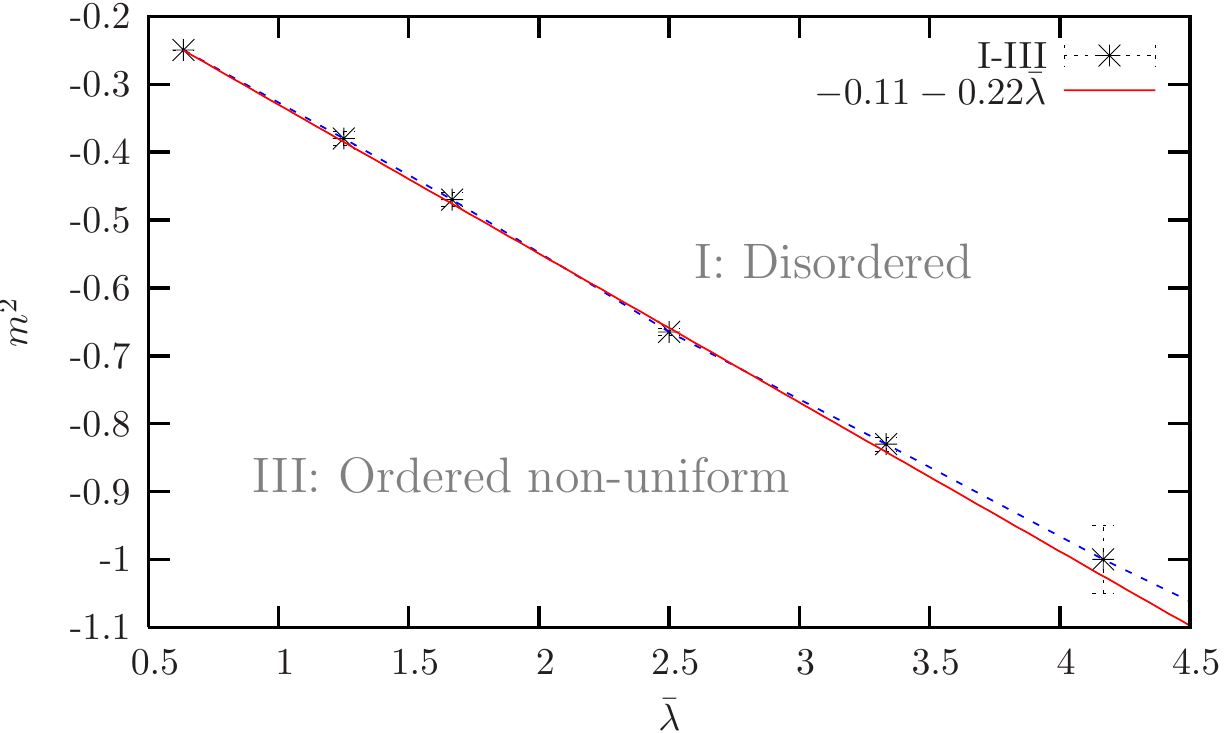}
 \end{center}
\begin{figure}[h]
 \vspace{-0.5in}
 \caption{An interval of the transition curve from the disordered phase to the ordered 
          non-uniform phase for $N=12,\bar{R}=8$.}
\label{tc-line-N12-R8}
\end{figure}
First we explore a linear fit for the transition curves. In the subsequent section 
\ref{considering-not-a-line} we will compare it with the results obtained for a fit using 
a polynomial of second degree.

The fit of {\bf figure \ref{tc-line-N12-R8}} is:
\be
  \bar{m}^2= -(0.11\pm0.02)  -(0.22\pm0.01 )\bar{\lambda}. \label{fit-tcline-1-3-N12-R8}
\ee

For general $N,\bar{R}$ we propose a linear fit for the transition curve from
the disordered phase to the ordered non-uniform phase of the form

\be
  \bar{m}^2= h_0(N,\bar{R})+    h_1(N,\bar{R})\bar{\lambda} \label{ffit_linear}
\ee
and we further make a factorisation ansatz
and $h_i(N,\bar{R})=h_i(N)\tilde{h}_i(\bar{R}) , i =0,1$.

First we concentrate on the scaling in $N$,
keeping $\bar{R}=8$ and varying $N=12,16,23,33$.

\pagebreak
\begin{center}
  \includegraphics[width=3.8in]{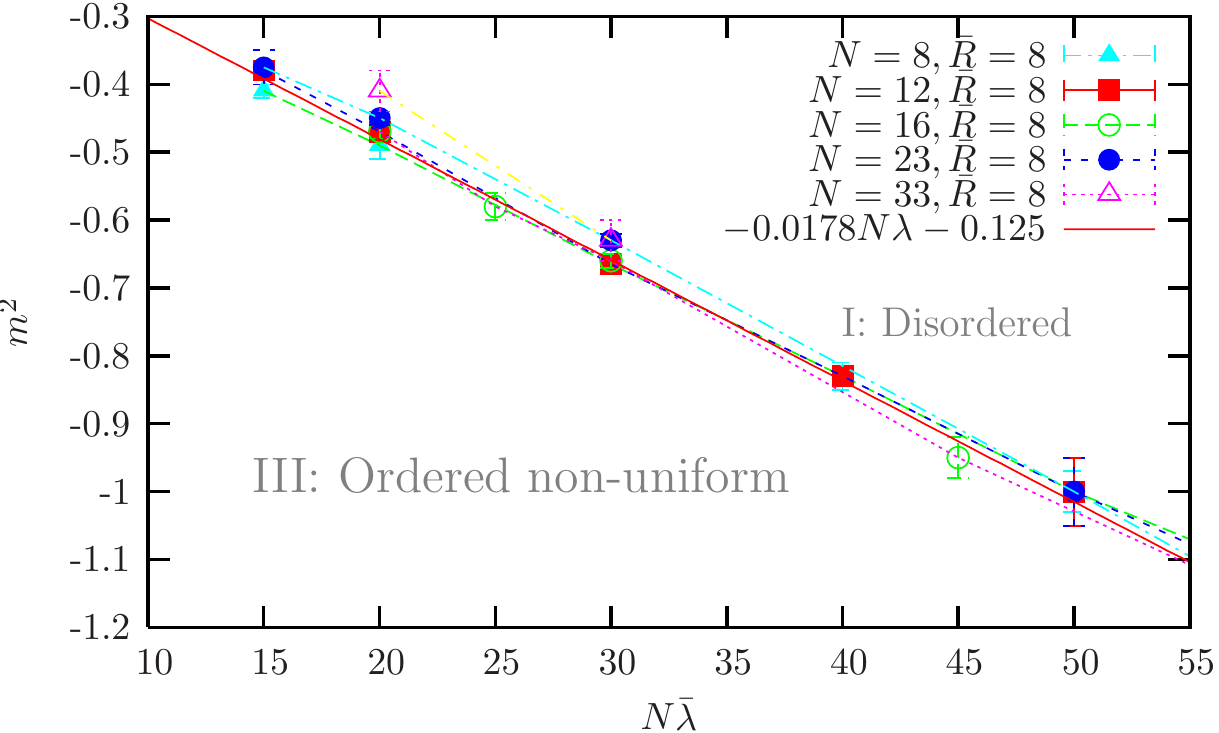}
 \end{center}
\begin{figure}[h]
 \vspace{-0.45in}
 \caption{Transition curves from the disordered phase to the ordered non-uniform phase for $\bar{R}=8$}
\label{1-coll-I-III-N-R8}
\end{figure}
The fit for {\bf figure \ref{1-coll-I-III-N-R8}  } is:
\be
   \bar{m}^2_c =   -{0.0178\pm 0.0022} N \bar{\lambda} - 0.125. \label{eqcoll-I-II-n-R8}
\ee
Here we conclude $h_1(N)\propto N$, $h_0(N)=const.$

Now we want the stabilise the triple point, at least the value of $\bar{\lambda}_T$ 
as\footnote{To stabilise the value of $\bar{m^2}_T$ it is not sufficient to re-scale the $y$-axis.} 
we did in the previous section. Then if we re-scale the $x$-axis in  {\bf figure \ref{1-coll-I-III-N-R8}  }
by a factor $N^{0.64}$ we get the {\bf figure \ref{coll-I-II-N-R8}}:
\begin{center}
  \includegraphics[width=4.5in]{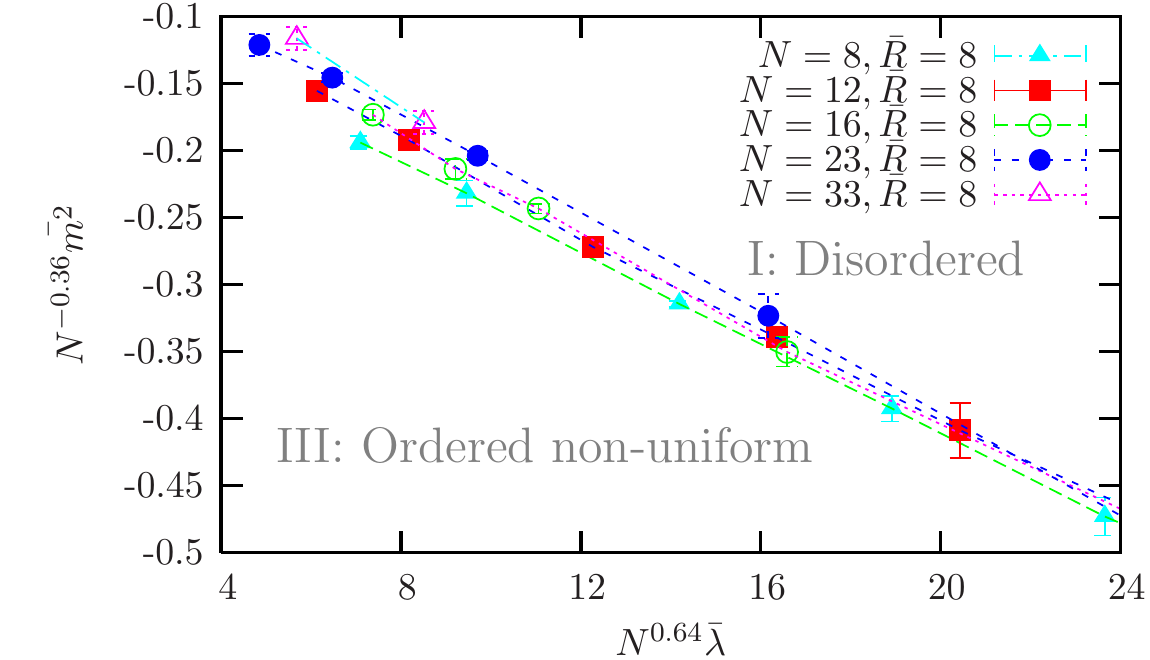}
 \end{center}
\begin{figure}[h]
 \vspace{-0.45in}
 \caption{Approximate collapse of the slope for the transition curves from the disordered phase to the ordered non-uniform phase for $\bar{R}=8$.}
\label{coll-I-II-N-R8}
\end{figure}

We fit the transition curves for $\bar{R}=4,8,32$ for different $N$ by
eq.~(\ref{ffit_linear}) in order to find the coefficient   $\tilde{h}_1(\bar{R})$. 

\begin{center}
  \includegraphics[width=4.3in]{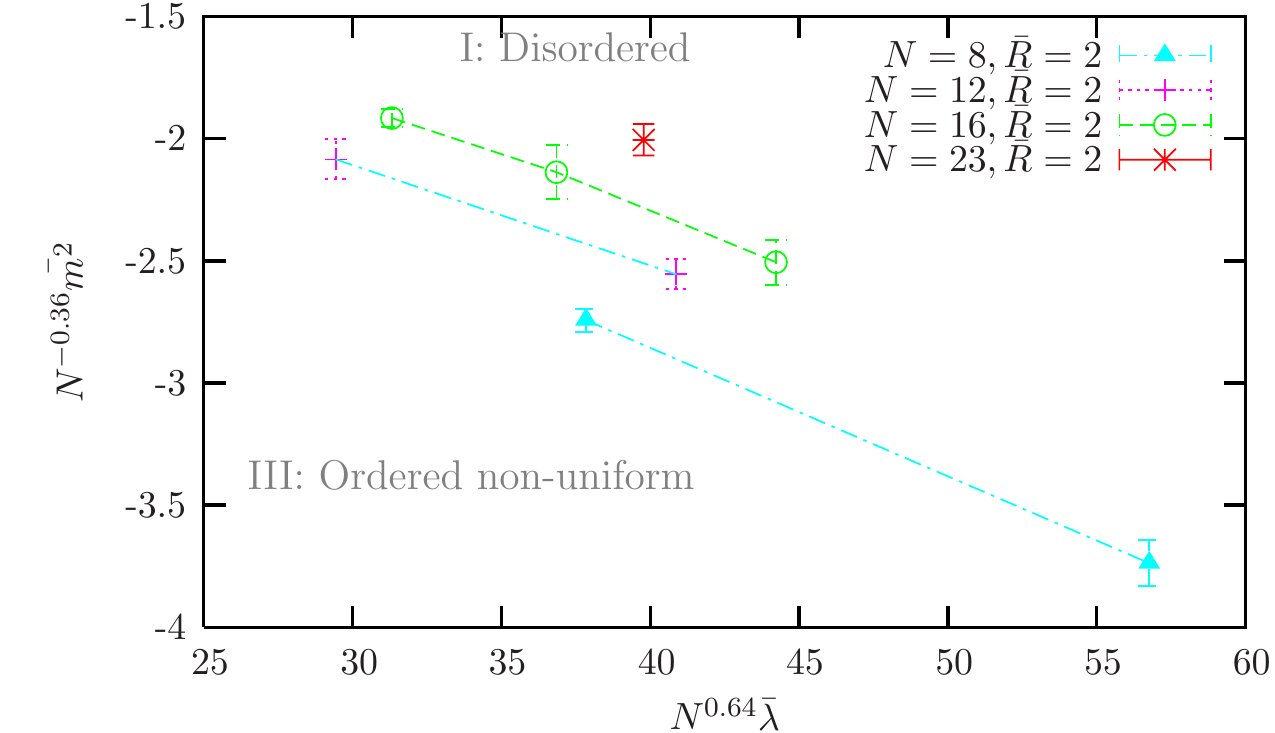}
 \end{center}
\begin{figure}[h]
 \vspace{-0.45in}
 \caption{Approximate collapse of the slope for the transition curves from the disordered phase to the ordered non-uniform phase for $\bar{R}=2$.}
\label{coll-I-II-N-R2}
\end{figure}

\begin{center}
  \includegraphics[width=4.8in]{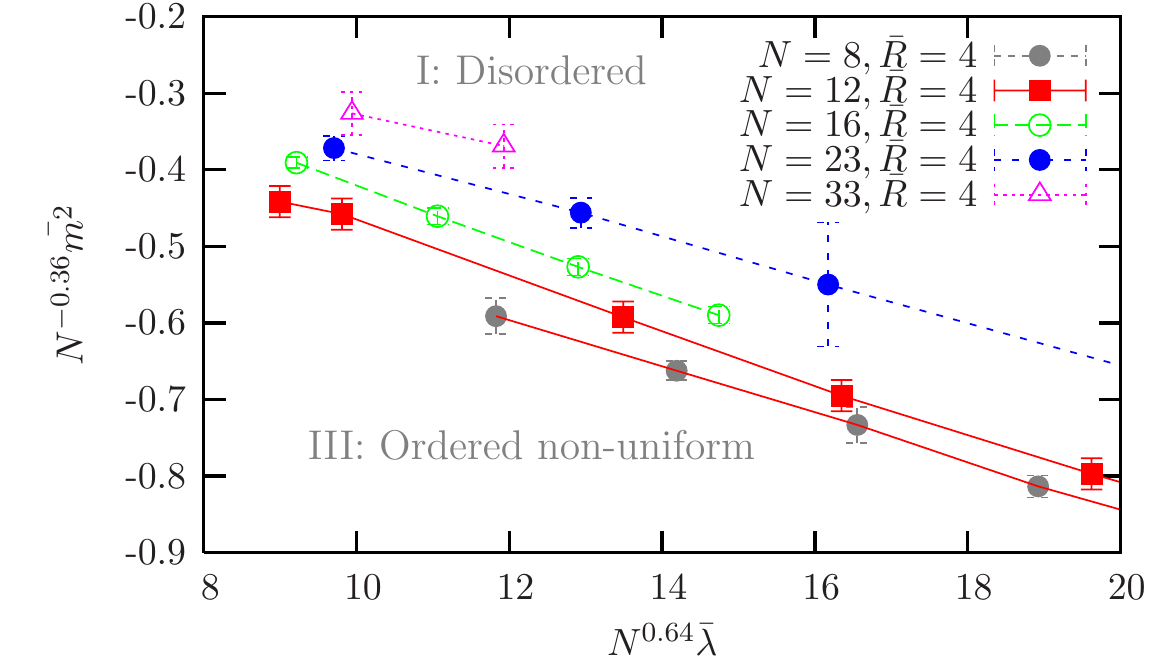}
 \end{center}
\begin{figure}[h]
 \vspace{-0.3in}
 \caption{Approximate collapse of the slope for the transition curves from the disordered phase to the ordered non-uniform phase for $\bar{R}=4$.}
\label{coll-I-II-N-R4}
\end{figure}

\begin{center}
  \includegraphics[width=4.8in]{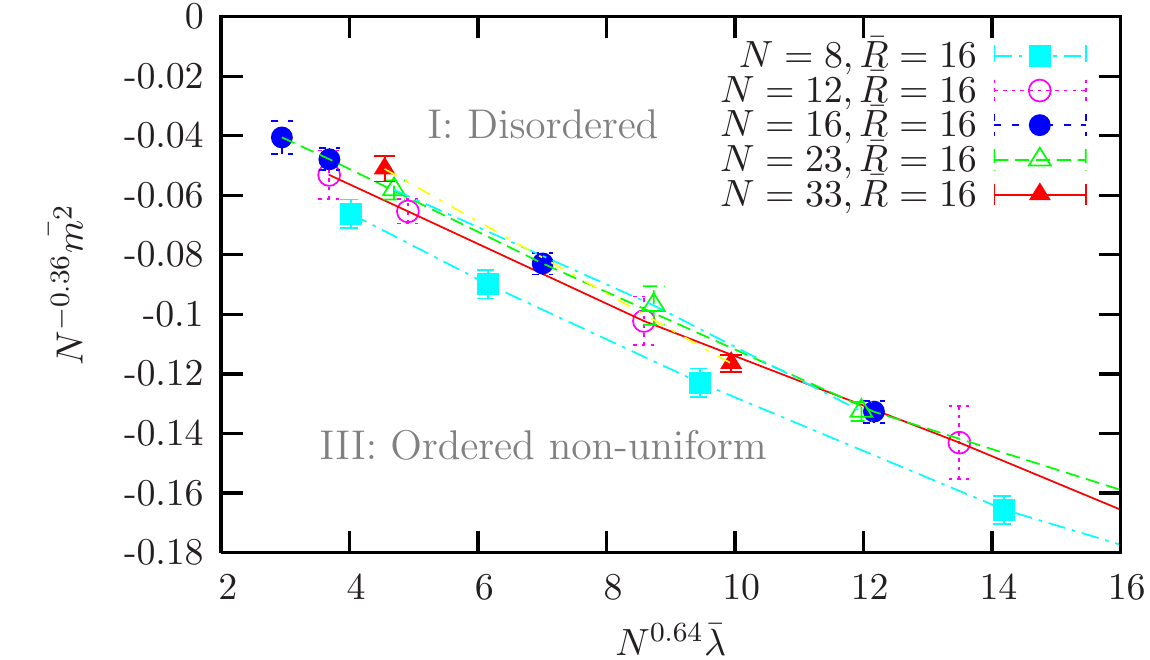}
 \end{center}
\begin{figure}[h]
 \vspace{-0.3in}
 \caption{Collapse of the slope for the transition curves from the disordered phase to the ordered non-uniform phase for $\bar{R}=16$.}
\label{coll-I-II-N-R16}
\end{figure}

\begin{center}
  \includegraphics[width=4.8in]{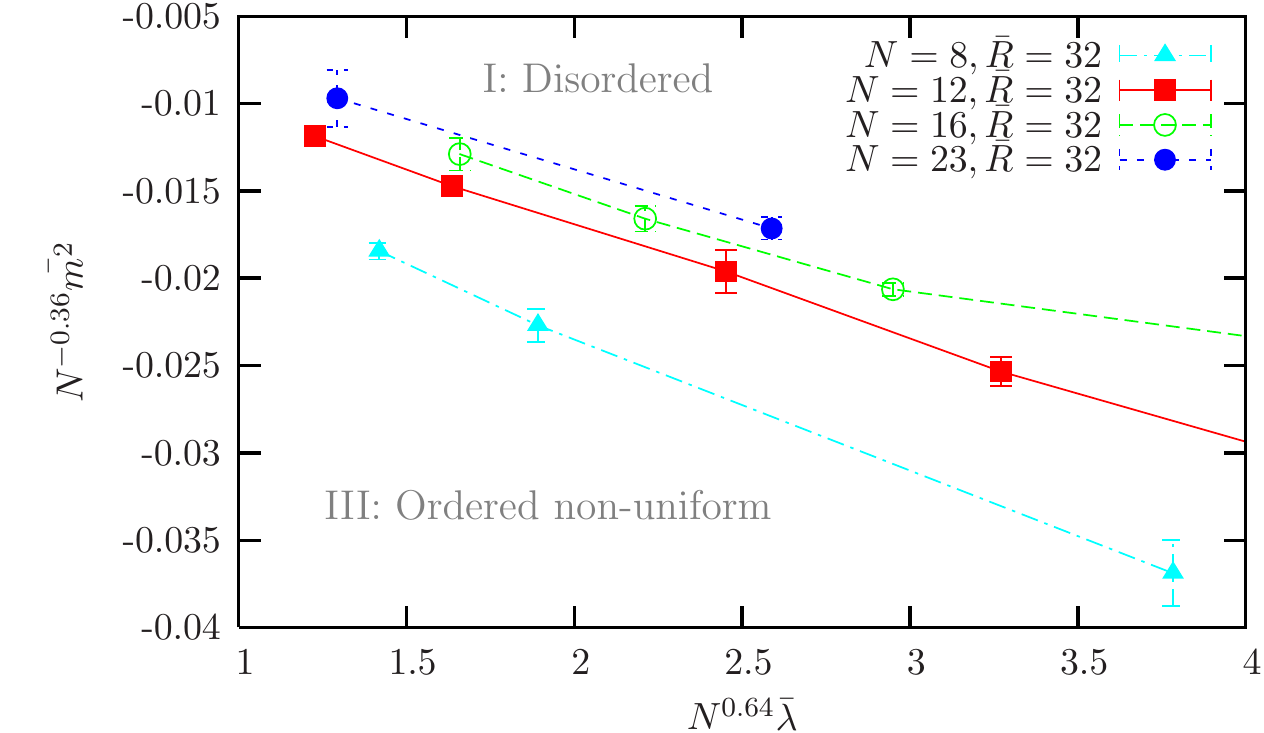}
 \end{center}
\begin{figure}[h]
 \vspace{-0.3in}
 \caption{Collapse of the slope for the transition curves from the disordered phase to the ordered non-uniform phase for $\bar{R}=32$.}
\label{coll-I-II-N-R32}
\end{figure}

Now that we get the collapse on $N$ as $h_1(N)\sim N$ we can determine
the coefficients, we propose:
\be
 {h}_1(N,\bar{R})= - N \tilde{h}_1(\bar{R})
\ee
{\bf Figure \ref{h1-on-R}} shows $\tilde{h_1}(\bar{R})$ and its fit.
\begin{center}
  \includegraphics[width=3.8in]{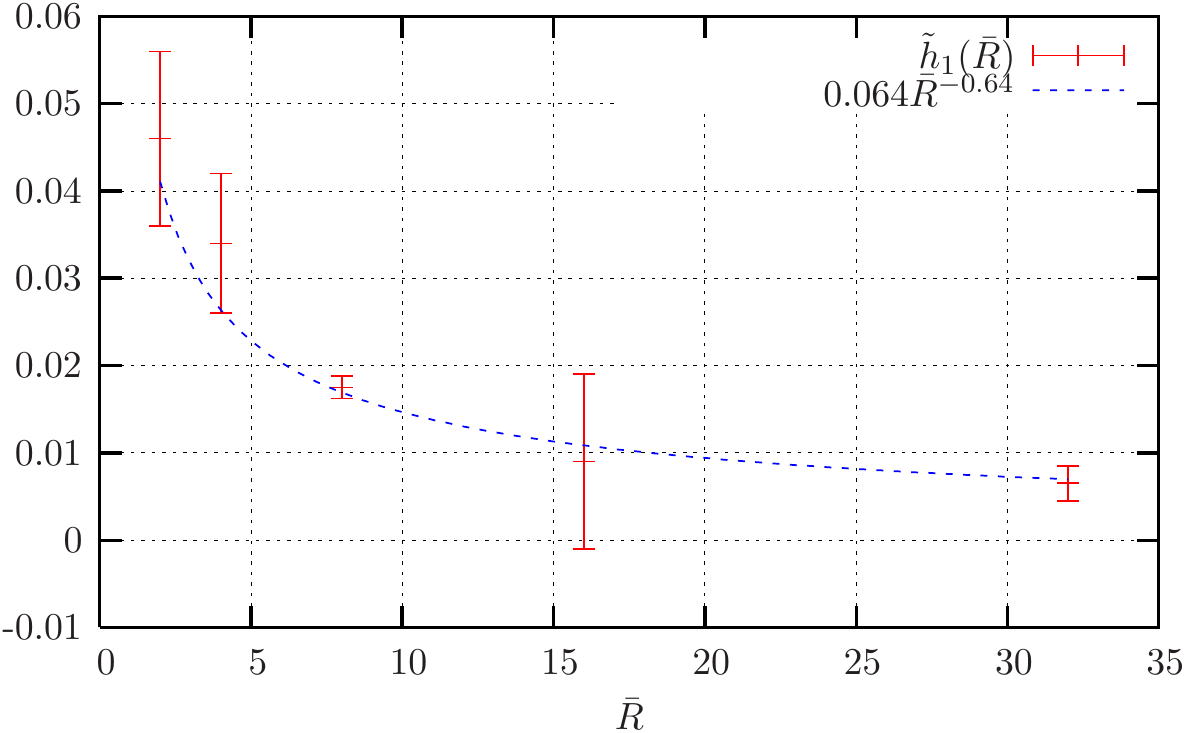}
 \end{center}
\begin{figure}[h]
 \vspace{-0.3in}
 \caption{Coefficients $\tilde{h}_1(\bar{R})$ for $\bar{R}=2,4,8,16,32$ and the fit
          $0.064\bar{R}^{-0.64}$.}
\label{h1-on-R}
\end{figure}

Then we have 
\be
 h_1(N,\bar{R})= - (0.064\pm 0.017) N \bar{R}^{-0.64\pm 0.1}. \label{slope-I-III}
\ee
Eq.~(\ref{slope-I-III}) is the slope of the coexistence line from the disordered phase 
to the ordered non-uniform phase. To solve the equation of this coexistence line we use 
the form:
\be
  \left(\bar{m^2} - \bar{m^2}_T \right)=h_1(N,\bar{R}) \left( \bar{\lambda} -\bar{\lambda}_T  \right).
\ee
If we want to collapse the transition line completely, it is not enough to re-scale the $y$-axis. 
It is necessary to shift $\bar{m^2}$, for example, by substituting 
$\bar{m^2} \longrightarrow \bar{m^2} + \frac{12.7}{\bar{R}^{1.92}}$ in {\bf figure \ref{coll-I-II-N-R8}}.
 This leads to {\bf figure \ref{v2-coll-I-II-N-R8}}.
\begin{center}
  \includegraphics[width=3.8in]{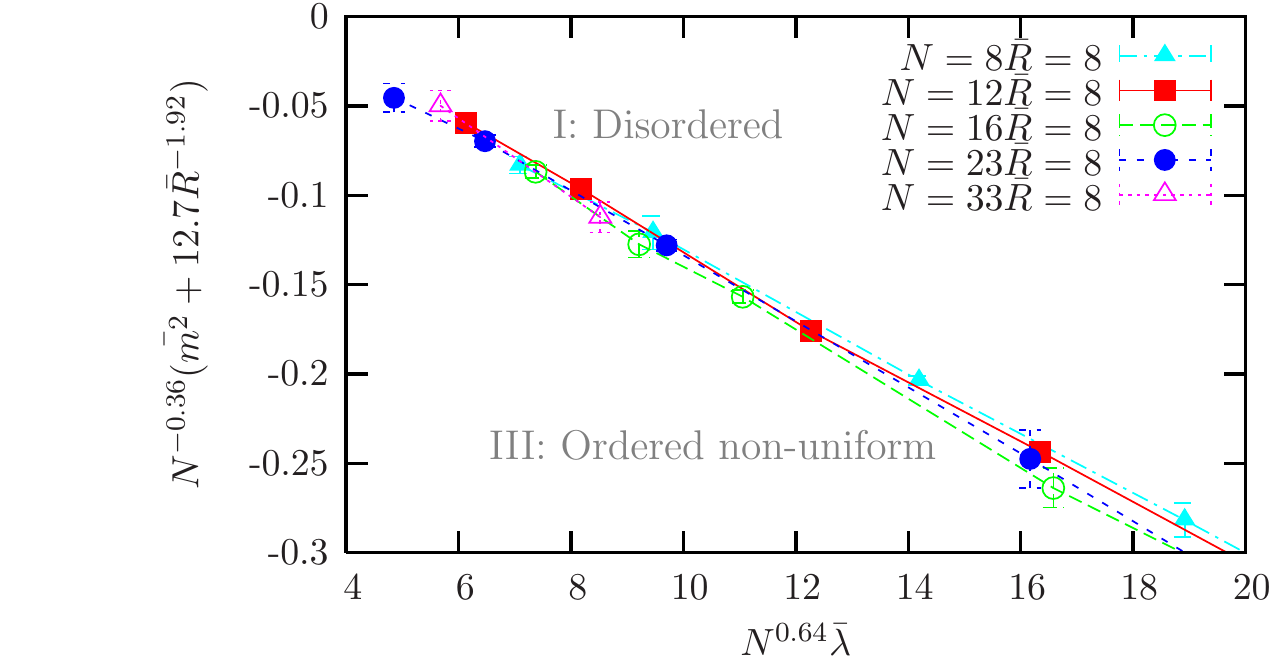}
 \end{center}
\begin{figure}[h]
 \vspace{-0.3in}
 \caption{Collapse of  transition curve from the disordered phase to the ordered non-uniform 
          phase for $\bar{R}=8$.}
\label{v2-coll-I-II-N-R8}
\end{figure}
\vspace{0.4cm}
 As another example for $\bar{R}=4$, form  {\bf figure \ref{coll-I-II-N-R4}} we obtain {\bf figure \ref{v2-coll-I-II-N-R4}}.

\begin{center}
  \includegraphics[width=4.8in]{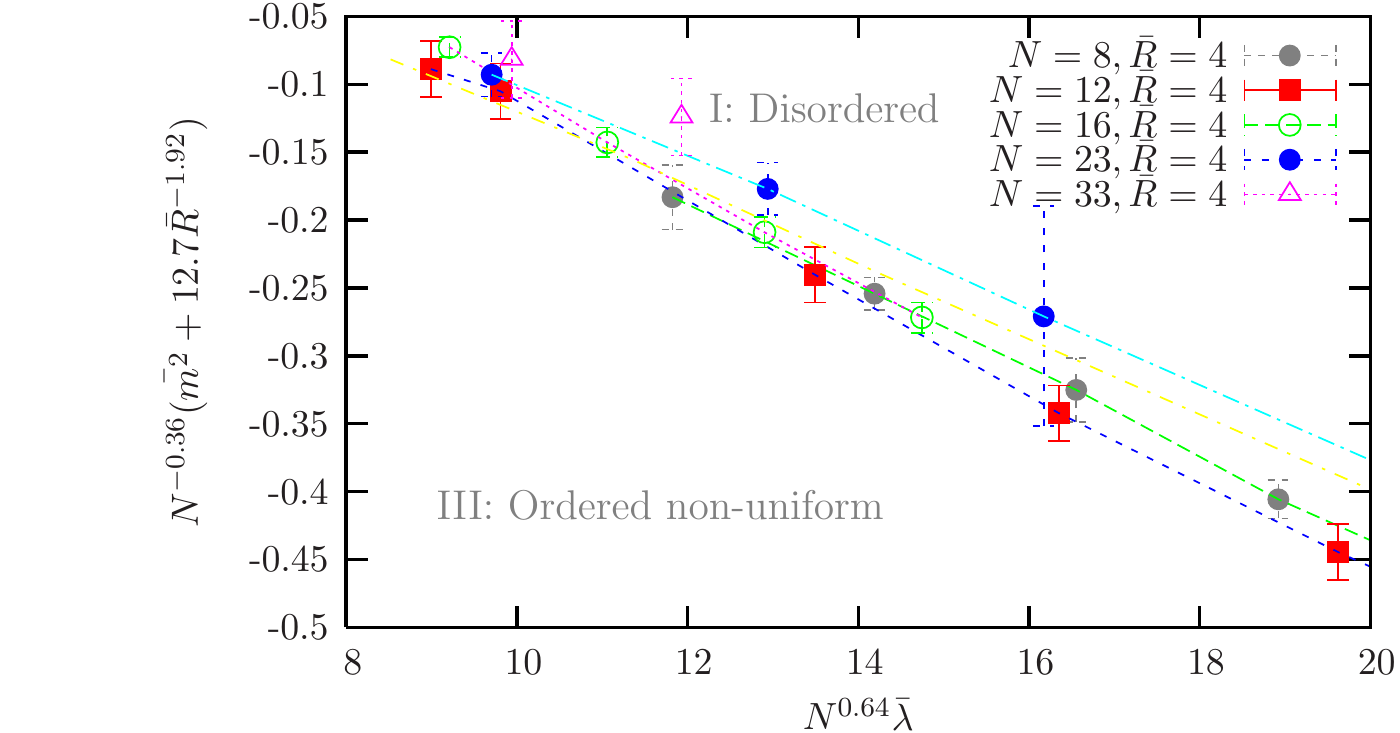}
 \end{center}
\begin{figure}[h]
 \vspace{-0.3in}
 \caption{Collapse of  transition curve from the disordered phase to the ordered non-uniform phase      for $\bar{R}=4$.}
\label{v2-coll-I-II-N-R4}
\end{figure}

We conclude that for $\bar{\lambda}$ around $\bar{\lambda}_T$ the transition curve
from the disordered phase to the ordered non-uniform phase obeys  eq.~(\ref{eq-collapse-uno-tres}).
\be
   \bar{m}^2_c =  -(0.064\pm0.017) {N} {\bar{R}}^{-0.64\pm0.1} \bar{\lambda} +\left(2.69 N^{0.41}-12.7  \right)\bar{R}^{-1.92}.
 \label{eq-collapse-uno-tres}
\ee

\section{Stabilising the triple point}
\label{estimation_triplepointsection}

As we mentioned in section \ref{ph-tr-dis-ord-uni} we have the freedom to
re-scale both axes of the phase diagram by a common factor. We want to
use this factor to fix the triple point, i.e.\ we want to find a
function of $N$ and $\bar{R}$, ${\mathit{k}}(N,\bar{R})$, such that ${\mathit{k}}(N,\bar{R})\bar{\lambda}_T=const.$ 

To identify the triple point $\left(\bar{\lambda}_T, \bar{m}^2_T \right)(\bar{R},N) $ our strategy is the following:
\begin{enumerate}
  \item First we extract the dependence on $\bar{R}$ varying $\bar{R}$ and keeping
  $N$ fixed. 
  What we would expect is of the form
  \bea
    \bar{\lambda}_T &=& Z(N) \bar{R}^{e_1(N)},  \\
    \bar{m}^2_T     &=& M(N) \bar{R}^{e_2(N)}.
  \eea  
  For each $N$ we estimate the exponents $e_1(N),e_2(N)$ and the coefficients $Z(N),M(N)$ 
  using the package ``gnu-plot''. 
  \item In general the exponents can depend on $N$, but we will show they do not and fix  
        $e_1,e_2$ to a certain value.
  Since our final expression of the triple point may strongly depend on the choice of  
  $e_1,e_2$, we will compare the expressions for two different sets of 
  exponents.\footnote{We will choose  $e_1=-1.28, e_2=-1.92$ vs. $e_1=-1.25, e_2=-1.89$ and its corresponding  coefficients $Z(N),M(N)$.} Both fits for $e_1$ in the case of $  \bar{\lambda}_T$ -- or $e_2$ in the case of $ \bar{m^2}_T $ -- will be presented in the corresponding figures. We include the case when we vary $e_1$ and $Z(N)$ -- or $e_2$ and $M(N)$ in the case of $ \bar{m^2}_T $.
  \item We extract the
  dependence on $N$ proposing 
  \bea
    Z(N) &=& const. \cdot N^{d_1}, \label{general-form-Z} \\
    M(N) &=& const. \cdot N^{d_2}.
  \eea  
\end{enumerate}
\paragraph{Step 0}
\ \newline

First we explore the dependence on $\bar{R}$. To do this we consider
$N=12$ and several values for $\bar{R}$ and  we estimate the intersection of
both transition curves. {\bf Figure \ref{inter-N16-R8} } is an example of this procedure.

 \begin{center}
  \includegraphics[width=4.5in]{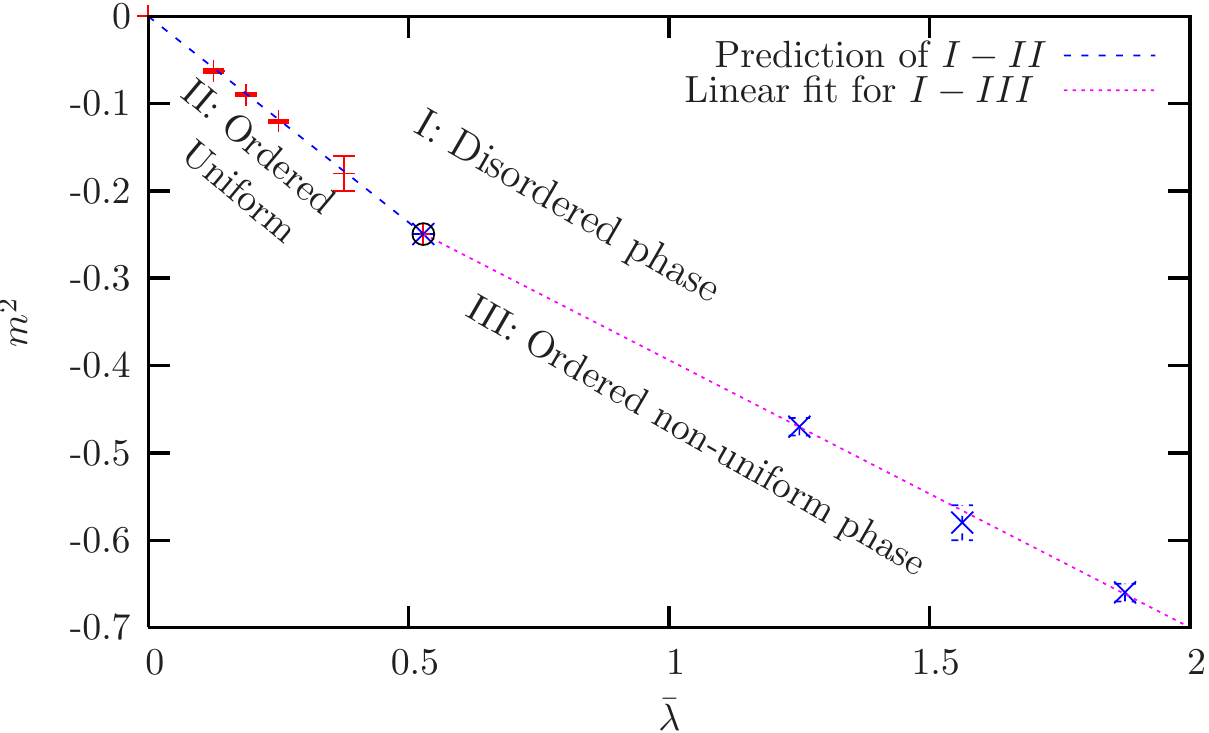}
 \end{center}
\begin{figure}[h]
 \vspace{-0.45in}
 \caption{Transition curves for $N=16, \bar{R}=8$.}
\label{inter-N16-R8}
\end{figure}
The linear fit for {\bf figure \ref{inter-N16-R8} } is given by the
equation:
\be
  \bar{m}^2= -(0.090\pm 0.0003) -(0.0191\pm 0.0013)\bar{\lambda}.
\ee
The intersection of both curves is estimated in $\bar{\lambda}_{T}= 0.528
\pm 0.02$, $\bar{m^2}_T=-0.2494 \pm0.003$.

\paragraph{Step 1}
\ \newline

{\bf Figure \ref{f-estimation-triple-N12}} shows the estimated values of  $\bar{\lambda}_{T}$ for different $\bar{R}$ fixing $N=12$ and a fit for these points:

 \begin{center}
  \includegraphics[width=4.6in]{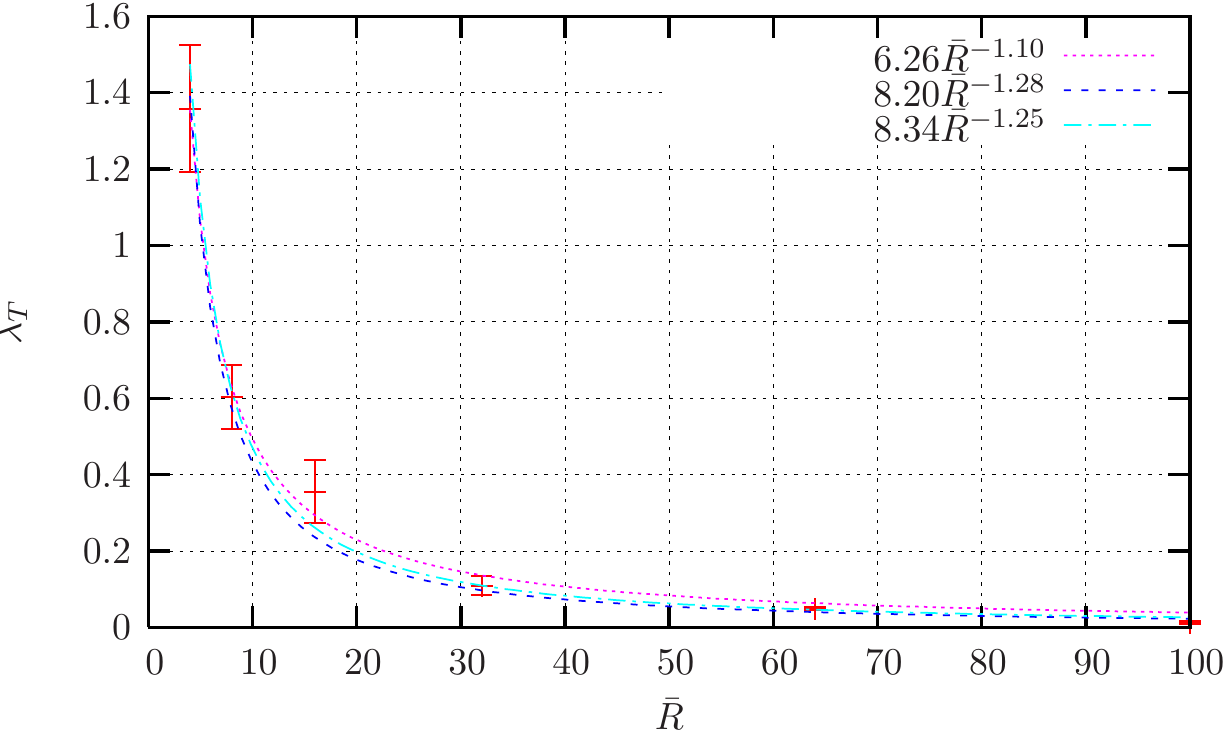}
 \end{center}
\begin{figure}[h]
 \vspace{-0.2in}
 \caption{Estimation of $\bar{\lambda}_{T}$ for  $N=12$ and three different fits. The fits were obtained via a function $\bar{\lambda}_{T}=  Z(N=12) \bar{R}^{e_1}$. In the first case we fit both parameters $Z(N=12),e_1$ and we got $\bar{\lambda}_T=(6.26 \pm 0.64)  \bar{R}^{(-1.10\pm0.063)}$. In the second and third case we fixed the exponent $e_1$ and vary $Z(N=12)$. For the second fit we chose $e_1=-1.28$ and we fit $Z(N=12)$ as in eq.~(\ref{lfit-triple-N12}) and finally in the third case we fixed  $e_1=-1.25$ to get $Z(N=12)=8.34\pm0.27$.}
\label{f-estimation-triple-N12}
\end{figure}

The fit for  {\bf Figure \ref{f-estimation-triple-N12}} is:
\footnote{ We tried another exponent and we got an acceptable fit for
$\bar{\lambda}_T= 9.11 \bar{R}^{-1.35}$.}
 
\be
   \bar{\lambda}_T=(8.20 \pm 0.31)\bar{R}^{-1.28 \pm0.1}. \label{lfit-triple-N12}
\ee

Then, if we multiply both axes in the phase diagram by a factor on $\bar{R}$ such that the $x$-axis
is $\bar{R}^{1.28} \bar{\lambda}$, we stabilise the triple point to the value of 
$\bar{\lambda}_T=8.20(31)$.

We can check the consistency of $\bar{m^2}_T$ with the eq.~(\ref{eq-collapse-uno-dos}):
 \begin{center}
  \includegraphics[width=4.6in]{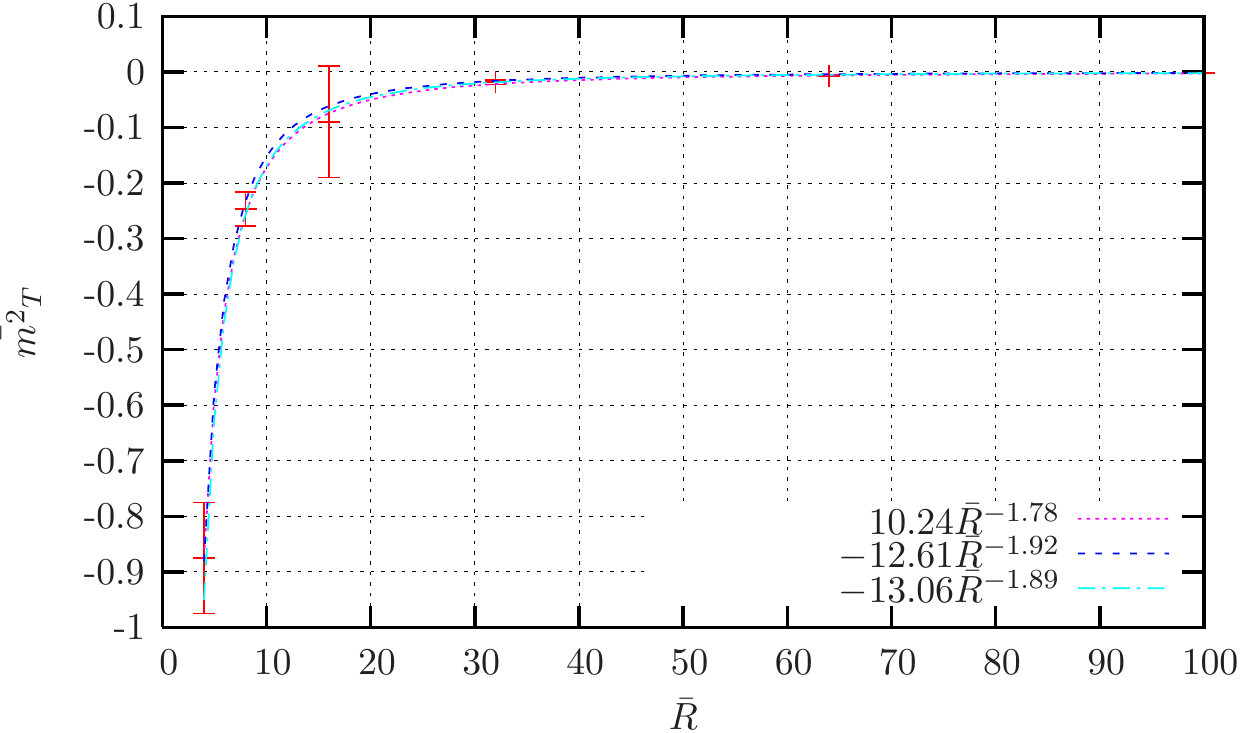}
 \end{center}
\begin{figure}[h]
 \vspace{-0.2in}
 \caption{Estimation of $\bar{m}^2_{T}$ for  $N=12$ and three different fits. 
         The fits were obtained via a function $\bar{\lambda}_{T}=  M(N=12) \bar{R}^{e_2}$. 
         In the first case we fit both parameters $M(N=12),e_2$ and we got 
         $\bar{m}^2_T=(-10.24 \pm0.66)\bar{R}^{-1.78\pm0.04}$.
         In the second case we fixed $e_2=-1.92$ and we fit $Z(N=12)$ as in 
         eq.~(\ref{mfit-triple-N12}) and finally we fixed  $e_2=-1.89$ to get $Z(N=12)=-13.06\pm0.33$.}
\label{f-estimation-triple-m-N12}
\end{figure}

The equation of the collapse is:

\be
  \bar{m}^2_T=(-12.61\pm0.20)\bar{R}^{(-1.92\pm0.1)}. \label{mfit-triple-N12}
\ee
Then, if we  re-scale the $y$-axis by a factor $\bar{R}^{1.92}$ we
stabilize the value of   $ \bar{R}^{1.92} \bar{m^2}_T=-12.61$. 

We conclude that for  $N=12$ the equation of the triple point reads
\bea 
  \bar{\lambda}_T(N=12)&=&(8.20 \pm 0.31)\bar{R}^{-1.28 \pm0.1} \label{fit-ltriple-N12}, \\
  \bar{m}^2_T(N=12)    &=&(-12.61\pm0.2)\bar{R}^{(-1.92\pm0.1)} \label{fit-mtriple-N12}.
\eea

Now, to estimate the dependence on $N$ we apply the same procedure as in
the case $N=12$, for $N=8,16$ and $N=23$.
As in {\bf figures \ref{f-estimation-triple-N12}}-{\bf \ref{f-estimation-triple-m-N12}} 
we will compare three different fits.

 \begin{center}
  \includegraphics[width=4.6in]{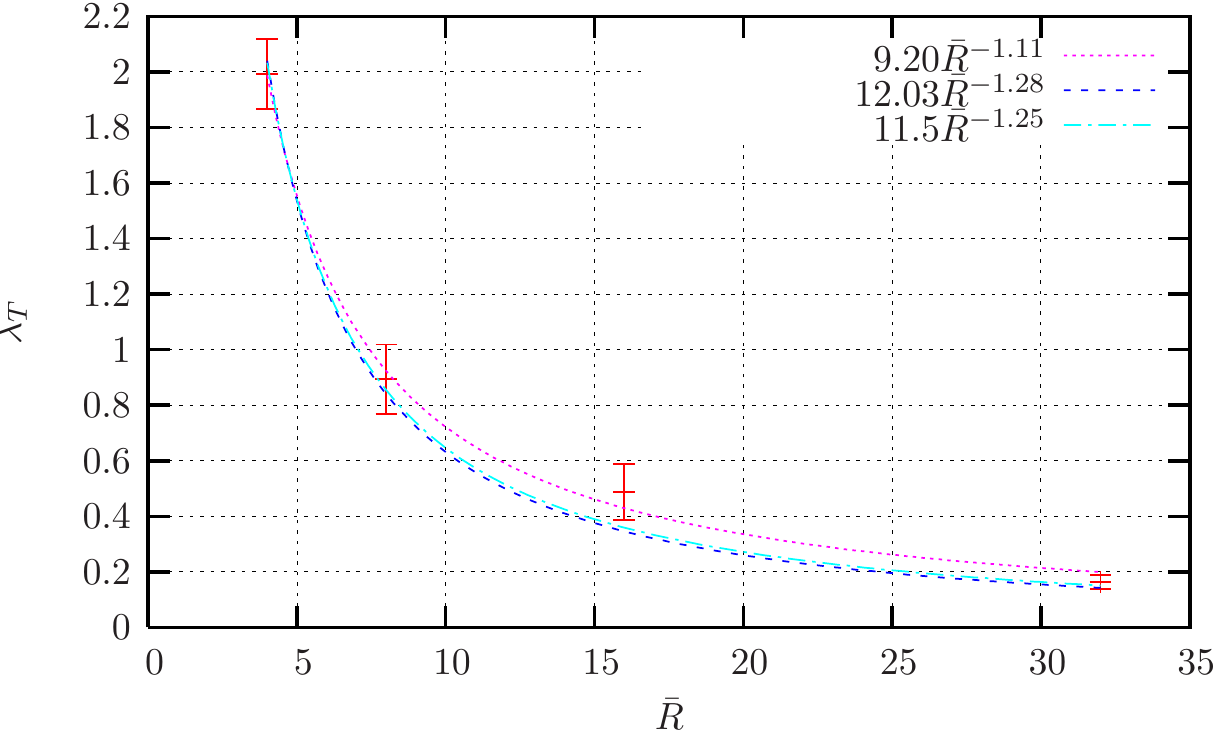}
 \end{center}
\begin{figure}[h]
 \vspace{-0.3in}
 \caption{Estimation of $\bar{\lambda}_{T}$ for  $N=8$ and three different fits. 
          The first fit is $\bar{\lambda}_T(N=8)=(9.20 \pm0.91)\bar{R}^{-1.11\pm0.061}$. 
          The second fit is given by eq.~(\ref{estimation-triple-N8}).}
\label{f-estimation-triple-N8}
\end{figure}
   \begin{center}
  \includegraphics[width=4.6in]{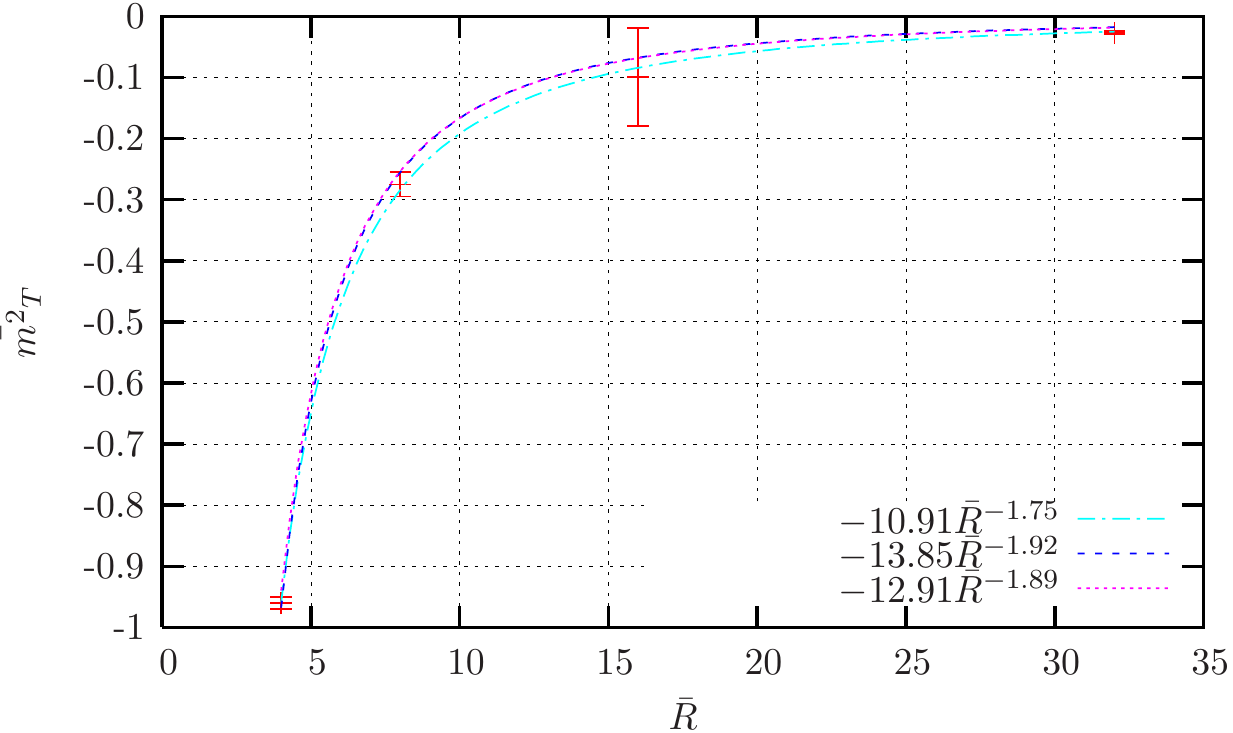}
 \end{center}
\begin{figure}[h]
 \vspace{-0.3in}
 \caption{Estimation of $\bar{m^2}_{T}$ for  $N=8$ and three different fits. 
          The first fit is  $\bar{m^2}_T(N=8)=-(10.91\pm0.88)\bar{R}^{-1.75\pm0.055} $. 
          The second fit is given by eq.~(\ref{estimation-triple-m-N8}).}
\label{f-estimation-triple-m-N8}
\end{figure}

We conclude that for  $N=8$ the equations of the triple point reads
\bea  
   \bar{\lambda}_T(N=8)&=& (12.03\pm 3 ){\bar{R}^{-1.28\pm0.17 }},  \label{estimation-triple-N8}\\
    \bar{m}^2_T(N=8) &=& (-12.91\pm 2) \bar{R}^{-1.92\pm0.22}. \label{estimation-triple-m-N8}
\eea

 \begin{center}
  \includegraphics[width=4.4in]{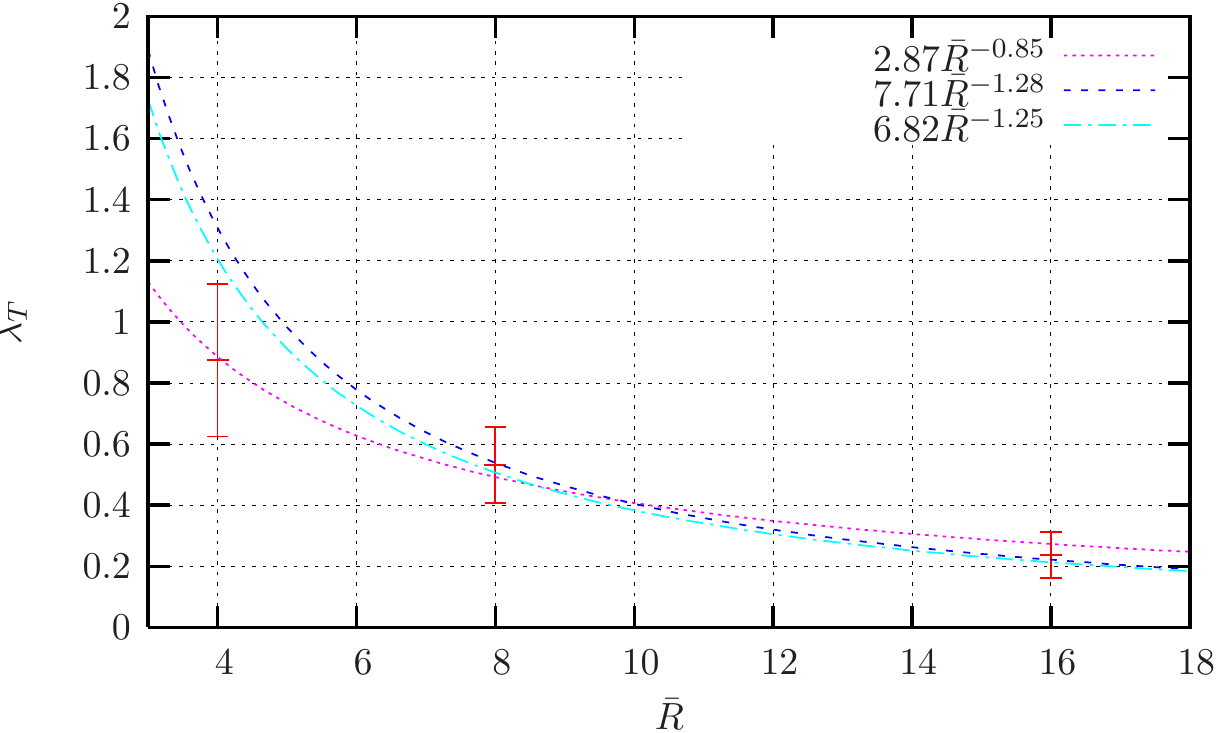}
 \end{center}
\begin{figure}[h]
 \vspace{-0.3in}
 \caption{Estimation of $\bar{\lambda}_{T}$ for  $N=16$. The first fit is $\bar{\lambda}_T(N=16)=(2.87\pm0.60)\bar{R}^{-0.85\pm0.125}$. The second fit is given by eq.~(\ref{estimation-triple-N16}).}
\label{f-estimation-triple-N16}
\end{figure}
   \begin{center}
  \includegraphics[width=4.6in]{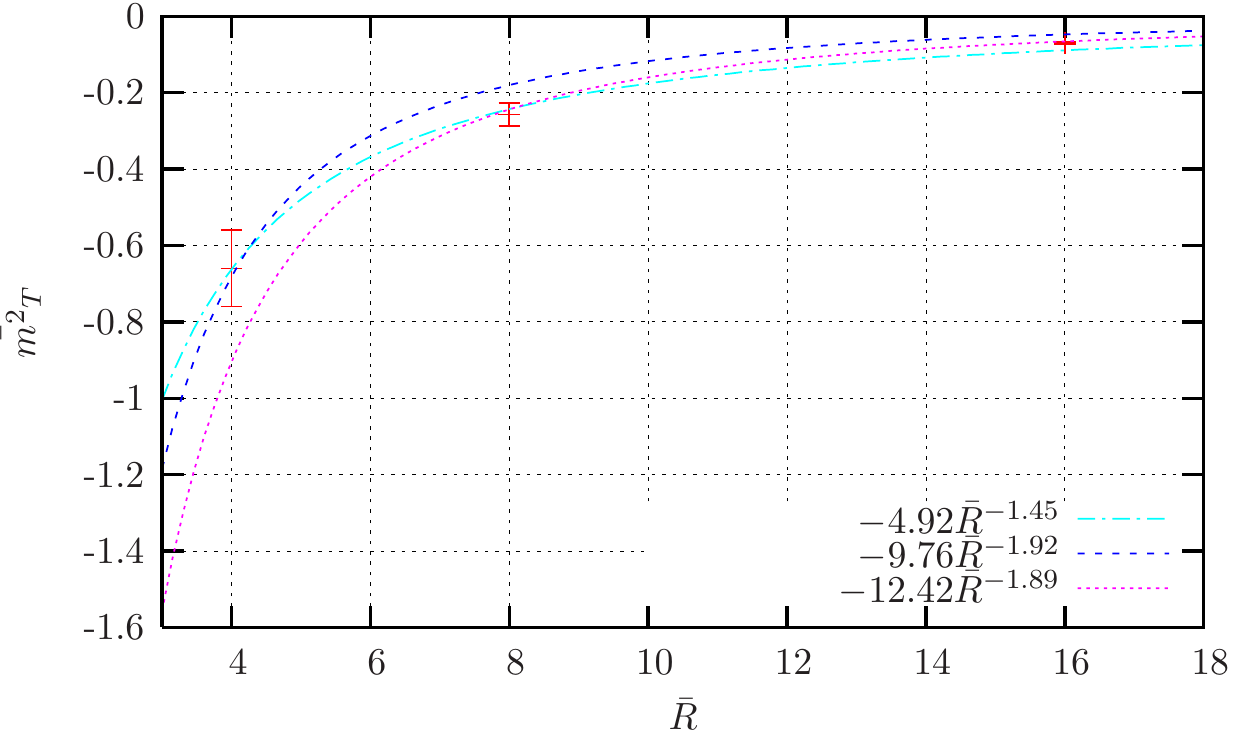}
 \end{center}
\begin{figure}[h]
 \vspace{-0.3in}
 \caption{Estimation of $\bar{m^2}_{T}$ for  $N=16$. The first fit is $\bar{m^2}_T(N=16)= -(4.9263\pm0.91)\bar{R}^{-1.45\pm0.125} $. The second fit is given by eq.~(\ref{estimation-triple-m-N16}).
}
\label{f-estimation-triple-m-N16}
\end{figure}

At $N=16$ the  equations of the triple point are 
\bea 
   \bar{\lambda}_T(N=16)&=& (7.71\pm5) {\bar{R}^{-1.28\pm 0.43}},  \label{estimation-triple-N16}\\
      \bar{m}^2_T(N=16) &=& -(9.76\pm5) \bar{R}^{-1.92\pm0.48}. \label{estimation-triple-m-N16}
\eea

 \begin{center}
  \includegraphics[width=4.6in]{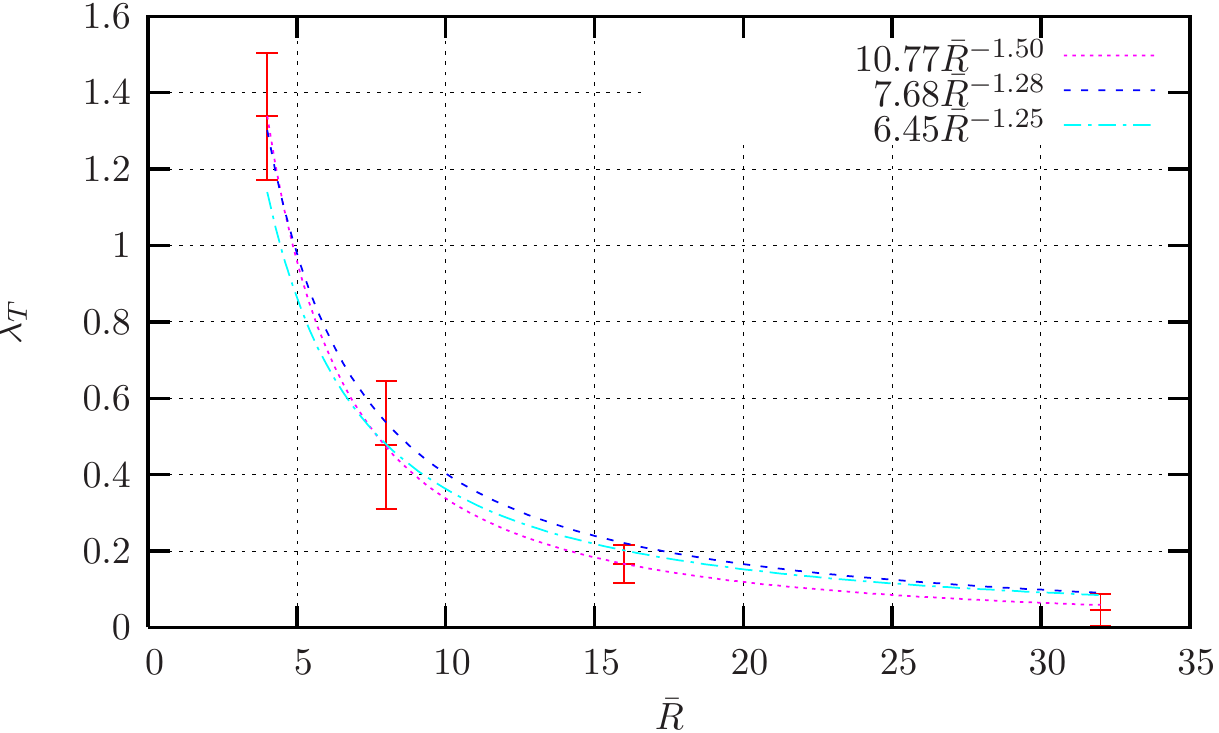}
 \end{center}
\begin{figure}[h]
 \caption{Estimation of $\bar{\lambda}_{T}$ for  $N=23$. The first fit is $ \bar{\lambda}_T(N=23)=
(10.77\pm0.39)\bar{R}^{-1.50\pm0.024}$. The second fit is given by eq.~(\ref{estimation-triple-N23}).}
\label{f-estimation-triple-N23}
\end{figure}
   \begin{center}
  \includegraphics[width=4.6in]{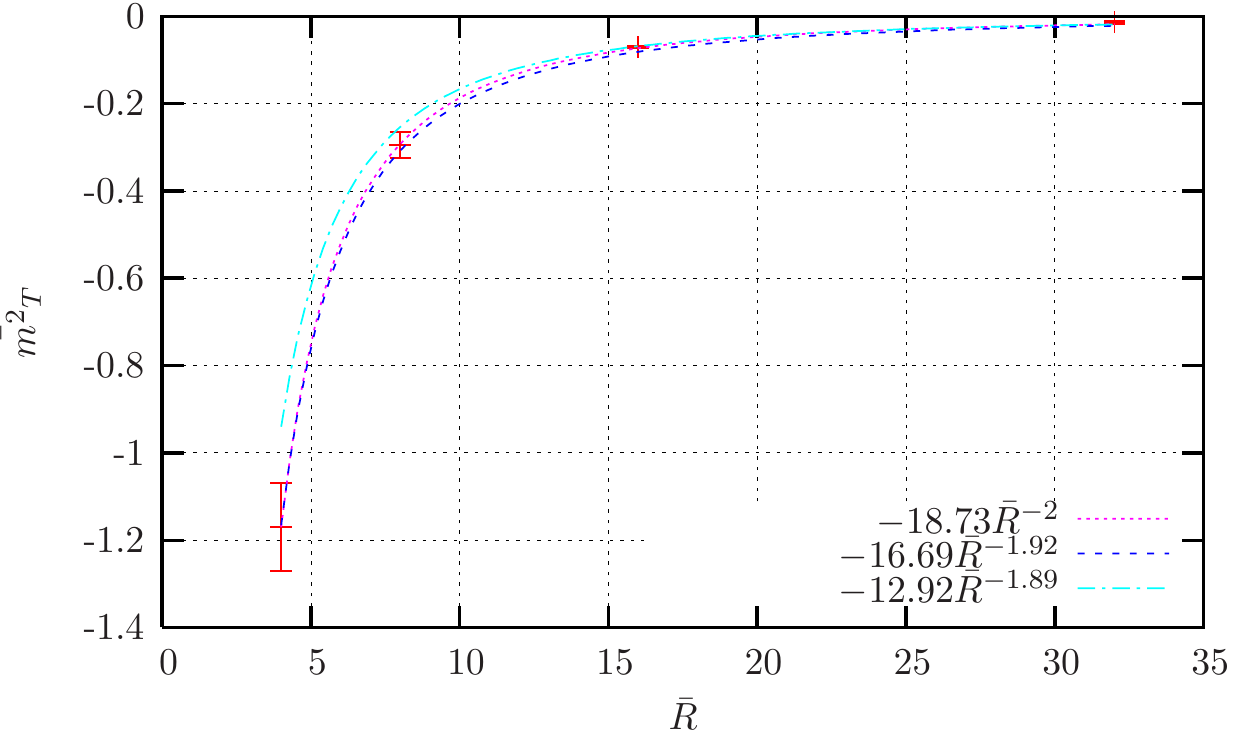}
 \end{center}
\begin{figure}[h]
 \caption{Estimation of $\bar{m^2}_{T}$ for  $N=23$.  The first fit is  
          $\bar{m^2}_T(N=23)= -(18.73\pm0.055)\bar{R}^{-2}$.
          The second fit is given by eq.~(\ref{estimation-triple-m-N23}).}
\label{f-estimation-triple-m-N23}
\end{figure}
At $N=23$ the  equations of the triple point are 
\bea 
      \bar{\lambda}_T(N=23)&=& (7.68\pm 3 ){\bar{R}^{-1.28\pm0.22}}, \label{estimation-triple-N23}\\
      \bar{m}^2_T(N=23) &=& -(16.69 \pm 2){\bar{R}^{-1.92\pm0.12}}. \label{estimation-triple-m-N23}
\eea
We summarise the coefficients $Z(N)$ obtained in eq.~(\ref{fit-ltriple-N12})-(\ref{estimation-triple-m-N23})
in  {\bf Table \ref{table-ZN-1}}.

\begin{table}[h]
    \caption{ Coefficients $Z(N)$ and $M(N)$ considering  \hspace{2cm}
              $\left( \bar{\lambda}_T(N), \bar{m^2}_T(N)\right) =\left( Z(N){\bar{R}^{-1.28}},  M(N)\bar{R}^{-1.92} \right)$. }
    \label{table-ZN-1}
\end{table}

\begin{center}
\begin{tabular}{rcc}\hline \hline
$N$   & $Z(N)$      &       $M(N)$ \\ \hline
$8$   & $12.03\pm3$ &     $-12.91 \pm 2$ \\
$12$  & $8.20\pm2$  &     $-12.61  \pm2$ \\
$16$  & $7.71\pm5$  &     $-12.42  \pm5$ \\
$23$  & $7.68\pm3$  &     $-12.92  \pm4$
\end{tabular}
\end{center}

{\bf Figure \ref{estimation-triple-on-N}} shows the values of $Z(N)$ in 
{\bf Table \ref{table-ZN-1}} and three different fits:
 \begin{center}
  \includegraphics[width=4.8in]{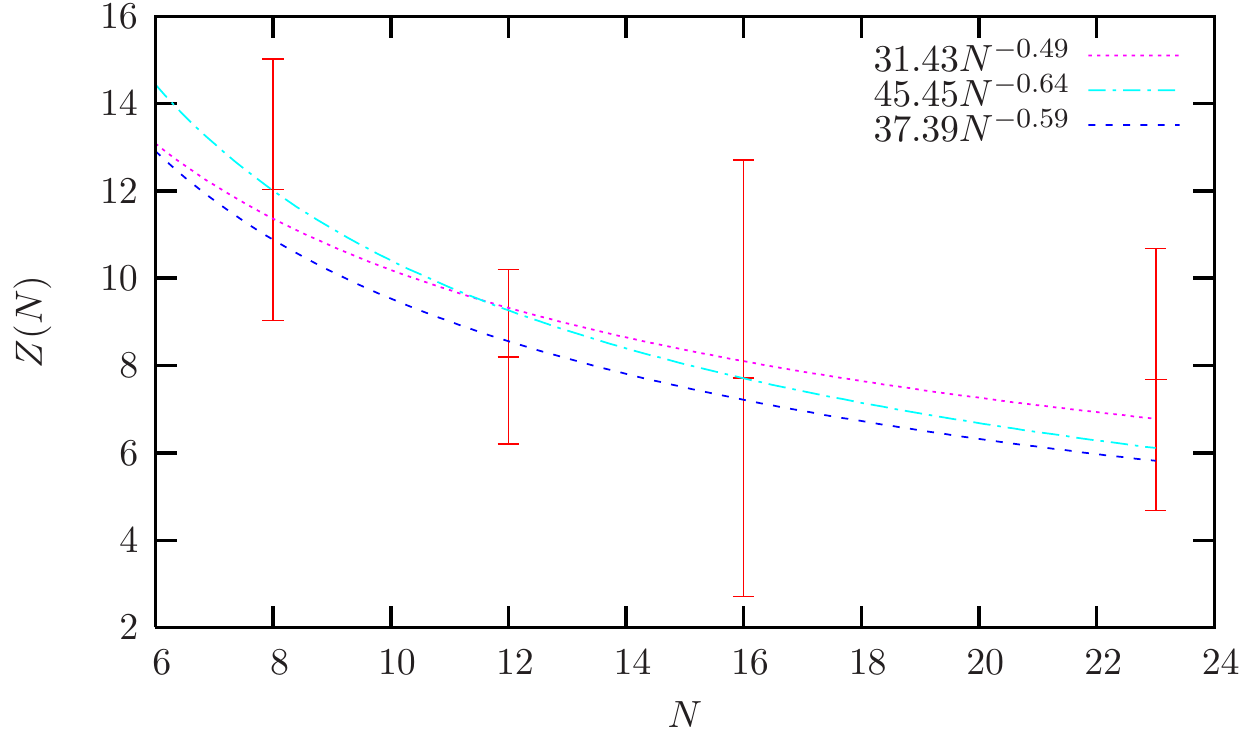}
 \end{center}
\begin{figure}[h]
 \vspace{-0.1in}
 \caption{Estimation of $Z(N)$ for  $N=8,12,16,23$.}
\label{estimation-triple-on-N}
\end{figure}

For $\bar{m^2}_T$ we  present a constant fit 
in {\bf figure \ref{estimation-mtriple-on-N}}.
\pagebreak

 \begin{center}
  \includegraphics[width=4.2in]{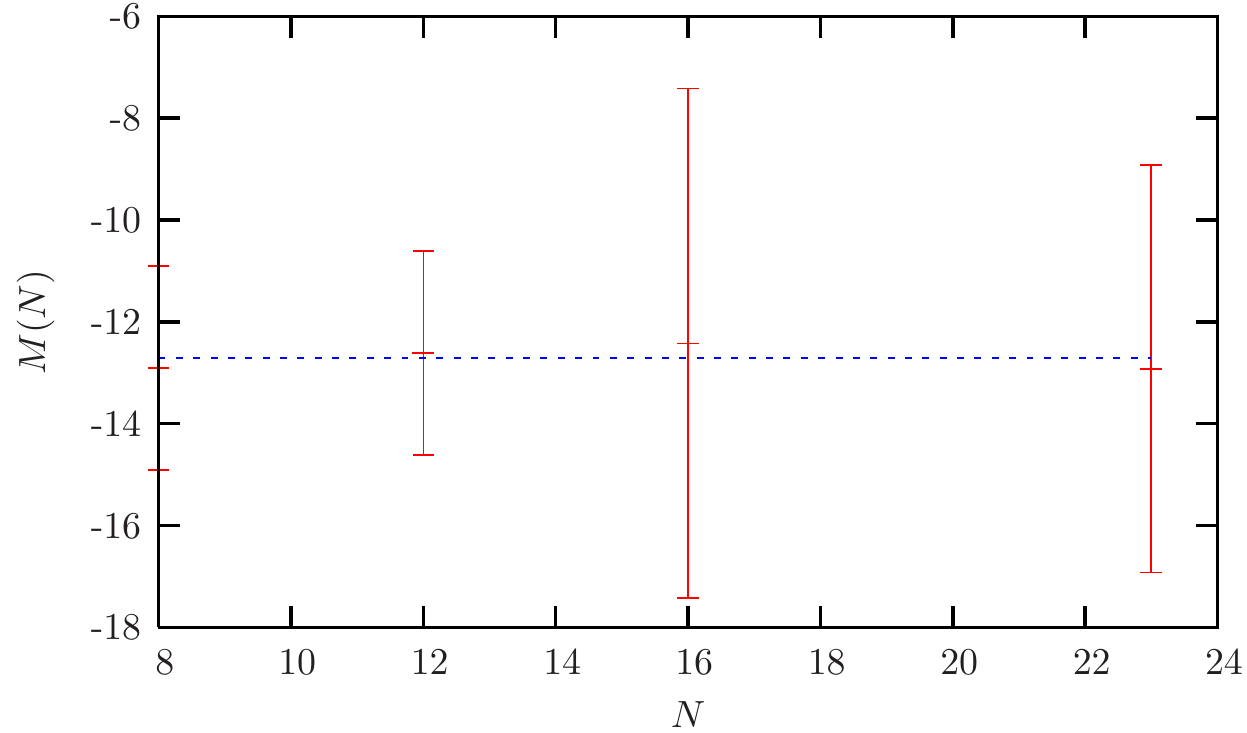}
 \end{center}
\begin{figure}[h]
 \vspace{-0.3in}
 \caption{Estimation of $M(N)$ for  $N=8,12,16,23$.}
\label{estimation-mtriple-on-N}
\end{figure}

Then we get as  estimation for the triple point the eqs.:
\bea 
  \bar{\lambda}_T &=& (45.45\pm15)N^{-0.64\pm0.21} {\bar{R}^{-1.28\pm0.2}}, \label{prediction-of-lambda-triple} \\
      \bar{m^2}_T  &=& -(12.7 \pm1){\bar{R}^{-1.92\pm 0.2 }}. \label{first-prediction-of-m-triple}
\eea
In eq.~(\ref{first-prediction-of-m-triple}) we observe that $\bar{m^2}_T$ seems not to depend on $N$.

Let us consider the expression for the triple point fixing the exponents to $e_1=-1.25, e_2=-1.89$. The coefficients  $Z(N)$and $M(N)$ are slightly different to those obtained with  $e_1=-1.28, e_2=-1.92$, they are summarise in the {\bf Table \ref{table-ZN-2}}.

\begin{table}[h]
    \caption{ Coefficients $Z(N)$and $M(N)$ considering  $\left( \bar{\lambda}_T(N), \bar{m^2}_T(N)\right) =\left( Z(N){\bar{R}^{-1.25}},  M(N)\bar{R}^{-1.89} \right)$. }
    \label{table-ZN-2}
\begin{center}
\begin{tabular}{rcc}\hline \hline
$N$   & $Z(N)$      &       $M(N)$ \\ \hline
$8$   & $11.0\pm3$ &     $-12.91 \pm 2$ \\
$12$  & $8.34\pm2$  &     $-13.06  \pm2$ \\
$16$  & $6.82\pm5$  &     $-12.42  \pm3$ \\
$23$  & $6.45\pm3$  &     $-12.92  \pm4$
\end{tabular}
\end{center}
\end{table}

The values of $Z(N)$ in {\bf Table \ref{table-ZN-2}} can be fitted via the function 
$Z(N)= (37.39\pm30)N^{-0.59\pm0.21}$, then the expression for $\bar{\lambda}_T$ reads
\be 
  \bar{\lambda}_T = (37.39\pm30)N^{-0.59\pm0.21} {\bar{R}^{-1.25\pm0.2}}. \label{prediction-of-lambda-triple-v2}
\ee
For $\bar{m^2}_T$  we observe the values of $M(N)$ in  {\bf Table \ref{table-ZN-2}} 
are practically the same that in  {\bf Table \ref{table-ZN-1}}. We take as prediction  
of  $\bar{m^2}_T$ eq.~(\ref{first-prediction-of-m-triple-v2}):
\be
      \bar{m^2}_T  = -(12.7 \pm1){\bar{R}^{-1.89\pm 0.2 }}. \label{first-prediction-of-m-triple-v2}
\ee

For simplicity from all the fits presented in this section we choose those where the exponents  
can be written as integer multiples  of $0.64$.
\be 
\framebox[1.1\width]{$ \displaystyle
  \left( \bar{\lambda}_T, \bar{m}^2_T \right) = \left( (41.91\pm15) N^{-0.64\pm0.20}
      {\bar{R}^{-1.28\pm0.25}}, - (12.7\pm1) \bar{R}^{-1.92\pm0.20}   \right).$} \label{prediction-of-triple-point}
\ee

Note eq.~(\ref{prediction-of-triple-point})  is in agreement with eq.~(\ref{eq-collapse-uno-dos}) since if we
substitute  $ \bar{\lambda}_T $ in  eq.~(\ref{prediction-of-lambda-triple-v2}) into eq.~(\ref{eq-collapse-uno-dos}), the expected $  \bar{m}^2_T$ turns out to be:

\be
      \bar{m^2}_T  =- 12.99 \bar{R}^{-1.92}, \label{prediction-of-m-triple-v2}
\ee
eq.~(\ref{first-prediction-of-m-triple-v2}) coincides with the expression for $\bar{m}^2_T$ in 
eq.~(\ref{prediction-of-triple-point}) within the errors.

\section{Testing the fit of the transition curve $I-II$}
\label{considering-not-a-line}
In this section we want to check the viability
of a linear fit for the disordered to ordered non-uniform phase transition curve
for values of $\bar{\lambda}$ slightly above  $\bar{\lambda}_{T}$.  

We study the significance of a quadratic term in $\bar{\lambda}$.
We propose 
\be
   \bar{m^2}_c = g_0(N,\bar{R})+ g_1(\bar{R},N)  \bar{\lambda} +g_2(\bar{R},N)  \bar{\lambda}^2. 
    \label{quadratic-fit}
\ee

From the previous section we adopt the assumption  $g_0(N,\bar{R}) \propto \tilde{g}_0(\bar{R}) $
and $g_1(N,\bar{R}) \propto N \tilde{g}_1(\bar{R}) $ and we propose $g_2(N,\bar{R})=  g_2(N)\tilde{g}_2(\bar{R})$.
 
We observe $g_2(N)= const$.
To find $\tilde{g}_i(\bar{R}), i =0,1,2$, we keep $N=12$ fixed and fit the curve (\ref{quadratic-fit}) for the set of data with different $\bar{R}$. {\bf Figure \ref{inter-N16-R8}} is an example for the procedure.

Now we present the plots and fits for $\tilde{h}_0(\bar{R}),\tilde{h}_1(\bar{R})$ 
and $\tilde{g}_i(\bar{R}), i =0,1,2$.

\begin{center}
  \includegraphics[width=3.8in]{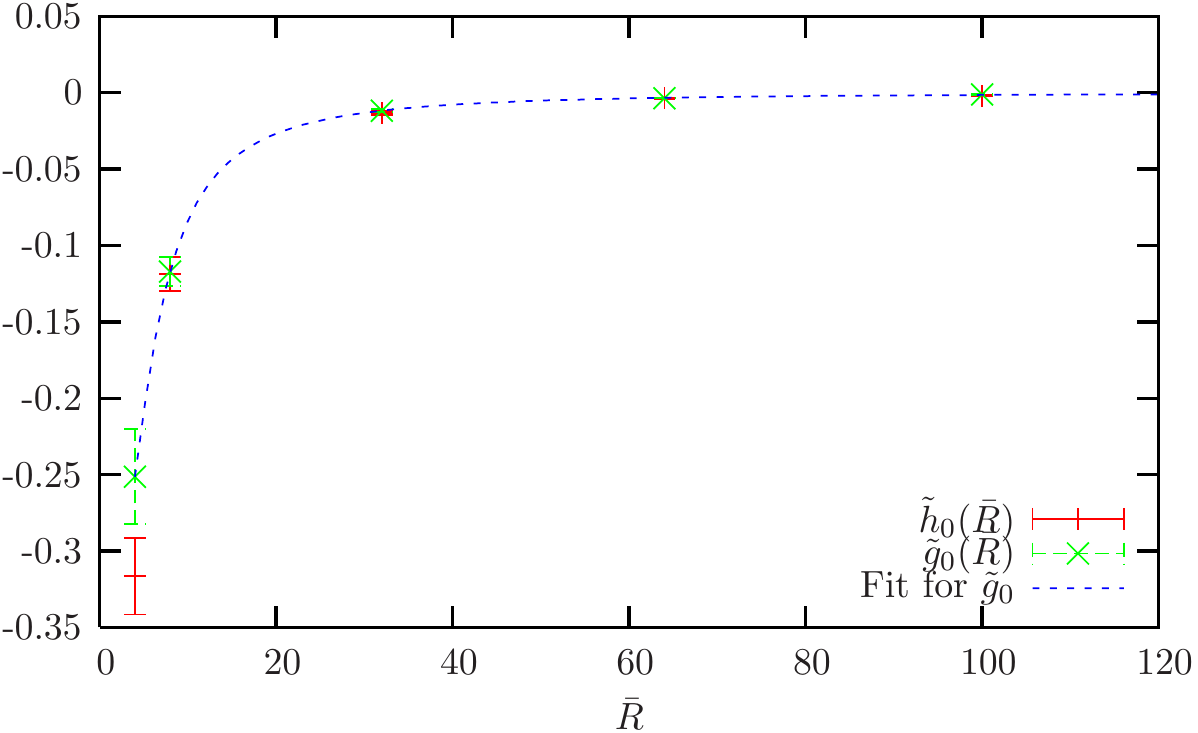}
 \end{center}
\begin{figure}[h]
 \vspace{-0.35in}
 \caption{Coefficients $\tilde{h}_0(\bar{R})$ defined in eq.~(\ref{ffit_linear}) and
          $\tilde{g}_0(\bar{R})$ in eq.~(\ref{quadratic-fit})
          for $\bar{R}=4,8,32,64,100$.
          We can observe  $\tilde{h}_0(\bar{R}) \approx \tilde{g}_0(\bar{R})$ for $\bar{R}>4$.
          The fit for $\tilde{g}_0(\bar{R})$ is given in eq.~(\ref{g0-fit-on-R}).}
\label{g0-0}
\end{figure}
The fit for {\bf figure \ref{g0-0}} is :
\be
   \tilde{g}_0(\bar{R})= 18.33 \bar{R}^{-2.63} -10.2576 \bar{R}^{-1.90877}. \label{g0-fit-on-R}
\ee
\begin{center}
  \includegraphics[width=3.8in]{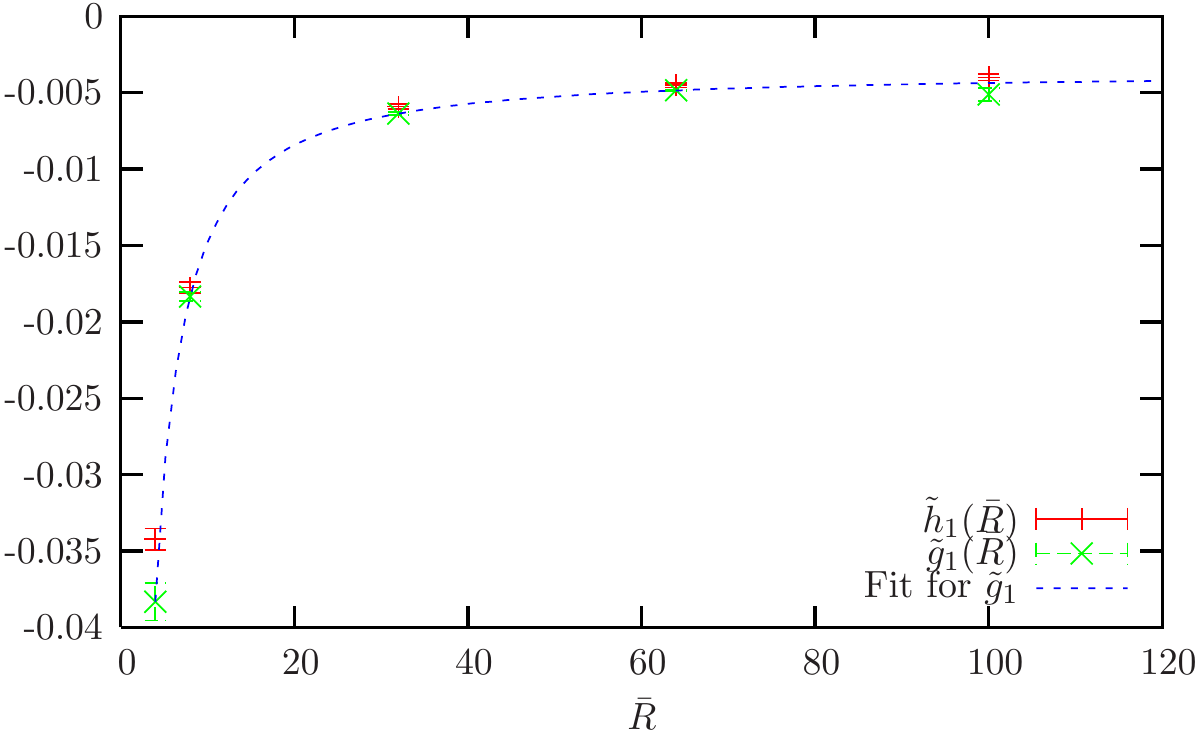}
 \end{center}
\begin{figure}[h]
 \vspace{-0.4in}
 \caption{Coefficients $\tilde{h}_1(\bar{R})$ defined in eq.~(\ref{ffit_linear}) and
          $\tilde{g}_1(\bar{R})$ in eq.~(\ref{quadratic-fit})  
          for $\bar{R}=4,8,32,64,100$.
          We can observe that for $\bar{R}>4$, $\tilde{h}_1(\bar{R}) \approx \tilde{g}_1(\bar{R})$. 
          The fit for $\tilde{g}_0(\bar{R})$ is given in eq.~(\ref{g1-fit-on-R}).}
\label{g1-1}
\end{figure}
The fit for {\bf figure \ref{g1-1}} is :
\be
   \tilde{g}_1(\bar{R})= -0.195 \bar{R}^{-1.25} -0.004 \bar{R}^{-0.015}. \label{g1-fit-on-R}
\ee

As we can observe the coefficients obtained by a fit with a polynomial of degree $1$ and 
$2$ are very similar. Thus the linear fit should be a good approximation for the region 
around the triple point.\footnote{We chose the fit of the form 
$\tilde{h}_0(\bar{R})=a_1 \bar{R}^{e1}+a_2 \bar{R}^{e2}$ since we could not find  a fit  
of the form $\tilde{h}_0(\bar{R})=a_1 \bar{R}^{e1}$ good enough to reproduce our set of data.}

\begin{center}
  \includegraphics[width=4.6in]{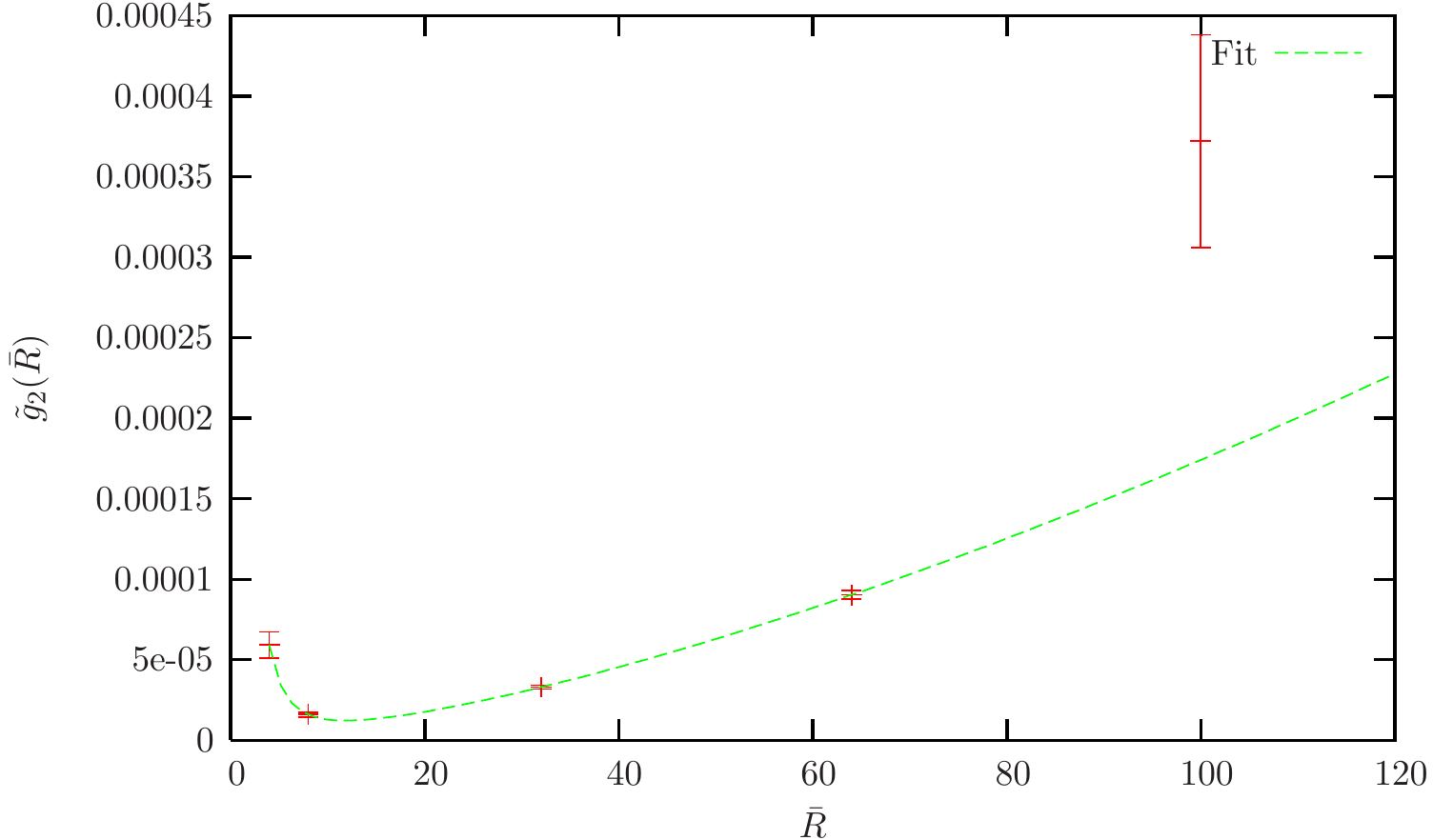}
 \end{center}
\begin{figure}[h]
 \vspace{-0.4in}
 \caption{Coefficients $g_2(\bar{R})$ for $\bar{R}=4,8,32,64,100$.  }
\label{g2-1}
\end{figure}
The fit for {\bf figure \ref{g2-1}} is :
\be
   \tilde{g}_2(\bar{R})= 0.0014\bar{R}^{-2.29}+ 1.97607 \cdot 10^{-7} \bar{R}^{1.47}. \label{fit-g2}
\ee

The final expression for this transition curve is:
\bea
  \bar{m^2}&=& (18.33 \bar{R}^{-2.63} -10.2576 \bar{R}^{-1.90877})-
               (0.195 \bar{R}^{-1.25} +0.004 \bar{R}^{-0.015})N\bar{\lambda} \nonumber 
  \\ & &+ 144 \cdot (0.0014\bar{R}^{-2.29}+ 1.97607 \times 10^{-7} \bar{R}^{1.47}) \bar{\lambda}^2 + O(\bar{\lambda}^3). 
\label{eq-collapse-uno-tres-2}
\eea

\section{Collapse of observables}
\subsection{Collapse for $\bar{\lambda} < \bar{\lambda}_T$}
For this range of  $\bar{\lambda}$ we find that rescaling the $x$-axis by the factor 
$\bar{\lambda}^{-1}$ the susceptibilities $\chi_0$ and $\chi_1$ 
collapse. We show an example in  {\bf figure \ref{sus-1-0-coll-N8-R8}}.

\centerline{
  \includegraphics[width=2.8in]{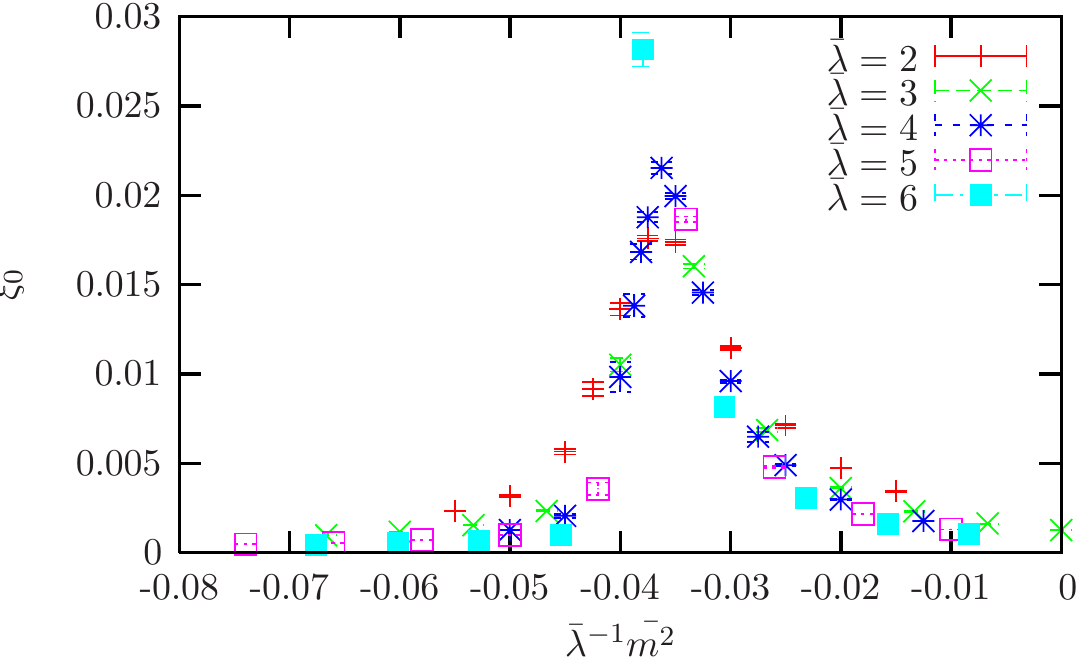}
  \includegraphics[width=2.8in]{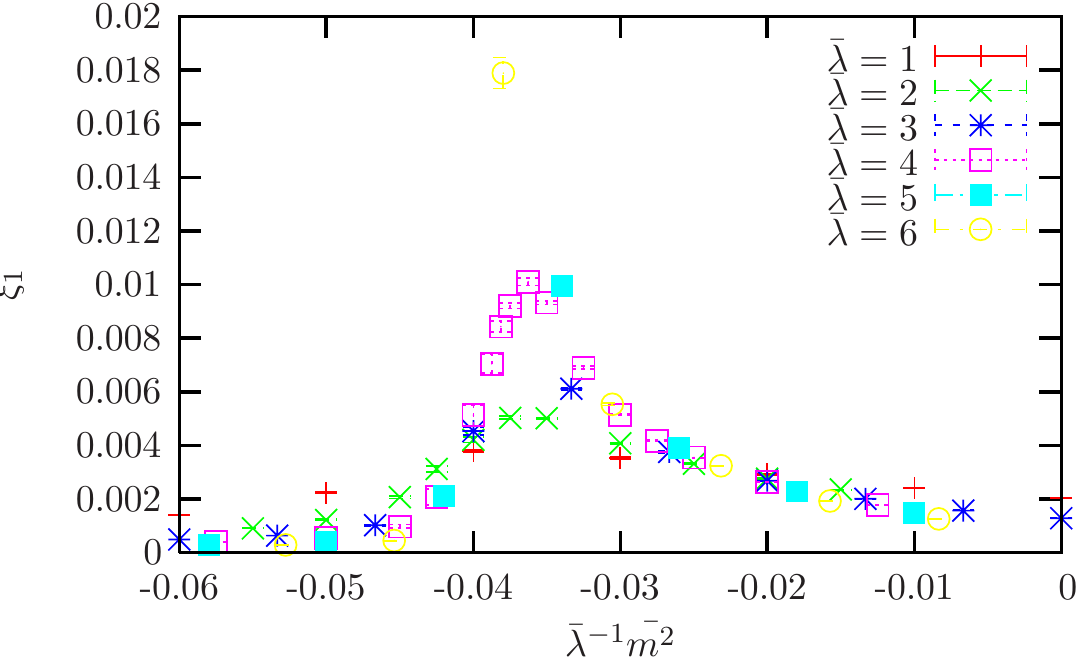}
}
\begin{figure}[h]
\vspace{-0.33in}
\caption{Collapse of $\chi_0$ and $\chi_1$ for $\bar{R}=8$, $N=8$.}
\label{sus-1-0-coll-N8-R8}
\end{figure}

The related graph of {\bf figure \ref{sus-1-0-coll-N8-R8}} for $\la \phi_{all}^2 \ra$ is:  
\begin{center}
  \includegraphics[width=3.5in]{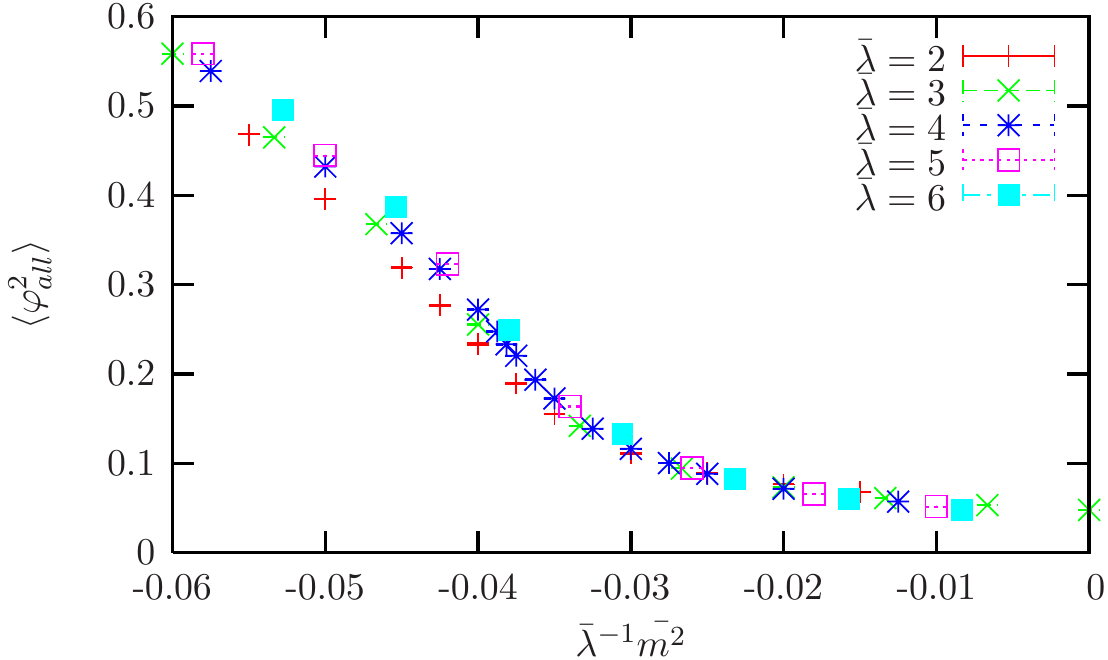}
 \end{center}
\begin{figure}[h]
\vspace{-0.33in}
\caption{Collapse of $\la \phi_{all}^2 \ra$ for $\bar{R}=8$, $N=8$.}
\end{figure}

\begin{center}
  \includegraphics[width=3.5in]{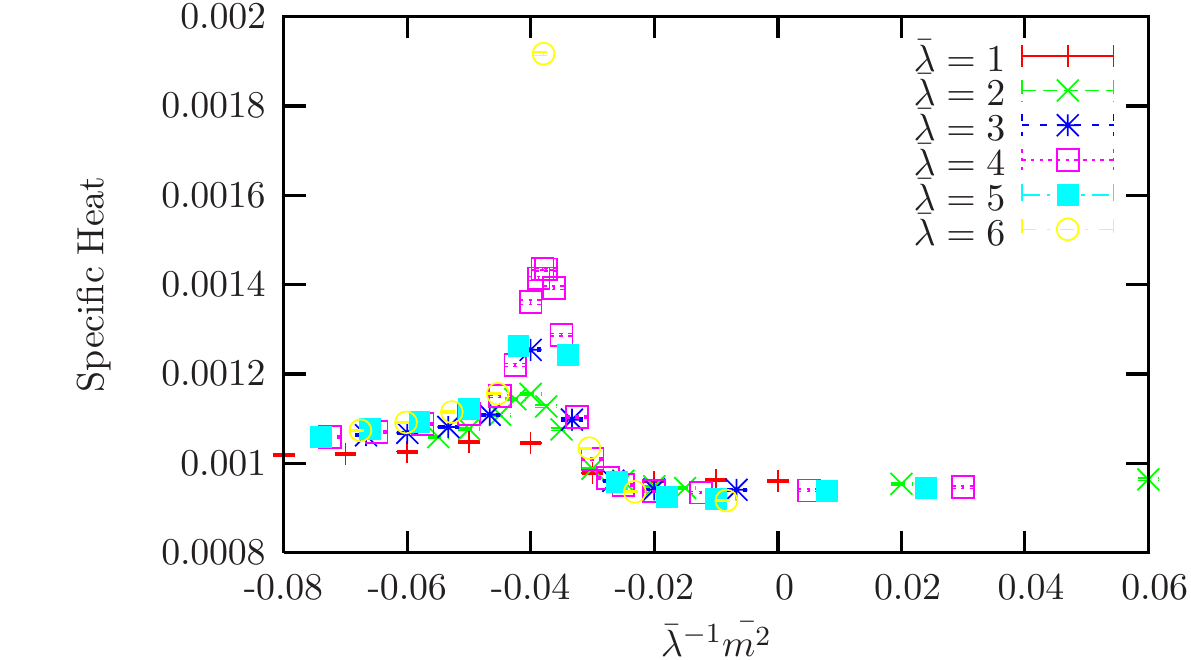}
 \end{center}
\begin{figure}[h]
\vspace{-0.33in}
\caption{Collapse of the Specific Heat in eq.~(\ref{specific_heat}) for $\bar{R}=8$, $N=8$.}
\label{sheat-1-0-coll-N8-R8}
\end{figure}
We can see from {\bf figures \ref{sus-1-0-coll-N8-R8}}-{\bf{\ref{sheat-1-0-coll-N8-R8}}} the equivallence  between the  Specific Heat  criteria and the susceptibilities criteria at this regime. 
This subject is more widely disccussed in appendix \ref{appendix-aside-results}.

\subsection{Collapse for $\bar{\lambda} \gg \bar{\lambda}_T$}
\label{collapse-large-lambda}
If  $\bar{\lambda}$ is sufficiently large we expect  that
the relevant contributions to the action in eq.~(\ref{action_two}) are those 
of the potential terms. Depending on the value of $\bar{R}$ one or both of the kinetic terms 
(fuzzy and time derivatives) can be negligible.

This leads us to {\em reduced models}, i.e.\ models that effectively depend on less parameters 
than those in our model in eq.~(\ref{action_two}).

If the time derivative terms are negligible we would have a chain of $N$-indepen\-dent fuzzy spheres.
If the fuzzy kinetic term is negligible the reduced model is a 
{\em chain of matrix models}, see Refs. \cite{BEynard}-\cite{Shimamune}.
If both kinetic terms are negligible two parameters are redundant, 
this lead us to the 1-matrix model (see Ref.~\cite{Matytsin}).

We devote this section to the study of our model for   $\bar{\lambda} \gg \bar{\lambda}_T$.
We concentrated our analysis in the collapse of the phase transition and some observables. 

The transition curve for this region of the space of parameters can be well 
fitted by the expression (\ref{large-lambda-tc-curve}):
\be
  \bar{m}^2_c= w_0(N,\bar{R})\bar{\lambda}^{w_1} \label{large-lambda-tc-curve}.
\ee
The exponent $w_1$ typically oscillates in  the range $[0.5,1]$ as it can be appreciated in  
Table {\bf \ref{table-critical-and-collapse-tcurve}} in appendix \ref{tables-appendix}.

As an example we present the transition curve disordered --- ordered non-uniform for
$N=8, \bar{R}=16$ and its proposed fit in {\bf{Figure \ref{tc-curve-N8-R16}}}:
\begin{center}
  \includegraphics[width=3.6in]{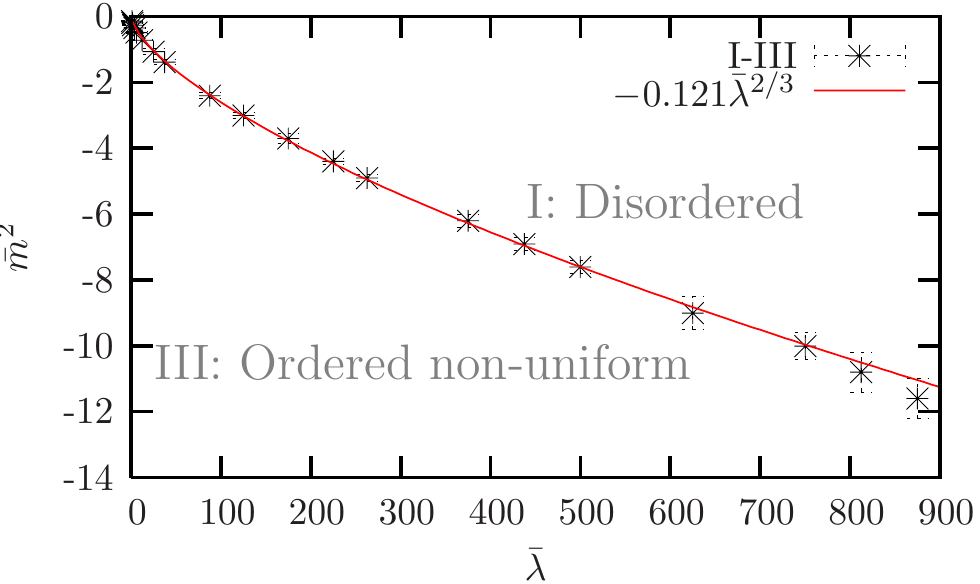}
 \end{center}
\begin{figure}[h]
 \vspace{-0.40in}
 \caption{Transition curve from the disordered phase to the ordered non-uniform phase for 
          $N=8,\bar{R}=16$.          
          The value of the exponent ${w_1}$ in eq.~(\ref{large-lambda-tc-curve}) was fixed to $\frac{2}{3}$.}
\label{tc-curve-N8-R16} 
\end{figure}
\pagebreak

A characteristical behaviour of the specific heat (\ref{specific_heat}) for the this region of 
large $\bar{\lambda}$ is shown in {\bf figure \ref{sus-heat-N8-l5000-R16}} for $\lambda=625$, the
last point in {\bf figure \ref{tc-curve-N8-R16}} ($N\bar{\lambda}=5000$): 
 \begin{center}
  \includegraphics[width=4.6in]{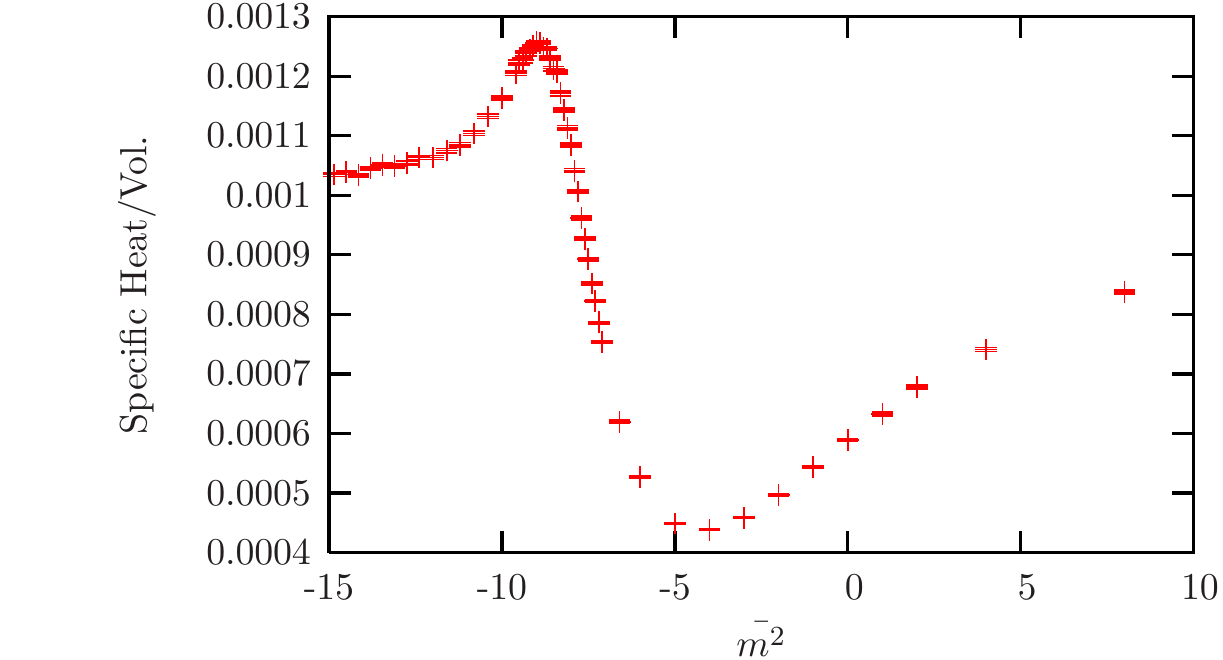}
 \end{center}
\begin{figure}[h]
 \vspace{-0.3in}
 \caption{ Specific heat at $\bar{\lambda}=625$, $\bar{R}=16$, $N=8$.
           The critical point is estimated  at $\bar{m}^2=-9\pm0.5$.}
\label{sus-heat-N8-l5000-R16}
\end{figure}

The corresponding internal energy (\ref{energy}) and its 
partial contributions (\ref{energy1})-(\ref{energy4}) to 
{\bf figure \ref{sus-heat-N8-l5000-R16}} are shown in {\bf figure \ref{contrib-N8-l5000-R16}}

\begin{center}
  \includegraphics[width=4.6in]{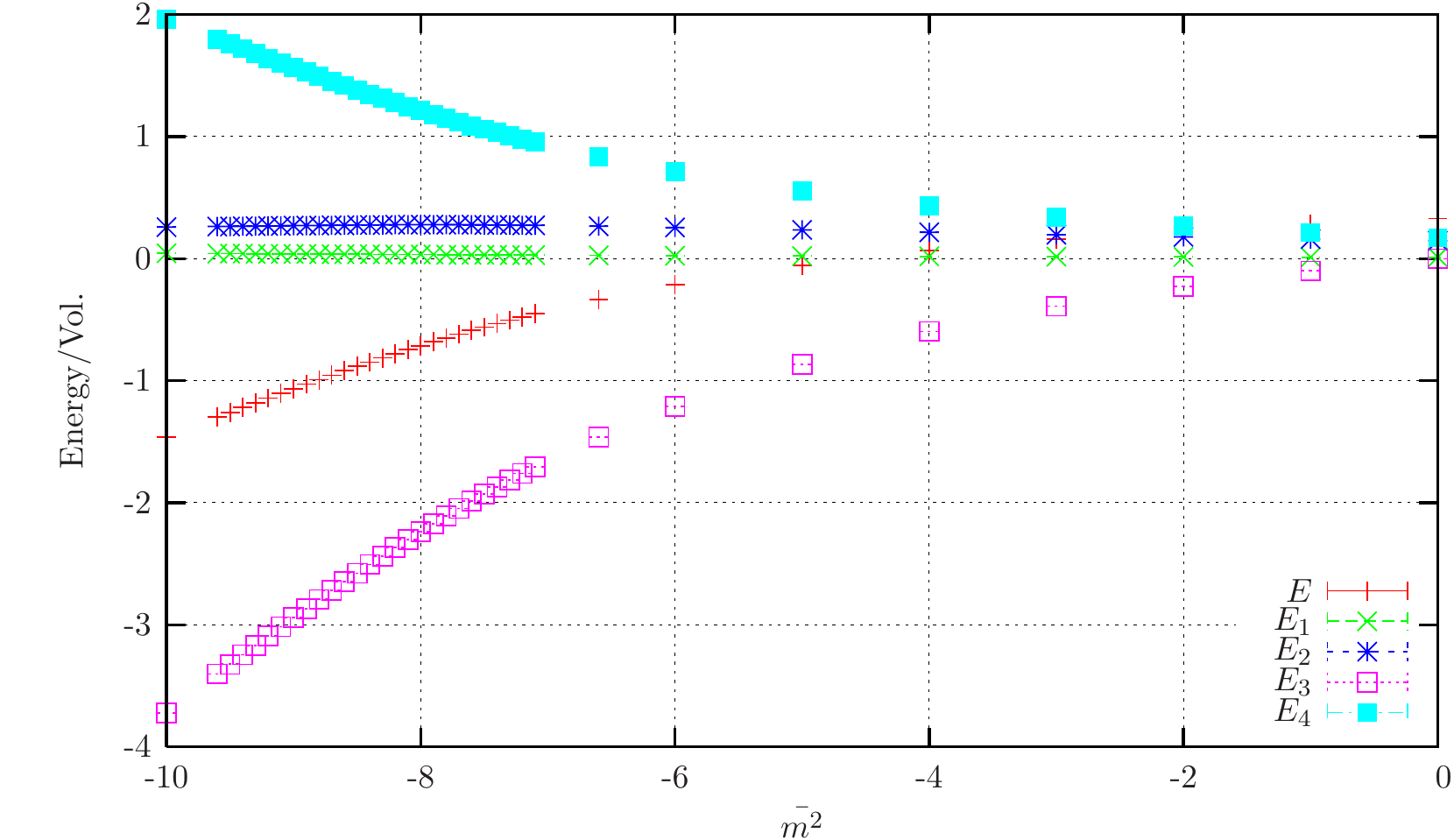}
 \end{center}
\begin{figure}[h]
 \vspace{-0.3in}
 \caption{Internal energy $E$ in eq.~(\ref{energy}) and its partial 
          contributions eqs.~(\ref{energy1})-(\ref{energy4}) at $\bar{\lambda}=625$, $\bar{R}=16$, $N=8$.}
\label{contrib-N8-l5000-R16}
\end{figure}

{\bf Figure \ref{contrib-N8-l5000-R16}} shows that the leading contributions to the 
internal energy is from the potential. Although small, the contributions of the kinetic 
terms are not negligible.

The next step would be to try to collapse in $N$ and $\bar{R}$ the transition curves in 
this region of large $\bar{\lambda}$. 
As we have mentioned, the exponent $w_1$ typically oscillates in the range $[0.5,1]$. 
The optimal value of $w_1$ strongly depends on the considered  region
of $\bar{\lambda}$ --see Table  {\bf \ref{table-critical-and-collapse-tcurve}} in  
appendix \ref{tables-appendix}.

In section \ref{section-transition-disordered-to-ordered-non-uniform} we focused in the region around the triple point, there we considered the transition curve a straight line, i.e.\ $w_1=1$. As we increase the range of $\bar{\lambda}$ the exponent $w_1=1$ decreases.

\subsubsection{The $N$-matrix model}
If the fuzzy kinetic term is negligible we arrive at a  {\em chain of matrix model}.

Ref.~\cite{Shimamune}  studied the large $N$ limit of this model, with a potential $g \phi^4$.
There  a phase transition was predicted at the critical value (\ref{critical_value_chain_matrix})
\be
g_c=\frac{(-\mu^2)^{3/2}}{3 \pi},\label{critical_value_chain_matrix}
\ee
where $g$ is the critical value of the coupling and $\mu^2$ is the squared mass parameter.
Under the appropiate translation to our parameters in eqs.~(\ref{R-prime})-(\ref{lambda-prime}), 
eq.~(\ref{critical_value_chain_matrix}) reads
\be
\framebox[1.1\width]{$ \displaystyle
  \bar{m}^2_c =-\left(\frac{3N^2}{16 \bar{R}^2}\right)^{2/3}\bar{\lambda}^{2/3}.$}
\label{prediction-trans-chain-matrix}
\ee

In terms of the coefficient $ w_0(N,\bar{R})$ and exponent $w_1$ in eq.~(\ref{large-lambda-tc-curve}) the prediction 
for the transition curve disordered ordered non-uniform at large $\bar{\lambda}$ is:
\be
 w_0(N,\bar{R})=-\left(\frac{3N^2}{16 \bar{R}^2}\right)^{2/3}, \quad w_1=\frac{2}{3}.
 \label{prediction_coeff_w}
\ee
The transition in {\bf figure \ref{tc-curve-N8-R16}} obeys eq.~(\ref{prediction-trans-chain-matrix}) (the fit in
{\bf figure \ref{tc-curve-N8-R16}} is $ \bar{m}^2_c=-0.121 \bar{\lambda}^{2/3}$ and the prediction from eq.~(\ref{prediction-trans-chain-matrix}) is $ \bar{m}^2_c=-0.130\bar{\lambda}^{2/3}$).

Some other examples that have the transition in eq.~(\ref{prediction-trans-chain-matrix}) are shown in 
{\bf figures \ref{tc-curve-N16-R16}}-{\bf{\ref{tc-curve-N23-R16}}}.
\begin{center}
  \includegraphics[width=3.8in]{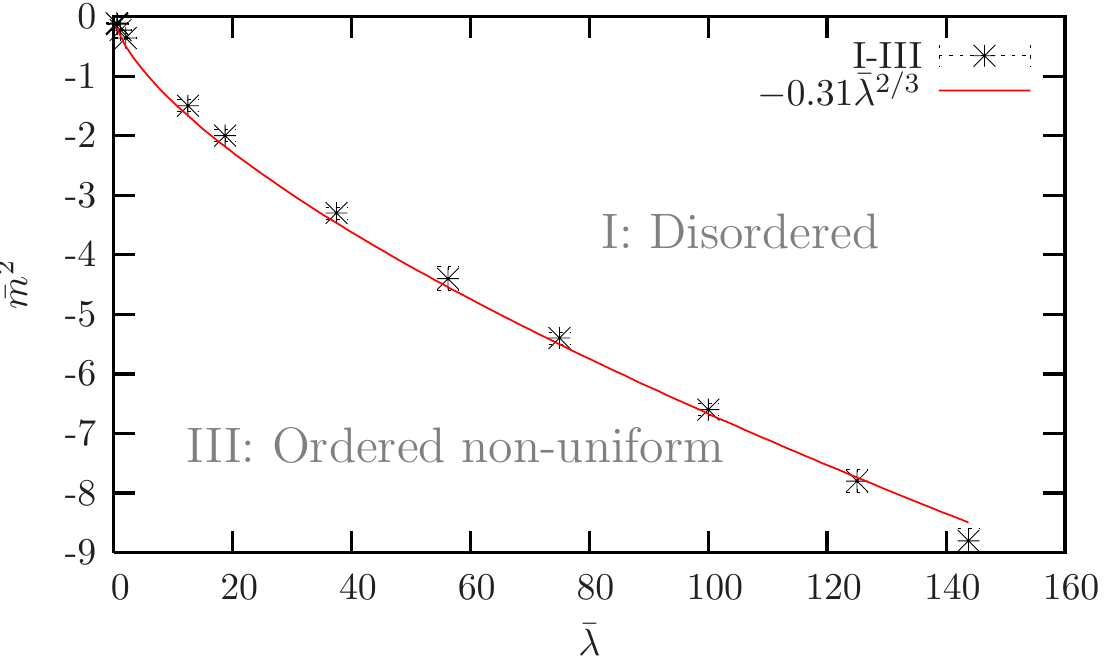}
 \end{center}
\begin{figure}[h]
 \vspace{-0.35in}
 \caption{Transition curve from the disordered phase to the ordered non-uniform phase for 
          $N=16,\bar{R}=16$.}                
\label{tc-curve-N16-R16} 
\end{figure}

\begin{center}
  \includegraphics[width=3.8in]{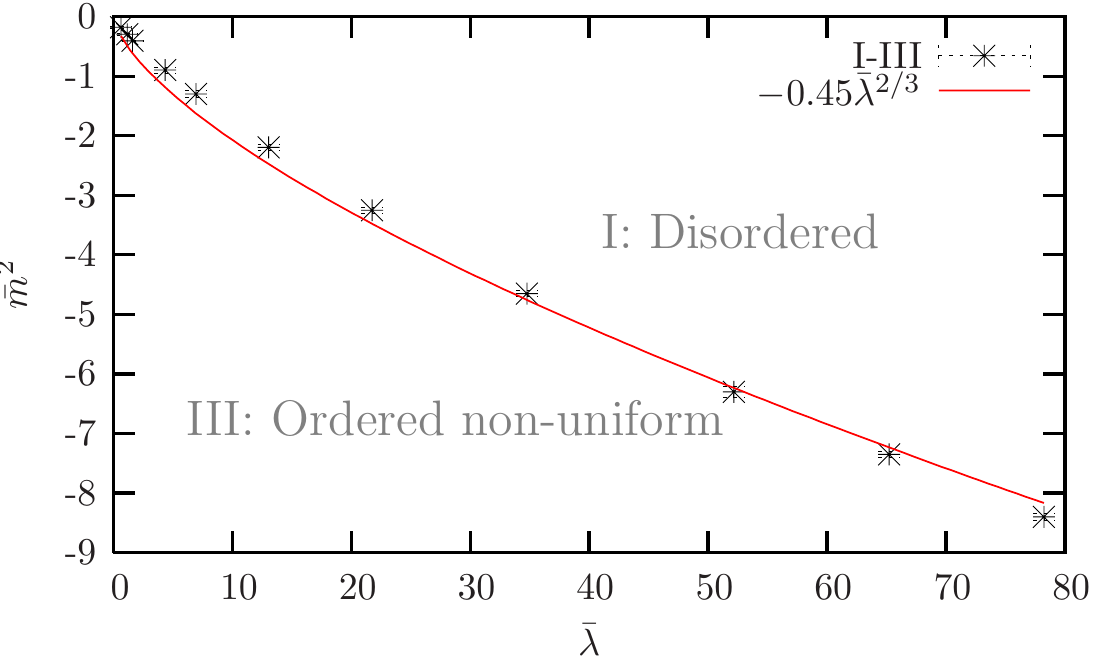}
 \end{center}
\begin{figure}[h]
 \vspace{-0.35in}
 \caption{Transition curve from the disordered phase to the ordered non-uniform phase for 
          $N=23,\bar{R}=16$.}                
\label{tc-curve-N23-R16} 
\end{figure}
In {\bf figure \ref{tc-curve-N23-R16}} we observe that  the fit  
$\bar{m}^2_c =-0.45 \bar{\lambda}^{2/3}$ works well for
$\bar{\lambda}>20$ (the prediction for $w_0(23,16)$ in eq.~(\ref{prediction_coeff_w}) is $-0.53$).
Therefore if we only consider (or measure)  $\bar{\lambda}>20$ for this case the estimated value
for $w_1$ would be $w_1>\frac{2}{3}$. 
This can explain the different estimated values for $w_1$ in {\bf Table {\bf \ref{table-critical-and-collapse-tcurve}}} in appendix \ref{tables-appendix}.
A second possibility is that for those cases in {\bf Table {\bf \ref{table-critical-and-collapse-tcurve}}}
where  $w_1<\frac{2}{3}$ is that we have a transition of a different  nature. 
This is analysed in the following section \ref{the-1-matrix model}.

\subsubsection{The 1-matrix model}
\label{the-1-matrix model}
In the 1-matrix model a transition is expected at 
\be
    m^2_c = -2 \sqrt{N \lambda}, \label{1-matrix-transition-curve}
\ee
according to the notation in  Ref.~\cite{xavier} where it was confirmed numerically, or 
\be
 b_c=-2\sqrt{Nc} \label{1-matrix-transition-curve-v2}
\ee
according to the notation in  Ref.~\cite{Garcia-Martin-OConnor} --see eq.~(\ref{convention-xavier}).

The question is if our model has such a transition. Theoretically for   
very large values of $\bar{\lambda}$ and {\em small} value of $\bar{R}$, 
the dominant term in the action is the potential. 
Then the model should effectively depend on less parameters.
Following Ref.~\cite{notes-models-strong-self-coupling} we obtain the relevant 
parameters in our model form those in the $2$-dimensional model:
\be
  m^2_{2d}=N m^2_{3d}, \quad \lambda_{2d}=N\lambda_{3d}.\label{prediction-trans-large-lambda-2d}
\ee
As a next step we verify  if our model has a transition at
\be
  \bar{m}^2_c \propto \sqrt{N \bar{\lambda}}.\label{prediction-trans-large-lambda}
\ee
With this intention we present {\bf figures \ref{large-lambda-sqrt(Nlambda)-R-8}-\ref{large-lambda-sqrt(Nlambda)-R-16}}.

 \begin{center}
  \includegraphics[width=4.2in]{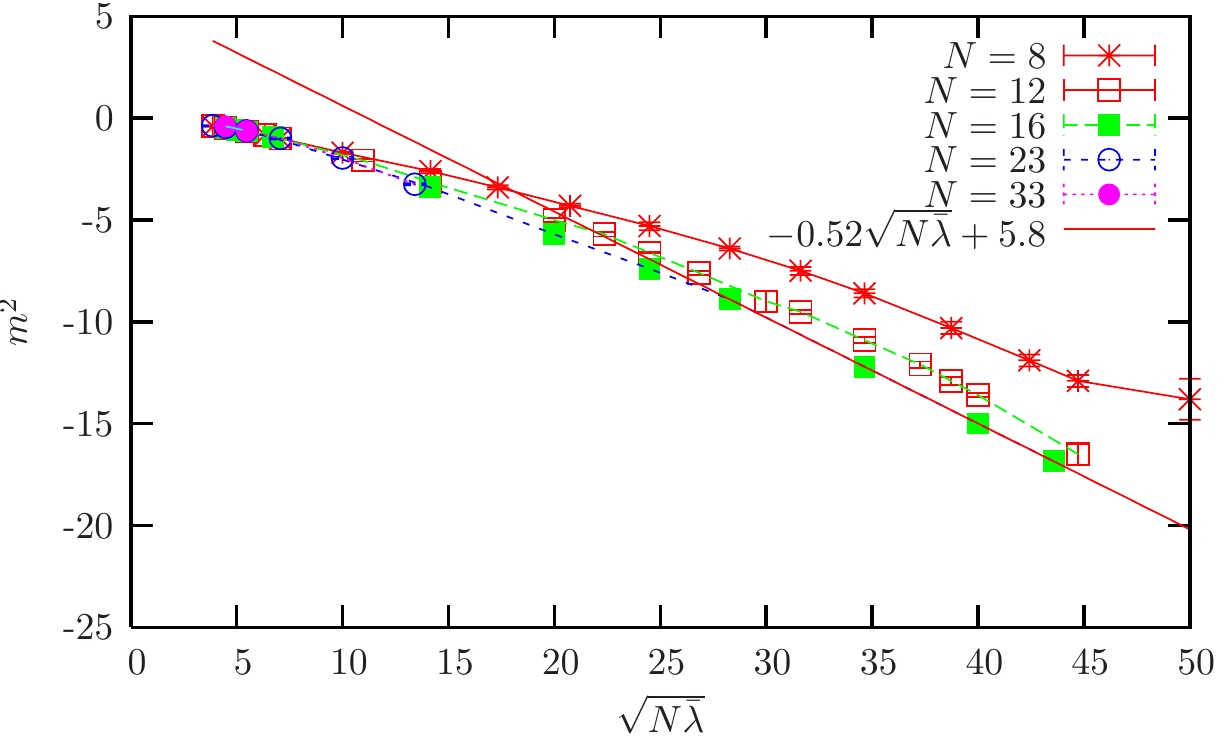}
 \end{center}
\begin{figure}[h]
 \vspace{-0.3in}
 \caption{Phase transition to the disordered phase for several values of $N$ and  $\bar{R}=8$.}
\label{large-lambda-sqrt(Nlambda)-R-8}
\end{figure}

 \begin{center}
  \includegraphics[width=4.2in]{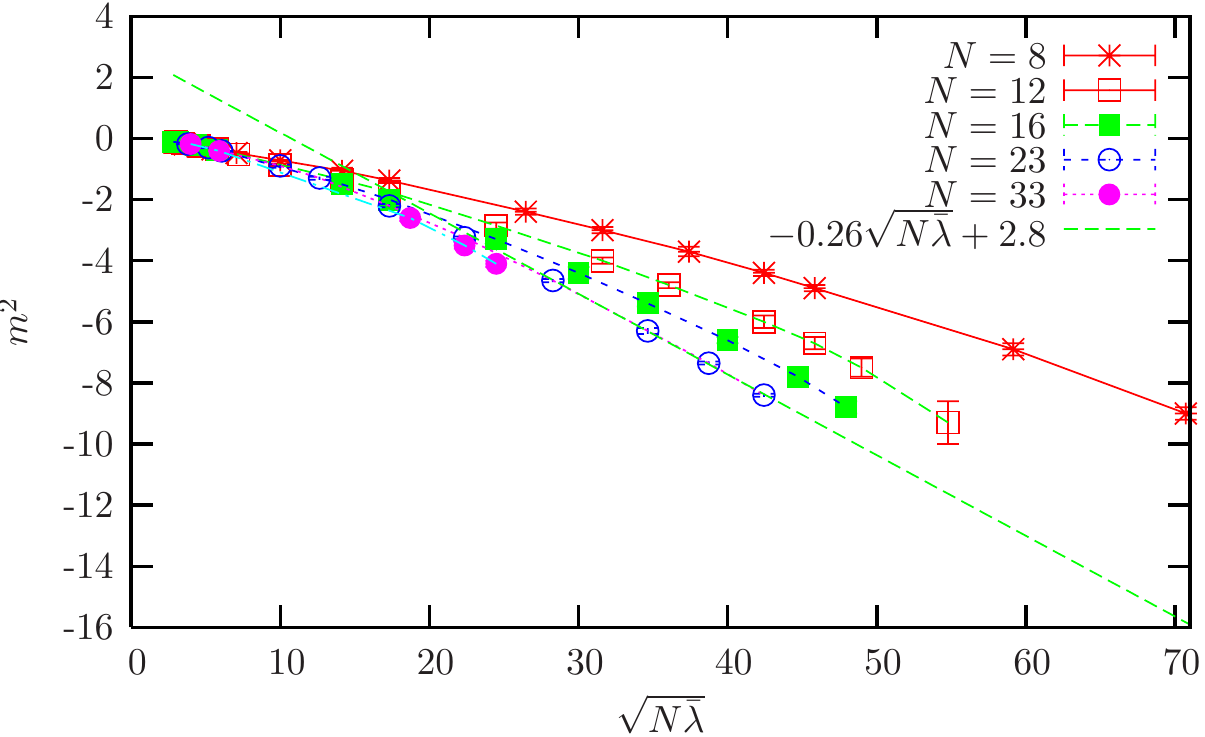}
 \end{center}
\begin{figure}[h]
 \vspace{-0.3in}
 \caption{Phase transition to the disordered phase for several values of $N$ and  $\bar{R}=16$.}
\label{large-lambda-sqrt(Nlambda)-R-16}
\end{figure}

For the points with larger $\bar{\lambda}$ in {\bf figure \ref{large-lambda-sqrt(Nlambda)-R-8}} the $N=16$ data can be fitted by a line eq.~(\ref{fit-of-figure-large-lambda-sqrt(Nlambda)-R-8}):
\be
 \bar{m}^2_c(\bar{R}=8)=-0.52\sqrt{N\bar{\lambda}}+5.8, \label{fit-of-figure-large-lambda-sqrt(Nlambda)-R-8} 
\ee
but a substantial difference is that 
it does not cross the origin as in eq.~(\ref{prediction-trans-large-lambda}).

A similar situation occurs for $\bar{R}=16$ in {\bf figure \ref{large-lambda-sqrt(Nlambda)-R-16}} 
for $N=16,23$ where the fit is given by eq.~(\ref{fit-of-figure-large-lambda-sqrt(Nlambda)-R-16}):  
\be
 \bar{m}^2_c(\bar{R}=16)=-0.26\sqrt{N\bar{\lambda}}+2.8. \label{fit-of-figure-large-lambda-sqrt(Nlambda)-R-16} 
\ee
Note that the coefficients in eq.~(\ref{fit-of-figure-large-lambda-sqrt(Nlambda)-R-8}) for $\bar{R}=8$
are approximately  doubled compared to eq.~(\ref{fit-of-figure-large-lambda-sqrt(Nlambda)-R-16})  
for $\bar{R}=16$.

We conclude that we cannot confirm the phase transition in 
eq.~(\ref{prediction-trans-large-lambda-2d}) for the $2$-dimensional model.
A possible explanation is that in the data obtained
we have not reached a {\em sufficiently large} $\bar{\lambda}$.

Now we focus on the collapse of other observables.

\vspace{.5in}
\subsubsection{Collapse of $\phi^2_{all}$}

In this section we investigate the collapse of the norm of the field in eq.~(\ref{full_power_field}) 
for large $\bar{\lambda}$. This quantity is of interest since for the ordered non-uniform phase for 
large $\bar{\lambda}$ the main contribution to  $\la \phi^2_{all} \ra$  comes form higher  modes,
 and the other quantities we are measuring are related to the lowest modes.
In addition it is known from the 1-matrix model studied in Ref.~\cite{xavier} that for a region where the kinetic term is negligible, $\la \phi^2_{all} \ra$ has the simple form:
\be
  \la  \phi^2_{all} \ra= -\frac{2 \pi r}{\lambda},
\ee
where $r$ is the squared mass parameter.
It would be interesting to check if our model, under the appropriate translation of parameters, 
can reproduce the 1-matrix result as a limiting case.

For $\bar{m^2} <  \bar{m^2}_c$ we found that the norm of the field,  $\la \phi^2_{all}\ra$ 
--- eq.~(\ref{full_power_field}) ---, can be well fitted by a line. A important difference from the 1-matrix model studied in Ref.~\cite{xavier} is that the line does not crosses the origin.
We propose eq.~(\ref{fit-norm-field}) 
\be
  \la  \phi^2_{all} \ra= v_0(N,\bar{R},\bar{\lambda} ) +v_1(N,\bar{R},\bar{\lambda} )\bar{m^2}.
   \label{fit-norm-field}
\ee

A concrete example is presented in  {\bf figure \ref{g-order-mix-N8-l5000-R16}} for the same parameters as in {\bf figures \ref{sus-heat-N8-l5000-R16}}- {\bf \ref{contrib-N8-l5000-R16}}.
 \begin{center} 
   \includegraphics[width=4.4in]{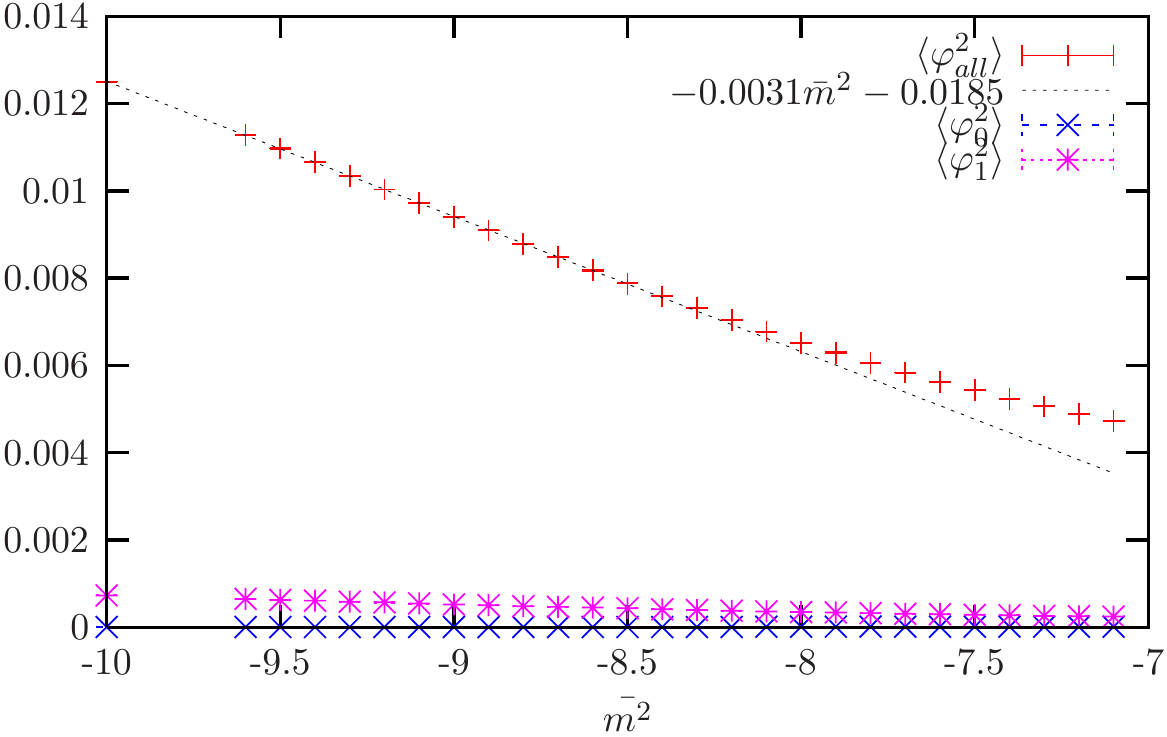}
\end{center}
\begin{figure}[h]
 \vspace{-0.35in}
 \caption{  $\la \varphi_{all}^2 \ra$ and a estimation  vs. $\bar{m^2}$ at $\bar{\lambda}=625$, 
    $\bar{R}=16$, $N=8$. We observe that the contributions from the zero and the first mode,
$\la \varphi_0^2 \ra$ and $\la \varphi_1^2 \ra$ respectively to $\la \varphi_{all}^2 \ra$ are small as we expected.
The critical value is  $\bar{m^2}_c=-9.0\pm0.5$. 
  }
\label{g-order-mix-N8-l5000-R16}
\end{figure}

\paragraph{Note:}
\label{note}
There are some technical difficulties in the measurement of $v_1$ and $v_0$:
The coefficients $v_1$ and $v_0$ must be measured on an appropriate range of $\bar{m}^2$, where $v_1$ and $v_0$ stabilise.
This typically happens for $ \bar{m}^2 < \bar{m^2}_c$, 
i.e.\ in the ordered non-uniform phase where we have thermalisation problems. For the
parameters in {\bf figure \ref{g-order-mix-N8-l5000-R16}} they appear for $\bar{m}^2 < -15$.
Then in {\bf figure \ref{g-order-mix-N8-l5000-R16}} the value
of $v_1$ and $v_0$ stabilise for $-15 <\bar{m}^2 < -8.3$.

In addition to the 1-matrix model, our model has another limiting case: the {\em chain of matrix models}.
In the {\em chain of matrix models} the fuzzy kinetic term should be negligible while in the  
1-matrix model both kinetic terms are negligible.

We conjecture that the non-vanishing coefficient $v_0(N,\bar{R},\bar{\lambda} )$ is related to 
the chain of matrix models and therefore should reduce to  zero for the limiting case of the 1-matrix model.
Therefore the coefficients $v_0(N,\bar{R},\bar{\lambda} )$ and $v_1(N,\bar{R},\bar{\lambda} )$ 
should depend differently on $N$ and $\bar{R}$.

Finally we present the attempts to collapse $\la \phi_{all}^2 \ra$ for  $\bar{\lambda}> \bar{\lambda}_T$ in 
{\bf figures \ref{g-order-coll-N8-R8}- \ref{v2-g-order-coll-N8-R8}}.
 
\begin{center}
  \includegraphics[width=3.5in]{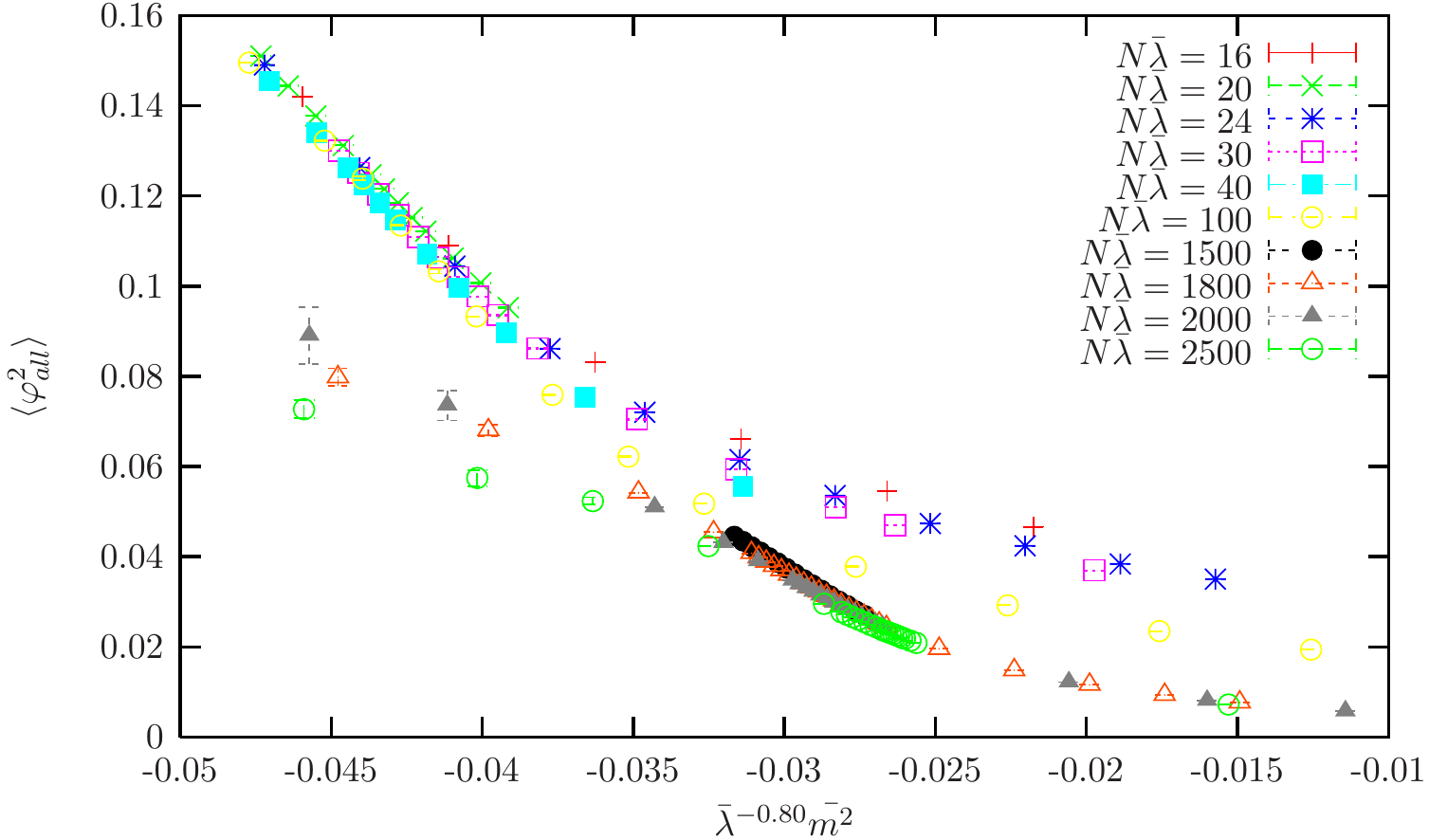}
 \end{center}
\begin{figure}[h]
\vspace{-0.33in}
\caption{Collapse of $\la \phi_{all}^2 \ra$ for $\bar{R}=8$, $N=8$. We re-scale the $x$-axis by the factor
 $\bar{\lambda}^{-0.8}$.  
  For values of $\bar{\lambda}$ slightly above  $\bar{\lambda}_{T}$ the collapse works but not for $\bar{\lambda}\ge1500$. }
\label{g-order-coll-N8-R8}
\end{figure}
In {\bf figure \ref{g-order-coll-N8-R8}} the $x$-axis is re-scaled by the factor $\bar{\lambda}^{\eta}$.
The optimal value of $\eta$ depends on the range of $\bar{\lambda}$.
$\eta=-0.8$ is optimal for $40\ge \bar{\lambda}>\bar{\lambda}_{T}$, 
it gives and acceptable collapse for $\bar{\lambda}<100$ in the range $\eta\in[-0.75,-0.85]$. 
For $2500>\bar{\lambda}>1500$ we choose $\eta=-1$ as in {\bf figure \ref{v2-g-order-coll-N8-R8}}. 
\begin{center}
  \includegraphics[width=3.5in]{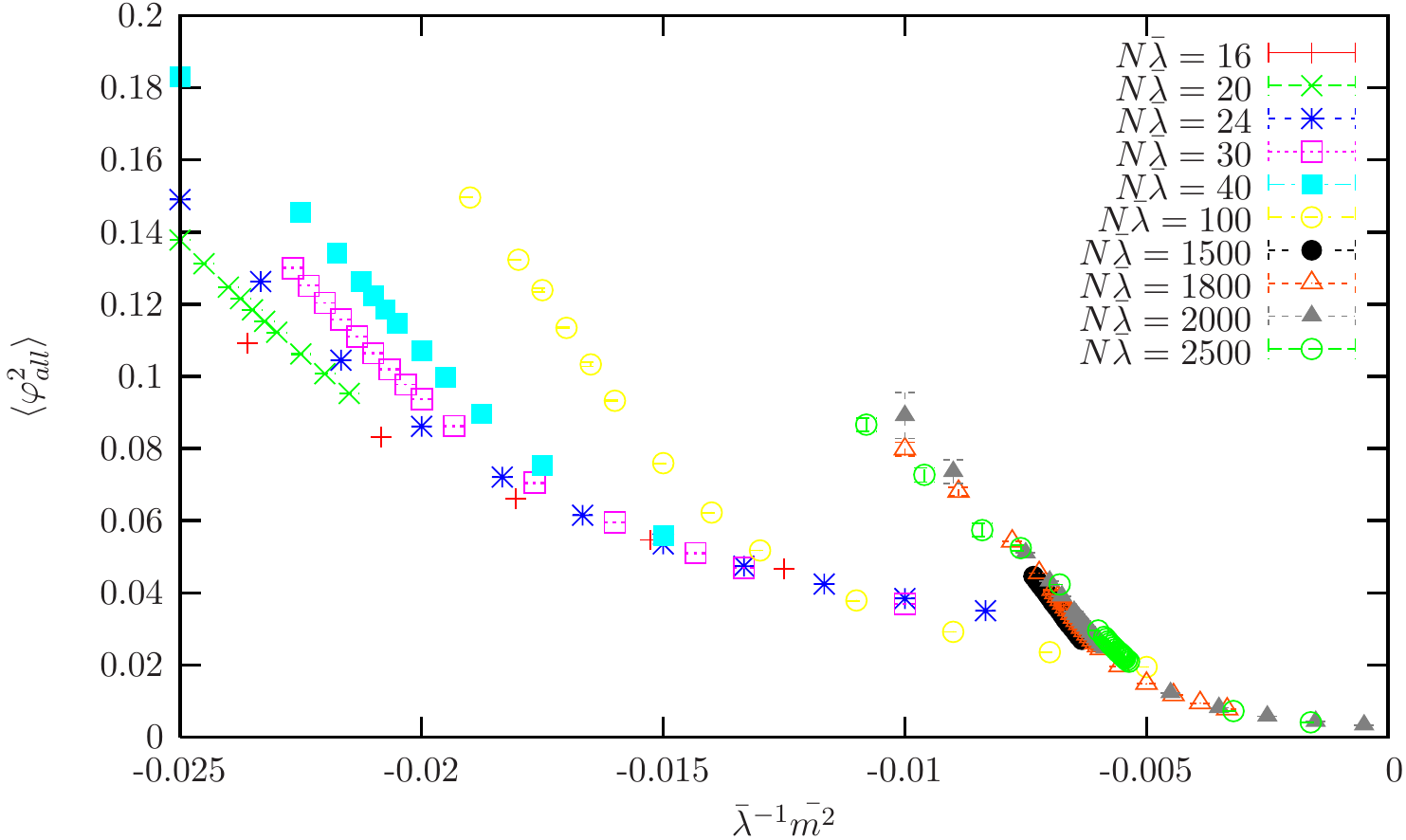}
 \end{center}
\begin{figure}[h]
\vspace{-0.33in}
\caption{Collapse of $\la \phi_{all}^2 \ra$ for $\bar{R}=8$, $N=8$. We re-scale the $x$-axis by the factor
 $\bar{\lambda}^{-1}$.  
  For values of $\bar{\lambda}\ge1500$ the collapse is valid. }
\label{v2-g-order-coll-N8-R8}
\end{figure}

We conclude that the collapse of $\la \phi_{all}^2 \ra$ depends on the range of $\bar{\lambda}$.
For the Specific Heat we have a similar situation a for  $\la \phi_{all}^2 \ra$.

For $\bar{\lambda} \le 100$ we have  in {\bf figure \ref{sheat-1-coll-N8-R8}} 
the collapse of the Specific Heat re-scaling the
$x$ axis by the factor $\bar{\lambda}^{-\eta}$ with $\eta=-0.8$ as in {\bf figure \ref{g-order-coll-N8-R8}}.

\begin{center}
  \includegraphics[width=3.5in]{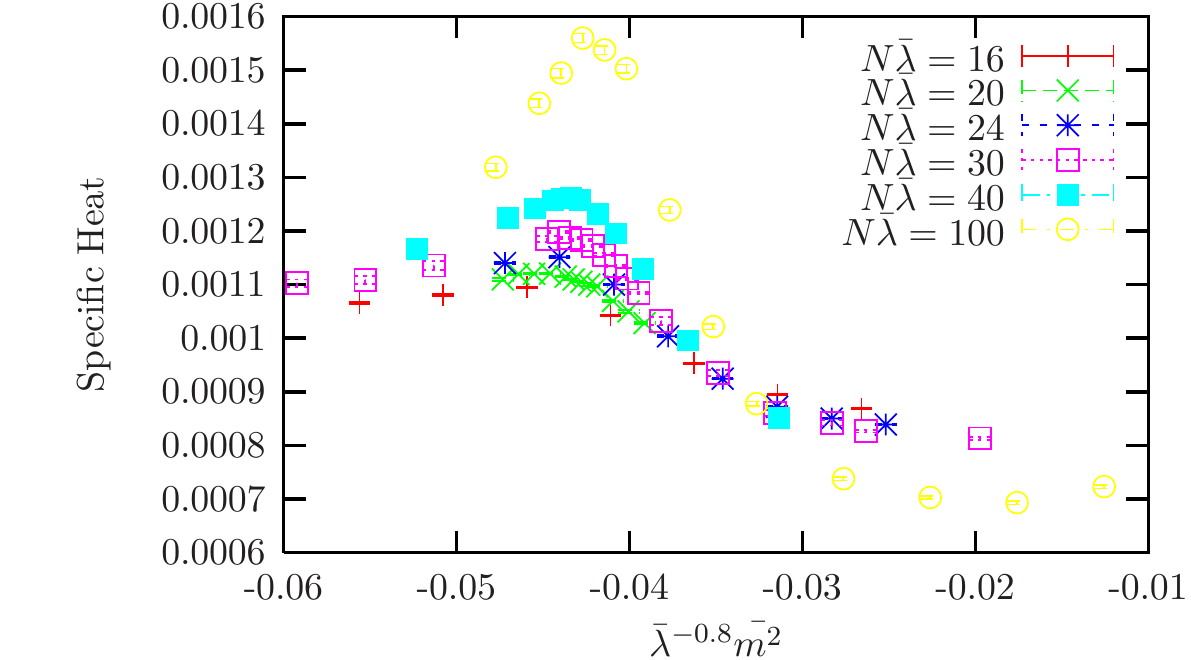}
 \end{center}
\begin{figure}[h]
\vspace{-0.33in}
\caption{Collapse of the Specific Heat in eq.~(\ref{specific_heat}) at $\bar{R}=8$, $N=8$ for
   $100\ge N\bar{\lambda}\ge16$.}
\label{sheat-1-coll-N8-R8}
\end{figure}
In {\bf figure \ref{v2-sheat-1-coll-N8-R8}} we present the collapse of the Specific Heat for the same data that in {\bf figure \ref{sheat-1-coll-N8-R8}}, but re-scaling the $x$ axis by the factor $N^{-2/3}\bar{R}^{4/3}\bar{\lambda}^{-2/3}$.

\begin{center}
  \includegraphics[width=4.5in]{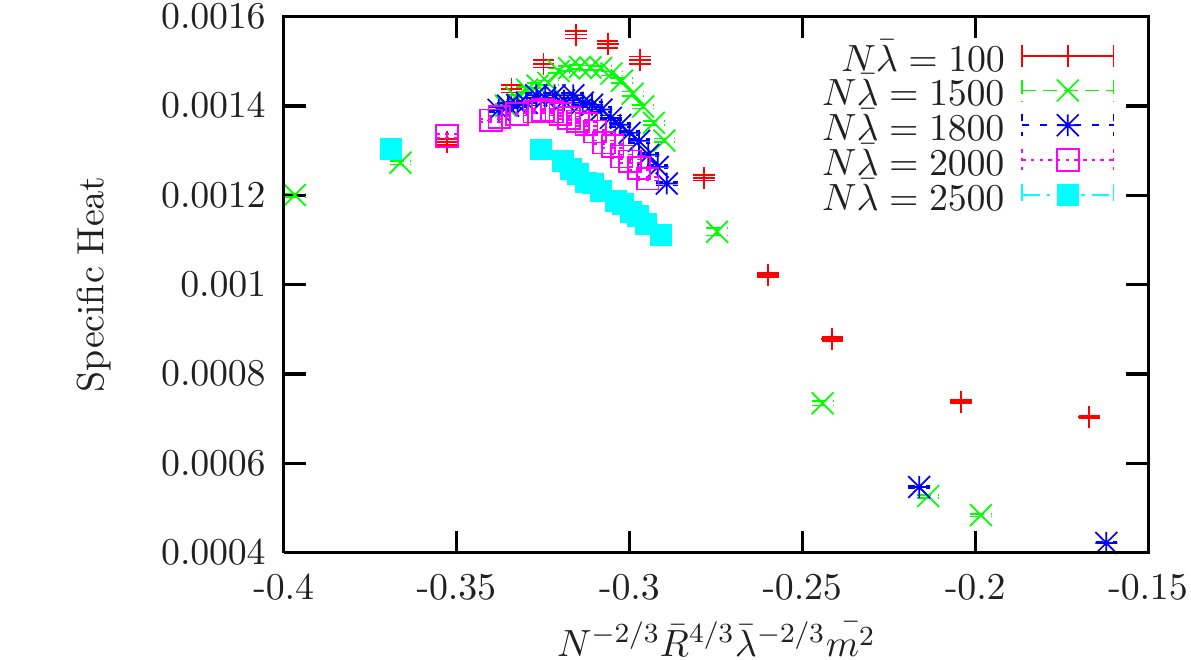}
 \end{center}
\begin{figure}[h]
\vspace{-0.33in}
\caption{Collapse of the Specific Heat in eq.~(\ref{specific_heat}) at $\bar{R}=8$, $N=8$ for
   $N\bar{\lambda}\ge100$.}
\label{v2-sheat-1-coll-N8-R8}
\end{figure}

\begin{center}
  \includegraphics[width=4.5in]{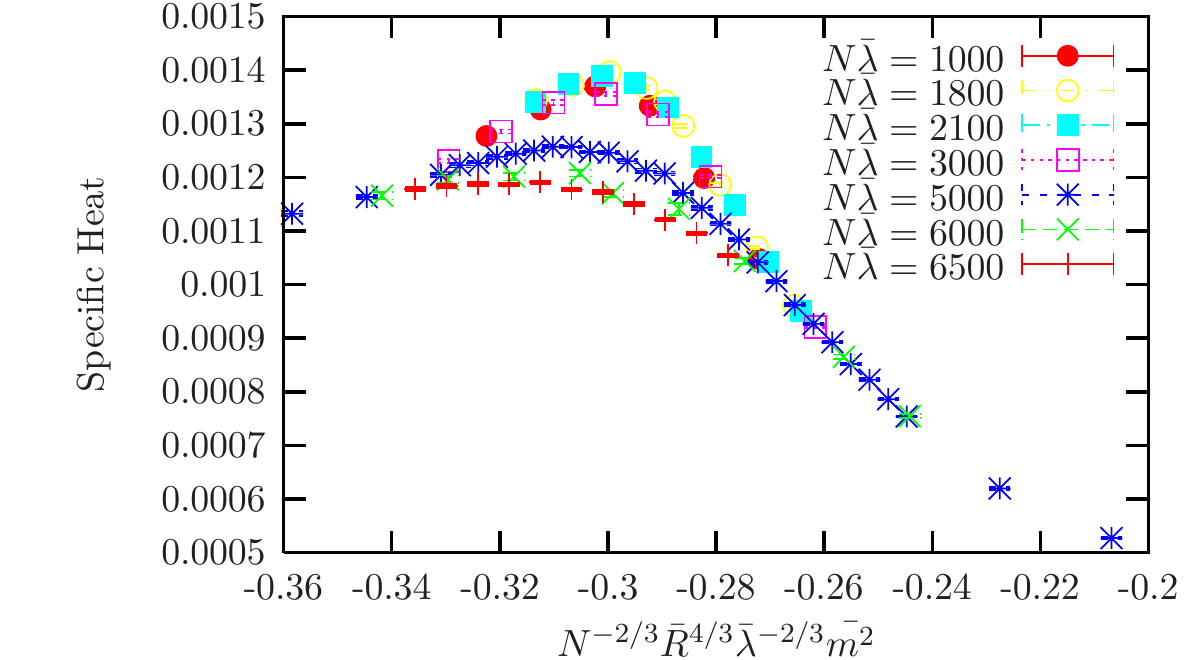}
 \end{center}
\begin{figure}[h]
\vspace{-0.33in}
\caption{Collapse of the Specific Heat in eq.~(\ref{specific_heat}) at $\bar{R}=16$, $N=8$ for
   $N\bar{\lambda}\ge1000$.}
\label{v2-sheat-1-coll-N8-R16}
\end{figure}
From {\bf figures \ref{v2-sheat-1-coll-N8-R8}-\ref{v2-sheat-1-coll-N8-R16}} we observe that re-scaling the $x$-axis by the factor \\  $N^{-2/3}\bar{R}^{4/3} \bar{\lambda}^{-2/3}$ we fix the critical value of 
$\bar{m}^2_c$ to
$N^{-2/3}\bar{R}^{4/3} \bar{\lambda}^{-2/3} \bar{m^2}\approx-0.031$. We have the same
case in {\bf figure \ref{v2-sheat-1-coll-N12-R16}} for $N=12$:

\begin{center}
  \includegraphics[width=4.5in]{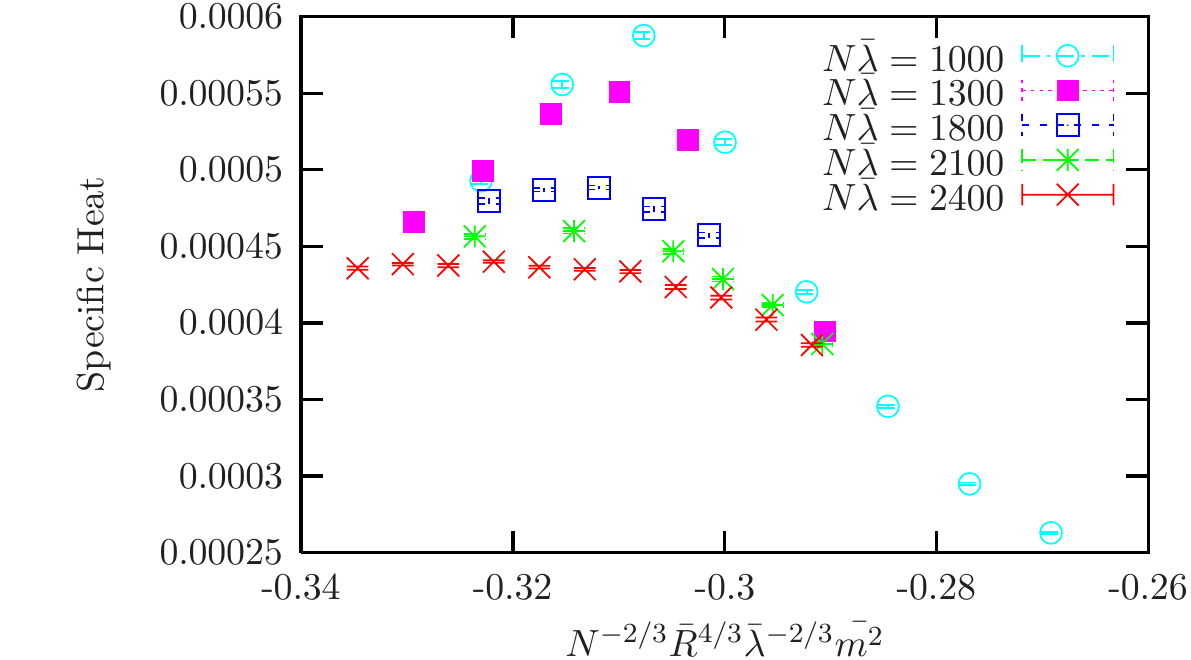}
 \end{center}
\begin{figure}[h]
\vspace{-0.0in}
\caption{Collapse of the Specific Heat in eq.~(\ref{specific_heat}) at $\bar{R}=16$, $N=12$ for
   $N\bar{\lambda}\ge1000$.}
\label{v2-sheat-1-coll-N12-R16}
\end{figure}

We conclude the collapse of observables and the transition curve 
for $\bar{\lambda}>\bar{\lambda}_{t}$ does not lead to the $1$-matrix model. 
Furthermore, the collapse of observables indicates that the exponent of collapse
--- $\eta$ for $\la \varphi_{all}^2\ra$ as in {\bf figures  \ref{g-order-coll-N8-R8}}-
{\bf \ref{v2-g-order-coll-N8-R8}}--- 
depends on the range of  $\bar{\lambda}$.

For the largest values of $\bar{\lambda}$ considered 
our results agree to those predicted for the chain of matrix models in Ref.~\cite{Shimamune} at large $N$, as 
it is shown in {\bf figures \ref{v2-sheat-1-coll-N8-R8}-\ref{v2-sheat-1-coll-N12-R16}}. 
\chapter{Discussion of the results}
\label{discussion}

As it was mentioned in section \ref{limits-of-the-model} we are interested in taking the limit 
$N \longrightarrow \infty$. If we scale $R$ in terms of $N$ we can access different 
limiting models.
The key point in our analysis is the behaviour of the triple point and the phase coexistence 
curves under those limits; it will decide which phases survive at the end.

If the triple point remains finite, it indicates that the three phases exist. 
If the triple point goes to infinity it would indicate the existence of the Ising type phases 
while the ordered non-uniform phases would disappear.

Now we proceed to analyse the model around the critical curves.
Coming back to the discretised model given by eq.~(\ref{action_two}), it
can be re-written as
 
\be
S \left[ \Phi\right] =  \sum_{\trm=1}^{N_{\trm}} 
                            \Tr \Big[
                            \cteA \Phi\left(t\right)
                             \lpl^2 \Phi\left(t\right) +  \cteD
                    \left( \Phi(t+\Delta t) -\Phi(t)  \right)^2
                    + \cteB \Phi^2(t)
                    +\cteC \Phi^4(t) \Big] \label{action_two_v2}
\ee
where we repeat for conveniece the 
definitions (\ref{cte-A})-(\ref{cte-C})
\bea
  \cteA &=& \frac{2  \pi \Delta t  }{N}, \label{cte-A-vv} \\
  \cteD &=& \frac{2  \pi R^2 }{N \Delta t}, \label{cte-D-vv} \\
  \cteB &=& \frac{2  \pi R^2 m^2  \Delta t }{N}, \label{cte-B-vv} \\
  \cteC &=& \frac{ \pi R^2 \lambda  \Delta t }{N}.  \label{cte-C-vv}
\eea

We have the freedom to re-scale the field $\Phi$ in the following way:
\be
   \Phi  \longrightarrow \Phi'= \frac{1}{\sqrt{z}}\Phi  \label{a-re-scaling-of field}
\ee 
and the constants (\ref{cte-A-vv}) - (\ref{cte-C-vv}) change to
\bea
  \cteA &=& \frac{2  \pi z \Delta t  }{N}, \label{cte-A-vv-re} \\
  \cteD &=& \frac{2  \pi R^2 z }{N \Delta t}, \label{cte-D-vv-re} \\
  \cteB &=& \frac{2  \pi R^2 m^2 z \Delta t  }{N}, \label{cte-B-vv-re} \\
  \cteC &=& \frac{ \pi R^2 \lambda  z^2\Delta t }{N}.  \label{cte-C-vv-re}
\eea

As we mentioned in section \ref{details_parameters} we can choose $z$ such that we fix 
one of the constants (\ref{cte-A-vv-re})-(\ref{cte-C-vv-re}). We fix  $A=2\pi$, i.e.\
\be
  z= \frac{N}{\Delta t} \label{fixing_scaling}.
\ee
Under the re-scaling (\ref{fixing_scaling})
the constants (\ref{cte-A-vv-re})-(\ref{cte-C-vv-re}) change to
\bea
  \cteA &=& 2  \pi  \label{cte-A-vv-z-fixed}, \\
  \cteD &=& \frac{2  \pi R^2 }{(\Delta t)^2}= 2 \pi \bar{R}^2,  \label{cte-D-vv-z-fixed} \\
  \cteB &=& 2  \pi R^2 m^2 =2 \pi \bar{R}^2 \bar{m^2}, \label{cte-B-vv-z-fixed} \\
  \cteC &=& \frac{ \pi R^2 \lambda  N}{\Delta t }= \pi N \bar{R}^2 \bar{\lambda}.  \label{cte-C-vv-z-fixed}
\eea

We re-write the obtained expression of the triple point of eq.~(\ref {prediction-of-triple-point}) 
making the substitutions of eqs.~(\ref{R-prime})-(\ref{lambda-prime}) to get
\be 
  \left( \Delta t \lambda_T, (\Delta t)^2 m^2_T \right) = \left(
  41.91\left(\frac{(\Delta t)^2}{ N R^2}\right)^{\gamma}
      , - 12.7 \left( \frac{\Delta t}{R}\right)^{3 \gamma}  \right) \label{f-point-R-function-N}
\ee
with $\gamma=0.64\pm 0.2$.

We scale $R=N^{\beta}$ for $\beta>0$, then $R\longrightarrow \infty$
as $N \longrightarrow \infty$. \\
Eq.~(\ref{f-point-R-function-N}) changes
to eq.~(\ref{f2-point-R-function-N})
\be 
  \left( \Delta t \lambda_T, (\Delta t)^2 m^2_T \right) = \left(
  41.91 (\Delta t)^{2\gamma}N^{-\gamma(2\beta+1)}
      , - 12.7 (\Delta t)^{3\gamma}N^{-3\gamma \beta}  \right), \label{f2-point-R-function-N}
\ee

\paragraph{Choice of $\Delta t$.}

We choose
\be
   \Delta t = \frac{1}{N^{\kappa}}, \label{delta_t_N_kappa}
\ee
with $\kappa > 0$ as an ansatz.

Then the time extension $T$ is 
\be
N_{\trm}\Delta t=N^{1-\kappa}
\ee
for $N_{\trm}=N$. \\ 
If $\kappa <1$ then
 $T\longrightarrow \infty$ as $N\longrightarrow \infty$.

Implementing eq.~(\ref{delta_t_N_kappa}) in eq.~(\ref{f2-point-R-function-N}) we get

\be 
  \left( \bar{\lambda}_T, \bar{m}^2_T \right) =
  \left( \Delta t \lambda_T, \Delta t^2 m^2_T \right) = \left(
  41.91 N^{-\gamma(2\beta+2\kappa+1)}
      , - 12.7 N^{-3\gamma(\beta+\kappa)}  \right). \label{f3-point-R-function-N}
\ee
We note that the dimensionless quantities in eq.~(\ref{f3-point-R-function-N}) 
goes to zero as $N\longrightarrow \infty$.

Eq.~(\ref{f3-point-R-function-N})  can be written as 
\be 
  \left(  \lambda_T,  m^2_T \right) = \left(
  41.91 N^{\kappa(1-2\gamma)-\gamma(1+2\beta)}
      , - 12.7 N^{2\kappa-3\gamma(\beta+ \kappa)}  \right). \label{f4-point-R-function-N}
\ee
To analyse the limit $N\longrightarrow \infty$ we focus on the exponents of $N$ in 
eq.~(\ref{f3-point-R-function-N}).
For  $\lambda_T$ the exponent is $\kappa(1-2\gamma)-\gamma(1+2\beta)$ and this is clearly 
negative if $1-2\gamma\le 0$. Although the error in $\gamma$ is quite large even taking 
the lower bound\footnote{ The lower bound would be $0.64-0.2=0.44$.}   
the exponent $\kappa(1-2\gamma)-\gamma(1+2\beta)$ is negative, which implies
\[
\lambda_T  \longrightarrow 0,     \quad       N\longrightarrow \infty.
\]

For $m^2_T$ the exponent of $N$ is ${2\kappa-3\gamma(\beta+ \kappa)}$, it is positive or negative 
depending on the values of $\beta$ and $\kappa$.

In any case what we would have is the disappearance of the ordered uniform phase and therefore
the phase diagram of the commutative theory cannot be recovered from the one of our studied
 model (\ref{action_two_v2}).

The tricritical action reads\footnote{We chose
  $\gamma\approx\frac{2}{3}$. } 
\bea
S_T \left[ \Phi,N,R\right]& \approx & \sum_{\trm=1}^{N} 
                             \Tr \Big[
                            \frac{2 \pi}{N} \Phi\left(t\right)
                             \lpl^2 \Phi\left(t\right) +  \frac{2
			       \pi}{N}\left(\frac{R}{\Delta t}  \right)^2
                    \left[ \Phi(t+\Delta t) -\Phi(t)  \right]^2
		    \nonumber \\ & &
                    \qquad -\frac{25.4 \pi}{N} \Phi^2(t)
                    +\frac{41.9\pi }{N} \left(\frac{R}{N \Delta t}\right)^{\frac{2}{3}} \Phi^4(t) \Big] .
     \label{action_triple}
\eea

We consider the particular case $\Delta t=\frac{1}{\sqrt{N}}$ and $R=N^{\beta}$,

\bea
S_T \left[ \Phi,N,R\right]& \approx & \sum_{\trm=1}^{N} 
                            \Tr \Big[
                            \frac{2 \pi}{N} \Phi\left(t\right)
                             \lpl^2 \Phi\left(t\right) +  2 \pi  N^{2\beta}
                    \left[ \Phi(t+\Delta t) -\Phi(t)  \right]^2
		    \nonumber \\ & &
                    \qquad -\frac{25.4 \pi}{N} \Phi^2(t)
                    +41.9\pi N^{\gamma(\beta-\med)-1 } \Phi^4(t) \Big]. \label{action_triple-v2}
\eea
The exponent of $N$ for the $\Phi^4$-term in eq.~(\ref{action_triple-v2}), $\gamma(\beta-\med)-1$,
is negative
for the values of $\beta$ considered in the limits in section \ref{limits-of-the-model} 
--see eqs.~(\ref{commutative-sphere-limit})-(\ref{commutative-flat-limit}).
We observe that for $N$ large the leading contribution to eq.~(\ref{action_triple-v2}) 
comes from the temporal kinetic term.\footnote{We will identify
$ \sum_t\Tr \left( \Phi\left(t\right)\lpl^2 \Phi\left(t\right) \right) /\sum_t \Tr \left[ \Phi(t+\Delta t) -\Phi(t)  \right]^2$.}

If we compare the powers on $N$ multiplying the kinetic terms in eq.~(\ref{action_triple-v2}), 
we see that the temporal kinetic term is $N^{2\beta+1}$ times larger than the fuzzy kinetic term.
Under the limit $N\longrightarrow \infty$ the fuzzy kinetic term is negligible and then, because in some sense 
the geometry of the sphere is {\em screened} by the temporal kinetic term, the model behaves more like a 
``matrix chain'' system interacting to first neighbours with the potential $\lambda \phi^4$ 
(see Ref.~\cite{BEynard}).

To maintain the uniform ordered phase in the limit $N\longrightarrow \infty$ it is necessary to 
reinforce the fuzzy kinetic term as in Refs.~\cite{matrix_phi4_models}-\cite{delgadillo_thesis}.  
We will come back to this point in the conclusions.

\section{Comparison with  other numerical studies}
\subsection{$\lambda \phi^4$ on the fuzzy sphere}
We found the $3$ phases present in the model in 
eq.~(\ref{eq:accion}) --see Refs.~\cite{xavier}-\cite{Garcia-Martin-OConnor}.

A substantial difference from the $2$-dimensional case
studied in Refs.~\cite{xavier}-\cite{Garcia-Martin-OConnor} is that the whole phase diagram
in {\bf figure \ref{fig:diagram}} collapses using the same
scaling function of $N$. We do not have  this situation since the scaling on $N$ for the 
the transition curves (\ref{eq-collapse-uno-dos}) (\ref{eq-collapse-uno-tres}) are different.
The collapse in the $2$ dimensional studies do not depend on the radius since it can
be absorbed in the couplings. This is in contrast to the $3$ dimensional case, where the radius plays an independent r\^ole.

In section \ref{collapse-large-lambda} we studied the collapse of observables and of the transition curve 
disordered to ordered non-uniform. We conclude that we do not observe the same scaling behaviour\footnote{This is under the appropriate translation of parameters from the $2$-dimensional model to the $3$-dimensional model, see eq.~(\ref{prediction-trans-large-lambda-2d}).} that as 
in the $2$-dimensional case. Furthermore, the collapse of the transition curve $I-III$ for large $\lambda$
-- see eq.~(\ref{prediction-trans-chain-matrix})--
indicates that in this regime our model behaves as a {\em chain of interacting matrices}, where the fuzzy kinetic term is neglected  Ref.~\cite{Shimamune}.

\subsection{Non-commutative lattice studies}
In this section we want to compare our results with those obtained in Refs.~\cite{BHN},\cite{hofheinz} and \cite{BHN-phase} where the following model was studied:

\bea
S \left[ \Phi\right]& =& N \Tr \sum_{t=1}^T 
                            \Big[
                            \frac{1}{2} \sum_i \left( \eater_i \Phi(t) \eater_i^{\dagger}- 
                                       \Phi(t) \right)^2 +
                            \med \left( \Phi(t+1) -\Phi(t)  \right)^2 \nonumber \\
                   & &    \hspace{1.7cm} + \frac{m^2}{2} \Phi^2(t)
                    +\frac{\lambda}{4} \Phi^4(t)
                           \Big] \ . \label{action_ek}
\eea
$\Phi$ is an hermitian matrix.
Eq.~(\ref{action_ek})  describes a scalar field $\phi$ living on a non-commutative torus and interacting 
under the $\lambda \phi^4$ potential. 

Comparing eq.~(\ref{action_ek}) to eq.~(\ref{action_two_v2}) we observe that, besides the different 
interpretation in each discretisation scheme, the substantial difference between the models is in the 
spatial kinetic term.
The first term on the right-hand-side  in eq.~(\ref{action_ek}) corresponds to the energy due to spatial translations 
on a squared $N\times N$ lattice with lattice spacing $a$, while the first term on the right-hand-side in   eq.~(\ref{action_two_v2}) corresponds to the energy due to translations (or rotations) over a discrete 
version of the sphere.
It is to be expected that in the regime where the spatial kinetic term is negligible, both models 
describe the same physics.
Now we want to compare the phase diagram for each model, but first we denote the constants of the model given by (\ref{action_ek}) as follows:
\bea
  \cteAA &=& \frac{N }{2}, \label{ek-cte-A} \\
  \cteDD &=& \frac{N}{2}, \label{ek-cte-D} \\
  \cteBB &=& \frac{N m^2  }{2}, \label{ek-cte-B} \\
  \cteCC &=& \frac{ N \lambda  }{4}.  \label{ek-cte-C}
\eea
In Refs.~\cite{BHN}-\cite{hofheinz} the phase diagram shows the existence of three phases:
\begin{itemize}
  \item Disordered phase
  \item Uniform phase
  \item Striped phase
\end{itemize}

This phase diagram stabilises taking the axes as $N^2 \lambda$ vs. $N^2 m^2 $, this is, under the re-scaling of the axes   $\lambda  \longrightarrow N^2 \lambda$ and $ m^2 \longrightarrow  N^2 m^2$  the transition lines collapse as follows:
\begin{itemize}
  \item Disordered phase --- Uniform phase coexistence line:
    \be
      N^2 m^2 \cong -0.88 N^2 \lambda. \label{tc-to-uniform}
    \ee
  \item Disordered phase --- Striped phase coexistence line:
    \be
      N^2 m^2 \cong -0.52 N^2 \lambda- 64, \label{tc-to-striped}
    \ee
\end{itemize}
and the triple point is given by 
\be
(N^2 \lambda, N^2 m^2) \cong (220,-150). \label{triple-point-ek}
\ee

In the model  (\ref{action_two_v2}) we found the existence of three phases:

\begin{itemize}
  \item I: Disordered phase
  \item II: Ordered uniform phase
  \item III: Ordered non-uniform phase
\end{itemize}
We conjecture that the striped phase of the model in eq.~(\ref{action_ek}) corresponds to  the 
ordered non-uniform phase in the model given by eq.~(\ref{action_two_v2}).

Coming back to the fuzzy model in eq.~(\ref{action_two_v2}),
their coexistence curves $I-II$ and $I-III$ stabilise under different scaling:
\begin{itemize}
  \item Disordered phase --- Ordered uniform phase:
    \be
      \bar{R}^{3\gamma} \bar{m^2} =  -0.31 {N}^{\gamma}\bar{R}^{2\gamma} \bar{\lambda}.
       \label{eq.-transition-I-II}
   \ee
  \item Disordered phase --- Ordered non-uniform phase:
    \be
       \bar{R}^{3\gamma}N^{\gamma-1} \left(\bar{m^2} + \frac{12.7}{\bar{R}^{3\gamma}} \right) = 
                 -0.064 ({N} \bar{R}^2)^{\gamma} \bar{\lambda} +2.69 N^{\gamma-0.59} \bar{R}^{3\gamma} 
       \label{eq.-transition-I-III}
  \ee
\end{itemize}
The triple point is given by the equations:
\be 
  \left( \bar{\lambda}_T, \bar{m^2}_T \right) = \left( 41.91 N^{-0.64\pm0.20}
      {\bar{R}^{-1.28\pm0.25}}, - (12.7\pm1) \bar{R}^{-1.92\pm0.20}  \right). \label{triple-point-eqs.}
\ee

In the  region around the transition curve $I-II$ (to the ordered uniform phase) the kinetic term 
is relevant.
Then,   because  of the difference  in  the nature   of the kinetic  term in the models in 
eq.~(\ref{action_ek}) and eq.~(\ref{action_two_v2}), we cannot expect that there exists a 
re-scaling such that both collapses are compatible. 
In the region around the transition  curve $I-III$ (to the ordered non-uniform phase) the leading 
term is the one due  to the potential,  then we could expect that there exists a  re-scaling such 
that both collapses are compatible for the transition curve  $I-III$.
Now we consider an special case of $R$ and $\Delta t $.

\subsubsection{Case $R=N$, $\Delta t=1$.}
We consider  $R=N$ and $\Delta t=1$.
Implementing the substitution in eq.~(\ref{eq.-transition-I-II}) we get
\be
      R^{3\gamma} m^2 =  -0.31 {N}^{3\gamma} \lambda.\label{eq.-transition-I-II-R=N}
\ee
Comparing eq.~(\ref{tc-to-uniform}) to eq.~(\ref{eq.-transition-I-II-R=N}) and considering $3\gamma \approx 2$, we conclude that the transition line $I-II$ for both models collapse with the same dependence in $N$. We have the same situation when we compare the triple points.
Implementing the substitutions $R=N$ and $\Delta t=1$ in eq.~(\ref{f-point-R-function-N}) we get
\be 
  \left( \lambda_T, m^2_T \right) = \left(
  41.91 \frac{1}{ N^{3 \gamma} }
      , - 12.7  \frac{1}{N^{3\gamma}}  \right) \approx \left(
  \frac{41.91}{ N^2 }
      , - \frac{12.7 }{N^2}\right). \label{tri-point-comparison}
\ee
The dependence  of the triple point on $N$ is the same as 
in Refs.~\cite{BHN}-\cite{hofheinz}.
The critical action is:
\bea
S_T \left[ \Phi,N\right]& \approx & \sum_{\trm=1}^{N} 
                            \Tr \Big[
                            \frac{2 \pi}{N} \Phi\left(t\right)
                             \lpl^2 \Phi\left(t\right) +  
			       2\pi  N
                    \left[ \Phi(t+\Delta t) -\Phi(t)  \right]^2
		    \nonumber \\ & &
                    \qquad -\frac{25.4 \pi}{N} \Phi^2(t)
                    +\frac{41.9\pi }{N}  \Phi^4(t) \Big]. \label{action_triple-v3}
\eea
From (\ref{action_triple-v3}) we observe that the dominance of the temporal kinetic term 
is even stronger than in eq.~(\ref{action_triple-v2}).
\begin{itemize}
\item {\bf Observation:} \\

Choosing $\beta=\med$ in eq.~(\ref{action_triple-v2}) we get the same critical action as in eq.~(\ref{action_triple-v3}), but the dependence on $N$ in the triple point is not the same.

The triple point (\ref{f3-point-R-function-N}) for $\beta=\med$ is:
\be 
  \left(  \lambda_T,  m^2_T \right) = \left(
  41.91 N^{\med-3\gamma}
      , - 12.7 N^{1-3\gamma}  \right). \label{f3-point-R-function-N-beta-05}
\ee
\end{itemize}

For the transition curve $I-III$ the predictions from models  (\ref{action_two_v2})-(\ref{action_ek})  
are different.
The main difference is that the transition curve in the model given by (\ref{action_two_v2}) is 
a curve as it is shown in {\bf figures \ref{tc-curve-N16-R16}}-{\bf{\ref{tc-curve-N23-R16}}}, whereas 
in the model given by eq.~(\ref{action_ek}) this  transition curve is a straight line. 
Nevertheless, we found that in a range of parameters around the triple point the transition curve $I-III$ 
in the model (\ref{action_two_v2}) could be approximated by a line as in  (\ref{action_ek}).
We conjecture that this is the range of parameters studied in Refs.~\cite{BHN}-\cite{hofheinz}.

Taking $R=N$ and $\Delta t=1$ in (\ref{eq.-transition-I-III}) we get
\be
  N^{4\gamma-1}\left(m^2 +\frac{12.7}{N^{3\gamma}}\right)=-0.064N^{3\gamma}\lambda+2.7N^{4\gamma-1.41}.  \label{eq.-transition-I-III-R=N}
\ee
If we consider just the terms in $m^2$ and $\lambda$ in (\ref{eq.-transition-I-III-R=N}),  we obtain 
$N^{4\gamma-1}m^2$ vs. $ N^{3\gamma}\lambda$. We observed $3\gamma\approx 2$ while $4\gamma-1= 1.67\pm 0.8$. 
We conclude that the dependence on $N$ in the eq.~(\ref{eq.-transition-I-III-R=N}) could be the same as in eq.~(\ref{tc-to-striped}).

\chapter{Conclusions from part \ref{simulation-part}}
\begin{itemize}
 \item We presented a numerical study of the $\lambda \phi^4$ model  on the $3$-dimensional Euclidean space which was regularised by means of:
   \begin{itemize}
     \item {\em the fuzzy sphere} $S_{F}^{2}$ for the spatial coordinates 
     \item a conventional lattice with periodic boundary conditions for the time direction.
   \end{itemize}
The obtained model was
{
\bea
S \left[ \Phi\right] = \frac{4  \pi R^2 }{N}  \Delta t \sum_{\trm=1}^{N_{\trm}} 
                            \Tr   \Big[&& \hspace{-5mm}
                            \frac{1}{2R^2} \Phi\left(t\right)
                             \lpl^2 \Phi\left(t\right) +  \med 
                    \left( \frac{\Phi(t+\Delta t)  -\Phi(t)}{\Delta t}  \right)^2 \nonumber \\ 
& & \hspace{-5mm}
                    + \frac{m^2}{2} \Phi^2(t)
                    +\frac{\lambda}{4} \Phi^4(t)
                           \Big]. \label{action_1_conclu}
\eea
}
where $\Phi (t) \in Mat_N$, for $\trm=1, \dots ,N_{\trm}$.

 \item We found the phase diagram of the model according to the specific heat.
       Following this criterion we determined  the critical values of $\lambda$ and $m^2$ for fixed $N$ and $R$
       denoted by an index ``c'', $\lambda_c$ and $m^2_c$.
       In addition we found  the tricritical point $(m_{T}^2, \lambda_{T})$.
  $\lambda_T$  
       divides the phase diagrams into two regions according to the behaviour of the observables.
   \begin{itemize}
     \item $\lambda_T > \lambda > 0$. In this domain we observe Ising type orderings.
       For $ m^2 > m^2_c$ we have a disordered phase.
      For $ m^2 <m^2_c$ we found a uniform ordering.
     \item $\lambda > \lambda_T $. For $ m^2 > m^2_c$ we have a disordered phase. 
      As in the two dimensional model discussed in chapter \ref{2-dim-model-chapter},
      we found   for     $ m^2 <m^2_c$  there is  a non-uniform ordering.
       For  $ m^2 \ll  m^2_c$ we encountered thermalisation problems,
       therefore we cannot conclude if there exists a boundary for the non-uniform ordered phase 
       as in the $2$-dimensional case \cite{Garcia-Martin-OConnor}.
   \end{itemize}
\item We compared our results with those obtained by other criteria (two point functions of the different modes) 
      and we conclude both criteria are qualitatively equivalent. We followed the specific heat criterion  which is 
      a more universal quantity: it captures the phase transition without taking into account which is  the dominant mode.
\item We found the existence of three phases:
 \begin{itemize}
   \item $I$: Disordered phase.
   \item $II$: Ordered Uniform phase.
   \item $III$: Ordered Non-Uniform phase. 
 \end{itemize}
   These three phases were also found in the 2-dimensional $\lambda \phi^4$ model on a fuzzy sphere 
   studied in Refs.~\cite{xavier}, \cite{Garcia-Martin-OConnor}.
\item The  phase  of non-uniform ordering is characterised by the dominance of several angular momenta for $l>0$. 
      In this phase the rotational invariance is broken.  
\item We get the transition curves of the model:
 \begin{itemize}
   \item $I-II$: It turns out to be a line given by the equation:
      \be
            (\Delta t)^2 m^2_c = ( -0.31\pm0.1) \frac{N^{0.64\pm0.3}}{\left(\frac{R}{\Delta t}\right)^{0.64\pm0.1}}
             \Delta t \lambda. 
            \label{eq-collapse-uno-dos-conclu}
      \ee
   \item $I-III$: We can observe that the transition curve shows curvature. The most natural fit we can propose is 
     a polynomial  where the coefficients are functions that depend on $R$, $N$ and $\Delta t$.
     Nevertheless, to predict the triple point we can concentrate on   values of $\lambda$ slightly above  $\lambda_T$.
     For this range of values the transition curve can be approximated by a line.
     We conclude that for $\lambda$ around $\lambda_T$ the transition curve
     from the disordered phase to the ordered non-uniform phase obeys the
     following equation:
     \bea
      (\Delta t)^2 m^2_c &=&  -(0.064\pm0.017) {N} \left(\frac{R}{\Delta t}\right)^{-0.64\pm0.1} \Delta t \lambda + 
        \nonumber \\
                        & &  \left(2.69 N^{0.41}-12.7  \right)\left(\frac{R}{\Delta t}\right)^{-1.92}.
     \label{eq-collapse-uno-tres-conclu}
     \eea
   \item We conjecture the existence of a transition curve $II-III$ as in Ref.~\cite{Garcia-Martin-OConnor}, 
     but due to thermalisation 
     problems it was not possible to measure it.
   \item The effective action has several minima, 
     and the thermalisation problems appear when it is not possible for the algorithm to tunnel between those minima.
     We sketched the main technical features of these thermalisation problems.

 \end{itemize}
\item We obtained the equation for the triple point:
{\footnotesize
  \be 
    \left(\Delta t \lambda_T, (\Delta t)^2 m^2_T \right) = \left( 41.91 N^{-0.64\pm0.20}
      {R^{-1.28\pm0.25}}, - (12.7\pm1) \left(\frac{R}{\Delta t}\right)^{-1.92\pm0.20}  \right).
    \label{prediction-of-triple-point-conclusions}
  \ee
}
\item Different limits of the fuzzy sphere can be obtained scaling $R$ as a function in $N$ and taking the limit 
      $N\longrightarrow \infty$,
 \begin{itemize}
  \item {Commutative sphere:   $R^2=const., N\longrightarrow \infty$ } 
  \item{Quantum Plane:   $R^2 \propto N$ }
  \item{Continuum flat limit}. It  requires:
  \be
    R^2 \sim N^{1-\epsilon}, \quad 1>\epsilon>0.
  \ee
  \end{itemize}
 We analysed the behaviour of the scalar model in (\ref{action_1_conclu}) under the different limiting spaces above. 
\item Our numerical results reveal that the triple point goes to zero under the limit $N\longrightarrow \infty$, 
   this is valid for all cases considered in the previous point. 
\item In other words, in the limit $N\longrightarrow \infty$ the non-uniform ordered phase dominates the phase diagram. 
      This result is as consequence of the UV-IR mixing:
      integrating out high energy in the loop produces non-trivial 
      effects at low external momenta \cite{steinacker}-\cite{Sachin}.

      A  perturbative analysis of the action 
      (our eq.~\ref{action_0} of chapter \ref{section2} ) :
\bea
S \left[ \Phi\right] &=& \frac{4  \pi R^2 }{N} \int_{S_1} d t \, 
                            \Tr \Big[
                            \frac{1}{2} \Phi\left(t\right)
                             \left( \frac{\lpl^2}{R^2}-\partial_t^2 \right)
                    \Phi\left(t\right) \nonumber \\  
                   & & \hspace{2.7cm} + \frac{m^2}{2} \Phi^2(t)
                    +\frac{\lambda}{4} \Phi^4(t)
                           \Big] \ , \label{action_0-conclusions}
\eea
is presented in Ref.~\cite{delgadillo_thesis}. It  shows that the action given by eq.~(\ref{action_0-conclusions})  does not reproduce the commutative continuum limit. This result was obtained from a  expansion to two loops. 
Our results show that  the UV-IR mixing is presented for all values of $\lambda$.
\item Following Refs.~\cite{matrix_phi4_models} and  \cite{delgadillo_thesis}\footnote{To be precise, in ref.~\cite{delgadillo_thesis} discusses $S^2_F \times \real$.} the action 
(\ref{action_0-conclusions}) should be corrected by modifying  the action.

This modification in the action is equivalent to reinforcing the fuzzy kinetic term and it
can be achieved adding a term $\Phi (\mathcal{L}^2)^2 \Phi/(\Lambda^2 R^4)$ where $\Lambda$ is a momentum cutoff.
\item We compare our results to those obtained in a numerical study of the $\lambda \phi^4$ potential on a non-commutative torus (see Ref.~\cite{BHN}-\cite{hofheinz}). 
Our results are in agreement if the parameters
$R$ and $\Delta t$ are scaled appropriately.
\item The axes cannot be chosen consistently for all regimes  to lead to a stable phase diagram for large $N$.
This is a significant difference to  the $2$-dimensional formulation which has
the ``privileged'' property of leading to a phase diagram.
\item We compare our results at large coupling $\lambda$ to those of the chain of matrices at the large $N$-limit
      in Refs.~\cite{Shimamune}-\cite{Gross-Klebanov} (where this model is known as the $c=1$ model in string theory). Our simulation results fully agree.
   The disordered to ordered non-uniform phase transition in this 
      regime obeys the predicted transition in the model $c=1$  in the large $N$-limit.
\end{itemize}                 
\part{Scalar Field Theory on $S^4$}
\label{analytical-part}
\chapter{ From fuzzy $\CP^3$ to a fuzzy $S^4_F$}
\label{construction-S4}
$S^4$ is an
especially important example since it is the most natural replacement
of ${\mathbb R}^4$ in studies of Euclidean quantum field theory. Therefore our motivation to propose a {\em fuzzy} approximation to  $S^4$, namely $S^4_F$.  
Nevertheless $S^4$ cannot be quantised in the strict sense since it is not a phase space. It is important to clarify how we can we obtain a matrix approximation to it. 

\section{$S^4_F$ in analogy to {$S^2_F$}}\label{s4_in_analogy_s2}
Having in mind the seminal example of fuzzy space, we search for five {\em matrix coordinates}, 
$X_a$, $a=1 \dots 5$, which fulfil a matrix equation of a $4$-sphere in ${\mathbb R}^5$:
\begin{equation}
\sum_{a=1}^5 X_aX_a=R^2\ID. \label{s4_F_defn}
\end{equation}
We propose 
\be
 X_a=\frac{R}{\sqrt{5}}\Gamma_a, \label{matrix_coordinates_4X4}
\ee 
with $\Gamma_a \in Mat_4$ the Dirac matrices including $\gamma_5$. They obey the algebra:
\be
  \{\Gamma_a, \Gamma_b  \}= 2 \delta_{ab}\ID. \label{Dirac_algebra}
\ee
Eq.~(\ref{s4_F_defn}) follows directly from  eqs.~(\ref{matrix_coordinates_4X4}) and (\ref{Dirac_algebra}).

The next step is to propose a sequence of matrices approximating $S^4$. 
To achieve this we note that $\Gamma_a$ (which are $4$-dimensional matrices) give
the representation $(\frac{1}{2},\frac{1}{2})$ of
$Spin(5)$.
 We can therefore consider the irreducible representation
obtained from the $L$ fold symmetric tensor product of this
representation i.e.\ the $Spin(5)$ representation
$(\frac{L}{2},\frac{L}{2})$. It will contain a set of five matrices:
$J_a$, $a=1 \dots5$ which can be realised as the symmetrisation of
$L$ copies of the $\Gamma$ matrices in the $Spin(5)$ fundamental
representation:

\begin{equation}
    J_a  =    \left( \underbrace{ \Gamma_a {\otimes} \mathbf{1}
                            {\otimes} \cdots {\otimes}
                            \mathbf{1}}_{L-terms}   +
                            \mathbf{1} {\otimes}  \Gamma_a
                            {\otimes} \cdots {\otimes}
                            \mathbf{1}   + \cdots +
                            \mathbf{1} {\otimes} \mathbf{1}
                            {\otimes} \cdots {\otimes}
                             \Gamma_a
                  \right)_{sym}.  \label{defJa}
\end{equation}
The subscript $sym$ indicates that we are
projecting onto the irreducible totally symmetrised representation.
The dimension of the matrices defined in eq.~(\ref{defJa}) is
\be
  d_L=\frac{(L+1)(L+2)(L+3)}{6}, \label{dimension_algebra_CP3}
\ee
and $J_a \in Mat_{d_L}$. They  satisfy the relation:
\be
 J_aJ_a=L(L+4)\ID. \label{J_a_squared}
\ee
Now we can generalise the definition of the {\em matrix coordinates} 
(\ref{matrix_coordinates_4X4}) to eq.~(\ref{in.2}):
\begin{equation}
   X_a=\frac{R}{\sqrt{L(L+4)}}J_a.
\label{in.2}
\end{equation}
The definition of the {\em matrix coordinates} (\ref{in.2}) guarantees 
that the matrix equation of a $4$-sphere (\ref{dimension_algebra_CP3}) is fulfilled for matrices of 
dimension $d_L$. 
In the limit $L \longrightarrow \infty$ the {\em matrix coordinates} in
eq.~(\ref{in.2}) commute. In this limit we recover  the algebra of function of continuous $S^4$,  $C^{\infty}(S^4)$ . 

However, a substantial difference to the fuzzy $2$-sphere case 
is that for  finite $L$ the {\em matrix coordinates} $X_a$ do not provide a complete
basis for the algebra of functions, i.e.\  $X_a$ cannot provide a basis for $Mat_{d_L}$.

To clarify this point we analyse the lowest approximation, $L=1$.
We define $\sigma_{ab}$ proportional to the commutators of $\Gamma_a$:
\be
\sigma_{ab}=\frac{1}{2i}[\Gamma_a,\Gamma_b].  \label{def_sigmas}
\ee

If $F$ is a matrix representing a function on $S^4$, it
will be of the form
\begin{equation}
F=F_0\ID+F_a\Gamma_a.  \label{function_L=1}
\end{equation}
However, a matrix product of two functions of the type in eq.~(\ref{function_L=1})
will involve  non-zero coefficients of $\sigma_{ab}$, the matrices in (\ref{def_sigmas}), 
\be
  F' =F'_0\ID+F'_a\Gamma_a+ F'_{ab}\sigma_{ab}.\label{expasion_product_function_L=1}
\ee 
The $10$ coefficients $F'_{ab}$  in eq.~(\ref{expasion_product_function_L=1})
have no corresponding counterparts in the expansion of functions on
commutative $S^4$. In the language of Statistical Physics these coefficients 
constitute a set of {\em extra degrees of freedom}.
On one hand, if we exclude them the involved algebra is not associative. 
On the other hand, if we included them, the approximated space is not exactly $S^4$.

The first option, advocated  by Ramgoolam
\cite{Ramgoolam_0105006}, is to project out such terms, in which case
one is left with a non-associative algebra. This involves additional
complications and does not seem particularly suited to numerical
work. In addition the necessary projector must be constructed. We will
return to this point in chapter  \ref{conclusions_analytical_part}   where we will, in fact,
give the projector.

An alternative is to include arbitrary coefficients of $\sigma_{ab}$
(demanding an associative algebra) and attempt to suppress such
coefficients of unwanted terms, by making their excitation improbable
in the dynamics. In this approach our algebra will be a full matrix
algebra and obviously associative.  

Including the extra degrees of freedom will lead us to work in a bigger space.
This bigger space is $\CP^3$, in  section \ref{review-contruction} 
will review its construction.

\section{Review of the construction of $\CP^3$}
\label{review-contruction} 
The fuzzy version of $\CP^{N-1}$ denoted by the subindex  ``$F$'', $\CP^{N-1}_F$, is a matrix approximation to the
continuous  $\CP^{N-1}$. In this section we review the construction of  $\CP^3_F$ following Ref.~\cite{starprod_CPN}.

A standard definition on $\CP^{N-1}$ is the space of all norm-1 vector in $\mathbb{C}^N$ 
modulo the phase. For any unit vector $\vert  \psi \rangle$ we can define a rank-one projector, 
\be
\proj(\psi):=\vert  \psi \rangle \langle \psi\vert.
\ee
 Then  $\CP^{N-1}$ can be defined as the space of all rank-one projectors 
\begin{equation}
 \CP^{N-1}:= \{ \proj \in Mat_N; \proj^{\dagger}=\proj, \proj^2= \proj,
               Tr \proj =1 \}.
\end{equation}

To construct the set of global coordinates for $\CP^{N-1}_F$ we need a set of $N^2$ hermitian matrices $\{ \identy, t_{\mu} \}$, $\mu=1 \dots N^2-1$. The set  
 $\{ t_{\mu} \}$ is a basis for the Lie algebra of $SU(N)$,  normalised as 
\be
 Tr \left( t_{\mu} t_{\nu}\right)= \delta_{\mu \nu}.
\ee
For our case of $\CP^3$ we start with  $Spin(6) \cong SU(4) $. Let $\{ \gen_{AB} \}$, $A,B=1 \dots 6$ be the $Spin(6)$ generators. They are equivalent to the generators $ t_{\mu}$ of $SU(4) $.
As a basis for our algebra we take the set $\{ \Lambda_{AB} \}$, $A,B=1 \dots 6$:
\be
 \Lambda_{AB}:= \frac{1}{\sqrt{2}}\gen_{AB},  \label{def-Lambdas}
\ee
their algebra is:
\begin{eqnarray}
    \Lambda_{AB} 
    \Lambda_{CD} &=& A_{AB;CD}\frac{\mathbf{1}}{4}+
                     \frac{1}{4\sqrt{2}}\epsilon_{ABCDEF} \Lambda_{EF} + \\
                 & & \frac{\imath}{2\sqrt{2}}
                     \left( \delta_{AC} \Lambda_{BD}  +
                             \delta_{BD} \Lambda_{AC}  -
                              \delta_{BC} \Lambda_{AD}  -
                                \delta_{AD} \Lambda_{BC}  
                     \right).  \label{algebra_lambdas}
\end{eqnarray}
$ A_{AB;CD}$ is defined as a  two indexes Kronecker symbol,
\be
  A_{AB;CD}=\med \left( \delta_{AC} \delta_{BD}- \delta_{AD} \delta_{BC}  
                 \right).
\ee

The algebra given by eq.~(\ref{algebra_lambdas}) admits representations of 
dimension \\
$d_L=\frac{(L+1)(L+2)(L+3)}{6}$ for $L$ integer. For the lowest 
non-trivial level $L=1$ we have the explicit form of the generators in appendix \ref{appendix-sigmas}.

A  projector $\proj$ can be expanded in terms of the basis given 
by eq.~(\ref{algebra_lambdas}):
\begin{equation}
 \proj = \frac{1}{4} \identy + \xi_{AB} \Lambda_{AB}, \label{proj}
\end{equation}
$\xi_{AB}$ are fifteen real coordinates.

$\proj^2=\proj$ implies  that $\xi_{AB}$   obeys the restrictions:
\begin{eqnarray}
  \xi_{AB}\xi_{AB}&=&\frac{3}{4}, \label{restriction1} \\
  \frac{1}{2\sqrt{2}}\epsilon_{ABCDEF}\xi_{AB}\xi_{CD}&=&\xi_{EF}. 
                                  \label{restriction2} 
\end{eqnarray}
Taking contractions of eqs.~(\ref{restriction1}) and (\ref{restriction2}) we get the 
three identities:
\begin{eqnarray}
  \frac{1}{2\sqrt{2}}\epsilon_{ABCDEF}\xi_{AB}\xi_{CD} \xi_{EF}&=&\frac{3}{4}, 
                                  \label{restriction3} \\
  \frac{1}{2\sqrt{2}}\epsilon_{ABCDEF}\xi_{AB}&=& 2\xi_{AB} \xi_{CD} -
                                    2\xi_{AC} \xi_{BD} +
                                    2\xi_{AD} \xi_{BC},
                   \label{restriction4} \\
       \xi_{AC}\xi_{CB} &=& -\frac{1}{8}\delta_{AB}.  \label{restriction5}
\end{eqnarray}
Eqs.~(\ref{restriction1})-(\ref{restriction2}) describe how  $\CP^3_F$ is 
embedded on $\mathbb{R}^{15}$. 
The global coordinates $\{ \xi_{AB} \}$ allow us to describe the geometry of 
$\CP^3_F$. 
Following Ref.~\cite{starprod_CPN} we have the geometrical structures
\begin{eqnarray}
 K_{AB;CD}^{\pm} & =& \med \left( P_{AB;CD} \pm \imath J_{AB;CD} \right), \label{K-projector} \\
 P_{AB;CD} & =& \med A_{AB;CD}+ \sqrt{2}  d_{ABCD}^{EF}\xi_{EF}
                -2\xi_{AB}\xi_{CD},   \label{P-projector} \\
 P^{\perp}_{AB,CD} & =& \med A_{AB;CD}- \sqrt{2}  d_{ABCD}^{EF}\xi_{EF}
                +2\xi_{AB}\xi_{CD},   \label{P-projector-perp} \\
 J_{AB;CD} & =& \sqrt{2}  f_{ABCD}^{EF}\xi_{EF}. \label{J-complex-structure}
\end{eqnarray}
$K_{AB;CD}$ is the {\em K\"ahler structure},
$J_{AB;CD}$ is the {\em complex structure} and $P_{AB;CD}$ is the {\em metric}.

From eq.~(\ref{algebra_lambdas}) we get the explicitly form of the  normalisation constants, 
eq.~(\ref{d_ABCDEF}), and  the structure constants, eq.~(\ref{f_ABCDEF}):
\begin{eqnarray}
  d_{ABCDEF} &= & \frac{1}{4} \epsilon_{ABCDEF},  \label{d_ABCDEF} \\
  f_{ABCDEF} &= &\med \big( \delta_{AC} A_{BD;EF}- \delta_{AD} A_{BC;EF}  \nonumber \\
                         &  & \hspace{0.4cm}+
                             \delta_{BD} A_{AC;EF}- \delta_{BC} A_{AD;EF}  
                       \Big). \label{f_ABCDEF}
\end{eqnarray}
From here to the end of the thesis, we used the Kronecker's delta to arise and 
low indexes unless the opposite is indicated.
A simplification for eqs.~(\ref{P-projector})-(\ref{J-complex-structure}) is the following:
\bea
   P_{AB;CD} & =&\med A_{AB;CD}-2\left( \xi_{AC}\xi_{BD}-\xi_{AD}\xi_{BC}\right), \\
   P^{\perp}_{AB;CD} & =&\med A_{AB;CD}+\left( \xi_{AC}\xi_{BD}-\xi_{AD}\xi_{BC} \right), \\
   J_{AB;CD} & =& \frac{1}{\sqrt{2}}\left( \delta_{AC}\xi_{BD} -\delta_{AD}\xi_{BC} +
                                           \delta_{BD}\xi_{AC} -\delta_{BC}\xi_{AD} 
\right).
\eea
\subsubsection{Some properties of the metric $P_{AB;CD}$ and the complex structure $J_{AB;CD}$.}
\bea
 P_{AB;CD} & =&P_{CD;AB}=-P_{BA;CD}, \label{properties-P-J-1}\\
 P^{\perp}_{AB;CD} & =&P^{\perp}_{CD;AB}=-P^{\perp}_{BA;CD}, \label{properties-P-J-2} \\
 J_{AB;CD} & =& -J_{CD;AB}= -J_{BA;CD}, \label{properties-P-J-3} \\
 P^2_{AB;CD} &:=&P_{AB;EF}P_{EF;CD}=P_{AB;CD}, \label{properties-P-J-4} \\
 \left(P^{\perp}_{AB;CD} \right)^2&:=&P^{\perp}_{AB;EF}P^{\perp}_{EF;CD}=P_{AB;CD}, \label{properties-P-J-5} \\
 J^2_{AB;CD} &:=&J_{AB;EF}J_{EF;CD}=- P_{AB;CD}, \label{properties-P-J-6}  \\
 J_{AB;EF}P_{EF;CD} &=& P_{AB;EF}J_{EF;CD}=J_{AB;CD}. \label{properties-P-J-7}
\eea

$P_{AB;CD}$, $P^{\perp}_{AB;CD}$ and $K_{AB;CD}$ are projectors, their ranks are 
given in eqs.~(\ref{rank-properties-P-J-1})-(\ref{rank-properties-P-J-3}):
\bea
   P_{AB;AB} & =&6, \label{rank-properties-P-J-1}\\
   P^{\perp}_{AB;AB} & =&9, \label{rank-properties-P-J-2}\\   
   K_{AB;AB}^{\pm} & =&3. \label{rank-properties-P-J-3}
\eea
$P_{AB;CD}$ projects on to the tangent space of $\CP^3$ and $P^{\perp}_{AB;AB}$ 
onto the orthogonal compliment in $\real^{15}$.

\subsection{$\CP^3$ as orbit under $Spin(6)$.  }
\label{CP3-orbit-spin6}
Now we want to perform an explicit construction for $\CP^3_F$, we will 
analyse the {\em induced line element}.
$\CP^3_F$ can be obtained taking one fiducial projector  $\proj^0$ and 
rotating it with and element of  $Spin(6)$:
\be
  P(\psi)=U(\psi) P^0 U^{-1}(\psi), \qquad U(\psi) \in Spin(6). 
                          \label{fiducial_1}
\ee
We choose $P^0$ as 
\begin{eqnarray}
  P^0&=& \frac{1}{4}\identy+ \xi_{AB}^0 \Lambda_{AB} \nonumber \\
     &=& \frac{1}{4}\identy+\frac{1}{\sqrt{2}}\left( \Lambda_{12}+ 
                              \Lambda_{34}+ \Lambda_{56} \right). 
        \label{north_pole_spin6}
\end{eqnarray}
Now we are interested on calculating the line element, $ds^2$, at the {\em north pole}, 
defined by eq.~(\ref{north_pole_spin6}).
\be
 ds^2:=\sum_{AB} d \xi_{AB}^2. \label{definition_metric_spin6}
\ee
$Spin(6)$ rotates $\Lambda_{AB}$ as a tensor, 
\be
  \xi_{AB}= R_{AC}R_{BD}\xi_{CD}^0.
\ee
The sum $d \xi_{AB}^2$ in eq.~(\ref{definition_metric_spin6})
can be written in terms of an infinitesimal rotation $R^{-1}d R$:
\be 
\sum_{A,B} d \xi_{AB}^2=-Tr \left[R^{-1}d R, \xi^0  \right]^2, \label{dxi_AB2}
\ee
where $ \xi^0$ is the matrix of coefficients $ \xi_{AB}^0$ and
\be
R^{-1}d R:=-e_{AB}L_{AB}. \label{Maurer-Cartan-spin6}
\ee

Eq.~(\ref{Maurer-Cartan-spin6}) is known as the {\em  Maurer-Cartan forms} for the group of rotations on $6$  dimensions.   $L_{AB}$ are the generators of this representation.

The line element turns out to be:
\bea
 ds^2&=& \med \big[ (e_{13}-e_{24})^2 + (e_{14}+e_{23})^2 +
                   (e_{15}-e_{26})^2 + (e_{16}+e_{25})^2 \nonumber \\
     & &          \hspace{0.3cm}  + (e_{35}-e_{46})^2 + (e_{36}+e_{45})^2 
            \big].  \label{line_element_cp3_spin6_symmetry}
\eea
From eq.~(\ref{line_element_cp3_spin6_symmetry}) we corroborate that $\CP^3$ is a $6$-dimensional space.

\subsubsection{Tangent forms}
We define the tangent forms as
\be
 e^{||}_{AB}= P_{AB;CD}  e_{CD}. \label{tangent-form}
\ee
At the north pole in eq.~(\ref{north_pole_spin6}) we have the non-vanishing coordinates are 
$\xi_{12}=\frac{1}{2\sqrt{2}}$,$\xi_{34}=\frac{1}{2\sqrt{2}}$,
$\xi_{56}=\frac{1}{2\sqrt{2}}$ and
permutation of them. The tangent forms at the north pole turn out to be:

\begin{eqnarray}
e^{||}_{12}&=&0,    \label{tangent-form_12}\\
e^{||}_{13}&=&\frac{1}{2}(e_{13}-e_{24})=-e^{||}_{24} , \label{tangent-form_13} \\
e^{||}_{14}&=&\frac{1}{2}(e_{14}+e_{23})=e^{||}_{23} ,  \label{tangent-form_14} \\
e^{||}_{15}&=&\frac{1}{2}(e_{15}-e_{26})=-e^{||}_{26} ,  \label{tangent-form_15}\\
e^{||}_{16}&=&\frac{1}{2}(e_{16}+e_{25})=e^{||}_{25} ,  \label{tangent-form_16} \\
e^{||}_{34}&=&0 ,                                       \label{tangent-form_34} \\
e^{||}_{35}&=&\frac{1}{2}(e_{35}-e_{46})=-e^{||}_{46},  \label{tangent-form_35} \\
e^{||}_{36}&=&\frac{1}{2}(e_{36}+e_{45})=e^{||}_{45},   \label{tangent-form_36} \\
e^{||}_{56}&=&   0 .  \label{tangent-form_56} 
\end{eqnarray}

We note from eq.~(\ref{line_element_cp3_spin6_symmetry})
\be
 ds^2= \sum_{A,B} \left( e^{||}_{AB} \right)^2. \label{ds2-in-tangent-forms}
\ee

Now we define the (anti)-holomorphic forms
\bea
e^{\pm}_{AB}&=&K^{\pm}_{AB,CD}e_{CD} \nonumber \\
 e^{\pm}_{AB}&=&\frac{1}{4}(e_{AB}+\epsilon_{AC}e_{CD}\epsilon_{DB}
\pm\frac{\imath}{2}(\epsilon_{AC}e_{CB}-e_{AC}\epsilon_{CB})) \nonumber \\
&=&\med e^{||}_{AB} \pm \frac{\imath}{\sqrt{2}}\left( \xi_{AC}e_{CB}-\xi_{BC}e_{CA} \right) \label{holomorphic-forms} 
\eea

At the north pole:
\begin{eqnarray}
e^{+}_{12}&=&0 \label{e-holo-12}\\
e^{+}_{13}&=&\frac{1}{4}\left(e_{13}-e_{24}+i(e_{23}+e_{14})\right), \label{e-holo-13}\\
e^{+}_{14}&=&\frac{1}{4}\left(e_{14}+e_{23}+i(e_{24}-e_{13})\right)=-\imath e^{+}_{13}, \label{e-holo-14} \\
e^{+}_{15}&=&\frac{1}{4}\left(e_{15}-e_{26}+i(e_{25}+e_{16})\right), \label{e-holo-15} \\
e^{+}_{16}&=&\frac{1}{4}\left(e_{16}+e_{25}+i(e_{26}-e_{15})\right)=-\imath e^{+}_{15}, \label{e-holo-16} \\
e^{+}_{23}&=&\frac{1}{4}\left(e_{23}+e_{14}+i(e_{24}-e_{13})\right)=-\imath e^{+}_{13}, \label{e-holo-23} \\
e^{+}_{24}&=&\frac{1}{4}\left(e_{24}-e_{13}-i(e_{14}+e_{23})\right)=- e^{+}_{13}, \label{e-holo-24}  \\
e^{+}_{25}&=&\frac{1}{4}\left(e_{25}+e_{16}+i(e_{26}-e_{15})\right)=-\imath e^{+}_{15}, \label{e-holo-25} \\
e^{+}_{26}&=&\frac{1}{4}\left(e_{26}-e_{15}-i(e_{25}+e_{16})\right)=-e^{+}_{15} , \label{e-holo-26} \\
e^{+}_{34}&=&0, \label{e-holo-34} \\
e^{+}_{35}&=&\frac{1}{4}\left(e_{35}-e_{46}+i(e_{45}+e_{36})\right), \label{e-holo-35} \\
e^{+}_{36}&=&\frac{1}{4}\left(e_{36}+e_{45}+i(e_{46}-e_{35})\right)=-\imath e^{+}_{35}, \label{e-holo-36}  \\
e^{+}_{45}&=&\frac{1}{4}\left(e_{45}+e_{36}+i(e_{46}-e_{35})\right)=-\imath e^{+}_{35}, \label{e-holo-45}  \\
e^{+}_{46}&=&\frac{1}{4}\left(e_{46}-e_{35}-i(e_{45}+e_{36})\right)=-e^{+}_{35}, \label{e-holo-46} \\
e^{+}_{56}&=&0. \label{e-holo-56}
\end{eqnarray}
From eqs.~(\ref{e-holo-12})-(\ref{e-holo-56}) we note that only $3$ holomorphic forms are independent:
\begin{eqnarray}
e^{+}_{13}&=&\frac{1}{2}\left(e^{||}_{13}+ie^{||}_{14}\right)=ie^{+}_{14},\\
e^{+}_{15}&=&\frac{1}{2}\left(e^{||}_{15}+ie^{||}_{16}\right)=ie^{+}_{16},\\
e^{+}_{35}&=&\frac{1}{2}\left(e^{||}_{35}+ie^{||}_{36}\right)=ie^{+}_{36}.
\end{eqnarray}
The line element can also be written as
\be
ds^2=e^{+}_{AB}e^{-}_{AB}, \label{ds2-in-holomorphic-forms}
\ee
where $e^{-}_{AB}=(e^{+}_{AB})^{*}$. 
At the north pole
\bea
ds^2&=& 2 \sum_{A<B}e^{+}_{AB}e^{-}_{AB},  \nonumber \label{ds2-in-holomorphic-forms-v2}\\
     &=& 8\left( e^{+}_{13}e^{-}_{13}+e^{+}_{15}e^{-}_{15}+e^{+}_{35}e^{-}_{35}\right).  
  \label{ds2-in-holomorphic-forms-v3}
\eea
Eq.~(\ref{ds2-in-holomorphic-forms-v3})
reduces exactly to eq.~(\ref{ds2-in-tangent-forms}).

\subsection{$\CP^3$ as orbit under $Spin(5)$.}\label{CP3-orbit-spin5}
Note that $\Lambda_{ab} \sim \gen_{ab}$, $a,b=1 \dots 5$ generate the $Spin(5)$ subalgebra of  $Spin(6)$ while  
$\Lambda_{a6}=\frac{1}{\sqrt{2}}\gen_{a6}$ transforms as a vector under $Spin(5)$. 

We define $\Lambda_a=\gen_{a6}$ so we can write the projector (\ref{proj}) as:
\be
  \proj=\frac{1}{4}\identy+ \xi_a \Lambda_a +\xi_{ab} \Lambda_{ab}.
\ee
The projector (\ref{fiducial_1}) takes the form:
\be 
  \proj^0=\frac{1}{4}\identy+ \med \Lambda_a +\frac{1}{\sqrt2}\left(
                    \Lambda_{12} +\Lambda_{34}\right). \label{proj-0-1}
\ee
Here we have an extra restriction: under $Spin(5)$ rotations the norm of the vector $\xi_a$ is not affected. From eq.~(\ref{proj-0-1}) we have $\sum_a \xi_a^2=\frac{1}{4}$.

The induced line element is:
\be
   ds^2= \alpha d\xi_a^2+ \beta d\xi_{ab}^2, \label{definition_metric_spin5}
\ee
the constants $\alpha,\beta$ are arbitrary numbers until know. We will come back to this point at the end of this chapter.

$Spin(5)$ rotates $\Lambda_{ab}$ as a tensor:
\be
  \xi_{ab}= R_{ac}R_{bd}\xi_{cd}^0
\ee
and it rotates  $\Lambda_{a}$ as a vector:
\be
  \xi_{a}= R_{ab}\xi_{b}^0
\ee
In analogy to eq.(\ref{dxi_AB2}),
$\sum_{a,b} d \xi_{ab}^2=-Tr \left[R^{-1}d R, \xi^0  \right]^2 $ but now the trace runs over the sub-indices $a,b=1 \dots 5$.
For the vector part, $d\xi_a$ we have:
\be
 \sum_a d\xi_a^2=-Tr\left( \left[ \vec{\xi}^{~0}\right]^t R^{-1}dR  R^{-1}dR \vec{\xi}^{~0}  \right), \label{dxi_a2}
\ee
 where $\xi^{0}$ represents the matrix of coefficients $\xi_{ab}^0$ and $ \vec{\xi}^{~0}$ the vector $\xi^{0}_{a}$.
Now we can use the Maurer-Cartan forms for the group of rotations on $4$-dimension: $R^{-1}d R:=-e_{ab}L_{ab}$  where $L_{ab}$ are the generators of this representation.
For more details of these calculations see appendix \ref{induced-metric-calculations}.

Finally, the line element at the {\em north pole} is:
\begin{equation}
  ds^2 = \frac{\alpha+\beta}{4}\left[ e_{15}^{2} +e_{25}^{2}+e_{35}^2+e_{45}^2 \right]+
\frac{\beta}{2}\left[ (e_{13}-e_{24})^2+(e_{14}+e_{23})^2 \right].
                                 \label{line-element-spin5}
\end{equation} 

Eq.~(\ref{line_element_cp3_spin6_symmetry}) reduces to eq.~(\ref{line-element-spin5}) if we ignore the rotations $e_{a6}$ and we choose $\alpha=1$, 
$\beta=1$. Then if we choose as a particular case $\alpha=\beta$ in eq.~(\ref{line-element-spin5}) 
 we recover the $Spin(6)$ symmetry.

From eq.~(\ref{line-element-spin5}) we obtain geometrical information of  $\CP^3$. We focus into 
the following two points:
\begin{enumerate}
 \item Isotropy group of the orbit. 
 \item Local form of $\CP^3$.
\end{enumerate}

\subsubsection{1) The isotropy group of the orbit}
To find the  isotropy group of the orbit, we have to identify the rotations that do not affect the 
projector given by eq.~(\ref{proj-0-1}).
Such rotations must be given by the orthogonal forms:
\be
 e^{\perp}_{AB}= P^{\perp}_{AB;CD}e_{CD}. \label{orthogonal-form}
\ee

\begin{eqnarray}
e^{\perp}_{12}&=&e_{12}, \label{e-perp-12} \\
e^{\perp}_{13}&=&\frac{1}{2}(e_{13}+e_{24})=e^{\perp}_{24}, \label{e-perp-13}\\
e^{\perp}_{14}&=&\frac{1}{2}(e_{14}-e_{23})=-e^{\perp}_{23},  \label{e-perp-14}\\
e^{\perp}_{15}&=&\frac{1}{2}(e_{15}+e_{26})=e^{\perp}_{26}, \label{e-perp-15}\\
e^{\perp}_{16}&=&\frac{1}{2}(e_{16}-e_{25})=-e^{\perp}_{25}, \label{e-perp-16} \\
e^{\perp}_{34}&=&e_{34}, \\
e^{\perp}_{35}&=&\frac{1}{2}(e_{35}+e_{46})=e^{\perp}_{46}, \\
e^{\perp}_{36}&=&\frac{1}{2}(e_{36}-e_{45})=-e^{\perp}_{45}, \\
e^{\perp}_{56}&=&e_{56}.  \label{e-perp-56}
\end{eqnarray}
From eqs.~(\ref{e-perp-12})-(\ref{e-perp-56}) we have to restrict to rotations on $4$-dimensions.
The forms $e^{\perp}_{a6}$, $a=1 \cdots 6$ are related to rotations exclusive on $5$-dimensions. 
 
From the remaining combination of coefficients we construct 
the corresponding generators of the isotropy of the orbit,
$T_{\mu}$, $\mu=1,2,3,4$ as follows
\begin{eqnarray}
    T_1 =  \frac{1}{\sqrt2} ( \Lambda_{23} - \Lambda_{14} )  
        =  \frac{1}{4}      ( \sigma_{23}  - \sigma_{14}  ),     \\
    T_2 = -\frac{1}{\sqrt2} ( \Lambda_{13} + \Lambda_{24} )  
        = -\frac{1}{4}      ( \sigma_{13}  + \sigma_{24}  ) ,    \\
    T_3 =  \frac{1}{\sqrt2} ( \Lambda_{12} - \Lambda_{34} )  
        =  \frac{1}{4}      ( \sigma_{12}  - \sigma_{34}  ) ,    \\
    T_4 =  \frac{1}{\sqrt2} ( \Lambda_{12} + \Lambda_{34} )  
        =  \frac{1}{4}      ( \sigma_{12}  + \sigma_{34}  ). 
\end{eqnarray}
The components $T_{\mu}$ do not affect the fiducial projector in eq.~(\ref{proj-0-1}) since:
\begin{equation}
     [ T_{\mu} , P_0 ] = 0,   \qquad    \mu=1,2,3,4.
\end{equation}

$T_i, i=1,2,3$ fulfil the $SU(2)$-subalgebra
\begin{equation}
    [ T_i , T_j ] = \imath \epsilon_{ijk}T_k. 
\end{equation}

$T_4 \propto \PP_0$ generates $U(1)$.

The orbit turns out  to be:
\begin{equation}
  \CP^3 \simeq Spin(5)/ [U(1) \times SU(2)].      \label{result-orbit}
\end{equation}

Note that both prescriptions, developed in sections \ref{CP3-orbit-spin6} and 
\ref{CP3-orbit-spin5}, give the same orbit, $\CP^3$, since we can rotate any point of 
$\CP^3$ as $Spin(6)$ orbit and then {\em unrotate} it using an element of $Spin(5)$.

If we follow the construction of  $\CP^3$ demanding a $Spin(6)$-symmetry we obtain
 a {\em rounded} version of $\CP^3$.
Demanding the less restrictive  $Spin(5)$-symmetry  we have a {\em squashed} version of  $\CP^3$.

\subsubsection{2) $\CP^3$ is locally of the form $S^4 \times S^2$.}
\label{CP3_locally}
As a second observation, in eq.~(\ref{line-element-spin5}) we distinguish that the line element is composed by two parts:
\begin{itemize}
\item A four dimensional part:  $ \sum_{a=1}^4 e_{a5}^2$. We identify it as a $S^4$-line element. 
      Then $\alpha+\beta$ is related to the square of the radius of the $S^4$, namely $R^2_{S^4}$.
      It can be re-written in terms of the (anti)-holomorphic forms as
      \[
          8 \beta \left(e^{+}_{15}e^{-}_{15} +e^{+}_{35}e^{-}_{35}\right).
      \]
\item A two dimensional part: $(e_{14} + e_{23} )^2 + (e_{13} - e_{24})^2$ or 
      in terms of the (anti)-holomorphic forms 
      \[
         4 \alpha e^{+}_{13}e^{-}_{13}
      \] 
Using the same procedure we used in the calculation of the isotropy group, now we look at the combination of indices $e_{ab}$ that appeared in eq.~(\ref{line-element-spin5}). We associated $\tau_i$ with the corresponding combination of generators:

\begin{eqnarray}
  (e_{14} + e_{23}) & \longrightarrow & \tau_1= \sqrt2 ( \Lambda_{14} 
                                                 + \Lambda_{23}), \nonumber \\
  (e_{24} - e_{13}) & \longrightarrow & \tau_2=\sqrt2 ( \Lambda_{24} 
                                                 - \Lambda_{13}). \nonumber
\end{eqnarray}
$\tau_3$ is obtained by demanding that $\tau_i$, $i=1,2,3$ obeys  
$  [ \tau_i , \tau_j ] = \imath \epsilon_{ijk}\tau_k $.

Summarising the above
\begin{eqnarray}
  \tau_1 &=& \sqrt2 ( \Lambda_{14}  + \Lambda_{23} ), \\
  \tau_2 &=& \sqrt2 ( \Lambda_{24}  - \Lambda_{13} ), \\
  \tau_3 &=& \sqrt2 ( \Lambda_{12}  + \Lambda_{34} ),
\end{eqnarray}
 generate a sphere $S^2$.

Then the constant $\beta$ in eq.~(\ref{line-element-spin5}) is related to the 
square of the radius of the $S^2$, $R^2_{S^2}$. $\alpha+\beta$ is proportional to the square of the radius of the $S^4$, $R^2_{S^4}$
\end{itemize}

\begin{equation}
  ds^2 = \frac{R^2_{S^4}}{R^2}\left[ e_{15}^{2} +e_{25}^{2}+e_{35}^2+e_{45}^2 \right]+
\frac{R^2_{S^2}}{R^2}\left[ (e_{13}-e_{24})^2+(e_{14}+e_{23})^2 \right].
                                 \label{line-element-spin5-v2}
\end{equation} 
In eq.~(\ref{line-element-spin5}) the radius of $\CP^3$, $R^2$ is $2$.
Summarising, eq.~(\ref{line-element-spin5-v2}) reflects that   locally $\CP^3\cong S^4 \times  S^2 $. 

It is known that $\CP^3$ is a fibre bundle with $S^2$ as a fibre and $S^4$ as a base space \cite{Nash_Sen}, in terms of diagrams we have
\[
\begin{array}{ccc}
       S^2       & \hookrightarrow & \CP          \\
                 &                 & \downarrow   \\
                 &                 & S^4
\end{array}
\]

The construction of $\CP^3$ as an orbit under $Spin(5)$ allows us to corroborate this fact.

As we mentioned in the introduction, in order to specify the geometry in fuzzy spaces we have to give a prescription for the Laplacian. 
In chapter \ref{geometry_for_CP3} we will get the exact relation between radius of the fibre
$S^2$ and a penalisation parameter $h$ appearing in the scalar theory of $S^4$.
\chapter{Decoding the geometry of the squashed $\CP^3_F$}
\label{geometry_for_CP3}

Ref.\  \cite{scalar-FT-S4}  presents a prescription for a scalar field theory on a  
fuzzy $4$-sphere.  \hspace{0.3cm}
However,  this theory depends on an additional parameter $h$ and it is defined on a 
{\em larger}\footnote{$\CP^3$ is a $6$-dimensional space.} space, $\CP^3$, instead $S^4$.

The motivation for this chapter is to explore the feature of fuzzy spaces,  to reflect the 
geometrical properties, and to explain  why the prescription in  \cite{scalar-FT-S4} works.

This chapter is divided into two parts: in section \ref{review-scalar-teory-s4} we present a review
of the results in \cite{scalar-FT-S4}.
In section \ref{geometry-in-laplacian}  we extract the geometrical information from the Laplacian via 
the tensor metric.

\section{Scalar field theory on fuzzy $4$-sphere}
\label{review-scalar-teory-s4}
In this section we will summarise the prescription for working with a
scalar field on $S^4_F$ following Ref.~\cite{scalar-FT-S4} . 
The action for the scalar field
on a  {\em rounded} $\CP^3$ is given by eq. (\ref{S_0.1}):
\begin{eqnarray}
S_0[\Phi] &=&\frac{R^4}{d_L} Tr \left( \frac{1}{4 R^2} 
[\gen_{AB},\Phi]^\dag [\gen_{AB},\Phi] + V[ \Phi ] \right),  
\label{S_0.1}
\end{eqnarray}
where $\gen_{AB}$ are the $Spin(6)$ generators defined in chapter \ref{review-contruction}. The constant in front of the trace, $ \frac{R^4}{d_L}$ represents the volume of $\CP^3_F$.

Since we are interested on retaining just the $Spin(5)$-symmetry, we  add to eq.~(\ref{S_0.1})
a  $SO(6)$ non-invariant but $SO(5)$ invariant term given by eq.~(\ref{S_I.1}):
\begin{eqnarray}
  S_I [ \Phi ] &=&\frac{R^4}{d_L} Tr \frac{1}{2R^2}\left( [\gen_{ab},\Phi]^\dag
[\gen_{ab} \Phi] - \frac{1}{2} [\gen_{AB}\Phi]^\dag[\gen_{AB}\Phi]\right)    \label{S_I.1}
\end{eqnarray}

so that we have the overall action for a {\em squashed} $\CP^3$ 
\be
  S [ \Phi ]= Tr \left(\Phi \Delta_h \Phi+ V[ \Phi ] \right).
\label{S_0.3}
\ee
The expression for the overall  Laplacian, $\Delta_h$ is 
\be
\Delta_h \cdot =  \frac{1}{2 R^2}\left(\frac{1}{2}[\gen_{AB},[\gen_{AB},
\cdot ]] + h ( [\gen_{ab}[\gen_{ab}, \cdot]]-\frac{1}{2} [\gen_{AB},[\gen_{AB},
\cdot]])\right).  \label{Delta_h}
\ee
or equivalently 
\begin{eqnarray}
 \Delta_h   & = & \frac{1}{2 R^2}\left( \CAS^{SO(6)}  
+ h ( 2\CAS^{SO(5)}-\CAS^{SO(6)}) \right)
\label{Delta_h_cas}
\end{eqnarray}
which gives a stable theory for all $L$ if $h \in (-1,\infty)$. 

This form (\ref{Delta_h_cas}) is an interpolation between $SO(5)$ and
$SO(6)$ Casimirs. As a particular case, the Laplacian is proportional to the $SO(6)$
Casimir for $h=0$ and the $SO(5)$ Casimir for $h=1$. 

The probability of any given matrix configuration has the form:
\begin{equation}
{\mathcal P}[\Phi]=\frac{{\rm e}^{-S_0[\Phi]-hS_I[\Phi]}}{Z}
\label{prob_of_config}
\end{equation}
where 
\begin{equation}
Z=\int d[\Phi] {\rm e}^{-S[\Phi]-hS_I[\Phi]}
\label{partition_fn}
\end{equation}
is the partition function of the model.

The values of $h$
of interest to us are those large and positive since in the
quantisation of the theory,  following Euclidean functional integral
methods, the states unrelated to $S^4$ then become highly
improbable.

Note that we have not specified the potential of the model since the 
above prescription is independent of the potential. The most obvious 
model to consider would be a quartic potential, since this is relevant
to the Higgs sector of the standard model.

\section{Analysing the geometry encoded in the Laplacian}
\label{geometry-in-laplacian}
\subsection{Mapping the fuzzy Laplacians to the continuum}

The non-commutative product of matrices induces a non-commutative product on functions, 
this is the {\em $\star$-product}. This useful tool allows us to access  the continuum limit.
Let $\widehat{M_1}$, $\widehat{M_2}$ be two matrices of dimension $d_L$, and  
 $M_1(\xi)$, $M_2(\xi)$ are the corresponding functions obtained by the mapping:
\be
  M_1(\xi):=Tr \left( {\proj}_L(\xi) \widehat{M_1} \right).
\ee
$ \proj_L(\xi) $ is called the projector at $L$-level and it is contracted taking 
the $L$-fold tensor product of $\proj$ defined in (\ref{proj}). 
$ \proj_L(\xi) $ carries the coordinates $\xi$. 

The definition of the {\em $\star$-product} is given by 
\be
  \left( M_1 \star M_2 \right) (\xi):=Tr \left( \proj_L(\xi) 
        \widehat{M_1} \widehat{M_2} \right). \label{star-product}
\ee
For $\CP^{N-1}$ the {\em $\star$-product} can be written as a sequences of derivatives 
on the coordinates $\xi$. For our proposes we will just follow the prescription 
given in Ref.\  \cite{starprod_CPN}.
Now we present the formal definitions of the set of coordinates  $\xi$ in eq.~(\ref{proj}):
\begin{equation}
  \xi_{AB}:= Tr\left( \proj_L \Lambda_{AB} \right),\label{def-global-coordinates}
\end{equation}
where $\Lambda_{AB}$ are proportional to $\gen_{AB}$ as in eq.~(\ref{def-Lambdas}).
Note eq.~(\ref{def-global-coordinates}) is consistent  with eq.~(\ref{proj}) via eq.~(\ref{algebra_lambdas}).

A first interesting fact is that the commutator of $\gen_{AB}$ maps into the covariant derivative:

\begin{eqnarray}
 \calL_{AB} M(\xi)&:=&Tr \left( \left[\gen_{AB},\widehat{M}\right] \right), \\
                  &= &\frac{\imath}{\sqrt{2}}J_{AB;CD}\partial_{CD}M(\xi).
\end{eqnarray} 

The quadratic Casimir operators are defined as:
\begin{eqnarray}
 [\gen_{AB},[\gen_{AB},\cdot ]]&=&C_2^{SO(6)}\cdot  \label{cc6}  \\   
 \left[\gen_{ab},[\gen_{ab}, \cdot ]\right]&=&C_2^{SO(5)} \cdot  \label{cc5}
\end{eqnarray}

What we want to calculate is the image of $\Delta_h$, eq.~(\ref{Delta_h_cas}), under the $\star$-product map. First we define the image of eqs.~(\ref{cc6})-(\ref{cc6}) in eqs.~(\ref{map-cas-spin6})-(\ref{map-cas-spin5}):

\begin{eqnarray}
   C_2^{SO(6)} \widehat{M} =\left[\gen_{AB},[\gen_{AB},\widehat{M}]\right] \longrightarrow 
             \mathcal{C}^{(6)} M,    \label{map-cas-spin6} \\ 
   C_2^{SO(5)} \widehat{M}=\left[\gen_{ab},[\gen_{ab},\widehat{M}]\right]
  \longrightarrow  \mathcal{C}^{(5)} M, \label{map-cas-spin5}
\end{eqnarray}
we can rewrite  $\mathcal{C}^{(6)}$ as
\begin{eqnarray} 
 \mathcal{C}^{(6)}&=&-\frac{1}{2} \kappa_6 \\
 \mathcal{C}^{(5)}&=&-\frac{1}{2} \kappa_5
\end{eqnarray}
where
\begin{eqnarray}
  \kappa_6 &=&  J_{AB,CD} \partial_{CD}\left( J_{AB,EF} 
                               \partial_{EF} \right)  \label{kappa6}      \\
                          &=& P_{CD;EF}\partial_{CD}\partial_{EF}+
                J_{AB;CD}(\partial_{CD} J_{AB;EF})\partial_{EF}  \\
             \kappa_5 &=&  J_{ab,CD} \partial_{CD}\left( J_{ab,EF} 
                               \partial_{EF} \right)    
\end{eqnarray}
A very useful expression to simplify the calculations is the contraction of the complex structure to the  partial derivative
\begin{equation}
  J_{AB;CD}\partial_{CD}=\sqrt{2}\left( \xi_{AC}\partial_{CB}- \xi_{BC}\partial_{CA} \right). 
  \label{J-contracted-derivative}
\end{equation}
We  are interested in extracting the metric tensor $\metric$ comparing the relevant continuous Laplacian with the general form:
\begin{eqnarray}
 -\calL^2&=& \frac{1}{\sqrt{G}}
              \partial_{\mu}\left(\sqrt{G}G^{\mu \nu} \partial_{\nu} \right) \\
        &=& G^{\mu \nu}  \partial_{\mu} \partial_{\nu} +
              \left(\partial_{\mu}G^{\mu \nu} \right)\partial_{\nu} +
              \frac{1}{\sqrt{G}}  G^{\mu \nu}
              \left( \partial_{\mu} \sqrt{G}  \right)\partial_{\nu}.
\end{eqnarray}
For the case when we retain the full $Spin(6)$-symmetry we have the Laplacian $\mathcal{C}^{(6)}$. The associated metric tensor is just $P_{AB;CD}$ as it can be verify from a  straightforward calculation for  $\kappa_6$:

\begin{equation}
   \kappa_6=\frac{1}{2 }\partial^2_{AB}-
            4\xi_{AC}\xi_{BD}\partial_{AB}\partial_{CD}-
            8\xi_{AB}\partial_{AB}.
\end{equation}

For the $Spin(5)$-symmetry case the corresponding image to the Laplacian eq.~(\ref{Delta_h}) is defined 
proportional to $\calL^2_h$  

\begin{equation}
  Tr\left(   \Delta_h  \widehat{M} \right):=\frac{1}{2R^2} \calL^2_h \label{map_deltah}
\end{equation}
then
\begin{equation}
  \calL^2_h= \kappa_6 +h\left(2\kappa_5-\kappa_6  \right).
\end{equation}
Our guess for the tensor metric related to $ \calL^2_h$ can be decomposed into a pure $Spin(6)$-symmetry part 
(given by $P^{AB;CD}$) plus a $Spin(5)$ invariant part denoted by $X^{AB;CD}$ as in eq.~(\ref{metric-tensor-covariant}):

\begin{equation}
 G^{AB;CD}=P^{AB;CD}+h X^{AB;CD}. \label{metric-tensor-covariant}
\end{equation}
$X$ is related  to $ \left(2\kappa_5-\kappa_6  \right)$, the term that breaks the $Spin(6)$ symmetry.

After a straightforward calculation we get:
\begin{equation}
 2\kappa_5-\kappa_6=\frac{1}{2}\partial_{ab}^2-4\xi_{ab}\xi_{cd}\partial_{ac}\partial_{bd}
                   - 8\xi_{a6}\xi_{b6}\partial_{ac}\partial_{bc}-  4\xi_{ab}\partial_{ab} \label{exterm}
\end{equation}
The tensor $X^{AB,CD}$ is obtained comparing the terms with second derivatives in eq.~(\ref{exterm}) to  $ G^{AB;CD}\partial_{AB}\partial_{CD} $
\begin{eqnarray}
   X^{ab;cd}&=& \frac{1}{2} A^{ab;cd}-2\left(\xi^{ac}\xi^{bd}-
                                             \xi^{ad}\xi^{bc}  
                                       \right)-  \nonumber \\ & &
                 2\left(\delta^{ac} \xi^{6b}\xi^{6d} -
                        \delta^{ad}   \xi^{6b}\xi^{6c}+ 
                        \delta^{bd}   \xi^{6a}\xi^{6c}-  
                        \delta^{bc}   \xi^{6a}\xi^{6d} \right)\\
   X^{a6;cd}&=& 0 = X^{ab;c6}= X^{a6;c6}
\end{eqnarray}

We can rewrite $X$ as:
\be
 X^{ab;cd}:= P^{ab;cd}-\med  M^{ab;cd},
\ee
with 
\be
 M^{ab;cd}=4\left(\delta^{ac} \xi^{6b}\xi^{6d} -
                        \delta^{ad}   \xi^{6b}\xi^{6c}+ 
                        \delta^{bd}   \xi^{6a}\xi^{6c}-  
                        \delta^{bc}   \xi^{6a}\xi^{6d} \right).
\ee
Some properties of $ M^{ab;cd}$ and $X^{ab;cd}$ are:
\bea
   M^{ab;cd}&=& M^{cd;ab} =- M^{ba;cd}, \\
   X^{ab;cd}&=& X^{cd;ab} =- X^{ba;cd}, \\
   M^2&=&M, \\
   X^2&=&X. 
\eea

Now the traces amount to

\begin{eqnarray}
   M^{ab;ab}&=& 8 \left( \delta_{aa}-1 \right) \xi^{6b}\xi^{6b}=4.\\
   X^{ab;ab}&=& \frac{1}{2} A^{ab;ab}+2\xi^{ab}\xi^{ba}-
                                     4(\delta_{aa}-1)   \xi^{6b}\xi^{6b}  
                     \nonumber                   \\
            &=& 5 +2\xi^{ab}\xi^{ba}-16  \xi^{6b}\xi^{6b}=2.
\end{eqnarray}
Notice $X^{ab;cd}$ is a rank-$2$ projector while $ M^{ab;cd}$is a rank-$4$ projector.
 
In order to invert the tensor metric we calculate the products  between projectors 
$P^{AB;CD}$,  $X^{ab;cd}$ and $ M^{ab;cd}$:
 
\begin{eqnarray}
   P^{ab;ef} P^{ef,cd}&=&   P^{ab;ef}- \frac{1}{4}M^{ef,cd}, \label{pro_PX-0} \\
   P^{ab;EF} X^{EF,cd}&=&   P^{ab;ef} X^{ef,cd}
                      = X^{ab;cd}. \label{pro_PX-1} \\
   P^{a6;EF} X^{EF,cd}&=&   P^{a6;ef} X^{ef,cd}
                      = 0, \label{pr0_PX-2} \\
   P^{ab;EF} M^{EF,cd}&=&   P^{ab;ef} M^{ef,cd}
                      = \med M^{ab;cd}. \label{pro_PM-1}\\
   P^{a6;EF} M^{EF,cd}&=&  P^{a6;ef} M^{ef,cd}
                      =  P^{a6;cd}, \label{pro_-2} \\
   X^{ab;ef} M^{ef,cd}&=&  0. \label{pro_XM}
\end{eqnarray}

The metric fulfils:
\begin{equation}
 G^{AB;CD}G_{CD;EF}=P^{AB;EF}.  \label{GGP}
\end{equation}

Now we assume the covariant tensor metric to take the form 
\be
 G_{AB;CD}= P_{AB;CD}+\alpha X_{AB,CD}, \label{propose_GGP_covariant}
\ee 
where $\alpha$ is a constant to determine.

From eq.~(\ref{GGP}) we distinguish two cases:
\begin{eqnarray}
   G^{ab;CD} G_{CD;ef} =
  &=&  (P+hX)^{ab;CD}(P+\alpha X)_{CD;ef} \nonumber\\
             &=&  P^{ab;ef}+(\alpha h +\alpha+h) X^{ab,ef}  \label{inverting_GGP_1}
\end{eqnarray}
Comparing eq.~(\ref{inverting_GGP_1}) to eq.~(\ref{GGP}) we get\footnote{We have to demand $h\neq -1$ in order to invert the tensor covariant metric tensor $G^{AB;CD}$.} $\alpha=\frac{h}{1+h}$. 
\be
{\mathbf{ \Rightarrow G_{ab;cd}= P_{ab;cd}-\frac{h}{1+h} X_{ab,cd}}}. \label{G_ab;cd} 
\ee
\begin{eqnarray}
  G^{a6;CD}G_{CD;ef}  &=&    (P+hX)^{a6;CD}(P+\alpha X)_{CD;ef} \nonumber \\
             &=&  P^{a6;ef}.  \label{inverting_GGP_2}
\end{eqnarray}
From eq.~(\ref{inverting_GGP_2}) we observe that for this case $G_{CD;ef}$ does not depend on $h$.
\be
 {\mathbf{ \Rightarrow  G_{a6;cd}= P_{a6;cd} }} \label{G_a6;cd} 
\ee

Finally we summarise the results in eqs.(\ref{G_ab;cd})-(\ref{G_a6;cd}) in eq.~(\ref{inverted_metric}):
\begin{equation}
\framebox[1.1\width]{$ \displaystyle
   G_{AB;CD}= P_{AB;CD}-\frac{h}{h+1} X_{AB,CD}.
$}  \label{inverted_metric}
\end{equation}

\subsection*{Observations}
Eq.~(\ref{pro_XM}) suggested  that $X_{ab;cd}$ and $M_{ab;cd}$ project onto {\em separated} spaces.
As we demonstrate in section \ref{CP3_locally}, $\CP^3$ is locally of the form $S^4 \times S^2$.

$X_{ab;cd}$ projects onto a 2-dimensional space that should be the fibre $S^2$ while 
$M_{ab;cd}$  projects onto a 4-dimensional that should be the base $S^4$. 
This information can be precisely obtained analysing the contributions to line element 
$ds^2$ due to the tensors $X_{ab;cd}$  and $M_{ab;cd}$. This will be achieved in the 
subsequent section \ref{line-element-ds2}.

Let us define $\Omega$, the symplectic structure as:
\be
  \Omega_{AB,EF}=G_{AB,CD}J^{CD}_{EF}
\ee
It is clear when $h=0$ (full $Spin(6)$-symmetry case) $\Omega_{AB,CD}=J_{AB;CD}$. In general  we have 

\begin{eqnarray}
   \Omega_{ab,c6} &=& P_{ab,c6},  \\
   \Omega_{ab;cd} &=& (1+h)P_{ab;cd}.
\end{eqnarray}

\subsection{The induced line element $ds^2$.}
\label{line-element-ds2}
The line element, {$ds^2$} at the north pole is defined as follows:
\be
  ds^2:=G_{AB;CD}^0e^{AB}e^{CD} \label{definition_ds2},
\ee
where $e^{AB}$ are defined as the coefficients of infinitesimal rotations under $Spin(6)$, i.e.\ the Maurer-Cartan forms, $R^{-1}dR=-e_{AB}L_{AB}$, where $L_{AB}$ are generators for rotations in 5 dimensions.

Note that in the case $h=0$ (full $Spin(6)$-symmetry) eq.~(\ref{definition_ds2}) reduces to 
\be
 ds^2= P_{AB;CD}e^{AB}e^{CD}
  \label{ds2-3}
\ee
as it was expected.

At the north pole we have
\bea
  M_{ab;cd}^0e^{ab}e^{cd}&=& 2(e^{c5})^2, \label{Mee-north-pole} \\
  X_{ab;cd}^0e^{ab}e^{cd}&=& \left( e^{13}-e^{24}\right)^2+
                               \left( e^{14}+e^{23}\right)^2. \label{Xee-north-pole}
\eea

From eqs.~(\ref{Mee-north-pole})-(\ref{Xee-north-pole}) we verify that $M_{ab;cd}$ 
projects to the base space $S^4$ and   $M_{ab;cd}$ projects to the fibre $S^2$.
The line element at the north pole depending on the parameter $h$ is
\be
  ds^2 \vert_{north pole}=\left( e_{15}^2 +e_{25}^2 +e_{35}^2 +e_{45}^2 \right)+
                         \frac{1}{1+h}\left[
                     \left( e_{13}-e_{24}\right)^2+ \left( e_{14}+e_{23}\right)^2  \right].\label{ds2-h}
\ee
We can compared eq.~(\ref{ds2-h}) to eq.~(\ref{line-element-spin5}).
We identify  the radius of  $S^2$ and $S^4$ as
\bea
  \frac{R^2_{S^2}}{R^2}&=&\frac{1}{1+h} \label{radius-s2},\\
  \frac{R^2_{S^4}}{R^2}  &=&1 \label{S4-radius}.
\eea

As we saw in section \ref{review-scalar-teory-s4}, $h$ takes values in the interval $(-1,\infty)$. 
We analyse some particular cases of eq.~(\ref{ds2-h}):
\begin{enumerate}
  \item $h=0$. We recover the $Spin(6)$ symmetry. The radius of the {\em rounded } $\CP^3$ is ${R^2=2}$.
  \item $h \longrightarrow \infty$. In this limit
        \[
          \frac{R^2_{S^2}}{R^2}\longrightarrow 0 
        \]
        then, the limit $h \rightarrow \infty$ corresponds to shrinking the $S^2$ 
        fibres to zero size. The radius of $S^4$ remains finite:
        \[
              \frac{R^2_{S^4}}{R^2}=1. 
        \]
  \item $h \longrightarrow -1$. Note that $h$ cannot take the exact value $1$ since for
        $h=1$ the tensor metric $G^{AB;CD}$ in eq.~(\ref{metric-tensor-covariant})
        cannot be inverted.
        For this limit we have        
        \[
          \frac{R^2_{S^2}}{R^2}\longrightarrow \infty, \qquad   \frac{R^2_{S^4}}{R^2}=1.
        \]
       This limit corresponds to making the size of the fibre infinitely large.
\end{enumerate}

\chapter{Conclusions from part \ref{analytical-part}}\label{conclusions_analytical_part}
\begin{itemize}
  \item We reviewed the prescription given in Ref. \cite{scalar-FT-S4} of the scalar field theory 
        on a {\em fuzzy} $4$-sphere.
  \item Since $S^4$ is not a phase space the construction of its {\em fuzzy} version has some 
        complications. The construction starts defining the matrix coordinates  $X_a$ as
        \begin{equation}
           X_a=\frac{R}{\sqrt{L(L+4)}}J_a, \nonumber
        \end{equation}
        where $J_a$ belong to the $(\frac{L}{2},\frac{L}{2})$ representation of Spin(5). They
        are matrices of dimension $d_L=\frac{(L+1)(L+2)(L+3)}{6}$ with $L$ an integer number.
        From $J_aJ_a=L(L+4)\ID$ follows an equation which holds for the matrix coordinates of 
        a $4$-sphere  
        \begin{equation}
          \sum_{a=1}^5 X_aX_a=R^2\ID. \nonumber \label{s4_F_defn-conclu}
        \end{equation}
        In the limit $L \longrightarrow \infty$ the  matrix coordinates commute,
        then  we have a matrix approximation to $S^4$ at algebraic level.
  \item The complications emerge for a finite $L$ where the five matrix coordinates $X_a$, $a=1,\dots,5$ 
        do not provide a 
        complete basis for the algebra of functions.
        To circumvent this problem we have two options:
        \begin{enumerate}
          \item To include more generators to complete the basis.
                This leads us to include more coefficients
                in the expansion of a function that are not related to 
                degrees of freedom of $S^4$, i.e.\ {\em{extra degrees of freedom}}.
                Then the constructed space is not $S^4_F$, it turns out to be $\CP^3_F$.
                To recover a scalar theory on  $S^4_F$. We implemented a {\em penalisation method}
                for all the non-$S^4_F$ modes in  $\CP^3_F$.
          \item A second alternative consists of projecting out the non-$S^4$ modes.   
                This leads us to deal with a non-associative algebra.
                We found that following the previous alternative we got the projector
                of the non-$S^4$ modes required in this second option.
        \end{enumerate}  
 \item The {\em penalisation method} introduced in the previous point consists of the following:
       \begin{itemize}
          \item We start defining an initial  action on $\CP^3_F$, $S_0[\Phi]$.

               Then we modify $S_0[\Phi]$ by adding a term  $S_I [ \Phi ]$.
               $S_I [ \Phi ]$ gives a positive value for those field configurations 
               associated to the non-$S^4$ modes and it is zero for those of  $S^4$.
         \item Then we construct the overall action as 
               \be   S[\Phi]= S_0[\Phi]+hS_I[\Phi] \label{modified-action}
               \ee
               where $h$ is a penalisation parameter in the interval $(-1,\infty)$.
         \item The probability of any given matrix configuration has the form:
               \begin{equation}
                 {\mathcal P}[\Phi]=\frac{{\rm e}^{-S_0[\Phi]-hS_I[\Phi]}}{Z}
                 \label{prob_of_config-conclu}
               \end{equation}
              where $Z=\int d[\Phi] {\rm e}^{-S[\Phi]-hS_I[\Phi]}$ is the partition function.
              For $h \longrightarrow \infty$ the states unrelated to $S^4$  become 
              highly improbable. 
     \end{itemize}
 \item The resulting action is
       \be
          S [ \Phi ]= Tr \left(\Phi \Delta_h \Phi+ V[ \Phi ] \right),
           \label{S_0.3-conclu}
       \ee
       where
       \begin{eqnarray}
           \Delta_h   & = & \frac{1}{2 R^2}\left( \CAS^{SO(6)}  
          + h ( 2\CAS^{SO(5)}-\CAS^{SO(6)}) \right)
        \label{Delta_h_cas-conclu}
       \end{eqnarray}
        is written in terms of 
       quadratic Casimir operator of the groups  $SO(5)$ and  $SO(6)$,
       $\CAS^{SO(5)}$ and $\CAS^{SO(6)}$ respectively.
       $R^2$ is the square of the radius of $\CP^3$.
 \item Since the defined action $S[\Phi]$ for $h$ large and positive contains only $S^4$ modes,
       the resulting model describes a scalar field theory for $S^4_F$.
 \item The {\em fuzzy} spaces can well retain the geometrical properties
       of the discretised space. This nice feature
       allowed us to provide a exact geometrical interpretation why the
       penalisation procedure described above works. The      
       Laplacian is given in eq.~(\ref{Delta_h_cas-conclu}).

 \item $\CP^3$ can be constructed either as a $Spin(6)$ orbit or as a $Spin(5)$ orbit.
      The Laplacian in eq.~(\ref{Delta_h_cas-conclu}) corresponds to $\CP^3_F$ with 
      $Spin(5)$ symmetry. We restore the $Spin(6)$ symmetry in a particular case of 
      eq.~(\ref{Delta_h_cas-conclu}) with $h=0$.
 \item It is known in the literature that $\CP^3$ is a non-trivial 
      fibre bundle over $S^4$ with $S^2$
      as the fibre. The construction of $\CP^3_F$ as $Spin(5)$ orbit  clarified this fact.
\item Using coherent states techniques we extracted the covariant tensor metric in 
      eq.~(\ref{Delta_h_cas-conclu}). Once we inverted the tensor metric
      we found the line element $ds^2$ in terms of the parameter $h$ as
     \be
       ds^2= \frac{1}{1+h}\sum_{a,b,c,d=1}^5 X_{ab;cd}e^{ab}e^{cd}+
             \med \sum_{a,b,c,d=1}^5 M_{ab;cd}e^{ab}e^{cd},
            \label{ds2-4-conclu}
     \ee
     where $e^{a,b}$ are the Maurer-Cartan forms for $Spin(5)$.
     $X_{ab;cd}$ is a rank-2 projector and $M_{ab;cd}$  is a rank-4 projector.
\item $X_{ab;cd}$ projects to the fibre $S^2$ and  $M_{ab;cd}$ projects to the 
     base space $S^4$. The line element in eq.~(\ref{ds2-4-conclu}) shows that
     locally $\CP^3=S^4\times S^2$.
\item We identified the radius of the fibre and base space as 
     \bea
         \frac{R^2_{S^2}}{R^2}&=&\frac{1}{1+h} \label{radius-s2-conclu},\\
         \frac{R^2_{S^4}}{R^2}  &=&1 \label{S4-radius-conclu}.
    \eea
\item As $h \longrightarrow \infty$ we have 
        \[
          \frac{R^2_{S^2}}{R^2}\longrightarrow 0.
        \]
        The meaning is that the $S^2$ fibres shrink to zero size while the radius of $S^4$ remains finite.
\item The suppression of the non-$S^4$ states in the field theory on $\CP^3$  corresponds to 
    a Kaluza-Klein type reduction of $\CP^3$ to $S^4$.
\item As we mentioned in section  \ref{s4_in_analogy_s2}, following our construction of $\CP^3$ as a $Spin(5)$ orbit we are able to
give the prescription for the projector to $S^4$ modes
\bea
{\cal P}_{S^4}&=&\prod_{n=1}^{L}\prod_{m=1}^{n}
\frac{\CAS^{SO(5)}-\lambda_{n,m}}{\med \CAS^{SO(6)}  -\lambda_{n,m}} \nonumber \\
      &=& \ID + \prod_{n=1}^{L}\prod_{m=1}^{n}
\frac{   \mathrm{C}_I }{ \CAS^{SO(6)}  -2\lambda_{n,m}}.
\label{projector_to_s4}
\eea

\end{itemize}                    
\chapter{General conclusions and perspectives}
\label{perspectives}
In this thesis we have presented a study of scalar field theory on
fuzzy spaces. The main motivation of our work was to explore the fuzzy
approach as a non-perturbative regularisation method of Quantum Field
Theories.  For this purpose we chose a hermitian scalar field theory
in a three dimensional space, with $\lambda\phi^4$ potential. This model
is perturbatively super-renormalisable (i.e.\ there are only a finite
number of perturbative diagrams that require renormalisation). Our
non-perturbative regularisation consisted of a fuzzy two sphere for
space and a lattice for Euclidean time. We performed Monte Carlo
simulations and obtained the phase diagram of the model. 
The study of such model via a standard lattice regularisation  leads to  a  phase
diagram consists of a disordered and a uniformly ordered phase
separated by a continuous second order phase transition that is
governed by the Ising universality class. In contrast
to the standard lattice regularisation \cite{LFT}, 
 in this new model we found
three phases, two are the disordered and uniformly ordered phases but
they are separated by a third new phase of non-uniform ordering.
This third phase is a property of the non-commutativity of the
regularised model and has arisen in other studies in the literature
and has variously been called a {\em striped phase} (Gubser and
Shondi \cite{Gubser:2000cd}, Ambj\o rn and Catterall \cite{caterall}, Bietenholz et al.\ \cite{BHN}-\cite{BHN-phase})
or a {\em matrix phase} by Martin \cite{xavier}.  We find that the three
phases meet at a triple point characterised by 
$\left( \bar{\lambda}_T, \bar{m^2}_T \right) = \left( (41.91\pm15) N^{-0.64\pm0.20}
      {\bar{R}^{-1.28\pm0.25}}, - (12.7\pm1) \bar{R}^{-1.92\pm0.20}   \right)$, 
see section \ref{discussion},  which inevitably runs
to the origin as the matrix size is sent to infinity. The implication
is that  the model we have studied in the end does not
capture the non-perturbative physics of the flat-space $\lambda\phi^4$
field theory. This was not unexpected since perturbative studies
\cite{delgadillo_thesis} found that this fuzzy sphere model suffers from
Ultraviolet-Infrared (UV/IR) mixing, and the non-uniformly ordered
phase is a non-perturbative manifestation of this phenomenon in the
neighbourhood of the ordering transition. It is conjectured (but
still has not been demonstrated in numerical studies) that the
introduction of another term into the action will suppress UV/IR
mixing and bring the model into the Ising Universality class. We have
not pursued this issue in this thesis as the introduction of this
additional term involves an additional parameter in the phase diagram
and it was necessary to proceed in steps. First it was essential to
understand the phase structure of this three parameter model before
studying how the phase diagram is deformed by the additional parameters.

The new model is naturally a non-commutative model and is of interest
also as it is a non-perturbative regularization of a non-commutative
field theory, that of a scalar field theory on the Moyal-Groenewold
plane. The new phase seems to be a characteristic feature of such
non-commutative theories. On the fuzzy sphere it is characterise by
the dominance of non zero angular momenta (i.e.\ $l>0$)  in the ground
state of the model and implies that even though rotational invariance
is preserved in the regularisation process it is spontaneously broken
by the ground state.

An advantage of the study is also  that different limiting spaces
(e.g. the commutative sphere, the Moyal plane and the commutative flat
space) can be obtained scaling $R$ as a power on $N$ and taking
$N\longrightarrow \infty$. Our analysis shows that in all the limits
we have  considered so far the phase of non-uniform ordering survives.
In particular (as mentioned above) the commutative flat model is not
in the Ising universality class. However, we expect that a normal
ordering in the vertex, i.e.\ the introduction of a counter-term to
cancel the tadpole diagrams, would return the model to the Ising
universality class.  This modification in the action is equivalent to
reinforcing the fuzzy kinetic term and can be achieved more simply
adding a term $\Phi (\mathcal{L}^2)^2 \Phi/(\Lambda^2 R^4)$ where
$\Lambda$ is a momentum cutoff to the action.  An analogous prescription 
of adding higher derivative terms to the quadratic terms should be 
applied to all fuzzy models, in such a manner that all diagrams are 
rendered finite, if the commutative theory is required in the infinite
matrix limit. 

This three dimensional model  is amenable to an alternative treatment as a 
Hamiltonian on the fuzzy sphere where the Hamiltonian would be
$${\cal{H}}=\frac{4\pi R^2}{N} \Tr \left(\frac{1}{2}\Pi^2+\frac{1}{2R^2}\Phi\hat{\cal
L}^2\Phi+\frac{1}{2}m^2 \Phi^2 + \frac{\lambda}{4}\Phi^4 \right)$$
with $\Pi= \dot{\Phi}$.

At strong coupling $\lambda$ it is expected that the model effectively depends on less parameters.
We found that our simulations in that regime reproduce the predictions for the chain of matrices \cite{Shimamune} where the contribution of the fuzzy kinetic term is neglected. 
This model is interesting from the String Theory point of view, it has been studied in Refs.~\cite{Gross-Klebanov}-
\cite{gromat} where it is known as $c=1$ {\em model}.

At the technical level, the simulations of scalar theories on fuzzy spaces
are slower than in their lattice counterparts since the fuzzy models
are intrinsically {\em non-local}. Nevertheless we have found that the 
simulations quickly converge as the matrix size is increased and 
we can capture the characteristic 
behaviour of the observables at very small matrix size.  It is, however,
expected that true advantages of the fuzzy approach only emerges with the
simulations involve a fermionic sector or the models of interest are 
supersymmetric.

A second, theoretical aspect of the thesis was the presentation of four
dimensional models on a round fuzzy approximation to $S^4$, via a
Kaluza-Klein reduction of $\CP^3$.  Though we did not perform numerical
studies in this case, from the studies that we have done so far, we 
are in a position to conjecture the structure of the phase diagram
for the $\CP^3$ and $S^4$ models. Since, the disordered-non-uniform
ordered transition line seems to be universal in the class of models
where space-time is modelled by a fuzzy space, we expect it to arise in
theses models also. It corresponds to the dominance of the potential
term and the models become pure potential one matrix models (with, in
our case, a $\phi^4$ potential).  The large $N$ limit of these models
are solved in terms of the density of eigenvalues. This density
undergoes a transition from a connected to a disconnected density as
the potential well is deepened by making the mass parameter, $m^2$,
more negative. It is this separation of the eigenvalue spectrum that
occurs at the disordered-non-uniform order transition line. In the
pure potential model (see eq.~(\ref{1-matrix-transition-curve-v2})) the transition curve is given by
$b_c=-2\sqrt{Nc}$ where $b$ is the total coefficient of the quadratic
potential term and $c$ that of the quartic term. This disordered phase
should give rise to a uniform ordered phase as the mass parameter is
furthered decreased. Again all these models should exhibit UV/IR
mixing.  The study of $S^4$ adds further complications due to the need
to squash $\CP^3$. A further complication is the need to introduce the
additional UV/IR suppressing term. However  the principal conclusion
of this thesis is that we see no insurmountable difficulty to the
implementation of the scheme implemented in this thesis and the
extensions outlined above as a regularisation scheme for quantum field
theories.  It has the distinct advantage that it is also a natural
regular method for non-commutative field theories.
         This constitutes an interesting numerical experiment for the Kaluza-Klein reductions 
         in the fuzzy context.

         Although there is still a lot to do to reach the high acceptance status 
         of Lattices Field Theories, to the date the fuzzy approach  have overcome the very 
         first tests. We may take the present work as a starting point to continue the  
         exploration of this possible alternative in future studies on fuzzy spaces. 
\appendix
\chapter{A small description of the Monte Carlo method}
\label{small-description-MC}

Let $\Phi$ be a configuration of the relevant field. The probability
for to this configuration is given by
\be
 \probability[\Phi]=\frac{e^{-S[\Phi]}}{\partition}, \label{def-probability}
\ee
where $S[\Phi]$ is the Euclidean action of the system in the configuration
$\Phi$. $\partition$ is called the {\em partition function},
\be
 \partition = \int D \left[\Phi\right] e^{-S \left[\Phi \right]}, \label{def-partition}
\ee
where $ \int D[\Phi]$ denotes the integration over all field configurations.
The expectation value of the observable $O$ is define by the expression:
\be
  \la O \ra=  \int D[\Phi] \frac{e^{-S[\Phi]}}{\partition} O[\Phi].\label{def-average}
\ee

The idea of the Monte Carlo method is to produce a sequence of
configurations $\{ \Phi_i \}$, $i=1,2,...,T_{MC}$\footnote{$T_{MC}$:
  Monte Carlo time} and evaluate the
 average of the observables over that set of configurations. In this
 way the expectation value is approximated as
\be
  \la O \ra \simeq  \frac{1}{T_{MC}} \sum_{i=1}^{T_{MC}} O_i,  \label{def-mean}
\ee
where $O_i$ is  the value of the observable $O$ evaluated in the $i$-sampled configuration,  
$\Phi_i$, i.e.\ $O_i=O[\Phi_i]$.

The sequence of configurations obtained by Monte Carlo have to be {\em
  representative} of the configuration space at the given parameters.

\section{The Metropolis algorithm}
\label{standard-mc-loop}
The concrete way that the Metropolis algorithm works is the following:
\begin{itemize}
  \item Start with a configuration $\Phi_{init}=\Phi_{0}$ and generate another configuration
    $\Phi_{test}$.
  \item Compute $\Delta S := S[\Phi_{test}] - S[\Phi_{0}]$.
    \begin{itemize}
       \item If  $\Delta S < 0$ then $\Phi_{test}$ is accepted,
	 i.e:
	 \[
         \Phi_1 = \Phi_{test}.
	 \]
       \item Otherwise the number $e^{-\Delta S }$ is compared to a
       random number, $ran\in[0,1]$; if  $e^{-\Delta S }>ran$ the
       configuration $\Phi_{test}$ is accepted, in the other case it is rejected, i.e.\  $\Phi_{1}=\Phi_0$.   
    \end{itemize}
    \item Set  $\Phi_{init}=\Phi_{1}$ and compare it with another
    configuration  $\Phi_{test}$.

\end{itemize}
Before we measure it is necessary to perform a number of steps,
in order to obtain  stable values for the observables of interest. 
This is called {\em thermalisation time}. In our study the thermalisation time was
estimated from the history of the action\footnote{ The picture
  obtained plotting the Monte Carlo step $i$ vs. the action at the configuration $\Phi_i$,
  $S[\Phi_i] $.}.

\subsubsection{The standard way to propose the following configuration}
\label{subsection2.1}
The variation of the configuration $\Phi(t)$ is performed element by element,
\be
  \Phi(t)_{ij} \longrightarrow  \Phi'(t)_{ij}= \Phi(t)_{ij}+ a_{ij}
\ee
where $a_{ij} \in \mathcal{C}$ is a random number.\footnote{In general the real and imaginary 
part are different random numbers.} Its real and imaginary part are in the
interval $\left[-N \sqrt{\vert \frac{m^2}{\lambda} \vert  },N
  \sqrt{\vert \frac{m^2}{\lambda} \vert  } \right]$ and the  $a_{ij}$ are chosen so that we preserve hermiticity of the field, i.e.\

$\Phi'(t)^{\dagger} = \Phi'(t)$.

One Monte Carlo step correspond 
to updating all entries of each $\Phi(t)$, $\trm=1, \dots ,N_{\trm}$ sequentially once.

\subsection{Modifying the Metropolis algorithm}
The standard Metropolis algorithm works well when 
the observable fluctuates around one value. In our  study we
found that the value of the observable fluctuates around several
values.\footnote{We call it {\em multilevel behaviour}, some
  examples of it appears in appendix \ref{appendix-aside-results}.} This was
  interpreted in the sense that the  effective potential of the system has several local minima.
It was also  observed that for certain values of the parameters there
is no tunnelling between the different minima for a very long history. Then the result depends
on which minimum the system falls in, and this  typically strongly depends on the
starting conditions. 
\subsubsection{Independent simulations}
As a first try to handle this dependence on the starting
conditions we perform many independent simulations. 
The algorithm described in \ref{standard-mc-loop} changes as follow:
\begin{itemize}
 \item Divide the $T_{MC}$-Metropolis steps into $sim$ parts.
 \item Choose a starting configuration  and {\em thermalise}.
 \item Perform the loop describe
    in \ref{standard-mc-loop}  $T_{MC}/sim$ times.
   \item Repeat the previous steps $sim$ times until you collect $T_{MC}$-configurations.   
\end{itemize}

This method is useful to check if the results do not depend on the starting conditions, 
but in some cases this method fails because the expectation values of the observables 
depend on the way that the $sim$ different initial conditions are chosen.

Let us characterise each minimum by the expectation value of the energy.

In {\bf figure \ref{ther-hot-m0p66-l15-N12-R8}} of chapter \ref{chapter-phases-characterisation} we showed that there is a clear difference between
results with {\em cold start} vs. {\em hot start}.
Between simulations with a {\em hot start}  no tunneling is observed, as 
 {\bf figure \ref{ther-hot-m24-l15-N12-R8}} illustrates.

\begin{center}
 \includegraphics[width=4.2in]{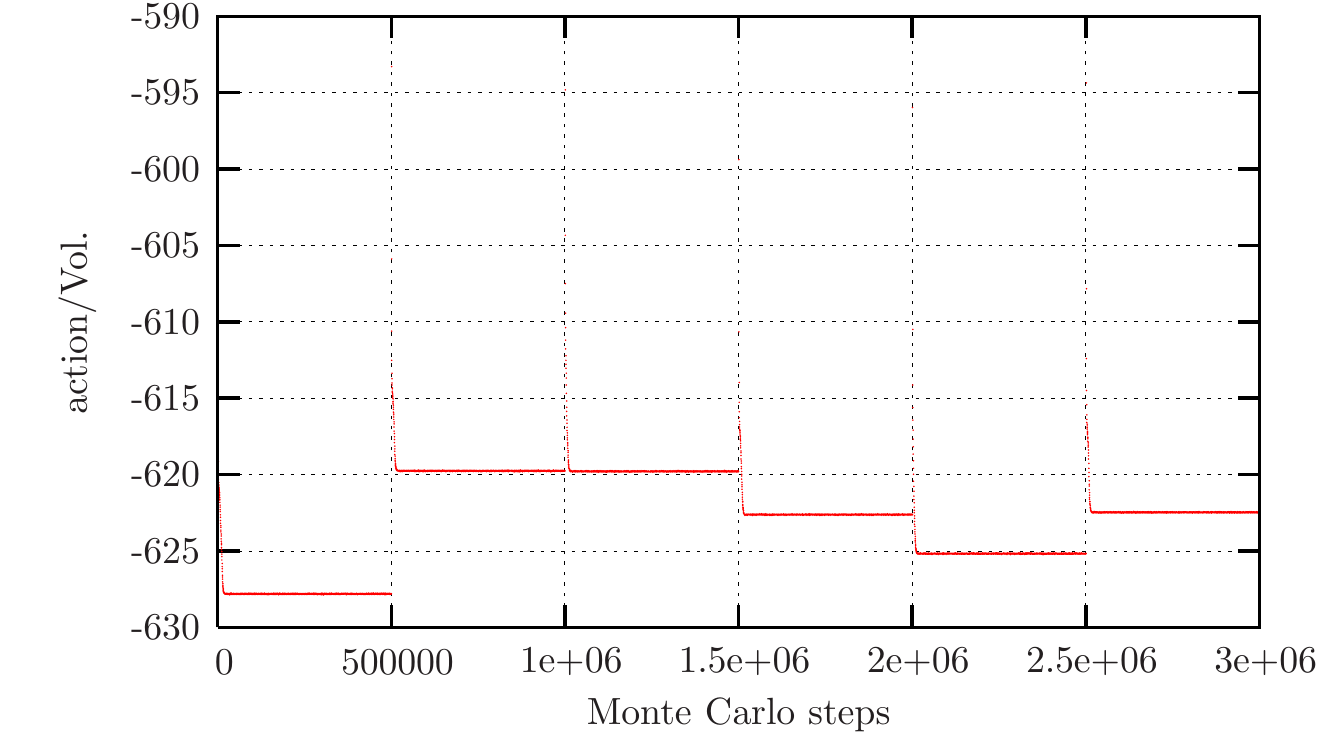}
\end{center}
\begin{figure}[h]
 \vspace{-0.1in}
 \caption{ History of the action  for $6$ different {\em hot} starting conditions 
  at $\bar{m^2}=-24$,
 $\bar{\lambda}=1.25$, $\bar{R}=8$, $N=12$. We take a new start every $500,000$ Monte Carlo steps. }
\label{ther-hot-m24-l15-N12-R8}
\end{figure}
In {\bf figure \ref{ther-hot-m24-l15-N12-R8}} we observe the history of the energy for 
$6$ independent simulations.
We observe  $4$ different mean values in the sense of  (\ref{def-mean}). This indicates that the space of 
configurations is divided into separated subspaces since we do not observe tunnelling in 
the same simulation.

According to the mean value of the energy, in the second and third simulation the 
system falls in the same minimum, characterised by  the central value $\la S \ra \approx -520$.
We have a similar situation for the fourth and sixth simulation where the expectation value of the energy
is  $\la S \ra \approx -622.5$.

If a history begins with a random hot start, the probability
to be trapped for a very long period in a specific action
minumum is what we denoted as the "size of the minimum" in
Chapter \ref{chapter-phases-characterisation}. 
In this situation it is obviously problematic to rely on a 
single run, or on the average over a few runs.

\subsection{An adaptive method for independent simulations}
A second attempt to sample the configurations takes into account the
{\em size of the minima}. To do this we {\em stick} the independent simulations in 
the following way:
\begin{itemize}
  \item Divide the $T_{MC}$-Metropolis steps into $sim$ parts
  \item Choose a starting configuration and then do the loop described
    in section \ref{standard-mc-loop} for $T_{MC}/sim$ times.
    Keep the last configuration, this will be the  $\Phi_{init}$ for
    the next loop.
  \item Choose a new starting configuration and thermalize to get
  $\Phi_{test}$, and test it as in \ref{standard-mc-loop} with
  $\Phi_{init}$ obtained from the previous independent simulation.
\end{itemize}

{\bf Figure \ref{many-sus-heat-N12-l22-R4}} shows the Specific Heat for $N=12$, $\bar{\lambda}=22/12$,
$\bar{R}=4$ obtained by the three different methods of measurement discussed in this appendix.
 \begin{center}
  \includegraphics[width=5.0in]{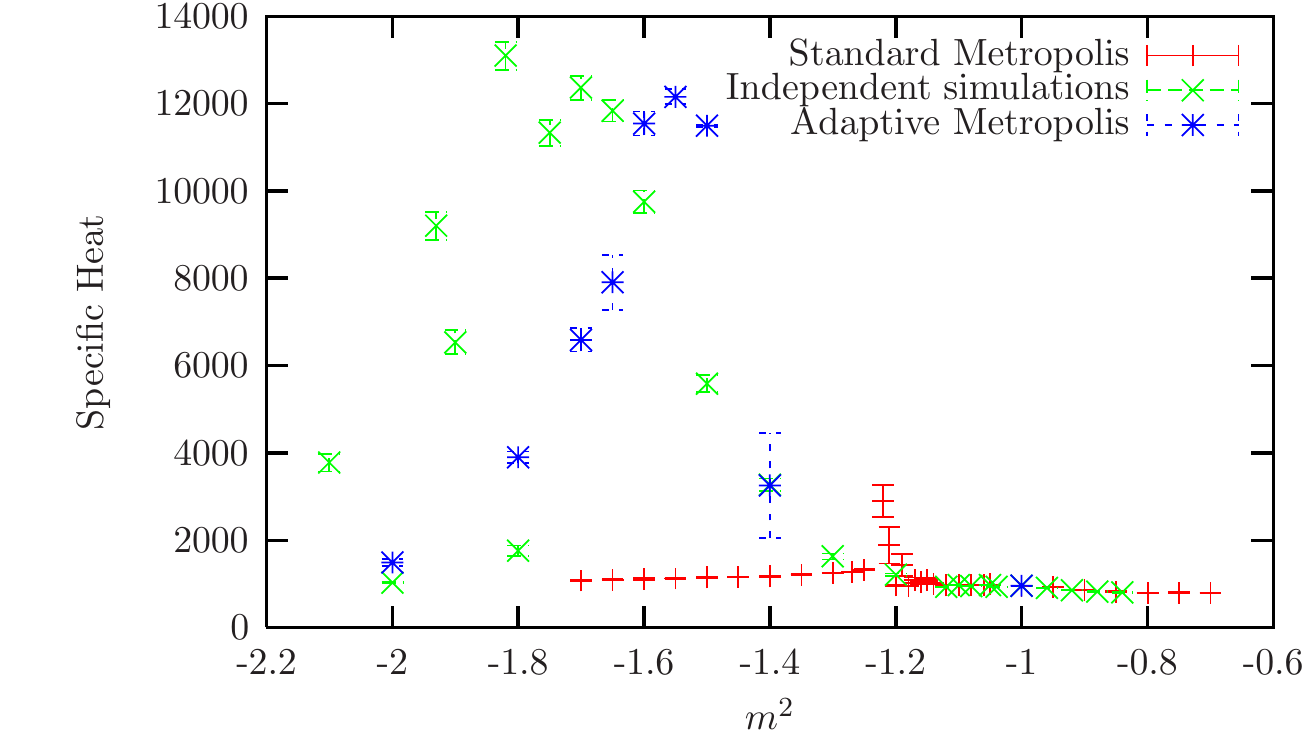}

 \end{center}
\begin{figure}[h]
 \vspace{-0.2in}
 \caption{The Specific Heat measured by the different methods at $\bar{\lambda}=1.83$, $\bar{R}=4$, $N=12$.}
\label{many-sus-heat-N12-l22-R4} 
\end{figure}
As we can observe, different methods give different
results and this is because the
value of the observables strongly depends on the way of
measuring. 
As an example of the statement above we present in {\bf figure \ref{ene-histos-N12-T12-m-1p7-l22-R4}} 
the histogram  for the action at $\bar{m}^2=-1.7$ of the
{\bf{figure} \ref{many-sus-heat-N12-l22-R4}}:
 \begin{center}
  \includegraphics[width=4.6in]{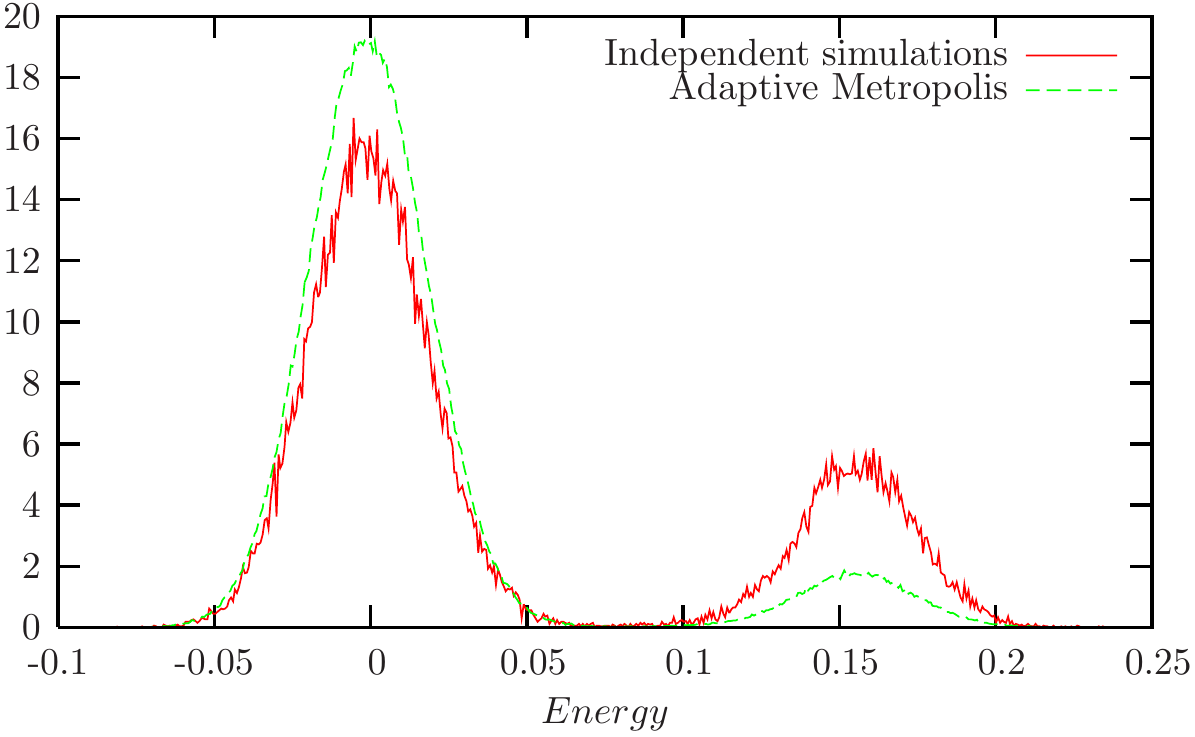}
 \end{center}
\begin{figure}[h]
 \vspace{-0.3in}
 \caption{ Histograms  for the action at
 $\bar{m}^2=-1.7$, $\bar{\lambda}=1.83$, $\bar{R}=4$, $N=12$.}
\label{ene-histos-N12-T12-m-1p7-l22-R4}
\end{figure}
{\bf Figure \ref{histos-N12-T12-m-1p7-l22-R4-v2} } shows the histogram for $\Tr(\Phi)$ with the
same setting as in   {\bf figure \ref{ene-histos-N12-T12-m-1p7-l22-R4}}.

\begin{center}
  \includegraphics[width=4.6in]{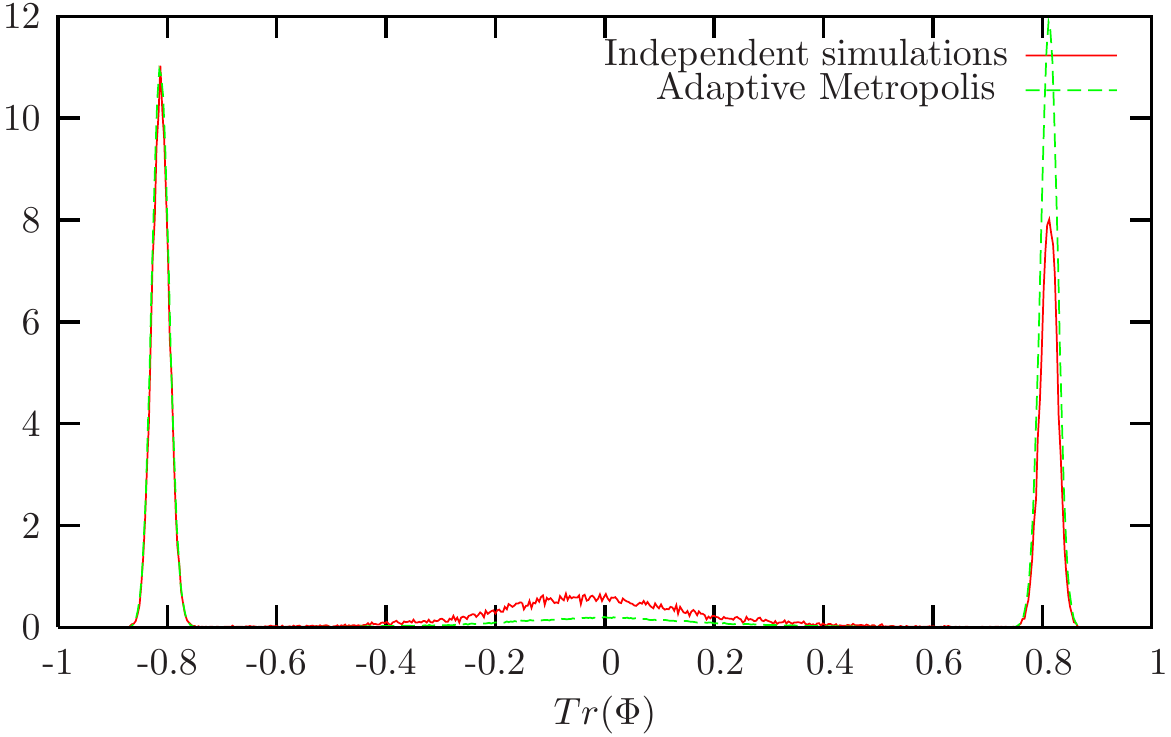}
 \end{center}
\begin{figure}[h]
 \vspace{-0.3in}
 \caption{ Histograms  for  $Tr(\Phi)$ at
 $\bar{m}^2=-1.7$, $\bar{\lambda}=1.83$, $\bar{R}=4$, $N=12$.}
\label{histos-N12-T12-m-1p7-l22-R4-v2}
\end{figure}

The thermalisation problems
reflect the fact that there exist a large potential barrier between the different subspaces.
As  consequence  we cannot sweep all the space of configurations, therefore the expectation values 
of the observables are not reliable.

\section{Methods to estimate the error}

In this section we give a brief explanation of the methods we used to estimate the statistical
errors in our simulations.

We start by defining the  mean $\bar{O}$ over a subset of $\{O_i \}_{i=1}^{T_{MC}}$ of $n$ elements ($n\leq T_{MC}$)

\be
   \bar{O}= \frac{1}{n} \sum_{i=1}^{n} O_i. \label{def-mean-v2}
\ee
The standard deviation is defined as
\be
  \sigma=\sqrt{\frac{\frac{1}{n} \sum_{i=1}^n \left(O_i-\bar{O} \right)^2   }{n-1}} 
        =\sqrt{\frac{1}{n-1}\left( \overline{O^2} -(\bar {O})^2\right)}. 
\label{standard_deviation}
\ee

If the samples are statistically independent eq.~(\ref{standard_deviation}) gives a good 
estimation of the error. If this is not fulfilled, then the correct expression for $\sigma$ 
is (see Ref.~\cite{Barkema}):
\be
 \sigma = \sqrt{\frac{1+\frac{2\tau}{\Delta n}}{n-1} \left( \overline{O^2}-(\bar{O})^2 \right) } 
          \label{sigma-correlated}
\ee
where $\tau$ is the autocorrelation time and $\Delta n$ is the Monte Carlo {\em time} interval 
at which the samples were taken. It is related to the total number of samples by the expression $n=\frac{T_{MC}}{\Delta n}$.
 
For large $n$ and $2 \tau \gg \Delta n$,  eq.~(\ref{sigma-correlated}) turns into eq.~(\ref{sigma-correlated-n-large}):
\be
   \sigma=\sqrt{\frac{2\tau}{T_{MC}}\left( \overline{O^2}-(\bar{O})^2 \right)  }.
   \label{sigma-correlated-n-large}
\ee

Now the problem is to estimate the autocorrelation time.
The formal expression for the autocorrelation time (see Ref.~\cite{Janke}) is 
\be
  \tau= \frac{1}{2} +\sum_{k=1}^n A(k) \left(1-\frac{k}{n} \right) ,
       \label{autocorrelation_time}
\ee
with
\be
  A(k)= \sum_{i\neq j}^{n}\frac{\la O_i O_{i+k}\ra -\la O_i\ra \la O_j\ra}{\la O_i^2\ra-\la O_i\ra\la O_j\ra}.
    \label{coefficient_A_k}
\ee

\subsection{Binning method}
This method is also called blocking method.
The idea behind this method is to 
divide the vector of measurements $\{O_i\}_{i=1}^{T_{MC}}$ into $n_b$ blocks (also called bin number), 
then we evaluate the observable for each block to obtain a new vector of measurements 
$\{\tilde{O}_j \}_{j=1}^{n_b}$. The error is estimated as if these new  measurements were statistically 
independent, then it is obtained via eq.~(\ref{standard_deviation}) replacing $n$ by $n_b$. 
The principal disadvantage of this  method is that the error may strongly depends on the choice of $n_b$.
Then one should test several values of  $n_b$ and keep the one where the error is maximal.
\subsection{Jackknife method}
This procedure can be consider a {\em re-sampling} method. It also starts by dividing the vector of measurements into $n_b$ blocks. Then
one forms $N_B$ large blocks cointaining all data but one of the previous binning blocks
\be
  O^{\rm jackknife}_j=\frac{T_{MC}\cdot \bar{O} - k \cdot \tilde{O}_j}{T_{MC}-j}, \quad j=1,...,N_B.
\ee
where $\tilde{O}_j$ is the average in the $j$ block.
$k$ is the number of samples in each block.
The error is then calculated as follows:
\be
 \sigma=\sqrt{\frac{N_B-1}{N_B}\sum_{j=1}^{N_B}\left(O^{\rm jackknife}_j- \bar{O}  \right)^2}.
\ee

\subsection{Sokal-Madras method}
This method is based on the estimation of the autocorrelation time given by 
eq.~(\ref{autocorrelation_time}), see \cite{SokalMadras}. 
The error is given by eq.~(\ref{sigma-correlated-n-large}).
In the case that the autocorrelation time turns out to be below $0.5$ (the samples are decorrelated), we take the standard error given by eq.~(\ref{standard_deviation}).

In our simulations we compare the error given by these three methods. In the case of the observables  
$\varphi_{all}^2$, $ \varphi_0$, $\chi_1$ and $E$ -- eqs.~(\ref{full_power_field}), (\ref{parameter0}), (\ref{phi_1}) and (\ref{energy}) respectively -- the errors given by the three methods are compatible.

For the Specific Heat (or the susceptibilities of the different modes) the most careful estimate 
was given by the Sokal-Madras method (largest errors).

\section{Technical notes}\label{tecnical-notes}The runs were performed on two different clusters:    
\begin{itemize}
 \item {\em Berlin cluster}:
       \begin{itemize}
         \item $450$MHz Athlon (pha4-pha9)
         \item $900$MHz (pha15-pha19) 
         \item 2.66GHz P-4 (pha30-pha33).
       \end{itemize}
 \item {\em Dublin cluster} 
       \begin{itemize}
         \item $3.06$GHz Intel Xeons-dual
              (cluster0-cluster15)
         \item $2.80$GHz Intel Xeon-dual (Gibbs) 
         \item  1.5Ghz P-4 
               (schrodinger, hamilton, oraifertaigh, lanczos)
       \end{itemize}
\end{itemize}

      The code is written in $C++$ and uses the Message Passing Interface (MPI).

\chapter{Polarisation tensors for $SU(2)$}
\label{app-polarisation}
In this section we want to describe some generalities of the
polarisation tensors  $\Yp_{l,m}$, which 
 form a basis for the matrices $\Phi$ in eq.~(\ref{expansion_field}). They are the
eigenvectors of the operator defined in eq.~(\ref{replace_fuzzy_5}) --the fuzzy version of
the angular momentum operator $\calL^2$
\be
 \lpl^2 \cdot=\sum_{i=1}^3
                    \left[L_i, \left[L_i, \cdot  \right]
                     \right]
\ee
where $L_i\in Mat_{L+1}$. So we have the set $\{ \Yp_{lm}\}_{l \le L,
  m\le l }$ such that
\be
 \lpl^2 \Yp_{lm}={l}({l} +1 ) \Yp_{lm},
\ee

\be
  \Yp_{l,m}^{\dagger}=(-1)^m \Yp_{l,-m}.    \label{yp_dagger}
\ee

Following Ref.~\cite{Varsha}, their algebra is
{\small
\bea
  \Yp_{l_1m_1}\Yp_{l_2 m_2}&=&\sqrt{\frac{L+1}{4\pi}}\sum_{l',m'}(-1)^{L+l'}
                  \sqrt{(2l_1+1)(2l_2+1)}  \nonumber \\
        & &       \hspace{2.2cm}  \times  \left\{ \begin{array}{ccc}
                               l_1  & l_2  & l' \\
                              L/2 & L/2 & L/2  \\
                        \end{array}
                  \right\} \clebsch^{l'm'}_{l_1 m_1 l_2 m_2}\Yp_{l'm'}
\eea
}
$ \left\{ \begin{array}{ccc}
           l_1  & l_3  & l' \\
           L/2 & L/2 & L/2  \\
          \end{array} \right\}$ are the {\em Wigner 3mj-symbols} 
--see Ref.\cite{Wigner-symbols}-\cite{Wigner-symbols-calculations} --
and $\clebsch_{l_1 m_1 l_2 m_2}^{l'm'}$ are the Clebsch-Gordan coefficients.

Their normalisation is\footnote{In fact the operators defined in \cite{Varsha}, $T_{lm}$,
are essentially  our polarisation tensors up to a factor: \[ T_{lm}=\sqrt{\cte} \Yp_{lm} \]} 
chosen as:
\be
   \frac{4\pi}{L+1} \Tr\left(\Yp_{lm}^{\dagger}\Yp_{l'm'}\right)=
      \delta_{ll'}\delta_{mm'}.   \label{orthonorma}
\ee

\section{Explicit form of  the generators of $SU(2)$ IRR of
  dimension $d_L$}
\label{section3.3.1}
As we mentioned in section \ref{fuzzy_sphere_section}, $L_i$ are the generators of
the $SU(2)$ irreducible representations (IRR) of dimension $N=L+1$. They satisfy the  Casimir constraint:
\be
   C^2_{SU(2)}:= \sum_{i=1}^3 L_i L_i= \frac{1}{4}(N^2-1) \cdot 1 \!\! 1 \ .
\ee

There is a well established procedure to construct these generators in an 
arbitrary representation. It operates in the {\em Cartan basis} where 
we work with $L_+:=L_1+\imath L_2$, $L_-:=L_1-\imath L_2$ and $L_z :=
L_3$.
\bea
   \left[L_z \right]_{ij} & = & 
\left\{ \begin{array}{cc} 
\med (d_L+1-2i) & \quad \mbox{if } i=j   \\
 0   & \quad \mbox{otherwise}  
\end{array} \right. \label{l.z} \\ 
\left[L_+ \right]_{ij} &=& 
\left\{ \begin{array}{cc}
\sqrt{i(d_L-i)} & \quad \mbox{if } i+1=j   \\ 
0 & \quad \mbox{otherwise}
\end{array} \right.  \label{l.+} \\ 
\left[L_- \right]_{ij} &=& 
\left\{ \begin{array}{cc}
\sqrt{j(d_L-j)}  & \quad \mbox{if } i-1=j \\
 0 & \quad \mbox{otherwise}
\end{array} \right. \label{l.-} 
\eea

Note that $(L_+)^{\dagger}=L_-$.

 The commutation relations read:
\bea
 \left[ L_z, L_+ \right] & =&  L_+, \nonumber \\
 \left[ L_z, L_- \right] & =& - L_-, \nonumber \\
 \left[ L_+, L_- \right] & =& 2 L_z \ .
\eea

\section{Explicit construction of the polarisation tensors}

$\Yp_{lm}$ can be constructed as traceless polynomials of order $l$ on $L_i$, for example, for $l=0$, 
$\Yp_{00} \propto 1 \!\! 1$. Demanding $\cte \Tr\left( \Yp_{00} \Yp_{00}\right)=1$,
 we arrive at 
 $\Yp_{00}   =  \frac{1}{\sqrt{4 \pi}} 1 \!\! 1  $ (also see Ref.~\cite{Yau}).

For $l=1$ we have $\Yp_{10} \propto L_z$, $\Yp_{1,1} \propto L_+$ and $\Yp_{1,-1} \propto L_-$. 
The normalisation can be found calculating the trace of the squared of $L_z,L_+,L_-$. 
Following our definitions (\ref{l.z})- (\ref{l.-}) 
we found $\Tr\left( L_z^{\dagger} L_z\right)=\frac{L(L+1)(L+2)}{12}$,
 $\Tr\left( L_+^{\dagger} L+\right)=\Tr\left( L_- L+\right)=\frac{L(L+1)(L+2)}{6}$. Then we choose $ \Yp_{1,+1} = e^{\imath \phi} \sqrt{\frac{3}{2 \pi}} \frac{1}{\sqrt{L(L+2)}} L_+ $ and so on for the remaining $m$'s. The phase $e^{\imath \phi}$ has to be fixed demanding eq.~(\ref{yp_dagger}) to hold.

Summarising:
\bea
   \Yp_{00}   &=& \frac{1}{\sqrt{4 \pi}} 1 \!\! 1_{l} \label{Yp00} \\
   \Yp_{1,+1} &=& \imath \sqrt{\frac{3}{2 \pi }} \frac{1}{\sqrt{L(L+2)}} L_+ \label{Yp11}\\
   \Yp_{10}   &=& \sqrt{\frac{3}{ \pi}} \frac{1}{\sqrt{L(L+2)}} L_z \label{Yp10}\\
   \Yp_{1,-1} &=& \imath \sqrt{\frac{3}{2 \pi}} \frac{1}{\sqrt{L(L+2)}}L_- \label{Yp1-1}
\eea
\chapter{Aside results}
\label{appendix-aside-results}
\section{Criteria to determine the phase transition}

To sketch the phase diagram we compare the following two criteria:
\begin{itemize}
\item {The criteria of the susceptibilities.}\label{criteria-1}
The first phase diagram  was revealed by searching for the values of $m^{2}$ 
where the two-point function of the zero mode  has its peak
for a given value of $\lambda$. 

\item {Specific Heat Criterion.}\label{criteria-2}
A second phase diagram   was found searching for the values of $m^{2}$ 
where the specific heat peaks.
\end{itemize}
We found that for values of $ \bar{\lambda} < \bar{\lambda}_T$ for $\bar{R}$ fixed
both  criteria roughly coincide, as it can be seen in the {\bf figure \ref{sus-order-N16-l7-R4}} 
for $N=16$, $\bar{\lambda}=0.44$, $\bar{R}=4$

\centerline{
  \includegraphics[width=3.0in]{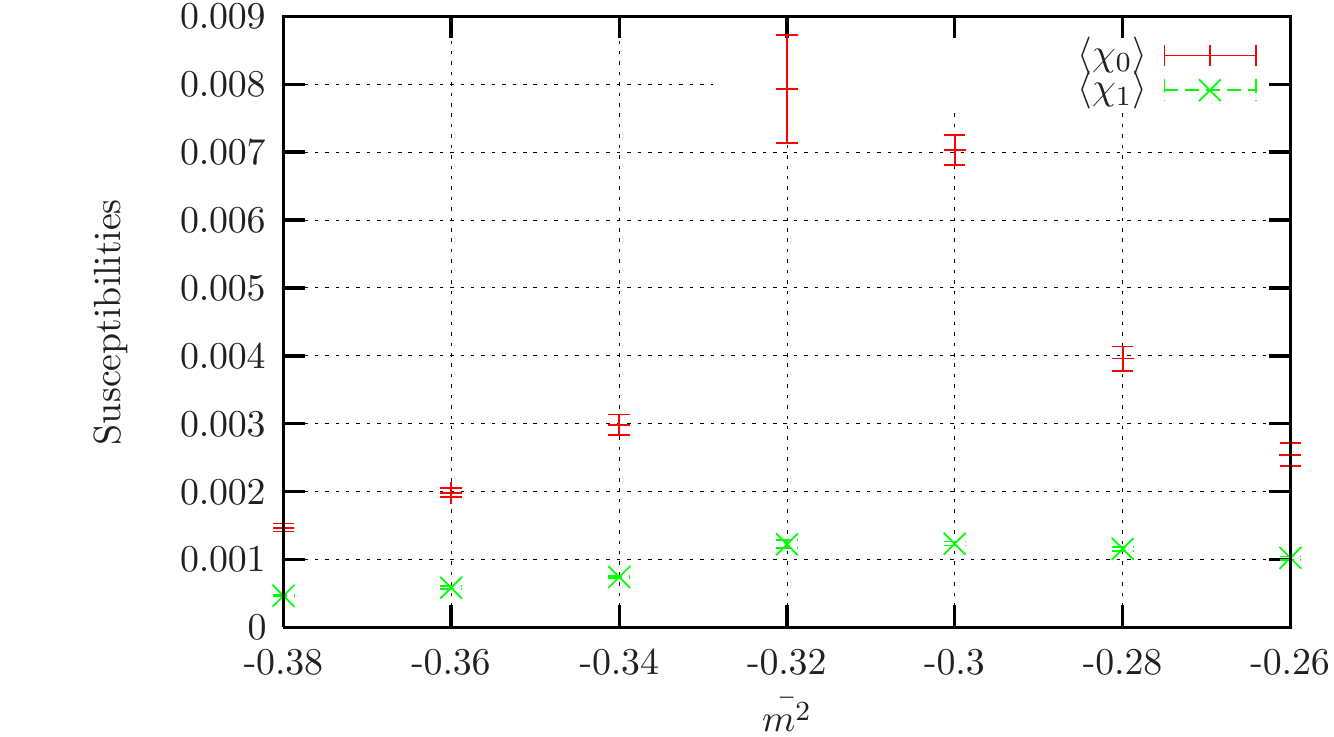}
  \includegraphics[width=3.0in]{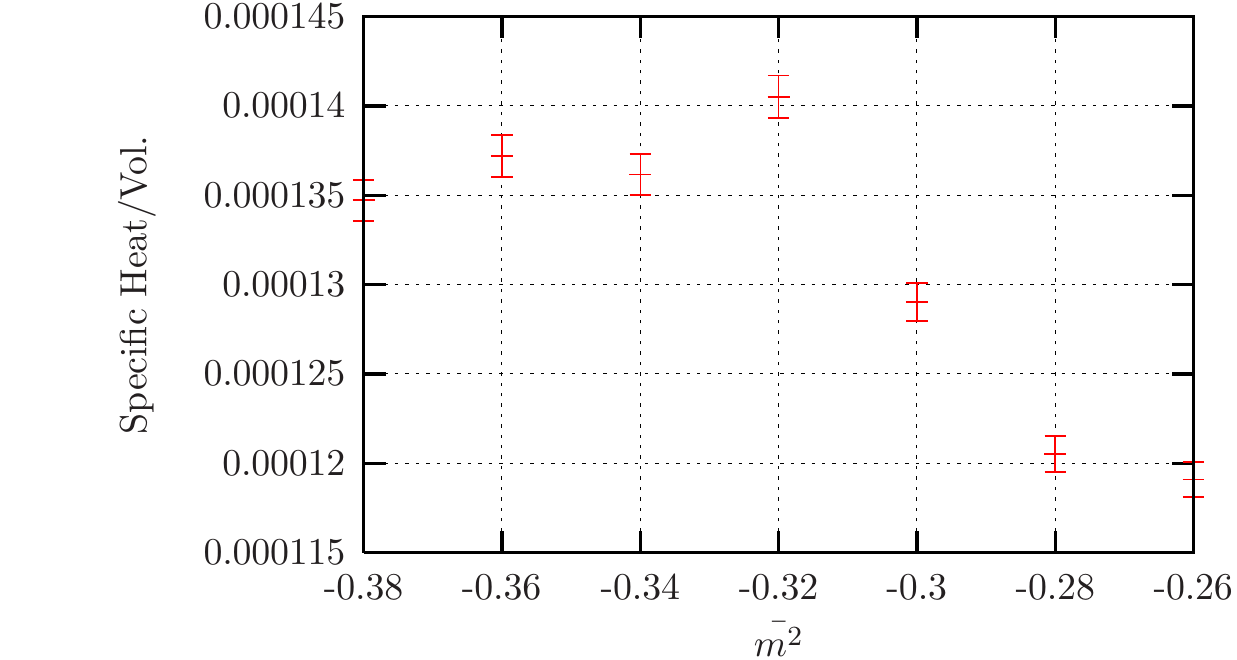}
}
\begin{figure}[h]
  \vspace{-0.3in}
  \caption{Susceptibilities $\chi_0$ and $\chi_1$ and the specific heat at $\bar{\lambda}=0.44$, $\bar{R}=4$, $N=16$.}
 \label{sus-order-N16-l7-R4}
\end{figure}

\pagebreak

We observe that the specific heat an the zero mode susceptibility $\chi_0$ in eq.~(\ref{sus_zero_mode}) 
roughly peak at the same value of $m^2$. This can be explainrd analysing the partial contributions to the action:

 \begin{center}
  \includegraphics[width=3.8in]{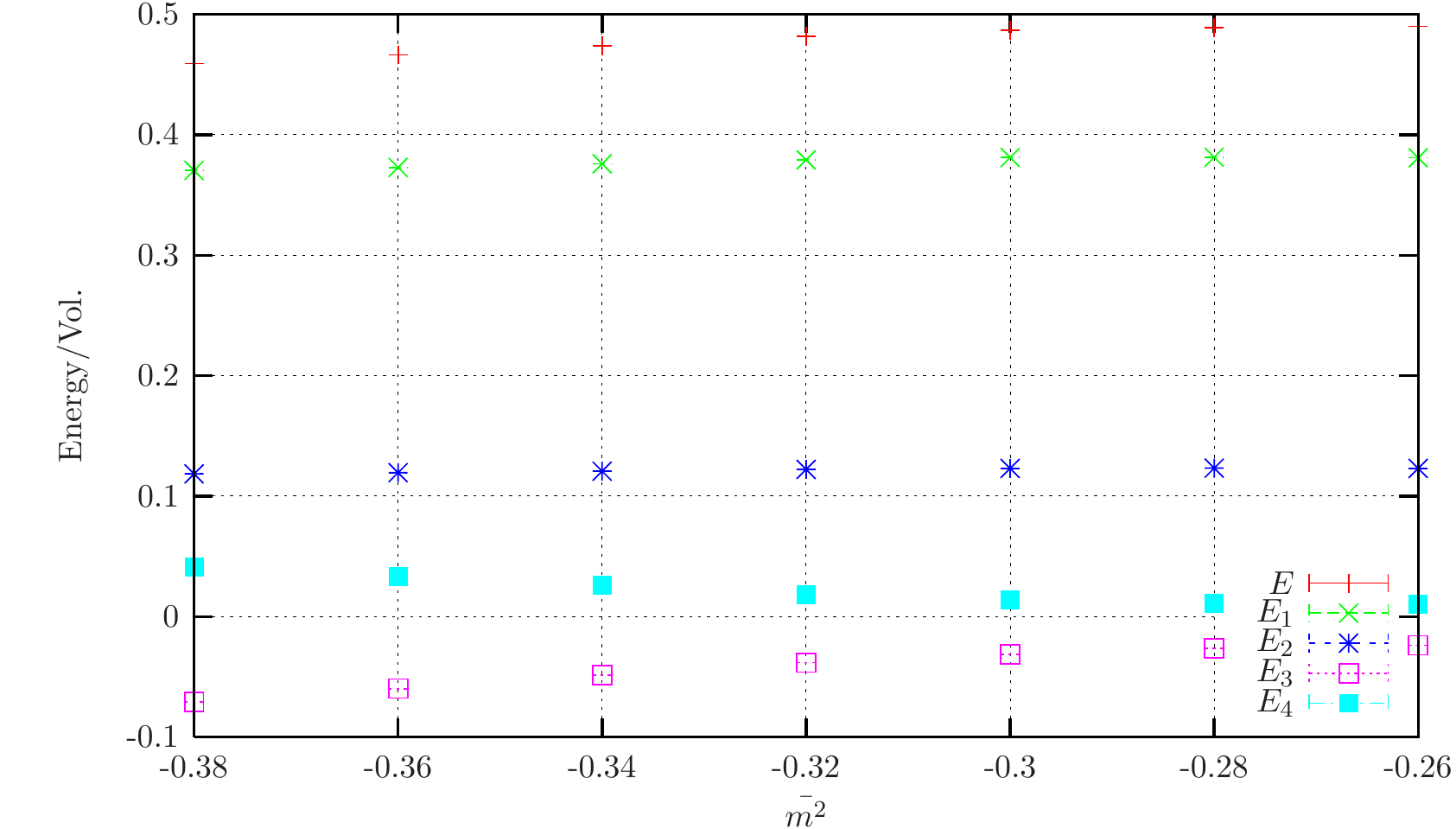}

 \end{center}
\begin{figure}[h]
 \vspace{-0.35in}
 \caption{ Partial contributions given by eqs.~(\ref{energy1})-(\ref{energy4}) to the internal energy eq.~(\ref{energy}) at $\bar{\lambda}=0.44$, $\bar{R}=4$, $N=16$}
\label{contrib-N16-l7-R4}
\end{figure}

We observe that the main contribution comes from the kinetics terms. The kinetic fuzzy term selects the configuration where the zero mode is leading.

 \begin{center} 
   \includegraphics[width=3.6in]{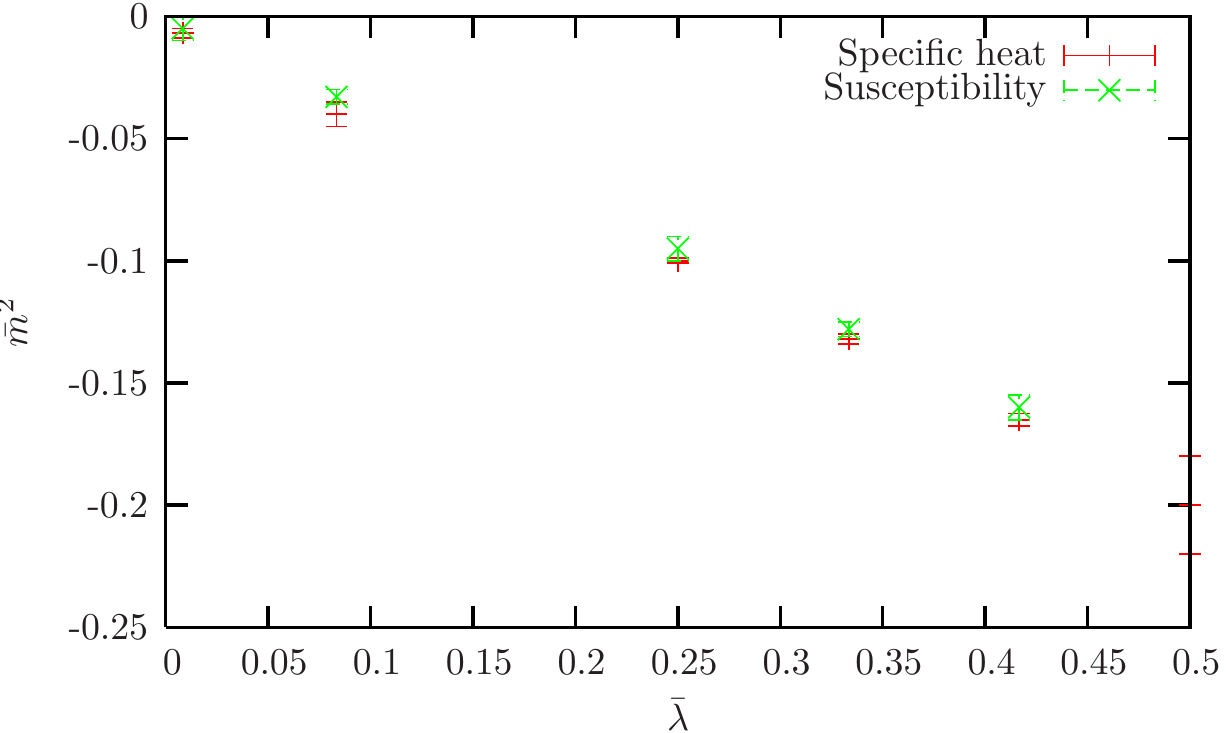}\end{center}
\begin{figure}[h]
 \vspace{-0.3in}
 \caption{ Comparison between the critical points obtained by the critera of the Specific Heat --eq.~(\ref{specific_heat})--
           and of the zero mode susceptibility in eq.~(\ref{sus_zero_mode}),   at
          $\bar{R}=8$, $N=12$. We  observe that the critical points  overlap within the error bars.}
\label{phase-diagram-N12-R8}
\end{figure}

\section{Free field results}
In the case $\lambda=0$, following Ref.~\cite{delgadillo_thesis} the expression for
the space correlator (\ref{space_correlator}) for $l=0$ is:
\be
  \la c_{00}^*(0) c_{00}(0) \ra :=  \la \varphi_0^2 \ra=  \frac{1}{4 \pi \bar{R}^2 \bar{m}^2}.
\ee

{\bf Figure \ref{free_field_result}} shows that our simulation results agree with this formula.
 \begin{center}
  \includegraphics[width=3.5in]{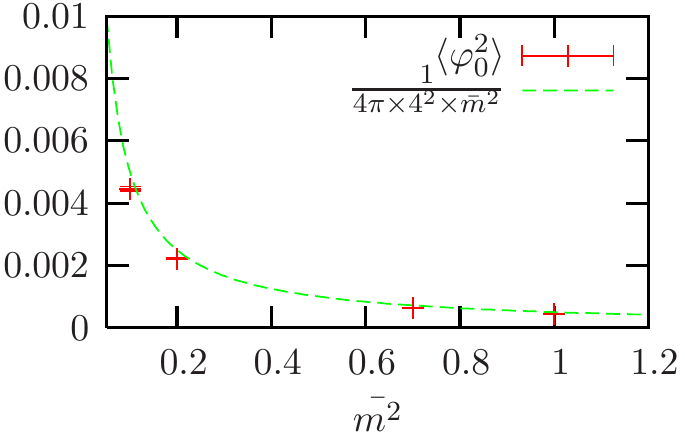}
 \end{center}
\begin{figure}[h]
\vspace{-0.33in}
\caption{$ \la \varphi_0^2 \ra$ for  $\bar{\lambda}=0$, $\bar{R}=4$, $N=12$.}
\label{free_field_result}
\end{figure}

\chapter{Tables}
\label{tables-appendix}

Table {\bf \ref{table-critical-and-collapse-tcurve}} is organised as follows:
\begin{enumerate}
\item For each pair $(N,\bar{R})$ it gives the maximal value of $\bar{\lambda}$ 
      simulated ---for several values of $\bar{m}^2$--- free of thermalisation problems
      around the phase transition. 
\item Next it indicates the critical value, $\bar{m}^2_c$. This is done with the purpose
      of having a reference of the magnitude of the simulated values of $\bar{m}^2$. 
\item Then it contains the corresponding  parameters $A,B,C$ and $D$ defined in 
      eqs.~(\ref{cte-D-vv-z-fixed})-(\ref{cte-C-vv-z-fixed})  
      (see chapter \ref{discussion} for more details).
\item Finally it present the proposed fit for the transition curve $I$-$III$ of the form
      (\ref{large-lambda-tc-curve}) (in some cases it indicates the range of $\bar{\lambda}$
      for the proposed fit). 
\end{enumerate}

{\footnotesize{
\begin{table}
  \caption{Critical values for the disordered-ordered non-uniform phase transition}
    \label{table-critical-and-collapse-tcurve}
\begin{center}
\footnotesize{
\begin{tabular}{|rrrr|r|}\hline \hline
 &Maximal &critical       & & Proposed fit for \\
 &parameters     & & & the transition curve \\
\hline
$N=8$ & $\bar{R}=4$ & $N \bar{\lambda}= 1300$ & $\bar{m}^2_c=-24\pm0.5$  &  
$\bar{m^2}_c= -0.099(N \bar{\lambda})^{0.76}$  \\  
$A=2\pi$ & $D=100.531$ & $B_c=-2412.74$ & $C= 65345.1$  &  \hspace{1.8in}  \\  
\hline  
$N=8$ & $\bar{R}=8$ & $N \bar{\lambda}= 2500$ & $\bar{m}^2=-13.8\pm1$  &  
$\bar{m^2}_c= -0.065(N \bar{\lambda})^{0.69}$  \\  
$A=2\pi$ & $D=402.124$ & $B_c=-5549.31$ & $C= 502655$  &  \hspace{1.0in}  \\  
\hline
$N=8$ & $\bar{R}=16$ & $N \bar{\lambda}= 5000$ & $\bar{m}=-9\pm0.3$  &  
$\bar{m^2}_c= -0.026(N \bar{\lambda})^{0.69}$ \\ 
$A=2\pi$ & $D=1608.5$ & $B_c=-14476.5$ & $C= 4.02124\times 10^6$  & \hspace{0.3in}  \\ 
\hline
$N=12$ & $\bar{R}=4$ & $N \bar{\lambda}= 1200$ & $\bar{m}^2=-24\pm0.5$  &  
$\bar{m^2}_c= -0.15(N \bar{\lambda})^{0.71}$  \\  
$A=2\pi$ & $D=100.531$ & $B_c=-2412.74$ & $C= 60318.6$  &  for  $N \bar{\lambda} \ge 400$ \\  
\hline
$N=12$ & $\bar{R}=8$ & $N \bar{\lambda}= 2000$ & $\bar{m}^2=-16.5\pm0.5$  & 
$\bar{m^2}_c= -0.05(N \bar{\lambda})^{0.76}$   \\  
$A=2\pi$ & $D=402.124$ & $B_c=-6635.04$ & $C= 402124$  &  for  $N \bar{\lambda}\in [600,2000]$ \\  
\hline
$N$=12 & $\bar{R}=16$ & $N \bar{\lambda}= 1800$ & $\bar{m}^2=-6\pm0.2$  &  
$\bar{m^2}_c= -0.037(N \bar{\lambda})^{0.68}$  \\  
$A=2\pi$ & $D=1608.5$ & $B_c=-9650.97$ & $C= 1.44765\times 10^6$  &  for $N \bar{\lambda} \ge 300$ \\ 
\hline  
$N=12$ & $\bar{R}=32$ & $N \bar{\lambda}= 1600$ & $\bar{m}^2=-2.2\pm0.2$  & 
$\bar{m^2}_c= -0.016(N \bar{\lambda})^{0.66}$  \\  
$A=2\pi$ & $D=6433.98$ & $B_c=-14154.8$ & $C= 5.14719\times 10^6$  &  $N \bar{\lambda}\ge 20$  \\  
\hline
$N=12$ & $\bar{R}=64$ & $N \bar{\lambda}= 60$ & $\bar{m}^2=-0.1125\pm0.006$  &  
$\bar{m^2}_c= -0.013(N \bar{\lambda})^{0.51}$  \\  
$A=2\pi$ & $D=25735.9$ & $B_c=-2895.29$ & $C= 772078$  &  for  $N \bar{\lambda} \ge 15$ \\  
\hline  
$N=12$ & $\bar{R}=100$ & $N \bar{\lambda}= 30$ & $\bar{m}^2=-0.0535\pm0.0025$  &  $\bar{m^2}_c= -0.005(N \bar{\lambda})^{0.64}$ \\  
$A=2\pi$ & $D=62831.9$ & $B_c=-3361.5$ & $C= 942478$  &  for $N \bar{\lambda} \ge5$  \\  
\hline
$N=16$ & $\bar{R}=2$ & $N \bar{\lambda}= 120$ & $\bar{m}^2=-6.8\pm0.25$  &  
$\bar{m^2}_c= -0.155(N \bar{\lambda})^{0.79}$  \\  
$A=2\pi$ & $D=25.1327$ & $B=-170.903$ & $C= 1507.96$  &    \\  
\hline
$N=16$ & $\bar{R}$=4 & $N \bar{\lambda}$= 1200 & $\bar{m}$=-27$\pm$0.5   &  
$\bar{m^2}_c= -0.084(N \bar{\lambda})^{0.81}$  \\ 
$A=2\pi$ & $D=100.531$ & $B=-2714.34$ & $C= 60318.6$   &  for $N \bar{\lambda} \ge 400$  \\ 
\hline
$N=16$ & $\bar{R}=8$ & $N \bar{\lambda}= 1600$ & $\bar{m}^2=-15\pm0.2$  & 
$\bar{m^2}_c= -0.081(N \bar{\lambda})^{0.71}$   \\  
$A=2\pi$ & $D=402.124$ & $B=-6031.86$ & $C= 321699$  &  $N \bar{\lambda}\in[200,1600]$  \\ 
\hline
$N=16$ & $\bar{R}=16$ & $N \bar{\lambda}= 2000$ & $\bar{m}^2=-7.8\pm0.2$  & 
$\bar{m^2}_c= -0.035(N \bar{\lambda})^{0.71}$   \\  
$A=2\pi$ & $D=1608.5$ & $B=-12546.3$ & $C= 1.6085\times 10^6$  &   \\  
\hline
$N=16$ & $\bar{R}=32$ & $N \bar{\lambda}= 400$ & $\bar{m}^2=-1.05\pm0.05$  &  
$\bar{m^2}_c= -0.012(N \bar{\lambda})^{0.75}$ \\  
$A=2\pi$ & $D=6433.98$ & $B=-6755.68$ & $C= 1.2868\times10^6$  &  for $N \bar{\lambda} \ge 4.5$ \\  
\hline
$N=23$ & $\bar{R}=4$ & $N \bar{\lambda}= 115$ & $\bar{m}^2=-3.5\pm0.1$  &  
$\bar{m^2}_c= -0.062(N \bar{\lambda})^{0.85}$  \\  
$A=2\pi$ & $D=100.531$ & $B=-351.858$ & $C= 5780.53$  &  for  $N \bar{\lambda} \ge 30$ \\  
\hline
$N=23$ & $\bar{R}=8$ & $N \bar{\lambda}= 180$ & $\bar{m}^2=-3.25\pm0.05$  &  
$\bar{m^2}_c= -0.029(N \bar{\lambda})^{0.91}$  \\  
$A=2\pi$ & $D=402.124$ & $B=-1306.9$ & $C= 36191.1$  &  for  $N \bar{\lambda} \ge 20$ \\ 
\hline
$N=23$  & $\bar{R}=16$     & $N \bar{\lambda}= 1500$ & $\bar{m}^2=-7.35\pm0.05$  & 
$\bar{m^2}_c= -0.032(N \bar{\lambda})^{0.74}$   \\  
$A=2\pi$ & $D=1608.5$ & $B=-11822.4$            & $C= 1.20637\times 10^6$  &  for $N \bar{\lambda} \ge 300$ \\ 
\hline
\end{tabular} 
}
\end{center}
\end{table}
}}
\chapter{Representations and Casimir operators}
\label{appendix-sigmas}
\section{Explicit form of the generators of $SO(6)$ in the $4$ dimensional IRR}
\label{appendix-sigmas-form}

$$
   \Gamma_1 =  \left( 
                 \begin{array}{rrrr}
                     0  &  0 &  0 & -1  \\ 
                     0  &  0 & -1 & 0  \\
                     0  & -1 &  0 & 0   \\
                    -1  &  0 &  0 & 0
                 \end{array}
                \right)  ; \quad
   \Gamma_2 =  \left( 
                      \begin{array}{rrrr}
                        0  &  0   & 0   &  \I  \\ 
                        0  &  0   & -\I &   0  \\
                        0  &  \I  &  0  &   0  \\
                      -\I  &  0   &  0  &   0
                      \end{array}
                     \right) ; \quad
   \Gamma_3 =   \left( 
                 \begin{array}{rrrr}
                     0 &  0 & -1 &  0   \\ 
                     0 &  0 &  0 &  1  \\
                    -1 &  0 &  0 &  0  \\
                     0 &  1 &  0 &  0
                 \end{array}
                \right)     ;
$$

$$
   \Gamma_4 =  \left( 
                 \begin{array}{rrrr}
                     0  &   0  & \I &  0  \\ 
                     0  &   0  &  0 & \I  \\
                   -\I  &   0  &  0 &  0  \\
                     0  & -\I  &  0 &  0
                 \end{array}
                \right)  ; \quad
   \Gamma_5 =  \left( 
                      \begin{array}{rrrr}
                        1  &  0   &  0  &   0  \\ 
                        0  &  1   &  0  &   0  \\
                        0  &  0   & -1  &   0  \\
                        0  &  0   &  0  &  -1
                      \end{array}
                     \right) ; \quad
   \sigma_{12} =  \left( 
                 \begin{array}{rrrr}
                     1 &  0 &  0 &  0   \\ 
                     0 & -1 &  0 &  0  \\
                     0 &  0 &  1 &  0  \\
                     0 &  0 &  0 & -1
                 \end{array}
                \right)     ;
$$

$$
   \sigma_{13} = \left( 
                  \begin{array}{rrrr}
                     0  & \I  &  0  &  0  \\ 
                   -\I  &  0  &  0  &  0  \\
                     0  &  0  &  0  & \I  \\
                     0  &  0  & -\I &  0
                  \end{array}
                \right)  ; \quad
   \sigma_{14} = \left( 
                      \begin{array}{rrrr}
                        0  &  1   &  0  &   0  \\ 
                        1  &  0   &  0  &   0  \\
                        0  &  0   &  0  &  -1  \\
                        0  &  0   & -1  &   0
                      \end{array}
                     \right) ; \quad
   \sigma_{15} =  \left( 
                 \begin{array}{rrrr}
                     0 &  0 &  0  & -\I   \\ 
                     0 &  0 & -\I &   0  \\
                     0 & \I &  0  &   0  \\
                    \I &  0 &  0  &   0
                 \end{array}
                \right)     ;
$$

$$
   \sigma_{23} = \left( 
                  \begin{array}{rrrr}
                     0  &  1  &  0  &  0  \\ 
                     1  &  0  &  0  &  0  \\
                     0  &  0  &  0  &  1  \\
                     0  &  0  &  1  &  0
                  \end{array}
                \right)  ; \quad
   \sigma_{24} = \left( 
                      \begin{array}{rrrr}
                        0  & -\I  &  0  &   0  \\ 
                       \I  &  0   &  0  &   0  \\
                        0  &  0   &  0  &  \I  \\
                        0  &  0   & -\I &   0
                      \end{array}
                     \right) ; \quad
   \sigma_{25} =  \left( 
                 \begin{array}{rrrr}
                     0 &  0 &  0  & -1  \\ 
                     0 &  0 &  1  &  0  \\
                     0 &  1 &  0  &  0  \\
                    -1 &  0 &  0  &  0
                 \end{array}
                \right)     ;
$$

$$
   \sigma_{34} = \left( 
                  \begin{array}{rrrr}
                     1  &  0  &  0  &  0  \\ 
                     0  & -1  &  0  &  0  \\
                     0  &  0  & -1  &  0  \\
                     0  &  0  &  0  &  1
                  \end{array}
                \right)  ; \quad
   \sigma_{35} = \left( 
                      \begin{array}{rrrr}
                        0  &  0   & -\I  &   0  \\ 
                        0  &  0   &  0   &  \I  \\
                       \I  &  0   &  0   &   0  \\
                        0  & -\I  &  0   &   0
                      \end{array}
                     \right) ; \quad
   \sigma_{45} =  \left( 
                 \begin{array}{rrrr}
                     0 &  0 & -1  &  0   \\ 
                     0 &  0 &  0  & -1   \\
                    -1 &  0 &  0  &  0  \\
                     0 & -1 &  0  &  0
                 \end{array}
                \right)     .
$$

The generators of $SO(6)$ in the fundamental representation are proportional to the former $\Gamma_a, \sigma_{a,b}$, 
$a,b=1,\cdots,5$ .
\be
  \gen_{a6}= \med\Gamma_a, \quad \gen_{ab}=\med \sigma_{ab}.   
\ee

\section{Gell-Mann matrices of $SU(4)$}
As $SO(6)\cong SU(4)$ we can find the relations between both bases. Before we give the matrix of transformation
between the bases we introduce the explicit form of the Gell-Mann matrices for $SU(4)$.

The fundamental representation of $SU(4)$ is given by the fifteen matrices 
$\{ \lambda_i\}_{i=1}^{15}$

$$
   \lambda_1 =  \left( 
                 \begin{array}{rrrr}
                     0  &  1 &  0 & 0   \\ 
                     1  &  0 &  0 & 0   \\
                     0  &  0 &  0 & 0   \\
                     0  &  0 &  0 & 0
                 \end{array}
                \right)  ; \quad
   \lambda_2 =  \left( 
                      \begin{array}{rrrr}
                        0  & -\I  & 0  &   0  \\ 
                       \I  &  0   & 0  &   0  \\
                        0  &  0   & 0  &   0  \\
                        0  &  0   & 0  &   0
                      \end{array}
                     \right) ; \quad
   \lambda_3 =   \left( 
                 \begin{array}{rrrr}
                     1 &  0 &  0 &  0   \\ 
                     0 & -1 &  0 &  0  \\
                     0 &  0 &  0 &  0  \\
                     0 &  0 &  0 &  0
                 \end{array}
                \right)     ;
$$
$$
   \lambda_4 =  \left( 
                 \begin{array}{rrrr}
                     0  &  0 &  1 & 0   \\ 
                     0  &  0 &  0 & 0   \\
                     1  &  0 &  0 & 0   \\
                     0  &  0 &  0 & 0
                 \end{array}
                \right)  ; \quad
   \lambda_5 =  \left( 
                      \begin{array}{rrrr}
                        0  &  0   & -\I  &   0  \\ 
                        0  &  0   &  0   &   0  \\
                       \I  &  0   &  0   &   0  \\
                        0  &  0   &  0   &   0
                      \end{array}
                     \right) ; \quad
   \lambda_6 =   \left( 
                 \begin{array}{rrrr}
                     0 &  0 &  0 &  0   \\ 
                     0 &  0 &  1 &  0  \\
                     0 &  1 &  0 &  0  \\
                     0 &  0 &  0 &  0
                 \end{array}
                \right)     ;
$$
$$
   \lambda_7 =  \left( 
                 \begin{array}{rrrr}
                     0  &  0 &  0  & 0   \\ 
                     0  &  0 & -\I & 0   \\
                     0  & \I &  0  & 0   \\
                     0  &  0 &  0  & 0
                 \end{array}
                \right)  ; \quad
   \lambda_8 =  \frac{1}{\sqrt{3}}
                  \left( 
                      \begin{array}{rrrr}
                        1  &  0   &  0  &   0  \\ 
                        0  &  1   &  0  &   0  \\
                        0  &  0   & -2  &   0  \\
                        0  &  0   &  0  &   0
                      \end{array}
                     \right) ; \quad
   \lambda_9 =   \left( 
                 \begin{array}{rrrr}
                     0 &  0 &  0 &  1   \\ 
                     0 &  0 &  0 &  0  \\
                     0 &  0 &  0 &  0  \\
                     1 &  0 &  0 &  0
                 \end{array}
                \right)     ;
$$
$$
   \lambda_{10} =  \left( 
                    \begin{array}{rrrr}
                     0  &  0 &  0 & -\I   \\ 
                     0  &  0 &  0 &  0   \\
                     0  &  0 &  0 &  0   \\
                    \I  &  0 &  0 &  0
                    \end{array}
                  \right)  ; \quad
   \lambda_{11} =  \left( 
                      \begin{array}{rrrr}
                        0  &  0   & 0  &   0  \\ 
                        0  &  0   & 0  &   1  \\
                        0  &  0   & 0  &   0  \\
                        0  &  1   & 0  &   0
                      \end{array}
                     \right) ; \quad
   \lambda_{12} =   \left( 
                 \begin{array}{rrrr}
                     0 &  0  &  0 &  0   \\ 
                     0 &  0  &  0 & -\I  \\
                     0 &  0  &  0 &  0   \\
                     0 &  \I &  0 &  0
                 \end{array}
                \right)     ;
$$

$$
   \lambda_{13} =  \left( 
                    \begin{array}{rrrr}
                     0  &  0 &  0 &  0   \\ 
                     0  &  0 &  0 &  0   \\
                     0  &  0 &  0 &  1   \\
                     0  &  0 &  1 &  0
                    \end{array}
                  \right)  ; \quad
   \lambda_{14} =  \left( 
                      \begin{array}{rrrr}
                        0  &  0   &  0  &   0  \\ 
                        0  &  0   &  0  &   0  \\
                        0  &  0   &  0  &  -\I  \\
                        0  &  0   & \I  &   0
                      \end{array}
                     \right) ; \quad
   \lambda_{15} = \frac{1}{\sqrt{6}}   
                 \left( 
                 \begin{array}{rrrr}
                     1 &  0  &  0 &  0   \\ 
                     0 &  1  &  0 &  0   \\
                     0 &  0  &  1 &  0   \\
                     0 &  0  &  0 & -3
                 \end{array}
                \right)     .
$$

Let $M$ be the matrix of transformation between Gell-Man matrices and $SO(6)$ generators such that:
{\footnotesize{
$$ M               
   \left( 
     \begin{array}{r}
           \Gamma_{1} \\ 
           \Gamma_{2} \\
           \Gamma_{3} \\
           \Gamma_{4} \\
           \Gamma_{5} \\
           \sigma_{12} \\
           \sigma_{13} \\
           \sigma_{14} \\
           \sigma_{15} \\
           \sigma_{23} \\
           \sigma_{24} \\
           \sigma_{25} \\
           \sigma_{34} \\
           \sigma_{35} \\
           \sigma_{45} 
       \end{array}
   \right) 
=
   \left( 
     \begin{array}{r}
           \lambda_{1} \\ 
           \lambda_{2} \\ 
           \lambda_{3} \\ 
           \lambda_{4} \\ 
           \lambda_{5} \\ 
           \lambda_{6} \\ 
           \lambda_{7} \\ 
           \lambda_{8} \\ 
           \lambda_{9} \\ 
           \lambda_{10} \\ 
           \lambda_{11} \\ 
           \lambda_{12} \\ 
           \lambda_{13} \\ 
           \lambda_{14} \\
           \lambda_{15}
       \end{array}
   \right) 
$$
}}

$M$ has the form:
{\footnotesize{
$$
 M = \left( 
       \begin{array}{rrrrrrrrrrrrrrr}
          0  &  0  &  0  &  0  &  0  & -1  &  0  &    0              & -1  &  0  &  0  &  0   &  0  &  0   &     0             \\
          0  &  0  &  0  &  0  &  0  &  0  &  1  &    0              &  0  & -1  &  0  &  0   &  0  &  0   &     0             \\ 
          0  &  0  &  0  & -1  &  0  &  0  &  0  &    0              &  0  &  0  &  1  &  0   &  0  &  0   &     0             \\ 
          0  &  0  &  0  &  0  & -1  &  0  &  0  &    0              &  0  &  0  &  0  & -1   &  0  &  0   &     0             \\ 
          0  &  0  &  0  &  0  &  0  &  0  &  0  &  \frac{2}{\sqrt3} &  0  &  0  &  0  &  0   &  0  &  0   & \frac{\sqrt6}{3}  \\ 
          0  &  0  &  1  &  0  &  0  &  0  &  0  & -\frac{1}{\sqrt3} &  0  &  0  &  0  &  0   &  0  &  0   & \frac{\sqrt6}{3}  \\ 
          0  & -1  &  0  &  0  &  0  &  0  &  0  &    0              &  0  &  0  &  0  &  0   &  0  & -1   &     0             \\ 
          1  &  0  &  0  &  0  &  0  &  0  &  0  &    0              &  0  &  0  &  0  &  0   & -1  &  0   &     0             \\ 
          0  &  0  &  0  &  0  &  0  &  0  &  1  &    0              &  0  &  1  &  0  &  0   &  0  &  0   &     0             \\ 
          1  &  0  &  0  &  0  &  0  &  0  &  0  &    0              &  0  &  0  &  0  &  0   &  1  &  0   &     0             \\ 
          0  &  1  &  0  &  0  &  0  &  0  &  0  &    0              &  0  &  0  &  0  &  0   &  0  & -1   &     0             \\  
          0  &  0  &  0  &  0  &  0  &  1  &  0  &    0              & -1  &  0  &  0  &  0   &  0  &  0   &     0             \\  
          0  &  0  &  1  &  0  &  0  &  0  &  0  &  \frac{1}{\sqrt3} &  0  &  0  &  0  &  0   &  0  &  0   & -\frac{\sqrt6}{3} \\  
          0  &  0  &  0  &  0  &  1  &  0  &  0  &    0              &  0  &  0  &  0  & -1   &  0  &  0   &     0             \\  
          0  &  0  &  0  & -1  &  0  &  0  &  0  &    0              &  0  &  0  & -1  &  0   &  0  &  0   &     0             \\  
                          \end{array}
                       \right) 
$$
}}

\section{$SO(N)$ Casimir operators}

Following Refs.~\cite{perelomov1}-\cite{barut}  the $p$-order Casimir operator is defined as
\begin{equation}
      C_p(m_1,m_2,...,m_n) = Tr(a^p E)
\end{equation}
$(m_1,m_2,...,m_n)$ denote the {\em highest weight vector} of the involved representation.
 $a_{ij}$ is a matrix associated to $m_i$.

For classical Lie groups we have
\begin{eqnarray}
   \CAS & = & 2 S_2 \\
   C_4 & = & 2S_4-(2 \alpha \beta + \beta -1)  \quad \mbox{where} \\
   S_k & = & \sum_{i=1}^n (l_i^k - r_i^k)
\end{eqnarray}
$l_i=m_i+r_i$, $\alpha, \beta$ are given in {\bf Table \ref{table-const-groups}}.

The fold symmetric representations of $O(2n+1)$ are labelled by {\em highest weight vector}
\[
 m=(f,0,0, \dots,0).
\]
their $p$-order Casimir operator is 
\begin{equation}
   C_p(f,0,...,0)=(f +2\alpha)^p +(-f)^p.
\end{equation}
The quadratic Casimir operator reads

\be
  \CAS  =  2f(f+2\alpha).
\ee

\begin{table}[h]
  \caption{The constants $\alpha,\beta$ for the classic groups.}
  \label{table-const-groups}
\begin{center}
\begin{tabular}{|c|c|c|c||c|||}
 
\hline\hline  
      &           &  & & \\
 Algebra & Group  &
    $\alpha$  &
    $\beta$  & $r_i$  \\
     &             &   &  &  \\
\hline 
      &             &   &  &  \\
       
 $A_{n-1}$ & $SU(n)$   & $\frac{n-1}{2}$  & $0$  & $\frac{n+1}{2}-i$   \\[.1cm]

   $B_n$   & $O(2n+1)$ & $n -\frac{1}{2}$ & $1$  & $(n+\frac{1}{2})\epsilon_i-i$                                                                             \\[.1cm]

   $C_n$   & $Sp(2n)$  & $n$              & $-1$ & $(n+1)\epsilon_i-i$ \\[.1cm]

   $Dn$    & $O(2n)$   & $n-1$            & $1$  & $n\epsilon_i-i$     \\[.1cm] 
\hline
\end{tabular}
\end{center}
where $\epsilon_i=1 $ for $i>0$, $\epsilon=-1 $ if $i<0$ and $\epsilon=0 $ if $i=0$.
\end{table}

\subsubsection{$SO(5)$ Casimirs operators}
$SO(5)$ is a rank-$2$ algebra.
For $O(5)$ we have $r=(r_1,r_2)=(\frac{3}{2},\frac{1}{2})$ 
\begin{eqnarray}
  \CAS(m_1,m_2) & = &  2 \left( m_1^2 +m_2^2 +3m_1+m_2 \right)=2S_2, \\
  C_4(m_1,m_2) & = &  2S_4-\frac{3}{2}\CAS \quad \mbox{where}\\
  S_4 & = & m_1^4 +m_2^4 +6m_1^3+2m_2^3 +\frac{27}{2}m_1^2 +\frac{3}{2}m_2^2
              \nonumber \\ & &
           +\frac{27}{2}m_1+\frac{1}{2}m_2.
\end{eqnarray}

\subsubsection{Fold symmetric case representations of $SO(5)$}

 For $O(2n+1)$ we have
\begin{eqnarray}
  \alpha = n-\frac{1}{2}, \\
  \beta  = 1.
\end{eqnarray}
For $O(5)$:
\begin{eqnarray}
  \CAS(f,0) & = & 2f(f+3), \\
  C_4(f,0) & = & 2f(f+3)[f^2 +3 f +3].
\end{eqnarray}

\subsubsection{$SO(6)$ Casimir operators}
\label{casi.so6}
$SO(6)$ is a rank-$3$.
For $SO(6)$ we have $r=(r_1,r_2,r_3)=( 2,1,0)$ .The Casimir operator reads
\be
  \CAS(m_1,m_2,m_3) =   2 \left( m_1^2 +m_2^2 +m_3^2 +4m_1+2m_2 \right). 
\ee
The involved representations in chapter \ref{construction-S4} are
$(n,n,0)$  $SO(6)$ IRR; their Casimir takes the value
\begin{equation}
  \CAS(n,n,0)  =   4n \left( n +3 \right). 
\end{equation}

\section{Dimension of representations of $SO(N)$ and $SU(N)$}

Following Ref.~\cite{fulton} 
we have

\subsection*{$SO(2n)$} 
\begin{equation}
  dim ( m_1, m_2,...,m_n )= \prod_{i<j}\frac{l_i^2 -l_j^2}{r_i^2 -r_j^2}
\end{equation}
with $ l_i = m_i +n-i$,$r_i=n-i$, $i=1,2,...,n$ .

\subsection*{ $SO(2n+1)$}
\begin{equation}
  dim ( m_1, m_2,...,m_n )= \prod_{i<j}\frac{l_i^2 -l_j^2}{r_i^2 -r_j^2}
               \cdot \prod_i \frac{l_i}{r_i}
\end{equation}
with $ l_i = m_i +r_i$,$r_i=n+\frac{1}{2}-i$, $i=1,2,...,n$ .

To calculate the dimension of IRR of $SU(N)$ we follow the {\em Young tableau} algorithm ---see Ref.~\cite{georgi}. As an example we have for the $(2n,n,n)$ IRR of $SU(4)$

\begin{equation}
 dim(2n,n,n)_{SU(4)}= \frac{1}{12}(2n+3)(n+1)^2(n+2)^2. \label{IRR-2n-n-n-SU4}
\end{equation}
The representation is denoted by the number of boxes at each line. The representation in the
 example in eq.~(\ref{IRR-2n-n-n-SU4})
 is shown  in {\bf figure \ref{Young-tab-2n-n-n}}.
{\large
\begin{figure}[hc]
\label{Young-tab-2n-n-n}
{\small\begin{eqnarray*}
 & \overbrace{\yng(8,4,4)}^{2n} & \\
 & \underbrace{ \hspace{.7in}}_{n}   \underbrace{\hspace{.7in}}_{n}    &
\end{eqnarray*}}
  \caption{IRR $(2n,n,n)$ of $SU(4)$.}   \label{young2n}
\end{figure}

}
Since  $dim(n,n,0)_{SO(6)}= \frac{1}{12}(2n+3)(n+1)^2(n+2)^2$ we conclude
\begin{equation}
 (2n,n,n) \mbox{ IRR } {SU(4)} \equiv
 (n,n,0)  \mbox{ IRR } {SO(6)}.
\end{equation}

\chapter{Calculation of the induced metric of $\CP^3$ as $Spin(5)$ orbit}
\label{induced-metric-calculations}

In chapter \ref{construction-S4} we introduce the expression for the induced line element of $\CP^3$ as a $Spin(5)$ orbit (see eq.~(\ref{definition_metric_spin5}):
\be
   ds^2= \alpha d\xi_a^2+ \beta d\xi_{ab}^2, \label{definition_metric_spin5-app}
\ee
where the constants $\alpha,\beta$ are arbitrary numbers.

We start the construction of the orbit choosing a fiducial projector $\proj^0$
\be 
  \proj^0=\frac{1}{4}\identy+ \med \Lambda_a +\frac{1}{\sqrt2}\left(
                    \Lambda_{12} +\Lambda_{34}\right). \label{proj-0-1-app}
\ee
We can place the coordinates of the  projector in eq.~(\ref{proj-0-1-app}) in a 
matrix of coefficients, $\xi^0$, for $\xi_{ab}^0$ and 
a vector $ \vec{\xi}^{~0}$ for $\xi_a^0$
$$
  \xi^0 =  \frac{1}{2\sqrt{2}}  \left( 
                                              \begin{array}{rrrrr}
                                               0 &  1 &  0 &  0  &  0  \\ 
                                              -1 &  0 &  0 &  0  &  0  \\
                                               0 &  0 &  0 & -1  &  0  \\
                                               0 &  0 &  1 &  0  &  0  \\
                                               0 &  0 &  0 &  0  &  0
                                              \end{array}
                                             \right), 
\qquad
  \vec{\xi}^{~0} =  \left( 
                      \begin{array}{c}
                         0  \\ 
                         0  \\
                         0  \\ 
                         0  \\
                         \frac{1}{2}
                      \end{array}
                     \right).        
$$

$Spin(5)$ rotates the coordinates of the fiducial projector as follows
\bea
  \vec{\xi}_{a} &=& R_{ab}      \vec{\xi}^{~0}_b, \nonumber \\
  \xi_{ab}&=& R_{ac}R_{bd}\xi_{cd}^0, \nonumber 
\eea
then
\begin{equation}
  d \xi_a = dR_{ab} [  \vec{\xi}^{~0} ]_b,   \qquad  
  d \xi_{ab} = dR_{ac} [\xi^0]_{cd} R^{-1}_{db} 
               + R_{ac} [\xi^0]_{cd} dR^{-1}_{db}. \nonumber
\end{equation}

\subsubsection{1. $(d \xi_a)^2$}
For  $(d \xi_a)^2$ we have
\bea
 (d \xi_a)^2 &=& \lbrack d \xi_a \rbrack^t  \lbrack d \xi_a \rbrack \nonumber \\
             &=& \sum_{a=1}^{5}  \left[ dR_{ab}  \vec{\xi}^{~0}_b \right]^t
                                   dR_{ac} \vec{\xi}^{~0}_c 
                               =  \sum_{a=1}^{5}  \left[  \vec{\xi}^{~0}_b \right]^t [dR^t]_{ba}
                                   dR_{ac} \vec{\xi}^{~0}_c. \\
             &=& Tr \left( [\xi^0]^t [dR]^t [dR] [\xi^0] \right) \nonumber \\
             &=& Tr \left( [\xi^0]^t [dR]^t  R R^{-1} [dR] [\xi^0] \right) \label{dxia-app1}
\eea 

Using  $R^{-1} = R^t$ and  $R^{-1}R = \mathbf{1}$ we have $dR^{t}R = dR^{-1}R = -R^{-1}dR$. 
In eq.~(\ref{dxia-app1}) we obtain

\begin{equation}
 (d \xi_a)^2 = - Tr \left( [\vec{\xi}^{~0}]^t [R^{-1} dR][R^{-1}dR] \vec{\xi}^{~0} \right). \label{dxia-app3}
\end{equation} 
A rotation in $4$-dimensions has the form
$ R= e^{\imath e_{ab}\theta_{ab}}$. 
$\theta_{ab}=\imath L_{ab}$ are the generators  and $e_{ab}$ the coefficients of the rotations as they were defined in 
eq.~(\ref{Maurer-Cartan-spin6}). $L_{ab}$ are the generators in the fundamental representation with the explicit form 

\be
  [ L_{ab}]_{ij}= \med \left( \delta_{ai}\delta_{bj}- \delta_{bi}\delta_{aj} \right). 
   \label{generators-so5-fundamental}
\ee 
The expression for $R^{-1}dR$ involves the {\em{Maurer-Cartan forms}}

\be
  [R^{-1}dR]_{ij}=-[e_{ab}L_ab]_{ij}=-e_{ij}. \label{def-R-1dR}
\ee
Substituting eq.~(\ref{def-R-1dR}) in eq.~(\ref{dxia-app3}) we obtain
\be
  \left(d\xi_a\right)^2=\frac{1}{4}\sum_{a=1}^4\left( e_{a5}\right)^2.
\ee

\subsubsection{2. $ d \xi_{ab}^2$}
For $ d \xi_{ab}^2$ we have

\begin{equation}
  d \xi_{ab}^2 = d \xi_{ab} d \xi_{ab} = -d \xi_{ab} d \xi_{ba}= -Tr [d \xi_{ab} ]^2
\end{equation}

\begin{eqnarray}
  {\left( d \xi_{ab} \right)}^2 & = &-\sum_{a=1}^5 
                                        \left( dR_{ac} \xi^0_{cd} R^{-1}_{dn} 
                                       + R_{ac} \xi^0_{cd} dR^{-1}_{dn} \right) \nonumber \\
                                  &   & \times
                                        \left( dR_{ne} \xi^0_{ef} R^{-1}_{fa}  
                                       + R_{ne} \xi^0_{ef} dR^{-1}_{fa} \right) \nonumber \\
                          & = & -Tr \Big( dR \xi^0 R^{-1} dR \xi^0 R^{-1}
                                           +  dR \xi^0 R^{-1} R \xi^0 dR^{-1}   \nonumber   \\                                        &   &    +  R \xi^0 dR^{-1}dR \xi^0R^{-1}
                                           +  R \xi^0 dR^{-1} R \xi^0 dR^{-1} \Big) \nonumber \\
                          & = & -2  Tr \Big([R^{-1}dR] \xi^0 [R^{-1}dR] \xi^0 
                                           -[R^{-1}dR] \xi^0\xi^0[R^{-1}dR]  \Big) \nonumber \\
                          & = &-Tr \left[R^{-1}dR, \xi^0\right]^2. \label{dxiab-app1}
\end{eqnarray}

Substituting eq.~(\ref{def-R-1dR}) in eq.~(\ref{dxiab-app1}) we get
\be
  {\left( d \xi_{ab} \right)}^2=-2\left( e_{ik} \xi^0_{km}e_{mn} \xi^0_{ni}-
                                         e_{ik} \xi^0_{km} \xi^0_{mn} e_{ni} \right). \label{dxiab-app2}
\ee
Some intermediate steps for eq.~(\ref{dxiab-app2}):
$$
 e_{ik} \xi^0_{kj} =  \frac{1}{2\sqrt{2}} \left(
                             \begin{array}{rrrrr}
                                      e_{12}  &      0   &  -e_{14} &  e_{13}  &  0  \\ 
                                      0       &  -e_{12} &  -e_{24} &  e_{23}  &  0  \\
                                      e_{23}  &  -e_{13} &  -e_{34} &    0     &  0  \\
                                      e_{24}  &  -e_{14} &    0     &  -e_{34} &  0  \\
                                      e_{25}  &  -e_{15} &  e_{45}  &  -e_{35} &  0 
                             \end{array}
                              \right)  ,
$$

\be
 e_{ik} \xi^0_{kj} e_{jm} \xi^0_{mi}=\frac{1}{4}
               \left(e_{12}^{2}+e_{34}^2-2e_{14}e_{23}+2e_{13}e_{24}  \right). \label{dxiab-app2-part1}
\ee
$$
  \left(\xi^0\right)^2 =  \frac{1}{8}  \left( 
                                              \begin{array}{rrrrr}
                                              -1 &  0 &  0 &  0  &  0  \\ 
                                               0 & -1 &  0 &  0  &  0  \\
                                               0 &  0 & -1 &  0  &  0  \\
                                               0 &  0 &  0 &  -1  &  0  \\
                                               0 &  0 &  0 &  0  &  0
                                              \end{array}
                                             \right). 
$$

$$
 e_{ik} \left(\xi^0 \right)_{kl} e_{lj}= -\frac{1}{8} \left(
                             \begin{array}{rrrrr}
                                         0    &   e_{12} &  e_{13} &  e_{14} &  e_{15}  \\ 
                                     -e_{12}  &     0    &  e_{23} &  e_{24} &  e_{25}  \\
                                     -e_{13}  &  -e_{23} &    0    &  e_{34} &  e_{35}  \\
                                     -e_{14}  &  -e_{24} & -e_{34} &    0    &  e_{45}  \\
                                     -e_{15}  &  -e_{25} & -e_{35} & -e_{45} &    0 
                             \end{array}
                              \right) 
\left(
                             \begin{array}{rrrrr}
                                         0    &   e_{12} &  e_{13} &  e_{14} &  0  \\ 
                                     -e_{12}  &     0    &  e_{23} &  e_{24} &  0  \\
                                     -e_{13}  &  -e_{23} &    0    &  e_{34} &  0  \\
                                     -e_{14}  &  -e_{24} & -e_{34} &    0    &  0  \\
                                     -e_{15}  &  -e_{25} & -e_{35} & -e_{45} &    0 
                             \end{array}
                              \right). 
$$
{\small
\be
 e_{ik} \left(\xi^0 \right)^2_{kl} e_{li}=\frac{1}{4}\left( 
       e_{12}^{2} +e_{13}^{2}+e_{14}^2+e_{23}^2+ e_{24}^{2}+e_{34}^2 \right)
       +\frac{1}{8}\left( e_{15}^2 + e_{25}^2 +e_{35}^2 +e_{45}^2 \right).
\label{dxiab-app2-part2}
\ee
}
Substituting
eqs.~(\ref{dxiab-app2-part1}) (\ref{dxiab-app2-part2})
in eq.(\ref{dxiab-app2}) we obtain

\be
  {\left( d \xi_{ab} \right)}^2=\med\left[ (e_{13}-e_{24})^2+(e_{14}+e_{23})^2 \right]
       +\frac{1}{4}\left( e_{15}^2 + e_{25}^2 +e_{35}^2 +e_{45}^2 \right).
  \label{dxiab-app3}
\ee

Finally we arrive at
\be
  ds^2= \frac{\alpha+\beta}{4}\left[ e_{15}^{2} +e_{25}^{2}+e_{35}^2+e_{45}^2 \right]+
\frac{\beta}{2}\left[ (e_{13}-e_{24})^2+(e_{14}+e_{23})^2 \right]. \nonumber
\ee                       

\thispagestyle{empty}
\chapter*{List of publications}
 \begin{itemize}
 \item J.\ Medina and D.\ O'Connor, \hspace{0.1cm}
      {\sl Scalar Field Theory on Fuzzy} $S^4$,
      {\JHEP}  0311 (2003) 051 \hspace{0.2cm} [{\hepth{0212170}}].
 \item J.\ Medina, W.\ Bietenholz, F.\ Hofheinz and D.\ O'Connor,
 {\sl Field Theory Simulations on a Fuzzy Sphere - an Alternative to the Lattice},
  {PoS(LAT2005)}263 (2005) \hspace{0.2cm} [{{\heplat{0509162}}}].
\end{itemize}                           

\end{document}